\documentclass[a4paper,11pt]{article}
\usepackage{subcaption}
\usepackage{jheppub} 
\usepackage{braket}
\usepackage{soul}
\usepackage{subcaption}
\usepackage{caption}
\usepackage{amsmath}
\usepackage{multirow}
\usepackage{graphicx}
 \usepackage{hhline}
 \usepackage[dvipsnames]{xcolor}
\usepackage{bm}

\newcommand{\IR}{\mathrm{IR}}
\newcommand{\UV}{\mathrm{UV}}
\newcommand{\IRcolour}[1]{{\color[rgb]{0.892054,0.000000,0.112536}{#1}}}
\newcommand{\UVcolour}[1]{{\color[rgb]{0.452659,0.131997,0.466922}{#1}}}

\newcommand{\muIR}{\IRcolour{\mu_{\IR}}}
\newcommand{\muIRsq}{{\IRcolour{\mu}}\color{black}^2_{{\IRcolour{\IR}}}}
\newcommand{\muUV}{\UVcolour{\mu_{\UV}}}
\newcommand{\muUVsq}{{\UVcolour{\mu}}\color{black}^2_{{\UVcolour{\UV}}}}

\newcommand{\Y}{\mathcal{Y}}

\newcommand{\eea}{\end{eqnarray}}
\newcommand{\beqa}{\begin{eqnarray}}
\newcommand{\eeqa}{\end{eqnarray}}
\newcommand{\be}{\begin{equation}}
\newcommand{\ee}{\end{equation}}
\newcommand{\beq}{\begin{equation}}
\newcommand{\eeq}{\end{equation}}

\DeclareMathOperator{\sing}{sing_\epsilon}
\DeclareMathOperator{\reg}{reg_\epsilon}

\allowdisplaybreaks

\title{\boldmath
Computing the soft anomalous dimension with massless particles using the method of regions}

\author{Einan Gardi and}
\author{Zehao Zhu}

\affiliation{Higgs Centre for Theoretical Physics, School of Physics and Astronomy,\\ The University of Edinburgh, Edinburgh EH9 3FD, Scotland, UK}

\emailAdd{Einan.Gardi@ed.ac.uk, Z.Zhu-37@sms.ed.ac.uk}

\abstract{ 
It is well known that soft singularities of massless amplitudes are significantly simpler than those of massive ones. However, the computation of the soft anomalous dimension (AD) using Wilson-lines correctors is only straightforward in the massive case, thanks to the multiplicative renormalizability of  correlators of non-lightlike Wilson lines. 
Instead, correlators involving lightlike lines, develop higher-order poles in dimensional regularization due to collinear singularities, on top of their ultraviolet divergences. 
We nevertheless show, using the method of regions, how correlators involving lightlike lines
can be interpreted and used in the computation of the multileg 
soft~AD.
As a case study, we compute the two-loop soft~AD for two massive and one massless particles.  To this end, we start with the correlator of three timelike Wilson lines and apply the method of regions in the limit where one of the lines becomes lightlike. A correlator involving a strictly lightlike line then emerges as the ``hard region'' in this expansion. Its collinear divergences are removed upon adding the remaining regions, recovering the correct ultraviolet pole corresponding to the sought-after AD. By applying the method of regions, we are able to disentangle between ultraviolet and infrared divergences appearing in the strict limit. We also discover new phenomena, such as hard-virtuality collinear modes whose presence reflects the rescaling symmetry of semi-infinite Wilson lines. Our approach generalizes to any combination of massive and massless particles at higher loop order.
}

\begin{document}
\maketitle

\section{Introduction}
\label{sec:Intro}
It is well known that on-shell QCD scattering amplitudes remain singular after ultraviolate (UV) renormalization of the parameters and fields in the Lagrangian. 
The remaining singularities are related to the long-distance interaction of on-shell initial- and final-state partons, and are therefore known as infrared (IR) singularities. 
These can be factorized from the renormalized amplitude  leaving behind a finite hard function~\cite{Polyakov:1980ca,Arefeva:1980zd,Sen:1982bt,Korchemsky:1985xj,Ivanov:1985np,Korchemsky:1988hd,Korchemsky:1991zp,Korchemskaya:1994qp,Catani:1998bh,Kidonakis:1998nf,Kidonakis:1997gm,Sterman:2002qn,Gardi:2005yi,Aybat:2006mz,Dixon:2008gr,Gardi:2009qi,Gardi:2009zv,Becher:2009cu,Becher:2009qa,Ma:2019hjq,Feige:2014wja,Agarwal:2021ais}. While the latter depends on all the degrees of freedom of the scattered particles, the former admit a relatively simple, universal structure. Furthermore, IR singularities exponentiate via a renormalization group equation with respect to the factorization scale at which the hard amplitude is defined.
Therefore, they can be generated by a path-ordered exponential of an integral over the so-called \emph{soft anomalous dimension} (soft AD).

The soft~AD itself is a finite quantity, which depends on the scale through the $D$-dimensional strong coupling (we set $D=4-2\epsilon$), as well as through simple single-logarithmic terms associated with the cusp anomalous dimension~\cite{Polyakov:1980ca,Arefeva:1980zd,Sen:1982bt,Korchemsky:1985xj,Ivanov:1985np,Korchemsky:1988hd,Korchemsky:1991zp}. Integrating the soft AD from zero momentum~\cite{Magnea:1990zb}, 
up to the factorization scale, is sufficient to generate all IR singularities in $\epsilon$ in dimensionally-regularized scattering amplitudes. In this way the soft~AD universally governs IR singularities in any on-shell amplitude, making it a central quantity in QCD~\cite{Agarwal:2021ais}. 

The soft~AD for any multileg massless amplitude is known  to three loops~\cite{Almelid:2015jia}. The three-loop corrections correlate up to four massless particles, though so-called quadrupole colour structures.
These represent the first corrections to the dipole formula, since in the massless case tripole corrections, correlating three partons, are forbidden owing to factorization and rescaling symmetry~\cite{Becher:2009cu,Gardi:2009qi,Becher:2009qa,Dixon:2009gx,Gardi:2009zv}.
In turn, the full result of the soft~AD involving any number of massless and massive particles is known to two loops~\cite{Ferroglia:2009ep,Ferroglia:2009ii}. 
The angle-dependent cusp anomalous dimension, which governs massive colour dipole terms, i.e. interaction between a pair of massive particles, has been computed to three loops~\cite{Grozin:2015kna},  with partial results  obtained at four loops~\cite{Grozin:2004yc,Grozin:2015kna,Grozin:2016ydd,Grozin:2018vdn,Bruser:2019auj,Bruser:2020bsh}. The most complicated contribution to the soft~AD of amplitudes with massive particles at two loops is the colour tripole term. Such a tripole structure survives if at least two of the coloured particles are massive. 

Thus, focusing on multileg amplitudes there is, at present, a clear gap between massless scattering, where IR singularities are known to three loops for any multileg amplitude, and scattering involving massive particles, where the state-of-the-art is still two loops, as it has been for well over a decade. The reason is that the computation of the multileg soft anomalous dimension in the general timelike case is technically challenging. 
Besides the theoretical interest in the structure of IR singularities in more complex amplitudes, precision computations, notably in the context of resummation of top-quark production cross sections,  require the knowledge of the soft~AD for multileg amplitudes involving both massless and massive particles at three loops (for related recent work see~\cite{Czakon:2013hxa,Angeles-Martinez:2018mqh,Shao:2025qgv,Buonocore:2025ysd,Liu:2024hfa,Kidonakis:2025eia,Guzzi:2024lol,Kidonakis:2024lht,Kidonakis:2023juy}). It thus becomes urgent to close this gap and determine the soft~AD at three loops for any combination of massive and massless lines.
Recently, the three-loop contribution to the soft~AD associated with one massive and two massless particles was computed~\cite{Liu:2022elt}.

The universality of the soft~AD implies that it should be computable without referring to a specific partonic process. 
For massive scattering amplitudes, IR singularities are generated by soft interactions, where the eikonal approximation is valid. This implies that the singularities are captured by a correlator of semi-infinite timelike Wilson lines, following the classical  trajectories of the external particles~\cite{Polyakov:1980ca,Arefeva:1980zd,Sen:1982bt,Korchemsky:1985xj,Ivanov:1985np,Korchemsky:1988hd,Korchemsky:1991zp}. 
Such a correlator was proved to be multiplicatively renormalizable~\cite{Polyakov:1980ca,Arefeva:1980zd,Dotsenko:1979wb,Brandt:1981kf}, namely its UV divergences can be removed by a universal factor which renormalizes the hard-interaction vertex where the Wilson lines meet. This underlies the standard procedure for computing the cusp AD and the soft~AD, see e.g.~\cite{Korchemsky:1985xj,Korchemsky:1987wg,Korchemskaya:1994qp,Kidonakis:1998nf,Kidonakis:1997gm,Aybat:2006wq,Becher:2009cu,Gardi:2009qi,Becher:2009qa,Dixon:2009gx,Gardi:2009zv,Ferroglia:2009ep,Ferroglia:2009ii,Grozin:2015kna,Grozin:2004yc,Gardi:2011yz,Grozin:2016ydd,Grozin:2018vdn,Bruser:2019auj,Bruser:2020bsh}.
Applying only dimensional regularization, the correlator is scaleless, implying that IR and UV poles exactly cancel. In a practical computation one therefore regularizes the IR singularities by introducing some auxiliary regulator, so dimensional regularization would be required only for the UV. 
The soft~AD can then be extracted from the computation of the UV divergences of the regularized correlator. 
The basic principle is that the regulator must not affect the UV singularities, and it better not violate the symmetries of the correlator, such as gauge invariance and rescaling symmetry. 
In this paper we follow the approach of ref.~\cite{Gardi:2011yz}, using an  exponential regulator in configuration space, which is consistent with these requirements.

Importantly, the standard method of computing the soft~AD using Wilson-line correlators does not immediately  generalize to the case of massless particles. This is  because  multiplicative renormalizability of correlators involving strictly lightlike lines is violated by collinear gluons. Depending on the regulator adopted,\footnote{An original method to handle the singularities of a cusp formed by lightlike lines in configuration space was proposed in ref.~\cite{Erdogan:2011yc}, which does not involve an extra IR regulator. This approach was further applied in a more complex Wilson-line configuration~\cite{Falcioni:2019nxk}, but it is not easy to extend to multileg scattering.} one therefore obtains a mix of singularities of different origin, which does not have a simple relation with the soft~AD. 

Despite this fundamental challenge, computations of the soft~AD for massless particles, have been successfully performed.\footnote{As mentioned above, recently~\cite{Liu:2022elt}, a three-loop computation was performed to obtain the soft~AD contribution associated with one massive particle two massless ones.
The soft~AD was extracted there from a cross-section level object, where IR singularities cancel through the phase-space integral. } 
In particular, as already mentioned, the three-loop soft~AD for purely massless scattering has been determined from a Wilson-line correlator in~\cite{Almelid:2015jia}.
In order to utilize multiplicative renormalizability, this computation was set up starting with a correlator of four \emph{timelike} Wilson lines. After deriving a Mellin-Barnes representation for the corresponding webs,  ref.~\cite{Almelid:2015jia} performed a simultaneous expansion of all four 
Wilson-line velocities near the lightlike limit. Upon performing this expansion the Mellin-Barnes representation drastically simplified, such that the computation could be completed, yielding an elegant closed-form polylogarithmic function. Clearly, the massless soft~AD is significantly simpler than the massive one. 
Proceeding to higher orders,  or to more complicated contributions to the three-loop soft~AD involving both massless and massive particles, it would be advantageous to make use of the simplification associated with the massless limit at the outset. This is the main motivation for the present study.

Here we develop a new method to compute the soft anomalous dimension involving both massive and massless partons. 
Starting with the usual set up of timelike Wilson lines, we perform an asymptotic expansion using the method of regions (MoR)~\cite{Beneke:1997zp,Smirnov:2002pj,Pak:2010pt,Jantzen:2011nz,Jantzen:2012mw,Semenova:2018cwy,Ananthanarayan:2018tog,Heinrich:2023til,Borowka:2017idc,Gardi:2022khw,Ma:2023hrt,Chen:2024xwt}. Owing to the non-commutativity of the expansion and the integration over the loop momenta, performing a Taylor expansion of the integrand would not yield the correct expansion. The MoR systematically accounts for this, by summing over a complete set of leading-region integrals. In this way the strict limit taken at integrand level appears as just one of several region integrals. This strict limit can also be interpreted as the contribution to a correlator involving strictly lightlike lines. Adding the other region integrals has the effect of eliminating the extra IR poles,  so as to recover the correct UV AD of the multiplicative-renormalizable correlator, in which the Wilson lines representing massless particles are \emph{near} the lightcone, rather than strictly on the lightcone. In this approach we are therefore making full use of the simplification associated with strictly lightlike Wilson lines, and of dimensional regularization,  while systematically eliminating the superfluous IR poles generated by taking this limit.

This approach is ideally suited for computing the soft~AD for amplitudes involving both massless and massive particles. In companion papers~\cite{GZ-PRL,GZ-TBP} soon to appear, we use the method advocated here to compute a new contribution to the three-loop soft~AD, namely the quadrupole terms associated with the interaction of a single massive particle and three massless ones. This, together with the result of~Ref.~\cite{Liu:2022elt}, provides a complete description of the singularities for any amplitude involving a single massive particle and any number of massless ones, through three loops.  The present paper focuses on the method itself. We perform a detailed study of the two-loop soft~AD, with two massive particles and a single massless one. We obtain the tripole contribution to this soft~AD using the MoR. To establish the new approach we separately compute each web through a sum of region integrals and compare it to the lightlike limit of the corresponding web computed with timelike lines. We use this example to study in detail the inner workings of the MoR in the lightcone expansion of Wilson-line correlators, analysing the modes and the regions in both parameter space and momentum space. Finally, we return to study the renormalization of  correlators involving  strictly massless lines, shedding light of the additional IR singularities they feature. 

This paper is organized as the follows. In section~\ref{sec:correlators}, we briefly review the structure of the soft~AD at one and two loops, and explain how it is obtained from the correlator of (timelike) Wilson lines. In section~\ref{sec:methodology}, we present our methodology, clarifying several aspects,  including the regulator we use, the concept of the webs, and the relevant background regarding the MoR. We also define there some new concepts, such as region functions, obtained through a sum over the contributions to  different webs from similar region integrals. To  demonstrate our method, in section~\ref{sec:twoloopcomp} we use the MoR to reproduce the soft~AD for the two-loop tripole terms with one lightlike and two timelike lines. We discuss there the modes and regions that arise in the lightcone expansion of Wilson-line correlators,  classifying them into IR, neutral and UV, depending on the loop-momentum virtuality from which they originate. We also show that  the presence of multiple neutral modes in this expansion is linked with the rescaling symmetry of semi-infinite Wilson lines. 
Finally, in section~\ref{sec:RenMixedCorr}, we study the singular structure of the mixed correlator, involving both timelike and lightlike lines, and explicitly show that the UV and IR singularities can be disentangled. In appendix~\ref{app:convfunc}, we provide the definitions of the transcendental functions appearing throughout the calculation. Appendix~\ref{app:oneloop} presents the computation of the region functions at one loop, while appendices~\ref{app:2loopIR},~\ref{app:2loopNe} and~\ref{app:2loopUV}, respectively, present the two-loop computation of IR, neutral and UV region functions.

\section{Infrared singularities from the correlator of timelike Wilson lines}
\label{sec:correlators}

\subsection{Infrared singularities of amplitudes}
It is well known that QCD scattering amplitudes feature both UV and IR singularities. For a UV renormalized scattering amplitude ${\cal M}$ with $N$ massless and $M$ massive external partons with momentum $\{p_i\},i=1,\ldots,N$ and $\{p_I\},I=1,\ldots,M$, we have the factorization form,
\begin{align}
\label{MZH}
\begin{split}
&\hspace{-40pt}\mathcal{M}\left(\left\{p_i,p_I\right\},\muUV,\alpha_s(\muUVsq),\epsilon_{\IR}\right)
    \\
&=Z\left(\left\{p_i,p_I\right\},{\mu},\alpha_s(\mu^2),\epsilon_{\IR}\right)\mathcal{H}\left(\left\{p_i,p_I\right\},
\muUV,\mu,\alpha_s(\muUVsq),\epsilon_{\IR}\right),
    \end{split}
 \end{align}
where $\muUV$ is the UV renormalization point while $\mu$ is the factorization scale introduced upon defining $Z$. The role of $Z$ is to captures all 
IR singularities in ${\cal M}$, rendering ${\cal H}$ free of IR poles in $\epsilon_{\IR}$. The function $Z$ satisfies the following Renormalization Group (RG) equation,
\begin{equation}
\label{RGZ}
\begin{aligned}
\begin{gathered}
    \frac{d}{d\log \mu}Z\left(\left\{p_i,p_I\right\},\mu,\alpha_s(\mu^2),\epsilon_{\IR}\right)
    =-\Gamma\left(\left\{p_i,p_I\right\},\mu,\alpha_s(\mu^2)\right)Z\left(\left\{p_i,p_I\right\},\mu,\alpha_s(\mu^2),\epsilon_{\IR}\right).
     \end{gathered}
 \end{aligned}
\end{equation}
In what follows we will be using minimal subtraction, keeping only poles in $\epsilon_{\IR}$ in $Z$. In turn, the hard function $\mathcal{H}$ in eq.~(\ref{MZH}) is by construction finite (note that it contains not only ${\cal O}(\epsilon_\IR^0)$ terms but also positive powers of $\epsilon_\IR$). 
The formal solution of eq.~(\ref{RGZ}) is a path ordered exponential,
\begin{equation}
\label{ZP}
  Z\left(\left\{p_i,p_I\right\},\mu,\alpha_s(\mu^2),\epsilon_{\IR}\right)=\text{P}\exp\left[\int_{0}^{\mu}\frac{d\tau}{\tau}\Gamma\left(\left\{p_i,p_I\right\},\tau,\alpha_s(\tau^2)\right)\right],
\end{equation}
where the path ordering ($\text{P}$) ensures that colour matrices in $Z$ multiply each other in line with the ordering in the scale $\tau$: the rightmost matrices are those defines at the hardest scale, $\tau=\mu$.
The strong coupling $\alpha_s(\mu^2)$ satisfies a similar RG equation,
\begin{equation}
\label{RunningCoupling}
    \frac{d}{d\log\mu^2}\frac{\alpha_s(\mu^2)}{4\pi}=-\epsilon_{\IR}\frac{\alpha_s(\mu^2)}{4\pi}-b_0\left(\frac{\alpha_s(\mu^2)}{4\pi}\right)^2+{\cal O}(\alpha_s^3(\mu^2)),
\end{equation}
where $b_0$ is one-loop QCD beta function,
\begin{align}
\label{betab0}
b_0=\frac{11}{3}C_A-\frac{4}{3}T_RN_f.
\end{align}
The finite function $\Gamma\left(\left\{p_i,p_I\right\},\mu,\alpha_s(\mu^2)\right)$ in eq.~\eqref{RGZ}, known as the soft~AD, depends on the scale $\mu$ in two way: through the argument of the coupling, and explicitly, through a term linear in $\log(\mu)$. The former dependence is sufficient in the case of massive particles, while the latter is needed to capture the overlapping soft and collinear double poles in massless scattering. The soft AD 
$\Gamma$ has been computed to two loops,
\begin{equation}
 \label{Gammadimension}
\begin{aligned}
\begin{gathered}
\\\Gamma\left(\left\{p_i,p_I\right\},{\mu},\alpha_s(\mu^2)\right)=
\sum_i\gamma_i(\alpha_s(\mu^2))
+
\Gamma^T(\alpha_s(\mu^2))
+\Gamma^L(\mu,\alpha_s(\mu^2))
,
\end{gathered}
\end{aligned}
\end{equation}
where the single-particle colour-singlet terms $\gamma_i$ arises from (hard) collinear singularities~\cite{Aybat:2006mz,Gardi:2009qi,Becher:2009qa,Falcioni:2019nxk}. The
$\gamma_i$ coefficients 
were determined up to four loops by computing form factors~\cite{vonManteuffel:2020vjv,Agarwal:2021zft,Moch:2005id,Moch:2005tm,Baikov:2009bg}.  $\Gamma^T$, where $T$ stands for \emph{timelike},  collects contributions involving \emph{only} massive particles 
\begin{align}
\label{GammaT}
\begin{split}
\Gamma^T(\alpha_s(\mu^2))
=\,\,& \sum_I\Omega_{(I)}(\alpha_s)+ 
\sum_{I<J}\mathbf{T}_I\cdot\mathbf{T}_J\Omega_{(IJ)}(\gamma_{IJ},\alpha_s)
\\&+
\sum_{I<J<K}\mathbf{T}_{IJK}\Omega_{(IJK)}(\{\gamma_{IJ},\gamma_{JK},\gamma_{IK}\},\alpha_s)+\ldots\,
,
\end{split}
\end{align}
while $\Gamma^L$, where $L$ stands for \emph{lightlike}, contains the remaining contributions that involve massless particles, including 
ones arising from the interaction between massive and massless ones, taking the form
\begin{align}
\label{GammaL}
\begin{split}
\Gamma^L(\mu,\alpha_s(\mu^2))
=\,\,&
\sum_{i<j}\mathbf{T}_i\cdot\mathbf{T}_j\gamma_{\text{cusp}}(\alpha_s)\log
\frac{\mu^2}{-2p_i\cdot p_j}
+\sum_{i,J}\mathbf{T}_i\cdot\mathbf{T}_J\gamma_{\text{cusp}}(\alpha_s)\log\frac{\mu\sqrt{p_J^2}}{-2p_i\cdot p_{J}}
\\&
+\sum_{I<J,k}\mathbf{T}_{IJk}\Omega_{(IJk)}\left(\left\{\gamma_{IJ},y_{IJk}\right\},\alpha_s\right)\,+\ldots ,
    \end{split}
\end{align}
where lower-case indices $i,j$ and $k$ run over the  massless particles, while upper-case ones run over massive ones. 
In both (\ref{GammaT}) and (\ref{GammaL}) we included ellipses to represent higher-order terms (three loop and beyond) which involve four or more coloured particles.
In these equations
we use the colour generator notation $\mathbf{T}_i$, where the representation is associated with the parton~$i$~\cite{Catani:1996jh}. In eq.~\eqref{GammaT} there are colour singet terms, $\Omega_{(I)}(\alpha_s)$, which involve no kinematic dependence and capture soft divergences of heavy partons; these are known to three loops \cite{Kidonakis:2009ev,Grozin:2014hna,Grozin:2015kna,Korchemsky:1987wg,Korchemsky:1991zp,Bruser:2019yjk}.  

A salient feature of the AD of massive particles is its rescaling invariance with respect to the momentum or velocity of each massive particle. To make this manifest we use 
rescaling-invariant variables, defined as follows:
\begin{equation}
\label{angleDef}
\gamma_{IJ}\equiv \frac{2\beta_I\cdot\beta_J}{\sqrt{\beta_I^2}\sqrt{\beta_J^2}}\,,
\end{equation}
where the velocity $\beta_I$ of a massive particle is a dimensionless four vector propotional to the four momentum, $\beta_I\sim p_I$. 
The variable $\gamma_{IJ}$ can be parametrized in the following way,
\begin{align}
    \label{sigmaIJ}\gamma_{IJ}=-2\cosh{\sigma_{IJ}},
\end{align}
where $\sigma_{IJ}$ is the Minkowski cusp angle formed between the four-velocities $\beta_I$ and $\beta_J$. The colour-dipole interaction $\mathbf{T}_I\cdot\mathbf{T}_J\Omega_{(IJ)}$ can be computed~\cite{Korchemsky:1987wg} as the AD associated with a pair of semi-infinite Wilson lines with  respective velocities $\beta_I$ and $\beta_J$  emanating from the hard interaction point, where they meet at an angle $\sigma_{IJ}$ as in~(\ref{sigmaIJ}),  forming a cusp; $\Omega_{(IJ)}$~is therefore known as the angle-dependent cusp AD. 

To represent the interaction 
associated with two massive particles and a massless one we
define a second rescaling-invariant kinematic variable
as follows:  
\begin{equation}
\label{yIJk}
 y_{IJk}\equiv
\frac{\beta_I\cdot\beta_k\sqrt{\beta_J^2}}{\beta_J\cdot\beta_k\sqrt{\beta_I^2}} =\lim_{\beta_K^2\to 0 } \frac{\gamma_{IK}}{\gamma_{JK}}\,,
\end{equation}
where we have used $\beta_k$ with a lower-case index $k$ to denote the velocity $\beta_K$ after having taken the lightlike limit,  $\beta_K^2\to 0$.

Besides the colour singlet term $\Omega_{(I)}$ and the colour dipole (cusp) term, starting at two-loops
eq.~\eqref{GammaT} contains also a colour-tripole terms,
arising form the interaction of three massive particles, 
involving a fully-connected antisymmetric colour factor defined by 
\begin{equation}
\label{fTTT}
\textbf{T}_{IJK}\equiv if^{abc}\textbf{T}_I^a\textbf{T}_J^b\textbf{T}_K^c\,,  
\end{equation}
and depending on the three cusp angles formed between their velocities. 

The kinematic dependence of the dipole terms is completely known up to three loops~\cite{Grozin:2015kna} (see section 5 there) and partially known at four loops \cite{Grozin:2004yc,Grozin:2015kna,Grozin:2016ydd,Grozin:2018vdn,Bruser:2019auj,Bruser:2020bsh}. 
Here we quote it at one loop:
\begin{subequations}
\begin{equation}
\Omega_{(IJ)}(\gamma_{IJ},\alpha_s)=\sum_{n=1}^{\infty}\left(\frac{\alpha_s}{4\pi}\right)^n\Omega_{(IJ)}^{(n)}\,, 
\end{equation}
where 
\begin{align}
\label{OneLoopCusp}
\Omega_{(IJ)}^{(1)}&=4\frac{1+\alpha_{IJ}^2}{1-\alpha_{IJ}^2}\log\left(\alpha_{IJ}\right)\,.
\end{align}
\end{subequations}
The variable $\alpha_{IJ}$  is defined by
\begin{equation}
\label{alphaIJ}
    \gamma_{IJ}=-\alpha_{IJ}-\frac{1}{\alpha_{IJ}}\,.
\end{equation}
It is introduced to rationalise the symbol alphabet (square roots appear upon using~$\gamma_{IJ}$).

The tripole contribution in eq.~(\ref{GammaT}) 
starts at two loops. It is given by~\cite{Ferroglia:2009ep}
\begin{equation}
\label{GammaIJK}
  \Omega_{(IJK)}^{(2)}=-4\epsilon_{IJK}\frac{1+\alpha_{IJ}^2}{1-\alpha_{IJ}^2}\log(\alpha_{IJ})\left[\frac{1+\alpha_{IK}^2}{1-\alpha_{IK}^2}M_{100}(\alpha_{IK})+2\log^2(\alpha_{IK})\right],
\end{equation}
where a sum over repeated indices is assumed and $\epsilon_{IJK}$ is the 3-dimensional Levi-Civita symbol. 
The functions $M_{abc}(\alpha)$, which are polylogarithmic functions with weight $a+b+c+1$, have been defined in ref.~\cite{Falcioni:2014pka}. In eq.~(\ref{GammaIJK}) we only require $M_{100}$, which is quoted in eq.~(\ref{M100def}).

Moving on to $\Gamma_L$, we can view the three types of terms in eq.~(\ref{GammaL}) as arising from 
$\Gamma_T$ in (\ref{GammaT}) in 
the limit where a subset of the particles ($k=1\ldots N$) become
massless. Specifically the terms in the first line of (\ref{GammaL}) can all be associated to the dipole (cusp) terms $\Omega_{(IJ)}$ in the limit where either one or both velocities $\beta_I$ and $\beta_J$ become lightlike, while the term in the second line arises form the tripole interaction $\Omega_{(IJK)}$ in the limit where \emph{one} of the three partons become massless. 
It is a well known fact that no contribution arises in the soft~AD from a situation where two or all of the three particles become massless -- in particular, massless scattering does not have a tripole contribution to the soft AD~\cite{Aybat:2006mz,Aybat:2006wq,Gardi:2009qi,Becher:2009qa,Ferroglia:2009ep,Ferroglia:2009ii}. This is because the two requirements~\cite{Becher:2009qa,Gardi:2009qi} of the rescaling invariance and the antisymmetry (which is induced by the colour structures), cannot be simultaneously satisfied with three massless particles or two massless and one massive particles.  

Note that the cusp contributions to $\Gamma_L$ involve an explicit $\log(\mu)$, which will generate a double pole in $\epsilon$ upon integration in eq.~\eqref{ZP}. This is a departure from $\Gamma_T$, where dependence on $\mu$ only occurs through the argument of the coupling constant, generating single poles. Physically, the double poles originate in the overlap between soft and collinear singularities, while the single poles have a purely soft or, alternatively, hard-collinear origin. 
Along with the double poles in eq.~\eqref{GammaL} ones find violation of rescaling symmetry -- and hence dependence on the particle momenta, rather than their velocities~\cite{Becher:2009qa,Gardi:2009qi}.

\subsection{Correlators of timelike Wilson lines and their lightlike limit}
\label{sec:MassiveMassless}

It is well known that $\Gamma^T$ can be obtained solely from a correlator of timelike Wilson lines $\left<\Phi_{\beta_I}\Phi_{\beta_J}\cdots\right>$, where $\Phi_{\beta_I}$ is a semi-infinite Wilson line~\cite{Polyakov:1980ca,Dotsenko:1979wb,Brandt:1981kf,Korchemsky:1985xj},
\begin{equation}
    \Phi_{\beta_I}=\text{P}\exp\left[ig_s\int^{\infty}_0ds\beta_I\cdot A(s\beta_I)\right],
\end{equation}
which describes the radiation from a coloured scalar particle moving along the classical trajectory with four velocity $\beta_I$.

\begin{figure}[thb]
    \centering
    \includegraphics[width=0.5\linewidth]{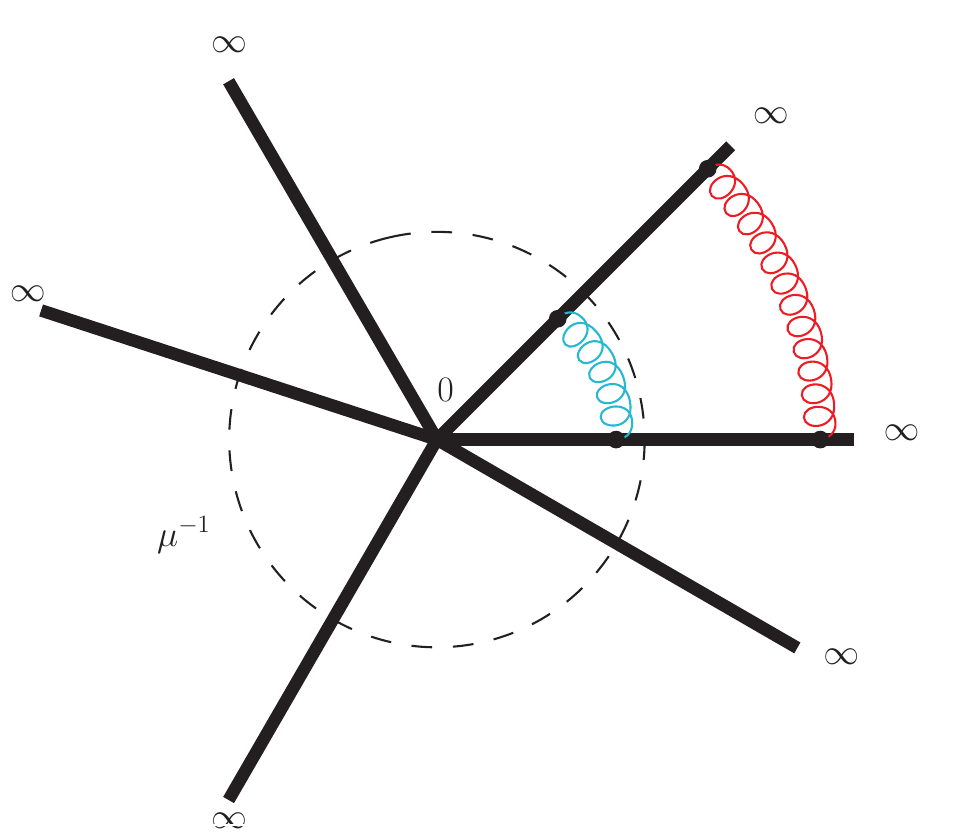}
    \caption{Soft and UV singularities are sketched in configuration space. The thick lines represent semi-infinite timelike Wilson lines, meeting at a single point, where the hard interaction takes place. }
    \label{fig:TimelikeRe}
\end{figure}

\begin{figure}
    \centering
    \includegraphics[width=0.5\linewidth]{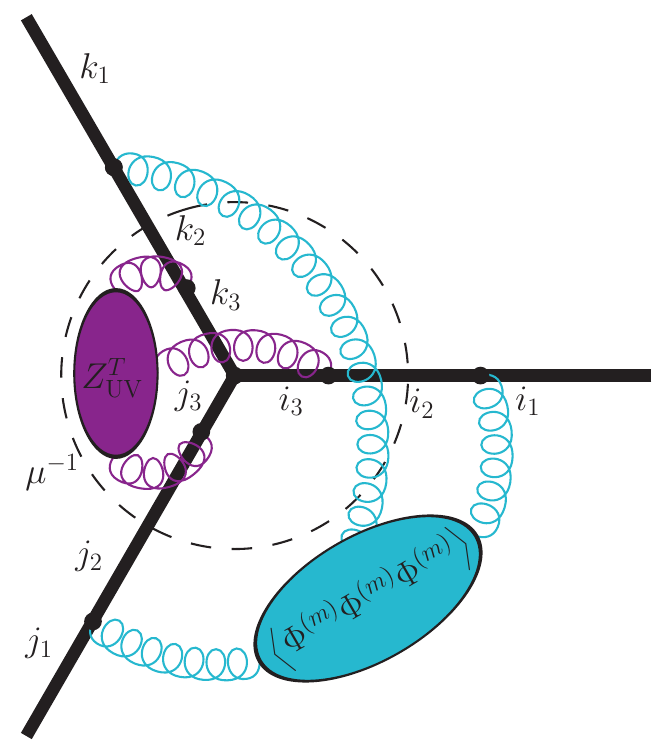}
    \caption{The factorization of the correlator of three (IR regularized) timelike Wilson lines sketched in configuration space. The indices $i_n$, $j_n$ and $k_n$ correspond respectively to the colour representations of the three lines, and the order of gluon attachments along the lines is consistent with the colour ordering in eq.~(\ref{MR2}). The UV renormalization scale~$\mu$ is labelled by the dashed line. }
    \label{fig:FactorizationT}
\end{figure}

A correlator of (timelike) Wilson lines can have both soft (IR) and UV divergences.  If all divergences are regulated solely by dimensional regularization, 
these two type of divergences will cancel each other, i.e. $\frac{1}{\epsilon_{\UV}}-\frac{1}{\epsilon_{\IR}}=0$, as all integrals are manifestly scaleless: they only involve velocities.
More globally, this implies a simple relation between the IR and the UV singularities of the correlator, which simply mirror each other, as illustrated in figure~\ref{fig:TimelikeRe}. This relation can be used to compute the IR singularities  indirectly as follows: one introduces 
an additional IR regulator $m$ so as to render  the integrals scaleful, and the correlator IR-finite but UV-divergent.\footnote{The specific way in which we introduce the regulator is discussed in the next section; see eq.~\ref{Phim} there.} The latter divergence, which cannot be affected by the regulator $m$, is associated with the renormalization of the vertex at which the Wilson lines meet, which physically corresponds to the hard interaction vertex \cite{Polyakov:1980ca,Arefeva:1980zd,Sen:1982bt,Korchemsky:1985xj,Ivanov:1985np,Korchemsky:1988hd,Korchemsky:1991zp,Dotsenko:1979wb,Brandt:1981kf,Korchemsky:1987wg,Kidonakis:1998nf,Kidonakis:1997gm,Becher:2009cu,Gardi:2011yz}.  
This way, the sought-after IR singularities are encoded in the UV renormalization factor $Z_{\text{UV}}^T$ defined by 
\begin{equation}
\label{MR2}
\left<\Phi^{(m)}_{\beta_I}\Phi^{(m)}_{\beta_J}\cdots\right>_{\text{ren.}(\mu)}=
\left<\Phi^{(m)}_{\beta_I}\Phi^{(m)}_{\beta_J}\cdots\right>Z_{\UV}^{T}(\{\alpha_{IJ}\},\alpha_s(\mu^2),\epsilon_{\UV})\,,
\end{equation}
where the angle brackets represent the expectation value of a time-ordered product of field operators, where the time-ordering of fields associated with different Wilson lines guaranties causality via Feynman's prescription. 
The superscript $(m)$ represents the IR regulator associated with  a given Wilson line,\footnote{In certain circumstances it is sufficient to associate such regulators to a subset of the emissions from a given Wilson line~\cite{Gardi:2021gzz} or even a subset of the Wilson lines~\cite{Henn:2023pqn}, leading to some simplification. In the present paper we shall keep the regulator on all emissions from every Wilson line, see section~\ref{sec:Regularised_WilsonLines} for details.} and where the subscript `$\text{ren.}$' identifies the renormalized correlator, which is finite. Equation~\eqref{MR2} is illustrated graphically in figure~\ref{fig:FactorizationT}.

The RG equation of the UV renormalization factor $Z_{\UV}^{T}$ is 
\begin{equation}
    \frac{dZ_{\UV}^{T}(\{\alpha_{IJ}\},\alpha_s(\mu^2),\epsilon_{\UV})}{d\log\mu}=-Z_{\UV}^{T}(\{\alpha_{IJ}\},\alpha_s(\mu^2),\epsilon_{\UV})
    \,
    \Gamma_{\UV}^{T}(\{\alpha_{IJ}\},\alpha_s(\mu^2))
    \,.
    \label{MR1}
\end{equation}
Equations~\eqref{MR1} and~\eqref{MR2} manifest the multiplicative renormalizability of the correlator~\cite{Brandt:1981kf}. 
The expression of $\Gamma_{\UV}^T$ up to two loops is given by eq.~(\ref{GammaT}).

In eq.~\eqref{MR2}, we regularize all the Wilson lines, such that soft divergences will not appear as $\epsilon_{\IR}$ poles. The setup where all Wilson lines are regularized will be referred to below as \emph{a complete regularization scheme}.
A new method was recently developed~\cite{Henn:2023pqn} to compute soft~ADs in an incomplete regularization scheme where only one Wilson line is regularized, which makes the computations simpler. In that scheme an additional subtraction is necessary for webs consisting of two or more connected subdiagrams. In the present paper, we will always work in the complete regularization scheme.

To clarify how multiplicative renormalizability of Wilson-line correlators facilitates the  calculation of the soft~AD, we follow \cite{Mitov:2010rp,Gardi:2013ita,Gardi:2011yz}     and define the exponential form of the correlator,
\begin{align}
    \begin{split}
       \label{nloopexp}
\left<\Phi^{(m)}_{\beta_I}\Phi^{(m)}_{\beta_J}\cdots\right>=\exp\left[\sum_{n}w^{(n)}\left(\frac{\alpha_s}{4\pi}\right)^n\right]=\exp\left[\sum_{n,l}w^{(n,l)}\left(\frac{\alpha_s}{4\pi}\right)^n\epsilon^l\right]. 
    \end{split}
\end{align}
Using the non-Abelian exponentiation theorem~\cite{Gardi:2013ita} (see also~\cite{Frenkel:1984pz,Gatheral:1983cz,Sterman:1981jc,Gardi:2010rn,Gardi:2011wa,Gardi:2011yz}), $w^{(n)}$ may be  expressed as the sum of webs containing 
only fully connected colour structures. 
We postpone further discussion of diagrammatic exponentiation to section~\ref{sec:RegularizeCorrelator} below. 

The renormalization factor $Z_{\UV}^{T}$  in eq.~(\ref{MR2}) can be written explicitly in terms of the expansion coefficients of the soft~AD,
\begin{align}
    \Gamma_{\UV}^{T}=\sum_{n}\left(\frac{\alpha_s}{4\pi}\right)^n\Gamma_{\UV}^{T,(n)}\,,
\end{align}
by  integrating (\ref{MR1}) as follows~\cite{Gardi:2011yz}:
\begin{align}
\label{PertExpGammaT}
    \begin{split}
        Z_{\UV}^{T}=\exp\left[\frac{\alpha_s}{4\pi}\frac{1}{2\epsilon}\Gamma_{\UV}^{T,(1)}+\left(\frac{\alpha_s}{4\pi}\right)^2\left(\frac{1}{4\epsilon}\Gamma_{\UV}^{T,(2)}-\frac{b_0}{4\epsilon^2}\Gamma_{\UV}^{T,(1)}\right)+{\cal O}(\alpha_s^3)\right]\,.
    \end{split}
\end{align}
Considering eq.~\eqref{MR2}, we insert the correlator expansion according to (\ref{nloopexp}) on the right-hand side, and require the left-hand side to be finite. Assuming minimal subtraction, this gives the following identities, 
\begin{subequations}
\label{Gamma_from_webs}
    \begin{align}
        \begin{split}
       \Gamma_{\UV}^{T,(1)}=\,\,&-2w^{(1,-1)},     \end{split}
    \\
         \begin{split}
         \label{twoloopGWeb}
       \Gamma_{\UV}^{T,(2)}=\,\,&-4w^{(2,-1)}-2\left[w^{(1,-1)},w^{(1,0)}\right].
       \end{split}
    \end{align}
\end{subequations}
The above procedure allows us to determine the soft~AD~$\Gamma^T$ by computing the webs to the relevant loop order (the generalization of eq.~(\ref{Gamma_from_webs}) to higher orders can be found in eq.~(2.15) of ref.~\cite{Gardi:2011yz}; we will not need it in the present paper). In what follows we will be looking closely at the tripole colour structure (\ref{fTTT}), which is the first instant where the commutator of one-loop webs in eq.~(\ref{twoloopGWeb}) plays a role. To see how it contributes assume that $w^{(1,-1)}$ and $w^{(1,0)}$ are each given by dipole terms of the form
\begin{align}   
w^{(1,l)}\equiv
\sum_{I<J<K}\mathbf{T}_I\cdot\mathbf{T}_J {\cal Y}^{(1,l)}_{(IJ)}
+\mathbf{T}_I\cdot\mathbf{T}_K {\cal Y}^{(1,l)}_{(IK)}
+\mathbf{T}_J\cdot\mathbf{T}_K {\cal Y}^{(1,l)}_{(JK)}\,,
\end{align}
where ${\cal Y}$ are the corresponding kinematically-dependent functions. 
It then follows that their commutator is
\begin{align}
\label{Commutator_of_Dipoles}
    \left[w^{(1,-1)},w^{(1,0)}\right]
    =-\sum_{I<J<K}\mathbf{T}_{IJK}\epsilon^{ABC}
    {\cal Y}^{(1,-1)}_{(AB)}
    {\cal Y}^{(1,0)}_{(BC)}
\end{align}
where $\{A,B,C\}$ take all the permutations of $\{I,J,K\}$.

A straightforward way to obtain mixed\footnote{We will use the terminology ``mixed" to represent the situation where there are both massive and massless particles, i.e. both timelike and lightlike Wilson lines.} soft~ADs in eq.~\eqref{GammaL}, such as $\Omega_{(IJk)}$ in $\Gamma^{L}$, 
is to start with the massive case and perform an expansion in the small Wilson-line virtualities. For example, at one loop, in the limit where one of the velocities gets close to the lightcone $\beta_{I}^2\rightarrow \beta_i^2=0$, we will get
logarithmical divergence,
\begin{equation}
\label{lliJ}
\Omega_{(iJ)}^{(1)}={\cal T}_{\lambda} \left[\Omega_{(IJ)}^{(1)}\bigg|_{\beta_{I}^2=t_i\lambda^2}\right]=-2\log\left(\frac{(2\beta_i\cdot\beta_J)^2}{t_i\lambda^2\beta_J^2}\right),
\end{equation} 
where $t_i$ is a finite positive parameter, a proportionality factor relating $\beta_I^2$ with $\lambda^2$, which encapsulates the way the limit is taken. The operator ${\cal T}_{\lambda}$ performs the asymptotic expansion in small $\lambda$, neglecting any power-suppressed terms. Notice that the result depends on $t_i\lambda^2$. 
Similarly, if we take the lightlike limit of both the Wilson lines around the cusp, we get
\begin{equation}
\label{GammaCuspij_one_loop}
\Omega_{(ij)}^{(1)}={\cal T}_{\lambda}\left[\Omega_{(IJ)}^{(1)}\bigg|_{\beta_{I}^2=t_i\lambda^2,\beta_{J}^2=t_j\lambda^2}\right]=-2\log\left(\frac{(2\beta_i\cdot\beta_j)^2}{t_it_j\lambda^4}\right)\,.
\end{equation}
The logarithmic divergence in~eqs.~(\ref{lliJ}) and~(\ref{GammaCuspij_one_loop}) is associated exclusively with cusp singularities, namely ones originating in overlapping UV and collinear singularities. 
In fact, this logarithmic divergence is an all-order property: it was proven~\cite{Korchemsky:1987wg} that the lightlike limit of the angle-dependent cusp AD $\Omega_{(IJ)}$ is always proportional to the logarithm. We therefore write down the lightlike limit of the cusp AD in the limit to all orders,
    \begin{subequations}
\label{dipole_small_lambda_lim}
    \begin{align}
    \begin{split}
    \label{GammaIk}
 \Omega_{(iJ)}=\,\,&{\cal T}_{\lambda} \left[\Omega_{(IJ)}\bigg|_{\beta_{I}^2=t_i\lambda^2}\right]=-\frac{1}{2}\gamma_{\text{cusp}}(\alpha_s)\log\left(\frac{(2\beta_i\cdot\beta_J)^2}{t_i\lambda^2\beta_J^2}\right),
 \end{split}
\\
 \begin{split}
 \label{GammaCuspij}
\Omega_{(ij)}=\,\,&{\cal T}_{\lambda} \left[\Omega_{(IJ)}\bigg|_{\beta_{I}^2=t_i\lambda^2,\beta_{J}^2=t_j\lambda^2}\right]=-\frac{1}{2}\gamma_{\text{cusp}}(\alpha_s)\log\left(\frac{(2\beta_i\cdot\beta_j)^2}{t_it_j\lambda^4}\right),
 \end{split}
 \end{align}
\end{subequations}
where the lightlike cusp AD is
\begin{equation}
    \label{llcusp}
    \gamma_{\text{cusp}}(\alpha_s)=\sum_{n=1}^{\infty}\left(\frac{\alpha_s}{4\pi}\right)^n\gamma_{\text{cusp}}^{(n)}\,.
\end{equation}
By comparing (\ref{GammaCuspij}) with (\ref{GammaCuspij_one_loop}) one finds that at one loop $\gamma_{\text{cusp}}^{(1)}=4$, consistently with the literature. The state-of-the-art knowledge of $\gamma_{\text{cusp}}$ in QCD is four loops~\cite{Moch:2004pa,Henn:2016men,Davies:2016jie,Henn:2016wlm,Lee:2017mip,Moch:2017uml,Grozin:2018vdn,Moch:2018wjh,Lee:2019zop,Henn:2019rmi,vonManteuffel:2019wbj,Henn:2019swt,vonManteuffel:2020vjv,Agarwal:2021zft}.

At this point it is natural to relate these results to the singularity structure of amplitudes.
By performing the transformation $\frac{\beta_i}{\sqrt{t_i}\lambda}\rightarrow \frac{p_i}{\mu}$ in eqs.~(\ref{GammaIk}) and~(\ref{GammaCuspij})
these logarithms directly 
generate the 
$\mu$-dependent dipole terms in~eq.~(\ref{GammaL}), i.e. they correspond to logarithmic factorization-scale-dependent terms in the soft~AD~(\ref{GammaL}), 
proportional to the lightlike cusp AD.
These dipole terms ultimately represents 
overlapping soft and collinear divergences in the factorized amplitude. They depend on the external momenta rather than the velocities, and hence they violate the rescaling symmetry present at the level of the eikonal Feynman rules. In fact, these terms are the sole reason for rescaling violation in the soft~AD~\cite{Becher:2009qa,Gardi:2009qi}, to any order in pertubation theory.

Moving on to the tripole terms, in the lightlike limit of $\beta_K$, eq.~(\ref{GammaIJK}) becomes:
\begin{align}
 \label{mixtwoloop}
\begin{split}
  \Omega_{(IJk)}^{(2)}= {\cal T}_{\lambda} \left[\Omega_{(IJK)}^{(2)}\bigg|_{\beta_{K}^2=c\lambda^2}\right]=-4\log(y_{IJk})
\left[\frac{1+\alpha_{IJ}^2}{1-\alpha_{IJ}^2}M_{100}(\alpha_{IJ})+2\log^2(\alpha_{IJ})+2\zeta_2\right]\,,
\end{split}
\end{align}
in line with ref.~\cite{Ferroglia:2009ep}.
We note that, in sharp contrast to the dipole terms of eq.~(\ref{dipole_small_lambda_lim}), no $\log(\lambda)$ terms arise in the lightlike limit of $\Omega_{(IJK)}$: the $\lambda\to 0$ limit in (\ref{mixtwoloop}) is smooth and the result directly corresponds to the tripole term $\Omega_{(IJk)}^{(2)}$ in the mixed soft~AD (\ref{GammaL}).
The same applies more generally. The expectation is that any non-dipole  term in the mixed soft~AD, i.e. any term that depends on more than two particles, some of which are massless, can be obtained 
directly by taking the lightlike limit of the corresponding timelike soft~AD.

Consider a correlator of $N+M$ timelike Wilson lines,  $\left<\Phi^{(m)}_{\beta_{1}}\cdots\Phi^{(m)}_{\beta_{_N}}\Phi^{(m)}_{\beta_{N+1}}\cdots\Phi^{(m)}_{\beta_{N+M}}\right>$
under the limit where $N$ velocities $\beta_K$ for $K\leq N$ get simultaneously close to the lightcone,  while the remaining $M$ remain timelike, i.e. $\beta_{K}^2\rightarrow \beta_{k}^2=0, k=1,\ldots,N$, the soft AD $\Gamma_{\UV}$ reads
\begin{align}
 \label{GammaUV}
\begin{split}
\\\Gamma_{\UV}(\alpha_s(\mu^2))&=\sum_{i}\tilde{\gamma}_{i}(\alpha_s)+\sum_{I}\Omega_{(I)}(\alpha_s)
\\&+\sum_{I<J}\mathbf{T}_{I}\cdot\mathbf{T}_{J}\Omega_{(IJ)}(\alpha_{IJ},\alpha_s)
\\&-\sum_{i}\sum_{I}\frac{1}{2}\mathbf{T}_{i}\cdot\mathbf{T}_{I}\gamma_{\text{cusp}}\left(\alpha_s\right)\log\left[\frac{(2\beta_{I}\cdot\beta_{i})^2}{t_i\lambda^2\beta_{I}^2}\right]
\\&-\sum_{i<j}\frac{1}{2}\mathbf{T}_{i}\cdot\mathbf{T}_{j}\gamma_{\text{cusp}}\left(\alpha_s\right)\log\left[\frac{(2\beta_{i}\cdot\beta_{j})^2}{t_it_j\lambda^4}\right]
\\&+\sum_{I<J<K}\mathbf{T}_{IJK}\Omega_{(IJK)}(\{\alpha_{IJ},\alpha_{JK},\alpha_{IK}\},\alpha_s)
\\
&+\sum_{I<J}\sum_{k}\mathbf{T}_{IJk}\Omega_{(IJk)}\left(\left\{\alpha_{IJ},y_{IJk}\right\},\alpha_s\right).
 \end{split}
\end{align}
We comment that the collinear AD~$\tilde{\gamma}_{i}$
appearing in the first line of eq.~\eqref{GammaUV} here, is distinct to the collinear AD~$\gamma_{i}$ in  eq.~\eqref{Gammadimension}, which governs collinear singularities in partonic amplitudes (see a related discussion in~\cite{Falcioni:2019nxk}).

As we have not taken the strict limit in eq.~(\ref{GammaUV}) (this is prohibited by the logarithmic singularity), multiplicative renormalizability still holds, 
\begin{equation}
\label{ZUV}
\begin{aligned}
\begin{gathered}
{\cal T}_{\lambda}\left[\left<\Phi^{(m)}_{\beta_{1}}\cdots\Phi^{(m)}_{\beta_{N}}\Phi^{(m)}_{\beta_{N+1}}\cdots\Phi^{(m)}_{\beta_{N+M}}\right>\bigg|_{\{\beta_{i}^2=t_i\lambda^2\}_{i=1}^N}\right]Z_{\UV}(\alpha_s(\mu^2))
\\={\cal T}_{\lambda}\left[\left<\Phi^{(m)}_{\beta_{1}}\cdots\Phi^{(m)}_{\beta_{N}}\Phi^{(m)}_{\beta_{N+1}}\cdots\Phi^{(m)}_{\beta_{N+M}}\right>_{\mathrm{ren.}}\bigg|_{\{\beta_{i}^2=t_i\lambda^2\}_{i=1}^N}\right],
\end{gathered}
 \end{aligned}
\end{equation}
where
\begin{equation}
\label{ZUVre}
    \frac{dZ_{\UV}(\alpha_s(\mu^2))}{d\log\mu}=-\Gamma_{\UV}(\alpha_s(\mu^2))Z_{\UV}(\alpha_s(\mu^2))\,,
\end{equation}
with the anomalous dimension $\Gamma_{\UV}$ the asymptotic expansion of $\Gamma_{\UV}^T$,
\begin{align}
    \Gamma_{\UV}(\alpha_s(\mu^2))={\cal T}_{\lambda}\left[ \Gamma_{\UV}^T(\alpha_s(\mu^2))\right]\,.
\end{align}
Notice that the dipole terms in eq.~\eqref{GammaUV} involving lightlike lines do not have a smooth $\lambda\to 0$ limit. Instead, they depend logarithmically on the combination $t_i\lambda^2$. In contrast, non-dipole contributions to the soft~ADs with lightlike indices ($k$), such as $\Omega_{(IJk)}$, are strictly $\lambda$-independent and are exactly the same as those in~$\Gamma^L$ in eq.~(\ref{GammaL}). 
Similarly to~$\Gamma_{\UV}^T$ of (\ref{GammaT}) -- and in contrast to~$\Gamma^L$ -- the AD $\Gamma_{\UV}$ depends on the renormalization scale~$\mu$ only through the strong coupling. 

In this section we briefly summarized the well-studied method to compute massive soft~ADs, which takes advantage of the multiplicative renormalizability of the correlator. We also reviewed how an expansion of the massive soft~AD near the lightcone, yields mixed terms, such as $\Omega_{(IJk)}$. However, while fully-massive results are known to two loops, they become hard to compute at higher loop orders. 
As mentioned earlier, at three loops the soft~ADs for multileg scattering has only been computed for the massless case~\cite{Almelid:2015jia,Gardi:2016ttq,Almelid:2016lrq}, starting with a correlator of timelike Wilson lines. To this end a high-dimensional Mellin-Barnes representation of the webs defined with timelike Wilson lines was first determined, and then an asymptotic expansion of the  Mellin-Barnes integrals near the simultaneous lightlike limit was performed, where major simplifications occur.  After the expansion, the result for these webs could be expressed as three-fold iterated sums, which were ultimately evaluated to obtain an elegant closed-form expression in terms of weight-5 generalised polylogarithms. 
This approach crucially relies on the vast simplification of the Mellin-Barnes representation in the simultaneous massless limit. It would be hard to apply it in the case where some of the Wilson lines remain timelike, let alone go to higher loop orders.   

If the central physics interest is in the massless or mixed soft~ADs, it may be more efficient to work with lightlike Wilson lines which significantly simplify the computations. Unfortunately, the relation between massless or mixed soft~ADs and correlators involving strictly lightlike Wilson lines is not clear due to the violation of  multiplicative renormalizability, a relation we will revisit in section~\ref{sec:RenMixedCorr} below. 

Mathematically, the  origin of the difficulty is the non-commutativity between the integration and the lightlike limit. This situation is of course commonplace in many QFT computations, and the standard method to address it is to perform asymptotic expansions using the Method of Regions (MoR)~\cite{Beneke:1997zp,Smirnov:2002pj,Pak:2010pt,Jantzen:2011nz,Jantzen:2012mw,Semenova:2018cwy,Gardi:2022khw,Ma:2023hrt}. The main goal of this paper is to develop a new strategy for computing  
mixed soft~ADs without evaluating the fully-massive integrals. In what follows we explain how the MoR can be used to achieve this goal.

\section{Methodology}
\label{sec:methodology}

In this section we present our new method to compute the soft~AD for amplitudes involving one or more massless particles. At the heart of this method stands the asymptotic expansion of Wilson-line correlators near the lightlike limit using the Method of Regions (MoR).

In section~\ref{sec:MassiveMassless}, we reviewed the conventional approach where one determines the soft AD by taking the lightlike limit of (the final result for) the correlator of timelike Wilson lines. 
Here we perform an asymptotic expansion of the Wilson-line correlator near  $\beta_K^2\to 0$. This leads, following the MoR, to the computation of several integrals associated with different approximations to the integrand, the so-called ``regions''. 
One of these contributions is the strict limit, dubbed the \emph{hard region}, which amounts to the computation of a ``mixed'' correlator where one or more Wilson lines are taken to be \emph{exactly} lightlike, $\beta_k^2=0$. To build up the correct asymptotic expansion of the correlator near $\beta_K^2\to 0$, starting with said hard region, one must include additional regions. We will show that adding these has the effect of removing all IR singularities generated in the strict limit, so as to recover the expanded correlator, which features a single UV pole. The coefficient of this pole is the sought-after soft AD. This will be exemplified in 
section~\ref{sec:twoloopcomp}, where we present the MoR computation of both the one- and two-loop soft~AD for the case where one Wilson line becomes lightlike.

The practical motivation for  developing this new strategy is that the computations of correlators with timelike Wilson lines become extremely hard as the loop order increases. Therefore, it would be very useful to build the connection between ``mixed soft~ADs'' -- that is ones involving both massless  and massive particles -- and ``mixed correlators'' -- that is ones involving strictly lightlike Wilson lines as well as timelike ones --  the computations of which are significantly simplified. 
It is also of theoretical interest to understand the singularity structure and the IR and UV renormalization of correlators involving strictly lightlike Wilson lines. This will be the subject of section~\ref{sec:RenMixedCorr}.

We set up the problem starting with a fully IR-regularized correlator of \emph{timelike} semi-infinite Wilson lines; this object defines the integrals to which we will be applying the asymptotic expansion. 
To this end, in section~\ref{sec:Regularised_WilsonLines} we specify the form of a regularized timelike Wilson line, which determines the eikonal Feynman rules. Then, in section~\ref{sec:RegularizeCorrelator}, we consider the correlator of any number of IR-regularized timelike Wilson lines. To manifest the exponentiation properties of this correlator we express it in terms of \emph{webs}~\cite{Frenkel:1984pz,Gatheral:1983cz,Sterman:1981jc,Gardi:2010rn,Gardi:2011wa,Gardi:2011yz,Gardi:2013ita}. We illustrate this for the particular case of tripole corrections at two loops, on which we shall focus in section~\ref{sec:twoloopcomp} below.
The final subsection~\ref{sec:MoR} summarizes the main steps we take in performing the asymptotic expansion of said correlator using the MoR, where a correlator involving strictly lightlike lines appears as the so-called hard region, and additional IR regions arise.

\subsection{Regularized Wilson lines}
\label{sec:Regularised_WilsonLines}

\begin{figure}
    \centering
    \includegraphics[width=0.8\linewidth]{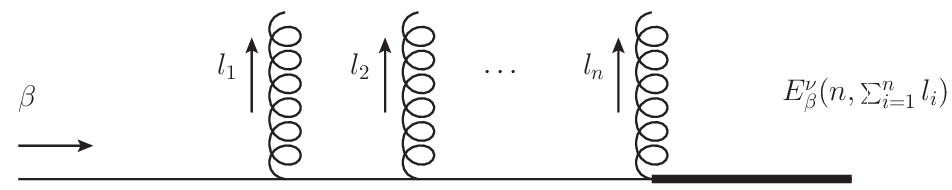}
    \caption{$n$ gluon emissions from an incoming Wilson line with four-velocity $\beta$. The eikonal propagator in eq.~(\ref{dimlessEikP})  is represented by the bold segment.
    }
    \label{fig:EikP}
\end{figure}

To compute the correlator and extract its UV divergences defining the soft~AD we should work with a suitable IR regulator. 
The key requirement is that the regulator would not alter the UV divergences, while removing all IR ones. Given that the relevant correlators have no inherent scale, and that we wish to use dimensional regularization for the UV divergences, the IR regulator must provide the only dimensionful scale. In this work, we will make use of the usual exponential IR regulator~\cite{Gardi:2011yz} in configuration space,\footnote{Various alternative regularizations may be used, see for example the early work in refs.~\cite{Korchemsky:1987wg,Korchemskaya:1994qp}. } which preserves rescaling symmetry, 
\begin{equation}
\label{Phim}\Phi^{(m)}_{\beta_{I}}=\text{P}\exp\left[ig_s\int^{\infty}_0ds\, e^{-ims\sqrt{\beta_{I}^2-i\varepsilon}}\, \beta_{I}\cdot A(s\beta_{I})\right].
\end{equation}
 In momentum space, the usual eikonal propagator is modified by a mass-like term. For example, the propagator after $n$ gluons are emitted from an incoming Wilson line with velocity $\beta$ (admitting $\beta^2>0$) takes the form:
\begin{align}
\label{EikP}
E^{\nu}_{\beta}\left(n,\sum_{i=1}^nl_i\right)\equiv\frac{(ig_s)\, i\beta^\nu}{-\beta\cdot \sum_{i=1}^nl_n-nm\sqrt{\beta^2}
+ i\varepsilon}\,.
\end{align}
This propagator corresponds to the bold segment in figure~\ref{fig:EikP}. For an outgoing line, one just flips the sign of $\beta$ in eq.~\eqref{EikP}.
Compared with an ordinary Feynman propagator, the denominator in (\ref{EikP}) is linear in the momentum $l_i$. 
The usual (unregularized) eikonal propagator can be obtained by setting $m=0$. 

In eq.~(\ref{EikP}) the rescaling symmetry of the velocity vector $\beta$ is explicit. In the course of the computation one may also use this freedom to rescale $\beta$ by the corresponding  (timelike) virtuality, i.e. 
\begin{equation}
\label{v_rescalled_beta}
 v^{\nu}\equiv {\beta^{\nu}}/{\sqrt{\beta^2}}\,.
\end{equation}
At the same time, one may also rescale the loop momentum~$l$ by the regulator $m$ to define the dimensionless propagator~$\tilde{E}^{\nu}$,
\begin{align}
\label{dimlessEikP}
E^{\nu}_{\beta}\left(n,\sum_{i=1}^nl_i\right)=\frac{-g_s v^\nu}{m}\left(\frac{1}{-v\cdot \sum_{i=1}^nq_i-n + i\varepsilon}\right)
\equiv\frac{-g_s v^{\nu}}{m}\tilde{E}_{v}\left(n,\sum_{i=1}^nq_i\right),
\end{align}
where the dimensionless loop momentum is $q_i^\mu={l_i^\mu}/{m}$. In eq.~\eqref{dimlessEikP} the dimension of the propagator $E^{\nu}_{\beta}$ is carried by the overall factor of ${1}/{m}$.

 In the strict lightlike limit, the regulator of eq.~(\ref{Phim}) is naturally removed as the exponential factor is replaced by unity,
 \begin{equation}
 \label{llWilsonLine}
     \phi^{(m)}_{\beta_{i}}= \phi_{\beta_{i}}=\text{P}\exp\left[ig_s\int^{\infty}_0ds\beta_{i}\cdot A(s\beta_{i})\right]\,.
 \end{equation}
This makes the computation of the mixed timelike-lightlike correlator relatively easy, but at the same time it introduces IR singularities. 

\subsection{Regularized correlators of timelike Wilson lines in terms of webs}
\label{sec:RegularizeCorrelator}

Our starting point is the correlator of timelike Wilson lines in a complete regularization scheme, where all Wilson line attachments are regularized as in eq.~(\ref{Phim}): 
\begin{equation}
\label{massiveCorr}
\left<\Phi^{(m)}_{\beta_I}\Phi^{(m)}_{\beta_J}\cdots\right> \,.
\end{equation}
This is precisely the object we renormalized in eq.~(\ref{MR2}), and it will be used to define the integrals to which we will be applying the asymptotic expansion  below.
We emphasize that there is no momentum conservation at the point where all Wilson lines meet (and hence no relation between the velocities), because other colourless particles in the scattering process also carry momentum. 

The exponential form of the correlator (\ref{massiveCorr}) and its perturbative expansion are given in eq.~\eqref{nloopexp}.
In section~\ref{sec:correlators} we mainly emphasized the relation between this exponential form and the renormalization-group equations. However, it is well known that there exists a  complementary, diagrammatic picture of non-Abelian 
exponentiation~\cite{Gardi:2013ita}, stemming from the pioneering work in the 1980s~\cite{Frenkel:1984pz,Gatheral:1983cz,Sterman:1981jc} which generalizes the Abelian case, albite in a rather non-trivial way. This has been extended to  correlators involving multiple Wilson lines in refs.~\cite{Gardi:2010rn,Gardi:2011wa,Gardi:2011yz,Dukes:2013wa,Dukes:2013gea,Gardi:2013ita,Gardi:2013saa,Falcioni:2014pka,Vladimirov:2015fea} (see also more recent work in~\cite{Becher:2019avh,Gardi:2021gzz,Agarwal:2020nyc,Mishra:2023acr,Agarwal:2024srg,Henn:2023pqn,Figueiredo:2025txe}). 
Here we follow this methodology and use diagrammatic exponentiation to simplify the computation of the singularities of the correlator. To this end we rewrite $w^{(n)}$, the $n$-th order expansion coefficient in the exponent of the correlator in~(\ref{nloopexp}) using webs, which manifest the fully-connected colour structure of the exponent~\cite{Gardi:2013ita},
 \begin{align}
 \label{Wij}
     \begin{split}
        \left(\frac{\alpha_s}{4\pi}\right)^n w^{(n)}=\sum_{i}{\cal C}_{i}\sum_{j}{\Y}_{ij}=\sum_{i,j}W_{ij}\,.
     \end{split}
 \end{align}
By fully-connected colour structure we mean a colour structure associated with a diagram which remains connected after removing all the Wilson lines.
In eq.~(\ref{Wij}) ${\cal C}_{i}$ is a fully connected colour structure indexed by $i$, while $j$ runs over all the configurations contributing to the colour structure ${\cal C}_i$ with ${\Y}_{ij}$ the corresponding kinematically-dependent function. The web $W_{ij}$ is defined as the product of the colour structure and the kinematic function, $W_{ij}\equiv{\cal C}_i{\Y}_{ij}$. Each web $W_{ij}$ is itself a combination of a set of Feynman diagrams distinguished by permutations of the order of attachments (gluon emissions) to each of the Wilson lines, as we illustrate below. 
 
In describing the computation it will be convenient to replace the abstract subscript~$ij$ in eq.~(\ref{Wij}) by an index that represents the web configuration; in ref.~\cite{Gardi:2010rn}, the subscript identifying a given web simply counts the number of attachments to each Wilson line. In the present paper, we will use a more explicit notation indicating the Wilson lines attached to by each 
connected subdiagram. For example, $W_{(IJK)}$ has one connected subdiagram, a three-gluon vertex with the three gluons attached to the Wilson lines $I$, $J$, and $K$ (as in eq.~(\ref{W_IJK}) below), while $W_{(IJ)(IK)}$ has two connected subdiagrams, one attached to lines $I$ and $J$ and the other to lines $I$ and~$K$ (eq.~(\ref{W_IJ_IK}) below); The latter is an example of the multiple-gluon-exchange class of webs analysed in detail in refs.~\cite{Gardi:2013saa,Falcioni:2014pka}.
A further notation we introduce here is the distinction between timelike and lightlike lines, which will be labelled respectively by upper-case or lower-case letters.

Let us now illustrate the concept of webs in the particular case on which we focus in the next section, namely those contributing to the tripole colour structure defined in eq.~(\ref{fTTT}), $\textbf{T}_{IJK}\equiv if^{abc}\textbf{T}_I^a\textbf{T}_J^b\textbf{T}_K^c$, at two loops. 
In this case the colour structure in eq.~(\ref{W_IJK})
is ${\cal C}_{(IJK)}=\textbf{T}_{IJK}$, 
are two types of webs to consider: the fully connected diagram,
\begin{align}
\label{W_IJK}
W_{(IJK)}^{(2)}=
{\textcolor{blue}{
\text{Col}\left[\vcenter{\hbox{\includegraphics[width=1.5cm]{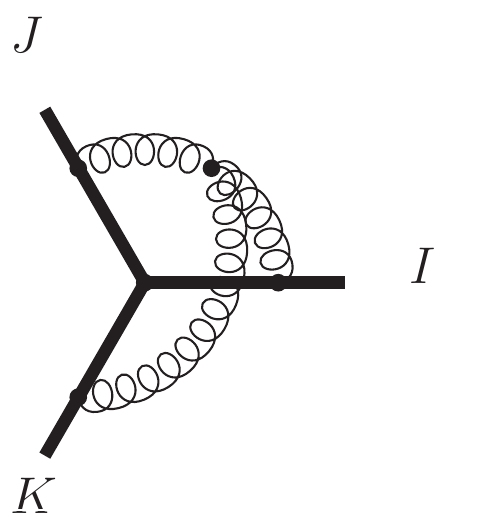}}}
\hspace*{-6pt}\right]}} \times 
{\textcolor{red}{\text{Kin}\left[
\vcenter{\hbox{\includegraphics[width=1.5cm]{fig/111IJK.pdf}}}
\hspace*{-6pt}\right]}}
=\,
{\textcolor{blue}{\textbf{T}_{IJK}}}
\times 
{\textcolor{red}{{\Y}_{(IJK)}}}
\end{align}
and the one consisting of two separate gluon exchanges between the three Wilson lines, 
\begin{align}
\label{W_IJ_IK}
\begin{split}
W_{(IJ)(IK)}^{(2)}=\,\,&\,
\left\{
{\textcolor{blue}{
\text{Col}\left[\vcenter{\hbox{\includegraphics[width=1.5cm]{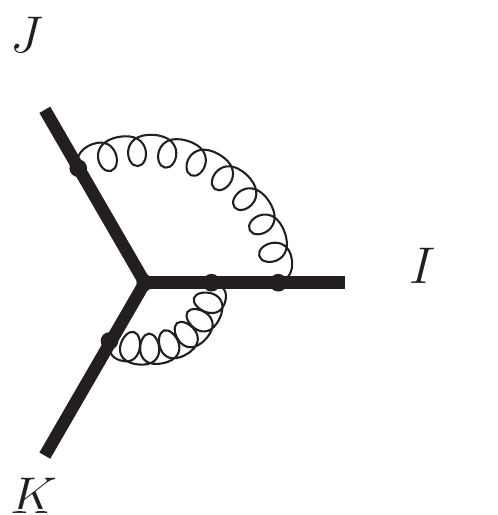}}}\hspace*{-6pt}\right]}} -
{\textcolor{blue}{\text{Col}\left[\vcenter{\hbox{\includegraphics[width=1.5cm]{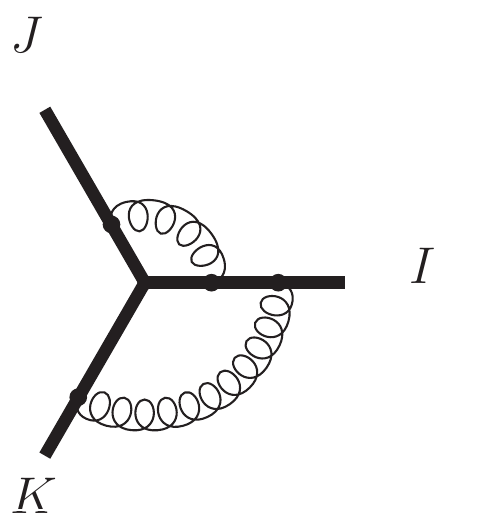}}}\hspace*{-6pt}\right]}}
\right\}\times
\frac12\left\{
{\textcolor{red}{
\text{Kin}\left[\vcenter{\hbox{\includegraphics[width=1.5cm]{fig/211IJKa.pdf}}}\hspace*{-6pt}\right]}} -
{\textcolor{red}{\text{Kin}\left[\vcenter{\hbox{\includegraphics[width=1.5cm]{fig/211IJKb.pdf}}}\hspace*{-6pt}\right]}}
\right\}
\\
\\
=\,\,&\,
\left\{
{\textcolor{blue}{
\textbf{T}_I^b\textbf{T}_J^b 
\textbf{T}_I^c\textbf{T}_K^c
- 
\textbf{T}_I^c\textbf{T}_K^c
\textbf{T}_I^b\textbf{T}_J^b 
}}
\right\}\times
{\textcolor{red}{{\Y}_{(IJ)(IK)}}}
=
{\textcolor{blue}{\textbf{T}_{IJK}}}
\times
{\textcolor{red}{
{\Y}_{(IJ)(IK)}}}\,,
\end{split}
\end{align}
where the operators ${\textcolor{blue}{
\text{Col}}}$ and ${\textcolor{red}{\text{Kin}}}$ extract, respectively, the colour and kinematic factors of the diagrams they act upon. In the last step in (\ref{W_IJ_IK}) we used 
the colour algebra for the generators on line $I$, namely, $[\textbf{T}_I^b,\textbf{T}_I^c]=if^{bca}\textbf{T}_I^a$, along with eq.~(\ref{fTTT}) to rewrite the colour structure of the web $W_{(IJ)(IK)}^{(2)}$ in terms of $\textbf{T}_{IJK}$, which is a connected colour factor.  
The two additional webs of the latter type, which can be obtained from eq.~(\ref{W_IJ_IK}) by permuting the Wilson-line indices,
namely $W_{(JI)(JK)}^{(2)}$ and 
$W_{(KI)(KJ)}^{(2)}$, also contribute to the correlator with the same connected colour factor.
The structure in the first line of eq.~(\ref{W_IJ_IK}), in which the colour and kinematic factors of different diagrams (differing just by the order of attachment of the gluons to the Wilson lines) combine is dictated by non-Abelian exponentiation.
A general algorithm to determine the relevant combinations of colour and kinematic factors for any web was proposed in ref.~\cite{Gardi:2010rn} using the replica trick (the specific example of eq.~(\ref{W_IJ_IK}) has been worked out in detail in
section 5.2 of ref.~\cite{Gardi:2010rn}).
The general properties of the resulting colour and kinematic factors of webs have subsequently been investigated, starting with the work of ref.~\cite{Gardi:2011wa}. In particular, it was shown that the colour factors which furnish the exponent are guaranteed to be 
fully-connected~\cite{Gardi:2013ita}. Many further studies and applications have followed and we refer the interested reader to the aforementioned references. In what follows we will only need the examples considered explicitly above. 

\subsection{The method of regions}
\label{sec:MoR}

For Feynman integrals, the operations of taking kinematic limits of interest 
and integrating over the loop momentum usually do not commute. 
This phenomenon is straightforward to understand. Consider for example the eikonal propagator $E^\nu_\beta(1,l)$ with timelike velocity  $\beta$,
\begin{align}
 \label{rad1eikonal}
 E^\nu_\beta(1,l)=\frac{-g_s \beta^\nu}{-\beta\cdot l-m\sqrt{\beta^2}+
 i\varepsilon }\,. 
\end{align}
By taking the lightlike limit, $\beta^\nu\rightarrow (\beta^+,\lambda^2\beta^-,\lambda\beta^\perp)$, where the lightcone coordinates\footnote{
We define lightcone coordinates such that a four vector $q^\mu$ with components $(q^0,q^1,q^2,q^3)$ is represented by $q=(q^+,q^-,q^\perp)$, such that 
$q^+=\frac{q^0+q^3}{\sqrt{2}}$, $q^-=\frac{q^0-q^3}{\sqrt{2}}$, so that its virtuality is~$q^2=2q^{+}q^{-}-|q^{\perp}|^2$.} are used, $\lambda$ tends to zero  ($\beta^2=\lambda^2(2                    \beta^+\beta^--|\beta^\perp|^2)\to 0$)  
and the propagator becomes
\begin{align}
\label{naiveLimitOfProp}
E^\nu_\beta(1,l)\rightarrow\frac{
-g_s \beta^\nu}{-\beta\cdot l +
 i\varepsilon}\,.
\end{align}
This (na\"ive) conclusion is based on considering the denominator of eq.~\eqref{rad1eikonal} with~$\beta\cdot l$ for generic (hard) loop momentum~$l\sim{\cal O}(\lambda^0)$. Under this assumption the regulator~$m$ is multiplied by $\sqrt{\beta^2}$, which is ${\cal O}(\lambda)$, and is therefore negligible compared with~$\beta\cdot l$ for generic~$l$, leading to eq.~(\ref{naiveLimitOfProp}). 
However, $l$ should be integrated over the entire, unrestricted momentum space. Thus, the replacement of (\ref{rad1eikonal}) by   (\ref{naiveLimitOfProp}) may be invalidated in certain regions. For example, suppose that all the components of $l$ are small, $l\equiv\lambda\tilde{l}=(\lambda\tilde{l}^+,\lambda\tilde{l}^-,\lambda\tilde{l}^\perp)$ (this is known as a soft loop-momentum mode). In this  region the regulator term $m\sqrt{\beta^2}$ is as important as $\beta\cdot l$ at small $\lambda$,
 \begin{align}
 \label{rad1eikonalSL}
 E^\nu_\beta(1,l)\rightarrow \frac{-g_s}{\lambda}\left(\frac{\beta^\nu}{-\beta\cdot \tilde{l}-m\sqrt{\beta^2}+
 i\varepsilon}\right)\,, 
\end{align}
and no simplification along the lines of (\ref{naiveLimitOfProp}) takes place.

This example illustrates that simply taking the limit at the integrand level (while naively regarding the integration variables as fixed) may alter the result of the integral already at leading power in the expansion. Instead, one must perform a proper asymptotic expansion, where all so-called \emph{leading regions}, such as the  soft region in the above example, are systematically considered and their contributions are summed over.     
 
The Method of Regions (MoR) \cite{Beneke:1997zp,Smirnov:2002pj,Pak:2010pt,Jantzen:2011nz,Jantzen:2012mw,Semenova:2018cwy,Gardi:2022khw,Ma:2023hrt} provides a strategy to perform asymptotic expansions of Feynman integrals. It systematically deals with the non-commutativity between taking kinematic limits and performing the integration.  
Firstly, given the kinematic limit of interest, i.e. the scaling law of the external momenta with the expansion parameter~$\lambda$, one should identify all leading regions. In every region contributing to the expansion, each loop momentum is in a certain ``mode'', that is, the corresponding momentum components scale in a particular way as the limit is approached, 
\begin{equation}
\label{l_n_scaling_law}
(l_n^+,l_n^-,l_n^\perp) \xrightarrow{\;{R}\;} 
(l_n^+\lambda^{a_+},l_n^-\lambda^{a_-},l_n^\perp \lambda^{a_\perp}),
\end{equation}
where ${a_+}$, ${a_-}$ and ${a_\perp}$ are three rational numbers, typically integers or half integers,  characterizing the behaviour of $l_n$ in region~$R$. 
In any region $R$ one then makes the scaling law manifest by applying (\ref{l_n_scaling_law}) to the integrand (and the integration measure) along with the scaling of the external momenta. With the scaling manifest, one performs a Laurent expansion of the integrand in the limit. 
In particular, the so-called hard region corresponds to retaining the loop momenta as they are (i.e. ${a_+}={a_-}={a_\perp}=0$) while scaling the external ones (this is equivalent to assuming that the loop momenta are hard, $l_n\sim (1,1,1)$ when $\lambda\to 0$).  
After expanding the integrand in each region, ones extends the integration over the loop momenta to the entire unrestricted space. The MoR asserts that upon performing the unrestricted integration of the expanded integrand in dimensional regularization based on the scaling associated with each region, and then adding them all up, one obtains the correct asymptotic expansion of the original integral. 

A key requirement for this remarkable statement to hold is the use of dimensional regularization in evaluating each of these ``region'' integrals, and in the course of doing that discard scaleless integrals (which  are separately divergent in both the UV and the~IR). Under certain conditions this guarantees the cancellation of the divergences that are generated in each of the region integrals  by extending the domain of integration to the entire space (well beyond the radius of convergence of the expansion in each of them).
This statement has been proven to hold if the integral satisfies a set of conditions~\cite{Jantzen:2011nz}. The main weakness of this strategy as a whole is that no general algorithm exists to identify all regions. 

The scaling law of the external kinematic variables $\{s_i\}$ defines the asymptotic expansion of interest. We therefore redefine the variables, making their scaling manifest, $s_i\rightarrow \lambda^{\sigma_i}\tilde{s}_i$, where $\sigma_i$ are rational numbers, typically integers or half integers. The asymptotic expansion~${\cal T}_{\lambda}$  of a Feynman integral ${\cal I}(\{s_i\})$ takes the general form
 \begin{align}
 \label{radieikonalSoft}
   {\cal T}_{\lambda} \left[{\cal I}\left(\left\{\lambda^{\sigma_i}\tilde{s}_i\right\}\right)\right]=\sum_R \lambda^{n_R\epsilon} {\cal I}^R_{\lambda}\left(\{\tilde{s}_i\}\right)\, ,
 \end{align}
 where~$R$ runs over all the regions, and ${\cal I}^R$ is the contribution of region~$R$ (integrated over the full, unrestricted space) with its non-analytic power behaviour, $\lambda^{n_R\epsilon}$, scaled out.  The exponent of the expansion parameter $\lambda$ is a rational number $n_R$,  usually an integer, or half integer,  multiplied by $\epsilon$. The presence of a non-analytic term for $\lambda\to 0$ explains the non-commutativity between the integration and the $\lambda$ expansion of the original integral.  
 Note that for the hard region, $n_R=0$.
 
Considering eq.~(\ref{radieikonalSoft}), it is straightforward to understand the origin of the non-analytic behaviour regions at $\lambda\to 0$  directly from the loop-momentum integration. Recall 
that while the integrand in a given region $R$ is a rational function of $\lambda$, thus admitting a Laurent expansion, the only factor in a loop integral that depend on the spacetime dimensions $D=4-2\epsilon$, is the measure of integration $d^D l_n$ for every loop momentum $l_n$, $n=1\ldots L$, where $L$ is the loop order. Since in what follows we will be using lightcone coordinates, $d^D l_n=dl_n^+dl_n^-d^{2-2\epsilon} l_n^\perp$, any non-analytic dependence in $\lambda$ at $\lambda=0$ is linked  directly to the scaling of the \emph{transverse} momentum components (which we assume to be homogeneous): according to eq.~(\ref{l_n_scaling_law}) the measure for loop $l_n$
scales as 
$
d^D l_n\xrightarrow{\;{R}\;} 
d^D l_n \lambda^{a_+(n)+a_-(n)+
(2-2\epsilon)a_\perp(n)}\,,
$
where the power of the factor $\lambda^{-2\epsilon}$ is governed by $a_\perp(n)$.
Summing over the loops $n=1\ldots L$ in the integral, one can therefore identify the non-analytic behaviour associated with region~$R$ in~eq.~(\ref{radieikonalSoft}) as 
\begin{equation}
\label{TransverseScaling_and_nR}
n_R=-2\sum_{n=1}^L
a_\perp^R(n)\,.
\end{equation}
Next we observe that regions that are characterized for $\lambda\to 0$ by vanishing transverse loop  momentum components -- so called \emph{IR regions} --  have $a_\perp>0$, so according to eq.~(\ref{TransverseScaling_and_nR}), the corresponding $n_R<0$,  while those that are characterized, for $\lambda\to 0$, by large transverse loop momentum components -- so called \emph{UV regions} -- have $a_\perp<0$, so 
the corresponding $n_R>0$. 
Following this, we will be classifying the regions into three broad categories as follows: 
\begin{align}
\label{n_IR_classification}
\left\{ 
\begin{array}{ll}
n_R<0  \qquad  & \text{IR Region}\\
n_R=0  \qquad & \text{neutral Region}\\
n_R>0  \qquad & \text{UV Region}\,.
\end{array}
\right.
\end{align}  
This classification will be essential in what follows in interpreting the regions and their singularities. 
 
A prototypical expansion, which is analogous to the one considered in this paper, 
is the on-shell expansion of wide-angle-scattering integrals, where the virtuality of some (or all) of the external particles, is taken small, serving as an expansion parameter~\cite{Gardi:2022khw,Ma:2023hrt,Gardi:2024axt}. 
The hard region (the strict on-shell limit) features additional IR poles in 
$\epsilon$, which are not present in the expansion of the original (off-shell) integral. These  originate in the extension of the domain of integration over the loop momenta towards low virtualities, where the ``hard'' approximation breaks down.    
In addition to the hard region, the on-shell expansion
also features so-called \emph{IR regions}, which are characterized by $n_R<0$. These give rise to UV poles in $\epsilon$, which originate in the extension of the domain of integration over the loop momenta towards high virtualities, beyond the validity of the IR-region approximation. 
In the sum of the integrated hard and IR regions, the superfluous poles in $\epsilon$ cancel, and logarithms in the external virtualities are generated, which reproduce the expansion of the original off-shell integral in the limit considered.      

In computations of Wilson-line correlators expanded near the lightlike limit, $\beta_K^2\to 0$, which we study in what follows, an analogous but somewhat more involved scenario is realised. In this case, for a given integral,  both positive and negative values of 
$n_R$ may arise, corresponding to UV and IR regions, respectively. Furthermore, in addition to the hard region, the expansion features other neutral regions with $n_R=0$.

The main hurdle in applying the MoR is the need to determine a priori the  complete set of regions. While this remains a major challenge in momentum space, a systematic region-finding algorithm has been devised in parameter space~\cite{Pak:2010pt,Jantzen:2012mw,Semenova:2018cwy,Gardi:2022khw,Ananthanarayan:2018tog,Heinrich:2023til,Borowka:2017idc,Chen:2024xwt} using a geometric construction based on the Symanzik graph polynomials. 

Let us then briefly introduce this powerful method.
Consider a Feynman integral~${\cal I}$ in Lee-Pomeransky representation~\cite{Lee:2013hzt} (see also~\cite{Weinzierl:2022eaz,Gardi:2022khw} for comparison with other parametric representations)
\begin{align}
\label{LeePom}
\begin{split}
   {\cal I}(\{s_i\})=\,\,&\,\, C_{D}(\{\nu_e\},L)\,\int_0^{\infty}\prod_{e}dx_ex_e^{\nu_e-1}\Big[{\cal P}(\{s_i\},\{x_j\})\Big]^{-\frac{D}{2}},
   \\
   &C_{D}(\{\nu_e\},L)\equiv\frac{\Gamma\left(\frac{D}{2}\right)}{\Gamma\left(\frac{(L+1)D}{2}-\nu\right)\prod_{e}\Gamma\left(\nu_e\right)}
   \end{split}
\end{align}
 where $e$ runs over all edges in the graph, $e\leq E$ ($E$ being the total number of internal edges), each of which is associated with a Lee-Pomeransky parameter $x_e$,  
 $L$ is the loop order and $D$ is the space-time dimension which is set to be $4-2\epsilon$ in this paper.  
 We use~$\nu_e$ to denote the power of the denominator of the corresponding propagator and $\nu$ is the sum of the powers $\nu=\sum_e\nu_e$. The set of kinematic variables is represented by $\{s_i\}$. The Lee-Pomeransky polynomial ${\cal P}$ is defined by  
 \begin{align}
\label{calPLeePom}
     {\cal P}(\{s_i\},\{x_j\})=  {\cal U}(\{x_j\})+ 
     {\cal F}(\{s_i\},\{x_j\}),
 \end{align}
 where ${\cal U}$ and ${\cal F}$ are the first and second Symanzik polynomials: ${\cal F}$ depends on the parameters $\{x_j\}$ with $1\leq j\leq E$ and on the kinematic variables $\{s_i\}$, while ${\cal U}$ only on the former.

 Parameteric representations are advantageous in systematically characterizing regions (as compared to momentum space) due to two fundamental properties.
First, they provides a Lorentz invariant way to characterize
the way in which 
different propagators in the graph approach their mass shell. Second, in parametric space many -- sometimes all -- of the regions  originate from endpoint Landau singularities  rather than from Landau singularities which satisfy a pinch condition inside the domain of integration.   
 
Regions associated with endpoint singularities in this parametric space are fully characterized by the scaling laws of the integration variables $\{ x_j\}$. 
In the Lee-Pomeransky representation, similarly to the Schwinger representation, 
\begin{equation}
\label{Schwinger}
    \frac{1}{P_j^{\nu_j}} =\frac{1}{\Gamma(\nu_j)}\,\int_0^\infty 
    \frac{d\bar{x}_j}{\bar{x}_j}\bar{x}_j^{\nu_j}
    \, e^{-\bar{x}_j P_j} \,,
\end{equation}
the scaling law of~$x_j$ with $\lambda$ is just the inverse of the scaling law of the denominator of the corresponding propagator~\cite{Engel:2022kde,Gardi:2022khw}. That is, if in a given region $R$ the momentum-space propagator behaves as $P_j\to \lambda^{-u_j^R} P_j$, then the corresponding Lee-Pomeransky parameter scales as $x_j\to \lambda^{u_j^R} x_j$. This rule allows us to map regions between parametric space and momentum space, as will be discussed below. 

We assume that when approaching the limit according to $s_i\rightarrow \lambda^{\sigma_{i}}\tilde{s}_i$, the scaling of the integration variables is \hbox{$x_j\rightarrow \lambda^{u_j}\tilde{x}_j$}, where $\{u_{j}\}$ is a vector of rational numbers. Thus, each region $R$ is characterized by a unique vector of exponents:
\begin{equation}
\label{regionVectors}
    {\bf{u}}^{R}
    \,=\{u_{1}^R,u_{2}^R,\ldots,u_{E}^R ,1\}\,,
\end{equation}
which we call the \emph{region vector}. Note that the last entry $1$ is introduced by convention.  

With the scalings implemented, we perform a Taylor expansion of the integrand, denoted by  $T_{\lambda}$,
 \begin{align}
 \label{rescaledLeePom}
\lambda^{n_R\epsilon}
{\cal I}^R_{\lambda}(\{\tilde{s}_i\})
\equiv\,
C_{D}(\{\nu_e\},L)
\,\int_0^{\infty}\prod_{e}d\tilde{x}_e\tilde{x}_e^{\nu_e-1}T_{\lambda}\left\{\lambda^{\nu}\Big[{\cal P}(\{\lambda^{\sigma_i}\tilde{s}_i\},\{\lambda^{u^R_j}\tilde{x}_j\})\Big]^{-\frac{D}{2}}\right\}\,,
\end{align}
defining the region integral 
${\cal I}^R_{\lambda}$ of eq.~\eqref{radieikonalSoft}, after scaling out the leading power behaviour.
Notice that the integration range of the new parameter $\tilde{x}_j$ extends over the whole positive axis, well beyond the radius of convergence of the expansion, similarly to what we have seen in momentum space. As discussed above, the resulting singularities are expected to cancel upon summing the regions. 

The asymptotic expansion of the original integral is obtained by summing over all region integrals, each characterized by a region vector $\mathbf{u}_R$:
 \begin{align}
 \begin{split}
 \label{pararesult}
   {\cal T}_{\lambda}\left[{\cal I}(\{\lambda^{\sigma_i}\tilde{s}_i\})\right]=\,\,&\,\,C_{D}(\{\nu_e\},L)\,
   \sum_{ \{n_{x_j}\}}\int_0^{\infty}\prod_{e}d\tilde{x}_e\tilde{x}_e^{\nu_e-1}T_{\lambda}\left\{\lambda^{\nu}\left[{\cal P}(\{\lambda^{\sigma_i}\tilde{s}_i\},\{\lambda^{u^R_{j}}\tilde{x}_j\})\right]^{-\frac{D}{2}}\right\}
   \\=\,\,&\sum_{R} \lambda^{n_R\epsilon}
   \, {\cal I}^R_{\lambda}(\{\tilde{s}_i\})\,.
   \end{split}
\end{align}
As already mentioned, the key feature of the parameter-space formulation is the existence of an algorithm to determine the region vectors. The idea is based on considering the space defined by the powers -- that is the exponents -- of the parameters $x_j$ in a given monomial in the polynomial $\cal P$ of eq.~(\ref{calPLeePom}). This forms a $(E+1)$ dimensional space, where the first~$E$ entries correspond to the powers of $x_j$ with $1\leq j\leq E$, while the $(E+1)$-th component corresponds to the scaling of each monomial with~$\lambda$ in the limit considered. Within this space one defines a Newton polytope, where each monomial in ${\cal P}$ is represented by a vertex, whose coordinates are the exponents. 
It can be shown that the region vectors $\mathbf{u}^R=\{u^R_j\}$ are precisely the (inwards pointing) normal vectors to the lower facets of this polytope~\cite{Pak:2010pt} (see also
\cite{Jantzen:2012mw,Semenova:2018cwy,Gardi:2022khw,Ananthanarayan:2018tog,Heinrich:2023til,Borowka:2017idc,Chen:2024xwt}). Here facets correspond to faces of codimension one, while lower facets are defined with respect to the $(E+1)$-th dimension: only these facets are relavant when expansing in positive powers of $\lambda$.  
An algorithm of finding all regions based on this geometrical construction  has been implemented in  several packages including  {\tt{Asy2}}~\cite{Jantzen:2012mw}, {\tt{ASPIRE}}~\cite{Ananthanarayan:2018tog}, {\tt{pySecDec}}~\cite{Heinrich:2023til,Borowka:2017idc} and {\tt{AmpRed}}~\cite{Chen:2024xwt}. In this work we use the last two. 

The geometric method to identify the complete set of regions relies on the assumption that they all originate in endpoint divergences in parameter space, and are hence fully described by their scaling law with respect to~$\lambda$. There exists a large class of integrals for which it is possible to prove this property, namely show that all Landau singularities that manifest themselves in the first Riemann sheet (in the limit about which we expand) are of the endpoint type. 
In this case the geometric algorithm is guaranteed to yield the complete set of regions.\footnote{We comment in passing that there exist important expansions of Feynman integrals which depart from this simple setting, and require additional case-by-case analysis to identify the complete set of regions. Classical examples are expansions involving potential or Glauber modes~\cite{Jantzen:2012mw}. Regions that do not correspond to facets of the Newton polytope appear due to pinch singularities in parameter space, and are referred to in the literature as ``hidden'' regions~\cite{Gardi:2024axt,Ma:2025emu,Becher:2024kmk,Becher:2025igg}. As oppose to facet regions, these depend on the signs of the monomials in the ${\cal F}$ polynomial and involve cancellation between terms of opposite signs. Hence, they cannot be fully characterized by the scaling of the individual $x_j$ parameters with $\lambda$.
Such regions will not be needed for the problem under consideration in this paper.} Specifically, expansions defined in a so-called Euclidean kinematic regime, belong to this class. A sufficient condition is that all monomials in the Symanzik ${\cal F}$ polynomial, and hence in the Lee-Pomeransky ${\cal P}$ polynomial in eq.~(\ref{pararesult}) have the same sign. We will use this,  and set up our computation of the Wilson-line correlator in the region where $\gamma_{IJ}<0$ for all $I$ and $J$ (see eq.~(\ref{angleDef})), which guarantees that all monomials in ${\cal P}$ are non-negative.

While the determination of the complete set of region vectors $\{\mathbf{u}_R\}$ is done in parameter space, it is useful to map these regions to momentum space, so as to characterize the region by the scaling law of the loop momentum components, in eq.~(\ref{l_n_scaling_law}).  This gives a clear physical interpretation of the regions. The computation of each region can be done by direct evaluation in either space or by setting up differential equations (so the computation itself certainly does not require a momentum space interpretation of the regions). 
Mapping a given region~$R$, with a region vector $\mathbf{u}_R$, to momentum space relies on the aforementioned observation~\cite{Engel:2022kde,Gardi:2022khw} that 
the $\lambda$ scaling of each parameter $x_j$ is inversely proportional to the scaling of the virtuality of the corresponding propagator denominator. However, this Lorentz-invariant information is not
sufficient by itself to fix the $\lambda$ scaling property of each momentum component according to 
eq.~(\ref{l_n_scaling_law}). Instead one must rely in addition on momentum conservation and on mild assumptions regarding the modes. This will be illustrated in the next section. 

Let us now to discuss the application of the MoR in the context of the expansion of a Wilson line correlator defined by eq.~(\ref{nloopexp}). 
Specifically we apply an asymptotic expansion to the kinematic functions contributing to the web ${\Y}_{ij}$ as in eq.~(\ref{Wij}), by summing over regions $R$ as follows,
\begin{align}
\label{ExpansionCalY}
   {\cal T}_{\lambda}\left[{\Y}_{ij}\right]=\sum_R\lambda^{n_R\epsilon} {\Y}^R_{ij}.
\end{align}
The asymptotic expansion of $w^{(n)}$ becomes
 \begin{align}
 \label{RegionFunction}
     \begin{split}
         \left(\frac{\alpha_s}{4\pi}\right)^n {\cal T}_{\lambda}\left[w^{(n)}\right]=\,\,&\sum_{i}{\cal C}_{i}\sum_{j}\left(\sum_{R}\lambda^{n_{R}\epsilon}{\Y}_{ij}^{R}\right)
         \\=\,\,&\sum_{i}{\cal C}_{i}\sum_{R}\lambda^{n_{R}\epsilon}\left(\sum_{j}{\Y}_{ij}^{R}\right)
         \\\equiv\,\,&\sum_{R}\lambda^{n_{R}\epsilon}\sum_{i}{\cal C}_{i}{\cal R}_{i}^{R}\,
    \\\equiv\,\,&\sum_{{n_R}}\lambda^{n_{R}\epsilon}\sum_{i}{\cal C}_{i}{\mathbb R}_{i}^{(n_R)}\,.
     \end{split}
 \end{align}
 In the second line, we reverses the order summation over webs ($j$) and over regions ($R$). 
 In the third line we collected the contributions from all webs associated with given colour factor ${\cal C}_i$ and given region $R$ into what we defined as \emph{region function}, ${\cal R}_{i}^{R}$.
 In the final line we further collected all region function contributions, into \emph{invariant region functions}, ${\mathbb R}_{i}^{(n_R)}$, with a unique characteristic scaling, $\lambda^{n_R\epsilon}$, defined by summing over all regions functions with a common value of $n_R$.
 Given that $w^{(n)}$ is the $L$-th order coefficient of a physical, gauge-invariant correlator, and that both the decomposition into colour factors and the decomposition into functions with distinct analytic behaviour $\lambda^{n_R\epsilon}$ are unique, it follows that the individual functions~${\mathbb R}_{i}^{(n_R)}$ defined by~eq.~(\ref{RegionFunction}) are themselves invariant with respect to any choice made in the computation, such as the choice of gauge or the choice of basis for the master integrals. Such choices might shuffle some contributions between distinct ${\cal R}^R$ region functions sharing the same behaviour $\lambda^{n_R\epsilon}$, but not between different functions~${\mathbb R}_{i}^{(n_R)}$.

\section{Computing the soft~anomalous dimension by the method of regions}
\label{sec:twoloopcomp}

In this section, we perform explicit computations of the soft~AD for amplitudes involving a single massless particle and any number of massive ones, using the MoR. Our aim is to  demonstrate how the method works using simple, pedagogical examples and explore the type of regions that are generated in this expansion, noting the special nature of semi-infinite Wilson lines. To this end we will examine the regions in both parameter space and momentum space.

We begin by considering the simplest example of a one-loop computation and then proceed to our main case study, namely the computation of the soft~AD $\Omega_{(IJk)}$, namely  tripole corrections arising from the interaction between two timelike Wilson lines ($I$ and $J$) and a single lightlike line ($k$), through two loops. The latter was first determined in ref.~\cite{Ferroglia:2009ep} by expanding the fully timelike result.

Given that two-loop webs can connect at most three Wilson lines, it will be sufficient to work with correlators of three lines at general angles (i.e. without imposing momentum conservation). We denote the lines $I$, $J$ and $K$, where line $K$ will be singled out as the one which approaches the lightlike limit, $\beta_K^2\to 0$. For convenience we associate the index $k$ to the corresponding velocity after taking the limit, i.e. $\beta_k^2=0$.

As explained above, we will initially set all three lines, $I$, $J$ and $K$, timelike, and then perform the asymptotic expansion in the lightlike limit $\beta_K^2\rightarrow \beta_k^2=0$. Importantly, we will perform the expansion in Euclidean regime $\{\alpha_{IJ}>0, \alpha_{IK}>0, \alpha_{JK}>0\}$  (recall the definition in eq.~\eqref{alphaIJ}) where the second Symanzik polynomials of all the integrals appearing in this section are positive definite. Besides the convenience of  computing integrals, the key advantage is that the geometric method of determining the regions is then guaranteed to give the complete set, i.e. no hidden regions exist. In the $\beta_K^2\to 0$ limit, the Euclidean regime becomes $\{y_{IJk}>0,\alpha_{IJ}>0\}$. In practice, we will use the following parametrization and approach the limit by sending $\lambda$ to be small, 
\begin{equation}
\label{limitalpha}
\alpha_{IK}=\frac{\lambda}{\sqrt{y_{IJk}}}, \qquad \quad \alpha_{JK}=\lambda\sqrt{y_{IJk}},
\end{equation}
while $\alpha_{IJ}$ remains ${\cal O}(\lambda^0)$, where we recall that the variables are defined in eq.~\eqref{alphaIJ} and eq.~\eqref{yIJk}. In terms of $\lambda^2$, the nearly lightlike  
$\beta_K^2$ becomes
\begin{equation}
\label{betaK2Lambda}
    \beta_K^2=\frac{(2\beta_I\cdot\beta_k)(2\beta_J\cdot\beta_k)}{\sqrt{\beta_I^2\beta_J^2}}\lambda^2+{\cal O}(\lambda^4)\,.
\end{equation}
This parametrization preserves the $(I,J)$-interchange symmetry as well as the rescaling invariance property of the separate velocity vectors $\beta_I$ and $\beta_J$.  A final comment regarding the parametrization is due in the context of our discussion in section~\ref{sec:correlators}. Recall that there we used $\lambda^2$ to denote our expansion parameter, while keeping its (leading-power)   proportionality coefficient with respect to the squared velocity unspecified, writing
$\beta_K^2=t_k\lambda^2$ (see eq.~\eqref{lliJ}). Here, in eq.~(\ref{betaK2Lambda}) this relation is specified, and we can therefore extract 
\begin{equation}
\label{tk}
t_k=\frac{(2\beta_I\cdot\beta_k)(2\beta_J\cdot\beta_k)}{\sqrt{\beta_I^2\beta_J^2}}\,.
\end{equation}

\subsection{Computation of the one-loop soft~anomalous dimension}
\label{sec:one-loop_computation}

At one loop, the correlator contains colour dipoles and singlets,
\begin{align}
\label{one_loop_webs}
    \begin{split}
        \frac{\alpha_s}{4\pi}w^{(1)}=\sum_{I<J}W_{(IJ)}^{(1)}+\cdots,
    \end{split}
\end{align}
where here and below ellipsis represent all colour structures that are not relevant for our discussion. Here they simply correspond to kinematic-independent singlet terms, which we ignore (these do not contribute to the computation of $\Omega_{(IJk)}$ below). The web $W_{(IJ)}^{(1)}$ contains both a colour factor and a kinematic function (see definition in section~\ref{sec:RegularizeCorrelator}):
\begin{align} W_{(IJ)}^{(1)}=\mathbf{T}_I\cdot\mathbf{T}_J{\Y}_{(IJ)}\,.
\end{align}
The subscript $(IJ)$ identifies the web configuration where the Wilson lines $I$ and $J$ are connected by an exchanged gluon. The kinematic function ${\Y}_{(IJ)}$ is expressed in terms of Feynman integrals containing eikonal propagators defined in section~\ref{sec:Regularised_WilsonLines}. Using the Feynman gauge we now write down the kinematic functions for the relevant webs that connects all pairs out of the three lines~$I$, $J$ and $K$,
\begin{subequations}
\begin{align}
\begin{split}
 \label{oneloopt}
{\Y}_{(IJ)}=&\,\,\vcenter{\hbox{\includegraphics[width=1.5cm]{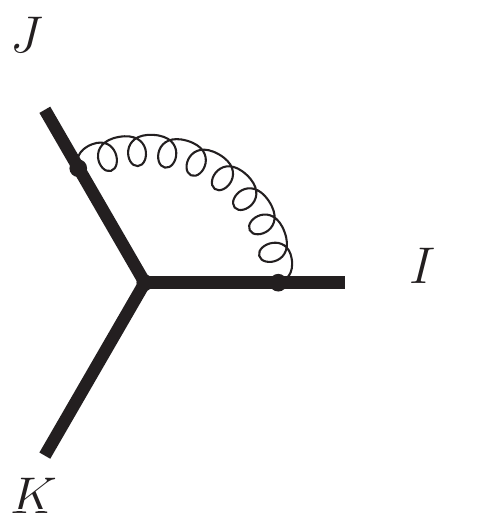}}}=\,\, \frac{\alpha_s}{4\pi} {\cal N} v_I\cdot v_J\int\left[{\cal D}q\right]Y_{(IJ)}(q)\,,
  \end{split}
 \\
\begin{split}
{\Y}_{(JK)}=&\,\,\vcenter{\hbox{\includegraphics[width=1.5cm]{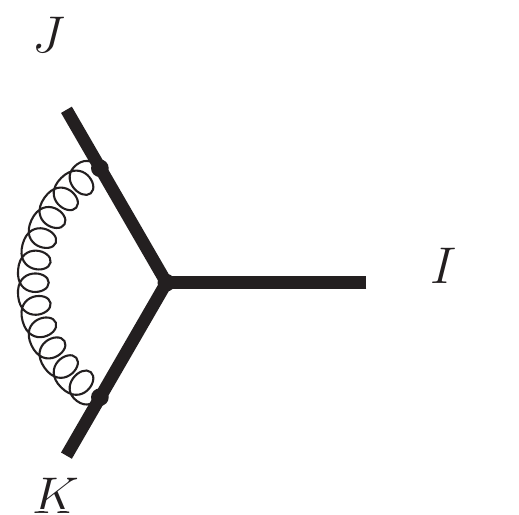}}}  =\,\,\frac{\alpha_s}{4\pi} {\cal N}v_J\cdot v_K\int\left[{\cal D}q\right]Y_{(JK)}(q)\,,
  \end{split}
\\
\begin{split}
{\Y}_{(IK)}=&\,\,\vcenter{\hbox{\includegraphics[width=1.5cm]{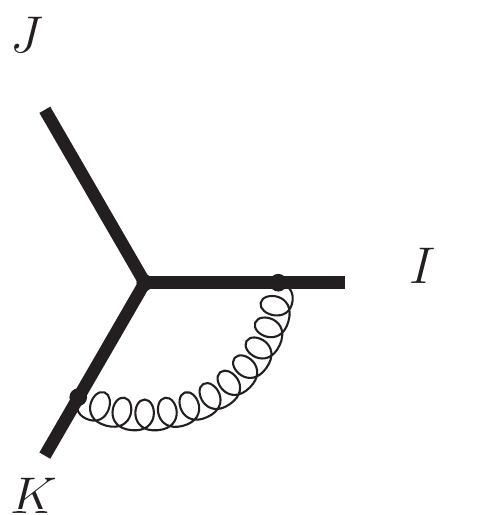}}} =\,\,\frac{\alpha_s}{4\pi} {\cal N} v_I\cdot v_K\int\left[{\cal D}q\right]Y_{(IK)}(q)\,,
  \end{split}
 \end{align}
 \end{subequations}
 where the normalization factor ${\cal N}$ and the one-loop integration measure $\left[{\cal D}q\right]$ are defined by
     \begin{align}
     \label{calNandDq}
     \begin{split}
         {\cal N}\equiv\left(\frac{4\pi \mu^2}{m^2}\right)^{\epsilon}\,,
     \qquad \quad 
       \left[ {\cal D}q\right] \equiv\frac{d^Dq}{i\pi^{\frac{D}{2}}}\,,
     \end{split}
     \end{align}
and the scalar integrand $Y_{(IJ)}$ is given by 
\begin{align}
\label{YIJdef}
         \begin{split}
            Y_{(IJ)}(q)\equiv\frac{1}{P_g}
            \frac{1}{P_I}
            \frac{1}{P_J}\,,
         \end{split}
     \end{align}
   with the propagators defined by
   $P_g=q^2+
 i\varepsilon$ and
     \begin{align}
     \label{P123}
     \begin{split} 
       P_I=\frac{1}{\tilde{E}_{v_I}\left(1,q\right)}\,=\,-v_I\cdot q-1
       +
 i\varepsilon\,,
       \qquad 
       P_J=\frac{1}{\tilde{E}_{v_J}\left(1,-q\right)}\,=\,v_J\cdot q-1
       +
 i\varepsilon\,,
     \end{split}
     \end{align}
where we used the rescaled eikonal propagator defined in eq.~(\ref{dimlessEikP}), expressed  in terms of $v^{\nu}$ defined in (\ref{v_rescalled_beta}) and in terms of the dimensionless momentum vector $q^\nu$. 
The integrands $Y_{(JK)}$ and $Y_{(IK)}$ can be obtained from eqs.~(\ref{YIJdef}) and (\ref{P123}) by permutations of the velocities. 

\begin{figure}[htb]
    \centering \includegraphics[width=0.4\linewidth]{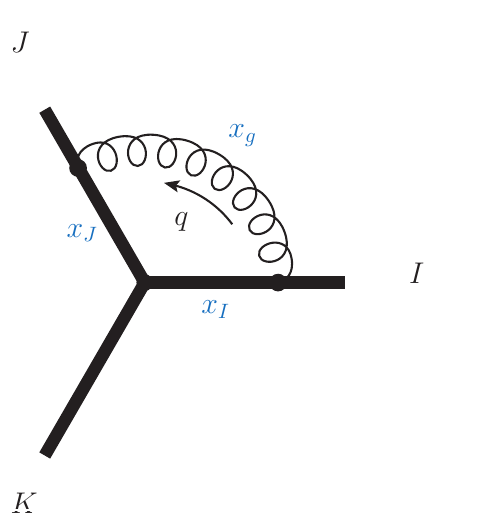}
    \caption{The propagators and the corresponding parameters for the integral ${\Y}_{(IJ)}$.}
    \label{fig:onelooppara}
\end{figure}

As discussed in section~\ref{sec:MoR}, we work in the Euclidean kinematic regime where $\alpha_{IJ}$, $\alpha_{JK}$ and $\alpha_{IK}$ are all positive. This allows us to reliably identify all regions in the MoR using the geometric algorithm in parametric space. We therefore proceed to write down the Lee-Pomeransky representation for ${\Y}_{(IJ)}$,
\begin{align}
\label{calYIJ}
\begin{split}
    {\Y}_{(IJ)}=&\,\, \frac{\alpha_s}{4\pi} {\cal N} 
    \frac{\Gamma (2-\epsilon )}{\Gamma (1-2 \epsilon )}
   v_I\cdot v_J\int_0^{\infty}
dx_g
\int_0^{\infty}
dx_I
\int_0^{\infty}
dx_J\left[{\cal P}_{(IJ)}\right]^{\epsilon-2}.
\end{split}
\end{align}
where the Lee-Pomeransky polynomial ${\cal P}_{(IJ)}$ is 
\begin{align}
\label{calPIJ}
   {\cal P}_{(IJ)}=
\frac{1}{4} x_I x_J  \left(\alpha _{IJ}+\frac{1}{\alpha _{IJ}}\right) 
+\frac{1}{4} x_I^2 
+\frac{1}{4} x_J^2 
+x_g x_I +x_g x_J+\textcolor{red}{x_g}\,,
\end{align}
where the correspondence between the parameters $x_i$ and the propagators is defined in figure~\ref{fig:onelooppara}. The red term ($\textcolor{red}{x_g}$) in ${\cal P}_{(IJ)}$ is the first Symanzik polynomial,\footnote{The absence of the parameters $x_I$ and $x_J$ in ${\cal U}_{(IJ)}$ follows from the absence of quadratic terms in the momenta in the corresponding (eikonal) propagators. } ${\cal U}_{(IJ)}$, while the rest of the expression is the second Symanzik polynomial, ${\cal F}_{(IJ)}$. One can confirm that all monomials in the Lee-Pomeransky polynomial (\ref{calPIJ}) are non-negative in the chosen kinematic regime, so no hidden regions can arise in the MoR.  

Using the package {\tt{pySecDec}}~\cite{Heinrich:2023til,Borowka:2017idc}, we get the region vectors associated with the expansion in small $\beta_K^2$. The scaling of the external variables can be found in eqs.~\eqref{limitalpha} and~\eqref{betaK2Lambda}, so we see that the expansion in $\beta_K^2$ corresponds to an expansion in $\alpha_{IK}$ and $\alpha_{JK}$, which are both of ${\cal O}(\lambda)$ while $\alpha_{IJ}={\cal O}(\lambda^0)$ is unaffected.
As an immediate consequence, one expects that the integral ${\Y}_{(IJ)}$ in eq.~(\ref{calYIJ}) has only a hard region in the $\lambda$ expansion, while those of ${\Y}_{(JK)}$ and ${\Y}_{(IK)}$ develop a non-trivial asymptotic expansion.
This is readily confirmed by examining the set of regions vectors $\mathbf{u}^R$ generated by {\tt{pySecDec}} for these integrals, which we present in table~\ref{tab:oneloopregion}.
Our notation
for the region vectors of the integral~${\Y}_{(IJ)}$ of eq.~(\ref{calYIJ}), corresponding to the diagram in figure~\ref{fig:onelooppara}, is:  $\mathbf{u}^{R}=\{u^R_{g},u^R_{I},u^R_{J},1\}$. Following eq.~(\ref{regionVectors}), this should be read as the following rescaling operation on the integration variables in the Lee-Pomeransky polynomial of eq.~(\ref{calPIJ}):
\begin{equation}
\label{YIJOneLoopScaing}
(x_g, x_I, x_J)\to
(x_g \lambda^{u_{g}}, 
x_I\lambda^{u_I}, 
x_J\lambda^{u_J})
\end{equation}
\begin{table}
\centering
\begin{subtable}[]{0.9\textwidth}
\centering
\begin{tabular}{|c|c|c|c|}
\hline
Integrand & Region vector $\mathbf{u}^R$ 
& $n_R$ & region $R$\\
& 
$\{u^R_{g},u^R_{I},u^R_{J},1\}$ &&\\
\hline
\hline
${\Y}_{(IJ)}$ & $\{0,0,0,1\}$ & $0$ & $\left\{H\right\}$ \\
\hline
\end{tabular}
\caption{Region vector for the integral ${\Y}_{(IJ)}$}
\label{tab:IJWebRegions}
\end{subtable}

\vspace{1em}

\begin{subtable}[]{0.9\textwidth}
\centering
\begin{tabular}{|c|c|c|c|}
\hline
Integrand & Region vector $\mathbf{u}^R$ 
& $n_R$ & region $R$\\
& 
$\{u^R_{g},u^R_{J},u^R_{K},1\}$ &&\\
\hline
\hline
\multirow{3}{*}{${\Y}_{(JK)}$} & $\{0,0,1,1\}$ & $0$ & $\left\{H\right\}$ \\
& $\{0,1,0,1\}$ & $0$ & $\left\{C_{N}\right\}$ \\
& $\{-1,0,0,1\}$ & $-1$ & $\left\{C_{\IR}\right\}$\\
\hline
\end{tabular}
\caption{Region vectors for the integral ${\Y}_{(JK)}$}
\label{tab:JKWebRegions}
\end{subtable}
\caption{The region vectors $\mathbf{u}^R$ and the coefficient $n_R$ identifying the overall scaling of each region, $\lambda^{n_R\epsilon}$ (see eq.~\eqref{radieikonalSoft}) of the one-loop integrals ${\Y}_{(IJ)}$ and ${\Y}_{(JK)}$. 
The regions of the remaining integral~${\Y}_{(IK)}$ can be obtained by upon replacing $J$ by $I$ in ${\Y}_{(JK)}$. 
The rightmost column describes each region in momentum space in terms of the modes defined in table~\ref{tab:ModesSum}, referring to the scaling of the lightcone momentum components of the gluon.
}
\label{tab:oneloopregion}
\end{table}

For ${\Y}_{(JK)}$ we use the same general convention, applying to ${\Y}_{(IJ)}$ the permutation $I\to J$ and $J\to K$ simultaneously.\footnote{We shall not discuss ${\Y}_{(IK)}$ explicitly: due to the $(I,J)$ symmetry, it can be obtained from the analysis of ${\Y}_{(JK)}$ below by replacing $J$ by $I$.}  
The scaling of the parameters of ${\Y}_{(JK)}$ in the small $\beta_K^2$ limit in a given region $R$ should therefore be read as follows:
\begin{equation}
(x_g, x_J, x_K)
\xrightarrow{\;{R}\;}
(x_g \lambda^{u_g^R}, 
x_J\lambda^{u_J^R}, 
x_K\lambda^{u_K^R})\,.
\end{equation}
The three region vectors of ${\Y}_{(JK)}$
in table~\ref{tab:JKWebRegions}
amounts to the rescaling: 
\begin{subequations}
\label{xscalingOneloop}
\begin{align}
(x_g, x_J, x_K)&\,\xrightarrow{\;{\{H\}}\;}
(x_g , x_J , x_K \lambda)\,,
\\
(x_g, x_J, x_K)&\,\xrightarrow{\;{\{C_N\}}\;}
(x_g , x_J \lambda, x_K)\,,
\\
(x_g, x_J, x_K)&\,\xrightarrow{\;{\{C_{\text{IR}}\}}\;}
(x_g \lambda^{-1}, x_J, x_K)\,,
\end{align}
\end{subequations}
respectively. The naming convention of the three regions in Table~\ref{tab:JKWebRegions} and in eq.~(\ref{xscalingOneloop}) will be explained below (see table~\ref{tab:ModesSum}).

Next, we would like to understand the physical interpretation of the three regions.\footnote{It is important to stress that the  integrals for the separate regions can be easily computed in parameter space (see, appendix~\ref{app:oneloop} below). Our motivation to express the regions in momentum space stems from our   interest in their interpretation. } 
To this end it is useful to revert to momentum space. The basic relation between 
the scaling of the parameter $x_e$ and that of the corresponding  propagator $P_e$ can be read of eq.~\eqref{Schwinger}, as discussed\footnote{We note that the Lee-Pomeransky parameters scale in the same as the Schwinger parameters (see appendix~B in~\cite{Gardi:2022khw}). } in ref.~\cite{Gardi:2022khw}, see eq. (2.45) there:
\begin{equation}
\label{scaling_law_conversion}
    x_e\sim \bar{x}_e\sim \frac{1}{P_e} \sim \lambda^{u_e^R}\,.
\end{equation} 
So the denominator of the propagator scales \emph{opposite} to the corresponding Lee-Pomeransky parameter. We stress that this applies to the denominator of the propagator as a whole. This of course makes a difference in converting the scaling law from parameter space to momentum space for linear propagators as compared to ordinary quadratic ones.

While the scaling law of the Lee-Pomeransky parameters in a given  region $R$ is of course Lorentz invariant, the scaling of the loop momentum components is Lorentz frame dependent. We will see that it can nonetheless be determined from the parameter-space region vectors using eq.~(\ref{scaling_law_conversion}), taking into account the full set of scaling laws of the propagators in the diagram, along with momentum conservation. 
One additional assumption we will use in making the transition from parameter space to momentum space is  that the loop momentum scaling should be consistent with the on-shell condition. 
The assumption that facet regions involve only on-shell loop-momentum modes is well motivated in the case of IR regions~\cite{Gardi:2022khw,Ma:2023hrt}, although it has not been proven. 
Our assumption here extends beyond the realm of IR modes; below we state it more precisely.

For what follows let us define the loop momentum in every one-loop diagram as the momentum~$q$ carried by the gluon. The generalization to two-loop will be discussed in section~\ref{sec:two-loop_computation}. 
We will decompose the momenta 
in lightcone coordinates, \hbox{$q=(q^+,q^-,q^\perp)$}, defined by the direction of the external (nearly) lightlike momentum $\beta_{K}\sim (1,\lambda^2,\lambda)$, such that the velocity in the strict limit is
$\beta_k=(\beta_k^+,0,0)$. 
Next, recall that we have defined the propagators (\ref{P123}) using rescaled velocities~$v_J^{\mu}=\beta_J^{\mu}/\sqrt{\beta_J^2}$ according to eq.~(\ref{v_rescalled_beta}). This means that while the timelike Wilson-line velocities admit $v_{J}\sim (1,1,1)$ (and similarly $v_{I}\sim (1,1,1)$), for the nearly lightlike Wilson line we have: $v_{K}\sim (\lambda^{-1},\lambda,1)$. 

Having fixed our conventions for the parametrization of the loop momentum components \hbox{$q=(q^+,q^-,q^\perp)$}, 
we can formulate more precisely the requirement above, that the loop momentum modes should be compatible with the on-shell condition.
If we assume, as in eq.~(\ref{l_n_scaling_law}), that in a given mode in region $R$, the scaling law is 
\[
(q^+,q^-,q^\perp) \xrightarrow{\;{R}\;} 
(q^+\lambda^{a_+},q^-\lambda^{a_-},q^\perp \lambda^{a_\perp})\,,
\]
where ${a_+}$, ${a_-}$ and ${a_\perp}$ are three rational numbers, then compatibility with the on-shell condition simply implies 
\begin{equation}
\label{on-shell_compatibility_of_modes}
{a_+}+{a_-}=2{a_\perp},
\end{equation}
so that an inverse propagator, 
\[
q^2 = 2q^+q^- - |p_{\perp}|^2\xrightarrow{\;{R}\;}  q^2 \lambda^{2{a_\perp}}
\]
admits \emph{homogeneous scaling} between the product of lightcone components and squared transverse components. Clearly, eq.~(\ref{on-shell_compatibility_of_modes}) is violated for potential or Glauber modes, which are associated with hidden regions in parameter space~\cite{Jantzen:2012mw,Gardi:2024axt}.
Recall that in the context of the present work we are working in the Euclidean regime, and hence we only have facet regions. In this context we will be assuming that the condition of eq.~(\ref{on-shell_compatibility_of_modes}) holds for every loop momentum mode (including UV modes!) and will provide further evidence for the validity of this assumption through the region analysis at one and two loops.

Let us now apply the relation of eq.~(\ref{scaling_law_conversion}) to infer the scaling of the  propagators corresponding to each of the three regions 
in eq.~(\ref{xscalingOneloop}):
\begin{subequations}
\begin{align}
\label{oneloopHardMode}
\left[q^2, \quad -v_J\cdot q-1,\quad  
v_K\cdot q-1\right]&\,
\xrightarrow{\;{\{H\}}\;}
\left[q^2,\quad  -v_J\cdot q-1, \quad 
(v_K\cdot q-1)\lambda^{-1}\right]\,,
\\
\label{oneloopNeutralMode}
\left[q^2, \quad -v_J\cdot q-1, \quad  
v_K\cdot q-1\right]&\,
\xrightarrow{\;{\{C_N\}}\;}
\left[q^2, \quad  (-v_J\cdot q-1)\lambda^{-1}, \quad 
v_K\cdot q-1\right]\,,
\\
\label{oneloopCollMode}
\left[q^2, \quad  -v_J\cdot q-1, \quad 
v_K\cdot q-1\right]&\,
\xrightarrow{\;{\{C_{\text{IR}}\}}\;}
\left[q^2 \lambda,\quad   -v_J\cdot q-1, \quad 
v_K\cdot q-1\right]\,.
\end{align}
\end{subequations}
Next, let us interpret these scaling laws by inferring the scaling of the separate lightcone momentum components. To this end we need to impose that $v_J\sim(1,1,1)$ and $v_K\sim(\lambda^{-1},\lambda, 1)$. 

Let us begin by considering (\ref{oneloopHardMode}) and (\ref{oneloopNeutralMode}). We notice that in both of these regions, the gluon propagator remain  ${\cal O}(\lambda^0)$, while in each one of the eikonal propagators scales by $\lambda^{-1}$.
Let us expand the propagators in lightcone coordinates. They are given respectively by
\begin{subequations}
\label{unscaledProps}
\begin{align}
q^2&\,=2q^+q^--|q^\perp|^2\,,
\\
\label{JinLightConeComp}
v_J\cdot q-1&\,=(v_J^+,v_J^-,v_J^\perp)\cdot 
(q^+,q^-,q^\perp) -1= 
v_J^+ q^-  + v_J^- q^+ - v_J^\perp \cdot  q^\perp -1\,,
\\
\label{KinLightConeComp}
v_K\cdot q-1&\,=(v_K^+,v_K^-,v_K^\perp)\cdot 
(q^+,q^-,q^\perp) -1= 
v_K^+ q^-  + v_K^- q^+ - v_K^\perp \cdot  q^\perp -1\,.
\end{align}
\end{subequations}
Since $v_J\sim(\lambda^0, \lambda^0, \lambda^0)$ and $v_K\sim(\lambda^{-1}, \lambda, \lambda^0)$
we see that the region of 
eq.~(\ref{oneloopHardMode})
corresponds to $(q^+,q^-,q^\perp)\sim (\lambda^{0},\lambda^0, \lambda^0)$
while that of eq.~(\ref{oneloopNeutralMode}) to
$(q^+,q^-,q^\perp)\sim (\lambda^{-1},\lambda, \lambda^0)$.   

The final region to interpret in momentum space is (\ref{oneloopCollMode}). Here we see that the gluon virtulaity scales as $\lambda^1$, while the two eikonal propagators remain $\sim\lambda^0$. An on-shell mode consistent with this is
$(q^+,q^-,q^\perp)\sim (1,\lambda^1, \sqrt{\lambda})$, which we call an IR collinear mode. 

The three loop-momentum modes defining these three regions are summarized in the first three rows of table~\ref{tab:ModesSum}. We notice that while the hard ($H$) and IR collinear ($C_{\text{IR}}$)  modes are commonplace in the on-shell expansion of wide-angle scattering non-eikonal integrals~\cite{Gardi:2022khw,Ma:2023hrt}, the mode~$C_N$   is rather special, as it involves both large and small momentum components. It is a feature of the eikonal approximation. We refer to it as the  \emph{neutral} collinear mode. This mode presents a potential ambiguity in disentangling IR and UV singularities, and we will discuss it further in section~\ref{sec:RenMixedCorr}.

With the region information given in table~\ref{tab:oneloopregion}, it is straightforward to write down the asymptotic expansion for kinematic functions,
\begin{subequations}
 \label{expansoneloop}
\begin{align}
\begin{split}
{\cal T}_{\lambda}  \left[{\Y}_{(IJ)}\right]=&\sum_{R}\lambda^{n_R\epsilon}{\Y}_{(IJ)}^{R}=\,\,{\Y}_{(IJ)}^{\{H\}}=\vcenter{\hbox{\includegraphics[width=1.5cm]{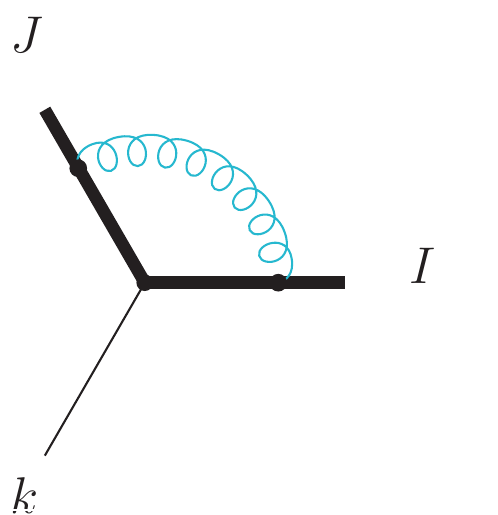}}},
\end{split}
\\
\begin{split}
  {\cal T}_{\lambda}  \left[{\Y}_{(IK)}\right]=&\sum_{R}\lambda^{n_R\epsilon}{\Y}_{(IK)}^{R}=\,\,{\Y}_{(Ik)}^{\{H\}}+{\Y}_{(Ik)}^{\{C_N\}}+\lambda^{-\epsilon}{\Y}_{(Ik)}^{\{C_{\IR}\}}
  \\=&\vcenter{\hbox{\includegraphics[width=1.5cm]{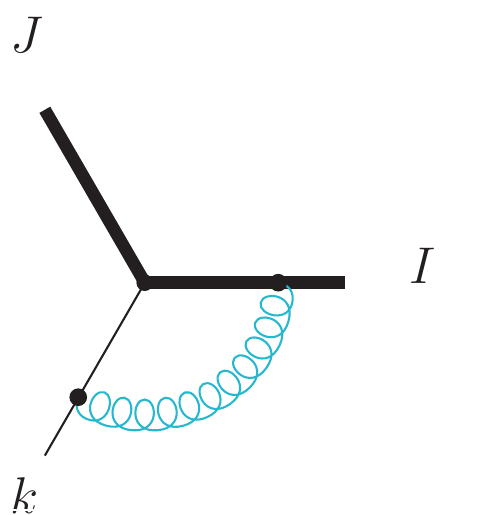}}} +\vcenter{\hbox{\includegraphics[width=1.5cm]{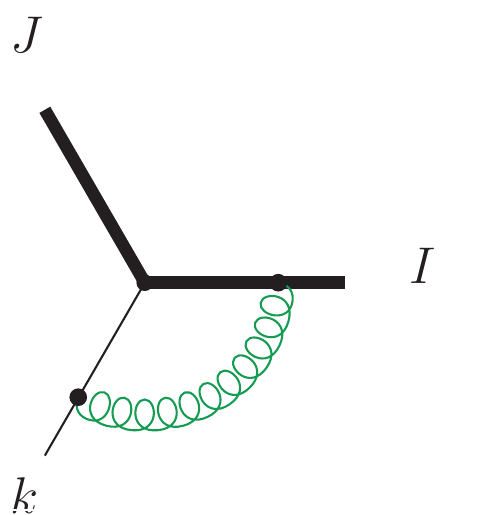}}} +\lambda^{-\epsilon}\vcenter{\hbox{\includegraphics[width=1.5cm]{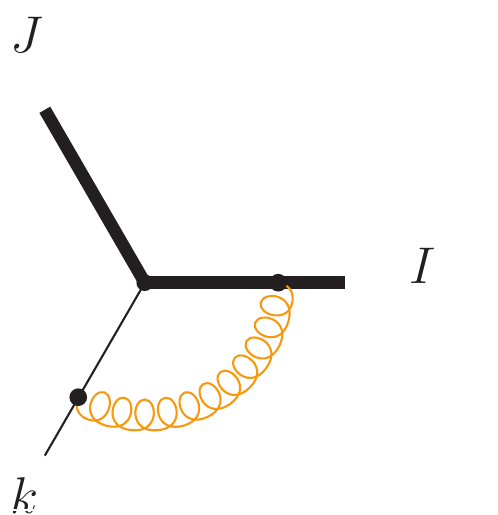}}},
  \end{split}
\\
\begin{split}
 {\cal T}_{\lambda}  \left[{\Y}_{(JK)}\right]=&\sum_{R}\lambda^{n_R\epsilon}{\Y}_{(JK)}^{R}=\,\,{\Y}_{(Jk)}^{\{H\}}+{\Y}_{(Jk)}^{\{C_N\}}+\lambda^{-\epsilon}{\Y}_{(Jk)}^{\{C_{\IR}\}}
 \\=&\vcenter{\hbox{\includegraphics[width=1.5cm]{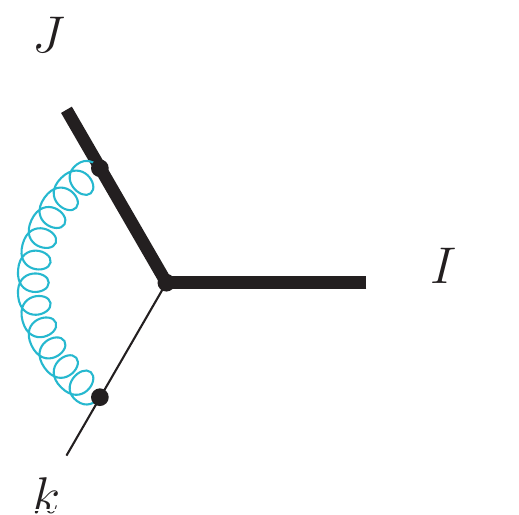}}} +\vcenter{\hbox{\includegraphics[width=1.5cm]{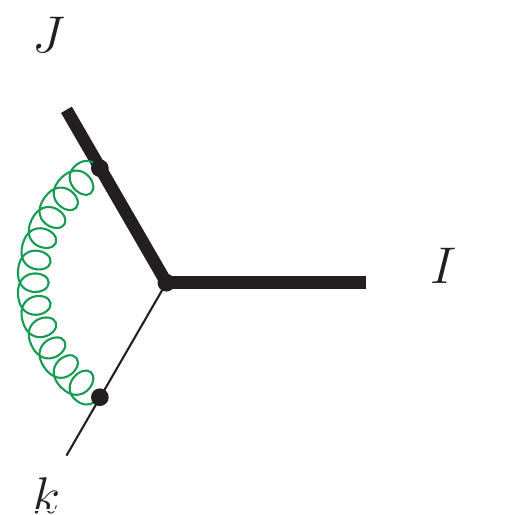}}}+\lambda^{-\epsilon}\vcenter{\hbox{\includegraphics[width=1.5cm]{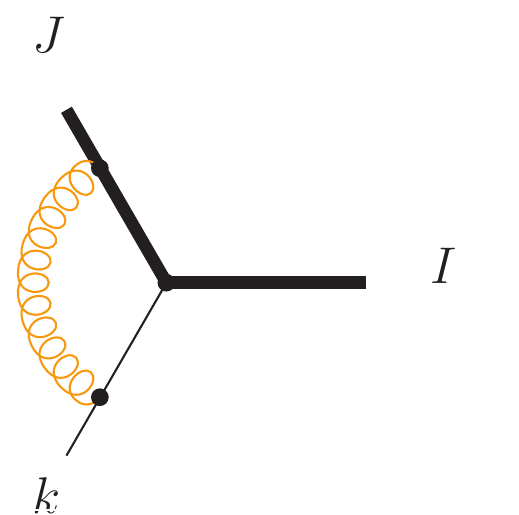}}} .
\end{split}
\end{align}
\end{subequations}
All three regions contribute at leading power in $\lambda$.

 \begin{table}
\centering
\begin{tabular}{|c|c|c|c|c|}
\hline
  Name & Notation & $n_R$ & Scaling $(q^+,q^-,q^\perp)$ & Colour Coding  \\
   \hline
   \hline
  IR Collinear & $C_{\IR}$ & $-1$& $(1,\lambda,\sqrt{\lambda})$ & \hbox{\includegraphics[width=1.5cm]{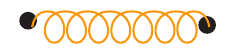}} \\
   \hline
  Neutral Collinear & $C_{N}$ & $0$ &
  $(\lambda^{-1},\lambda,1)$ & \hbox{\includegraphics[width=1.5cm]{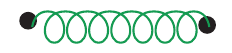}} \\
\hline
  Hard & $H$ & $0$ & $(1,1,1)$ & \hbox{\includegraphics[width=1.5cm]{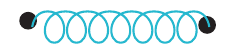}} \\
 \hline
  UV Collinear & $C_{\UV}$ & $2$ &
  $(\lambda^{-2},1,\lambda^{-1})$ & \hbox{\includegraphics[width=1.5cm]{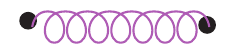}} \\
   \hline
  UV Hard & $H_{\UV}$ & $2$ &
  $(\lambda^{-1},\lambda^{-1},\lambda^{-1})$ & \hbox{\includegraphics[width=1.5cm]{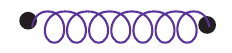}} \\
   \hline
 \end{tabular}
  \caption{Loop momentum modes in the region analysis in terms of lightcone coordinates $(q^+,q^-,q^\perp)$, where the external (nearly) lightlike line $K$ has velocity $\beta_{K}\sim(1,\lambda^2,\lambda)$, or equivalently, normalized velocity~$v_K\sim(\lambda^{-1},\lambda, 1)$. Out of the five modes listed here, only the first three -- IR and neutral modes -- appear in one-loop integrals, see table~\ref{tab:oneloopregion}, while the remaining two -- UV modes -- appear first at two loops, see tables~\ref{tab:twolooptopregionv} and~\ref{tab:121WebRegions} below.
 }
    \label{tab:ModesSum}
 \end{table}

By definition, the leading power of the hard region corresponds to the computation of the correlator with a  strictly lightlike Wilson line defined by eq.~\eqref{llWilsonLine}. We therefore single out the hard region by splitting the asymptotic expansion of the correlator (at any given loop order $n$) into two parts,
\begin{equation}
\label{correlatorsplit}
    {\cal T}_{\lambda} \left[w^{(n)}\right]=h^{(n)}+r^{(n)},
 \end{equation}
 where $h^{(n)}$ contains the hard region only, while~$r^{(n)}$ contains all the remaining regions. The contributions to $h^{(n)}$ and $r^{(n)}$ can be further decomposed by colour. 
 In $r^{(n)}$ each colour structure is multiplied by an asymptotic expansion consisting of \emph{region functions}, as in eq.~(\ref{RegionFunction}). Each such region function may in general receive contributions from different webs, i.e. different ${\Y}_{ij}^R$.
 
 At one loop, $h^{(1)}$ and $r^{(1)}$ may be  written in terms of the region functions ${\cal R}$ as follows:
\begin{subequations}
     \begin{align}
     \begin{split}
\frac{\alpha_s}{4\pi}h^{(1)}=\,\,&\sum_{I<J}\frac{\alpha_s}{4\pi}h^{(1)}_{IJ}+\sum_{I}\frac{\alpha_s}{4\pi}h^{(1)}_{Ik}+\cdots
\\=\,\,&\sum_{I<J}\mathbf{T}_{I}\cdot\mathbf{T}_{J}{\cal R}_{(IJ)}^{\{H\}}+\sum_{I}\mathbf{T}_{I}\cdot\mathbf{T}_{k}{\cal R}_{(Ik)}^{\{H\}}+\cdots,
\end{split}
\\
 \begin{split}
     \frac{\alpha_s}{4\pi}r^{(1)}=\,\,&\sum_{I}\frac{\alpha_s}{4\pi}r^{(1)}_{(Ik)}+\cdots
     \\=\,\,&\sum_{I}\mathbf{T}_{I}\cdot\mathbf{T}_{k}\left\{{\cal R}_{(Ik)}^{\{C_N\}}+\lambda^{-\epsilon}{\cal R}_{(Ik)}^{\{C_{\IR}\}}\right\}+\cdots.
 \end{split}
 \end{align}
 \end{subequations}
  Note that only the colour structures involving the lightlike line $k$ will enter $r^{(1)}$, because others will not be influenced by the expansion, so they are fully contained in $h^{(1)}$. While in the multi-loop context the region functions ${\cal R}^R$ with a given colour structure are combinations of several kinematic functions~${\Y}_{ij}^R$, at one loop, there is just one web (in fact, one diagram), hence one function ${\Y}^R$, contributing to each region function~${\cal R}^R$, as follows:
 \begin{align}
 \label{oneloopRegions}
 \begin{split}
 {\cal R}_{(IJ)}^{\{H\}}\,&= {\Y}_{(IJ)}^{\{H\}}=\vcenter{\hbox{\includegraphics[width=1.5cm]{fig/11IJH.pdf}}},
 \qquad
 {\cal R}_{(Ik)}^{\{H\}}= {\Y}_{(Ik)}^{\{H\}}=\vcenter{\hbox{\includegraphics[width=1.5cm]{fig/11IkH.pdf}}},
 \qquad 
  {\cal R}_{(Jk)}^{\{H\}}= {\Y}_{(Jk)}^{\{H\}}=\vcenter{\hbox{\includegraphics[width=1.5cm]{fig/11JkH.pdf}}},
 \\
     {\cal R}_{(Ik)}^{\{C_N\}}&\,=  {\Y}_{(Ik)}^{\{C_N\}}=\vcenter{\hbox{\includegraphics[width=1.5cm]{fig/11IKCN.pdf}}},
 \qquad
    {\cal R}_{(Jk)}^{\{C_N\}}={\Y}_{(Jk)}^{\{C_N\}}=\vcenter{\hbox{\includegraphics[width=1.5cm]{fig/11JKHCN.pdf}}},
 \\
     {\cal R}_{(Ik)}^{\{C_{\IR}\}}&\,={\Y}_{(Ik)}^{\{C_{\IR}\}}=\vcenter{\hbox{\includegraphics[width=1.5cm]{fig/11IKCIR.pdf}}},
\qquad
     {\cal R}_{(Jk)}^{\{C_{\IR}\}}= {\Y}_{(Jk)}^{\{C_{\IR}\}}=\vcenter{\hbox{\includegraphics[width=1.5cm]{fig/11JKHCIR.pdf}}}.
    \end{split}
 \end{align}
The invariant functions $\mathbb{R}^{(n_R)}$ are then simply the sum of the region functions ${\cal R}^R$ having a common $n_R$, namely 
\begin{align}
    \begin{split}
        \mathbb{R}_{(IJ)}^{(0)}=\,\,&{\cal R}_{(IJ)}^{\{H\}}\,,
        \\
        \mathbb{R}_{(JK)}^{(0)}=\,\,&{\cal R}_{(JK)}^{\{H\}}+{\cal R}_{(JK)}^{\{C_N\}}\,,
        \qquad
        \mathbb{R}_{(JK)}^{(-1)}={\cal R}_{(JK)}^{\{C_{\IR}\}}\,,
        \\
        \mathbb{R}_{(IK)}^{(0)}=\,\,&{\cal R}_{(IK)}^{\{H\}}+{\cal R}_{(IK)}^{\{C_N\}}\,,
        \qquad
        \mathbb{R}_{(IK)}^{(-1)}={\cal R}_{(IK)}^{\{C_{\IR}\}}\,.
    \end{split}
\end{align}

The one-loop region functions at leading order in $\lambda$ and through finite terms in $\epsilon$ are summarized below. Detailed computations can be found in appendix~\ref{app:oneloop}, where explicit results are provided through ${\cal O}(\epsilon^1)$. Keeping the  ${\cal O}(\epsilon^1)$ contribution at one loop is necessary, since  they contribute to the renormalization of the two-loop functions.

We note that all three regions contributing to the $(Jk)$ web (and similarly to the $(Ik)$ web), namely the hard, the neutral collinear and the IR collinear regions, have double poles in $\epsilon$. We shall see that these cancel in the sum of regions, as they must do to reproduce the expansion of the fully-timelike web $(JK)$.

It is also interesting to note that the neutral region functions, ${\cal R}_{(Jk)}^{\{C_N\}}$ and ${\cal R}_{(Ik)}^{\{C_N\}}$, are independent with any kinematic variables; see eq.~\eqref{oneloopNeutC}. This observation will become important in  section~\ref{sec:RenMixedCorr} below, where we will discuss the IR versus UV origin of the various contributions, referring to the special role of the neutral collinear mode. 

The contributions of the hard region to the one-loop correlator with a strictly lightlike like $k$ and two timelike lines $I$ and~$J$ is: 
\begin{subequations}
\label{onelooph}
\begin{align}
\begin{split}
h^{(1)}_{(IJ)}=\,\,&\textbf{T}_I\cdot\textbf{T}_J\left(\frac{\bar{m}^2}{\mu^2}\right)^{-\epsilon}\frac{1+\alpha _{IJ}^2}{1-\alpha _{IJ}^2}\bigg[-\frac{2}{\epsilon}\log(\alpha_{IJ})+V_1(\alpha_{IJ})+{\cal O}(\epsilon)+{\cal O}(\lambda)\bigg],
\end{split}
 \\
\begin{split}
h^{(1)}_{(Ik)}=\,\,&\textbf{T}_I\cdot\textbf{T}_k\left(\frac{\bar{m}^2}{\mu^2}\right)^{-\epsilon}\left[-\frac{1}{\epsilon^2}-\frac{5 \pi ^2}{12}+{\cal O}(\epsilon)+{\cal O}(\lambda)\right],
\end{split}
 \\
\begin{split}
h^{(1)}_{(Jk)}=\,\,&\textbf{T}_J\cdot\textbf{T}_k\left(\frac{\bar{m}^2}{\mu^2}\right)^{-\epsilon}\left[-\frac{1}{\epsilon^2}-\frac{5 \pi ^2}{12}+{\cal O}(\epsilon)+{\cal O}(\lambda)\right],
\end{split}
\end{align}
\end{subequations}
where 
$V_1(\alpha)=-4 \text{Li}_2(-\alpha)+\log ^2(\alpha)-4 \log (\alpha+1) \log (\alpha)-\frac{\pi ^2}{3}$ (an equivalent expression in terms of Goncharov polylogs is provided in appendix~\ref{app:convfunc}). The scale $\bar{m}$ is defined as 
\begin{align}
    \bar{m}\equiv m\sqrt{\frac{ e^{\gamma_{\text{E}}}}{\pi}}\,.
\end{align}
Similarly, the contributions of the remaining (non-hard) regions to the one-loop correlator with a lightlike like $k$ and two timelike lines $I$ and $J$ are
\begin{subequations}
 \label{oneloopr}
 \begin{align}
\begin{split} 
\label{onelooprIk}
r^{(1)}_{(Ik)}=\,\,&\textbf{T}_I\cdot\textbf{T}_k\left(\frac{\bar{m}^2}{\mu^2}\right)^{-\epsilon}\bigg\{\frac{1}{\epsilon^2}-\frac{1}{\epsilon}\log\left(\frac{\lambda^2}{y_{IJk}}\right)+\frac{1}{4}\left[\log^2\left(\frac{\lambda^2}{y_{IJk}}\right)+\frac{\pi^2}{3}\right]+{\cal O}(\epsilon)
\\
&\,\hspace*{100pt}+{\cal O}(\lambda)\bigg\},
    \end{split}
 \\
\begin{split} 
\label{onelooprJk}
r^{(1)}_{(Jk)}=\,\,&\textbf{T}_J\cdot\textbf{T}_k\left(\frac{\bar{m}^2}{\mu^2}\right)^{-\epsilon}\bigg\{\frac{1}{\epsilon^2}-\frac{1}{\epsilon}\log\left(\lambda^2y_{IJk}\right)+\frac{1}{4}\left[\log^2\left(\lambda^2y_{IJk}\right)+\frac{\pi^2}{3}\right]+{\cal O}(\epsilon)
\\
&\,\hspace*{100pt}+{\cal O}(\lambda)\bigg\},
    \end{split}
 \end{align}
 \end{subequations}
As one expects, the IR region contributions include logarithms of the expasnion parameter. In the way we performed the asymptotic expansion (see eq.~\eqref{limitalpha}), the variables $\frac{\lambda^2}{y_{IJk}}$ and $\lambda^2y_{IJk}$ in eqs.~\eqref{onelooprIk} and~\eqref{onelooprJk} are, respectively,  $\alpha_{IK}^{2}$ and $\alpha_{JK}^{2}$, which are considered infinitesimally small in the limit. 

Next we note that all double poles in $\epsilon$ cancel, for each colour structure,  between the hard region, $h^{(1)}$, and the other regions, $r^{(1)}$. 
Upon summing up all the regions, we get the correlator in the asymptotic expansion,
\begin{align}
\begin{split}
\label{oneloopw}
   {\cal T}_{\lambda} \left[w^{(1)}\right]=\,\,&\sum_{I<J}\textbf{T}_I\cdot\textbf{T}_J\left(\frac{\bar{m}^2}{\mu^2}\right)^{-\epsilon}\frac{1+\alpha _{IJ}^2}{1-\alpha _{IJ}^2}\bigg[-\frac{2}{\epsilon}\log(\alpha_{IJ})+V_1(\alpha_{IJ})+{\cal O}(\epsilon)+{\cal O}(\lambda)\bigg]
   \\&+\sum_{I}\textbf{T}_I\cdot\textbf{T}_k\left(\frac{\bar{m}^2}{\mu^2}\right)^{-\epsilon}\bigg\{\frac{1}{\epsilon}\log\left(\frac{(2\beta_I\cdot\beta_k)^2}{t_k\lambda^2\beta_I^2}\right)
   \\&+\frac{1}{4}\log^2\left(\frac{(2\beta_I\cdot\beta_k)^2}{t_k\lambda^2\beta_I^2}\right)-\frac{\pi^2}{3}+{\cal O}(\epsilon)+{\cal O}(\lambda)\bigg\}
   \\&+\cdots\,,
   \end{split}
 \end{align}
where we have restored the coefficient $t_k$ of eq.~(\ref{tk}).

 To obtain the soft~AD, we write explicitly the terms in the Laurent expansion in $\epsilon$ for each order of $h^{(n)}$ and $r^{(n)}$ as follows, 
\begin{subequations}
\label{singreg}
    \begin{align}
h^{(n)}=\,\,&\sum_{l}h^{(n,l)}\epsilon^l=\sing\left({h}^{(n)}\right)
+\reg\left({h}^{(n)}\right),
    \\
r^{(n)}=\,\,&\sum_{l}r^{(n,l)}\epsilon^l=\sing\left({r}^{(n)}\right)
+\reg\left({r}^{(n)}\right),
\end{align}
\end{subequations}
where we defined the singular (sometimes called principal part)  and regular part operators acting on the Laurent expansion of a function $f(\epsilon)$ such that 
\begin{align}
\label{defineSingReg}
\sing(f(\epsilon))=\sum_{l\leq -1}f^{(l)}\epsilon^l,\qquad 
\reg(f(\epsilon))=\sum_{l\geq 0}f^{(l)}\epsilon^l\,,
\end{align}
and $\sing(f(\epsilon))+\reg(f(\epsilon))=f(\epsilon)$. 
Finally, using eq.~\eqref{twoloopGWeb}, the soft~ADs are given by,
\begin{align}
\label{Gamma1sr}
\begin{split}
 \frac{1}{\epsilon}\Gamma_{\UV}^{(1)}=\frac{1}{\epsilon}{\cal T}_{\lambda}\left[-2w^{(1,-1)}\right]=\,\,&-\frac{2}{\epsilon^2}\left(h^{(1,-2)}+r^{(1,-2)}\right)-\frac{2}{\epsilon}\left(h^{(1,-1)}+r^{(1,-1)}\right)
\\=\,\,&-2\sing\left(h^{(1)}+r^{(1)}\right)\,.
 \end{split}
\end{align}\\
Upon substituting in the functions, the double pole vanishes owing the an exact cancellation between the three regions, and the soft~AD of eq.~\eqref{GammaIk} is exactly reproduced. 

\subsection{Computation of the two-loop two-mass tripole }
\label{sec:two-loop_computation}

Starting from two loops, colour tripoles contribute to correlators of Wilson lines, and hence to the soft~AD, provided that at least two of the particles are massive. In what follows we compute the two-loop correlator of three Wilson lines in the limit where one of them becomes lightlike. We will use the MoR, and follow the same steps we took above at one loop, focusing now exclusively to the tripoles  contributions~$\mathbf{T}_{IJK}$ in the limit $\beta_K^2\to 0$.

The correlator 
of a product of any number of timelike Wilson lines
at two loops takes the form 
\begin{align}
\label{Tripole_webs}
\begin{split}
   \left( \frac{\alpha_s}{4\pi}\right)^2 w^{(2)}=\,\,&\sum_{I<J<K}\left[W_{(IJK)}^{(2)}+W_{(IJ)(IK)}^{(2)}+W_{(JK)(JI)}^{(2)}+W_{(KI)(KJ)}^{(2)}\right]+\cdots
   \\
=\,\,&\sum_{I<J<K}\mathbf{T}_{IJK}\, \left[{\Y}_{(IJK)}^{(2)}+{\Y}_{(IJ)(IK)}^{(2)}+{\Y}_{(JK)(JI)}^{(2)}+{\Y}_{(KI)(KJ)}^{(2)}\right]+\cdots,
\end{split}
\end{align}
where we identified the four webs that contribute with the colour structure $\mathbf{T}_{(IJK)}$, and where ${\Y}$ represent the corresponding kinematically dependent function, as discussed in section~\ref{sec:RegularizeCorrelator}. The most complicated of these is ${\Y}_{(IJK)}$, where the three Wilson lines $I$, $J$ and $K$ are connected by a three-gluon vertex. 

\subsubsection{The fully connected two-loop tripole web}
\label{sec:connectedweb}

The fully-connected tripole web was computed in refs.~\cite{Ferroglia:2009ii,Ferroglia:2009ep} (see also~\cite{Mitov:2009sv}) for the general case of timelike Wilson lines, as part of the computation of the soft~AD quoted in eq.~(\ref{GammaIJK}) above. Here we proceed to compute it using the MoR, expanding in~$\beta_K^2$.

The kinematic function of this fully-connected web is 
\begin{align}
\label{Tripole_connected_web}
\begin{split}
{\Y}_{(IJK)}=&\vcenter{\hbox{\includegraphics[width=1.5cm]{fig/111IJK.pdf}}}=\,\,\left( \frac{\alpha_s}{4\pi}\right)^2{\cal N}^2v_I^{\mu}v_J^{\nu}v_K^{\rho}\int\left[{\cal D}q\right]^2L_{\mu\nu\rho}(q_I,q_J,q_K)\times Y_{(IJK)}(q_I,q_J,q_K),
  \end{split}
 \end{align}
where the two-loop measure $\left[{\cal D}q\right]^2$ and the numerator $L_{\mu\nu\rho}$ related to the three-gluon vertex are    
\begin{align}
    \label{twoloopmeasure}
         \left[{\cal D}q\right]^2\equiv&\,\,\left(\frac{1}{i\pi^{\frac{D}{2}}}\right)^2
         \, d^Dq_I \, d^Dq_J\, d^Dq_K\, \delta^D(q_I+q_J+q_K),
    \\
      L_{\mu\nu\rho}(q_I,q_J,q_K)\equiv& \,\,g_{\mu\nu}\left(q_I-q_J\right)_{\rho}+g_{\nu\rho}\left(q_J-q_K\right)_{\mu}+g_{\rho\mu}\left(q_K-q_I\right)_{\nu}.
  \end{align}
  The scalar integrand $Y_{(IJK)}$ is defined by
  \begin{align}
  \label{connIntegrand}
  \begin{split}
     Y_{(IJK)}(k_I,k_J,k_K)\equiv&\,\, \frac{1}{P_{g_I}}\frac{1}{P_{g_J}}\frac{1}{P_{g_K}}\frac{1}{P_{I}}\frac{1}{P_{J}}\frac{1}{P_{K}},
  \end{split}
    \end{align}
with the six propagators: 
\begin{align}
\label{3gprop}
  \begin{split}
     P_{g_I}=&\,\,q_I^2+i\varepsilon\,,
     \qquad
      P_{I}=\frac{1}{\tilde{E}_{v_I}(1,q_I)}=-v_I\cdot q_I-1+i\varepsilon\,,
      \\
     P_{g_J}=&\,\,q_J^2+i\varepsilon\,,
     \qquad
      P_{J}=\frac{1}{\tilde{E}_{v_J}(1,q_J)}=-v_J\cdot q_J-1+i\varepsilon\,,
      \\
     P_{g_K}=&\,\,q_K^2+i\varepsilon\,,
     \qquad
     P_{K}=\frac{1}{\tilde{E}_{v_K}(1,q_K)}=-v_K\cdot q_K-1+i\varepsilon\,.
  \end{split}
    \end{align}
From these definitions, it is straightforward to check that the scalar integrand $Y_{(IJK)}$ is symmetric under the interchange of any pair of the of external lines $I$, $J$ and $K$, while the numerator $v_I^{\mu}v_J^{\nu}v_K^{\rho}L_{\mu\nu\rho}(q_I,q_J,q_K)$ is totally antisymmetric with respect to such an interchange. 
Thus, the kinematic function ${\Y}_{(IJK)}$ is consistent with the antisymmetry of the colour factor $\mathbf{T}_{(IJK)}$, and the Bose-symmetric nature of the web $W_{(IJK)}$ as a whole.

\begin{figure}
    \centering \includegraphics[width=0.4\linewidth]{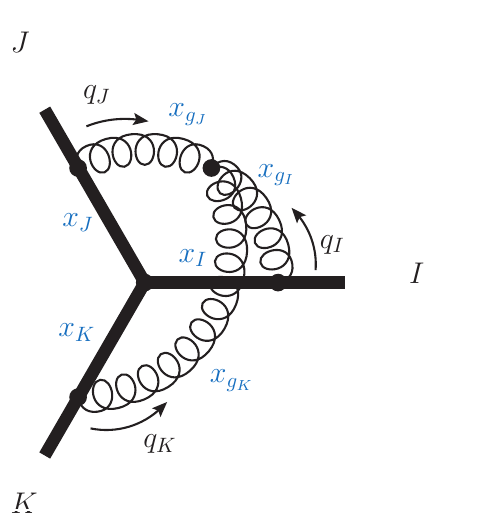}
    \caption{The propagators and the corresponding parameters for the integral $\tilde{{\Y}}_{(IJK)}$.}
    \label{fig:twolooptoppara}
\end{figure}

Because the numerator $L_{\mu\nu\rho}(q_I,q_J,q_K)$ depends on loop momenta, the kinematic function ${\Y}_{(IJK)}$ is a combination of several scalar integrals, written in Lee-Pomeransky representation. However, the full set of regions is determined by the denominator, while the numerator only suppresses or enhances certain regions. To figure out the region structures, we analyze the following scalar integral which contains all propagators in eq.~(\ref{3gprop}) and is represented by a single Lee-Pomeransky integrand
\begin{align}
\label{ScalarTripole}
    \tilde{{\Y}}_{(IJK)}\equiv\,\,\int\left[{\cal D}q\right]^2Y_{(IJK)}(q_I,q_J,q_K).
\end{align}
The Lee-Pomeransky representation of this integral is 
\begin{align}
\tilde{{\Y}}_{(IJK)}=\frac{\Gamma (2-\epsilon )}{\Gamma (-3 \epsilon )} \int_0^{\infty}dx_{g_I}\int_0^{\infty}dx_{g_J}\int_0^{\infty}dx_{g_K}\int_0^{\infty}dx_{I}\int_0^{\infty}dx_{J}\int_0^{\infty}dx_{K} \left[{\cal P}_{(IJK)}\right]^{\epsilon-2} 
\end{align}
with its polynomial ${\cal P}_{(IJK)}$ given by
\begin{align}
\label{TripoleGraphLeePomPoly}
    \begin{split}
{\cal P}_{(IJK)}=\,&
        \frac{x_{g_K}}{4} x_{J} x_{I} \left(\alpha _{IJ}+\frac{1}{ \alpha _{IJ}}\right)+\frac{x_{g_J}}{4}  x_{K} x_{I} \left(\alpha _{IK}+\frac{1}{\alpha _{IK}}\right)+\frac{x_{g_I}}{4}  x_{J} x_{K} \left(\alpha _{JK}+\frac{1}{\alpha _{JK}}\right)
    \\&+\frac{x_{g_K}}{4}  \left(x_{I}^2+x_{J}^2\right)+\frac{x_{g_J}}{4} \left(x_{I}^2+x_{K}^2\right)+\frac{x_{g_I}}{4} \left(x_{J}^2+x_{K}^2\right)
    \\&+\left(x_{g_J} x_{g_K}+x_{g_I} x_{g_J}+x_{g_I} x_{g_K}\right) \left(x_{I}+x_{J}+x_{K}\right)+\textcolor{red}{x_{g_I} x_{g_J}+x_{g_K} x_{g_J}+x_{g_I} x_{g_K}}\,.
    \end{split}
\end{align}
The correspondence between the parameters and the propagators is displayed in figure~\ref{fig:twolooptoppara}. The red monomials in eq.~(\ref{TripoleGraphLeePomPoly}) originate in the first Synamzik polynomial, ${\cal U}_{(IJK)}$. These monomials only involve the parameters associated with gluon propagators, as expected. 

The region vectors of the 
small $\beta_K^2$ expansion of the scalar integral $\tilde{\Y}_{(IJK)}$ of eq.~(\ref{ScalarTripole}), and hence of the original connected web ${\Y}_{(IJK)}$ of eq.~(\ref{Tripole_connected_web}), computed using {\tt{pySecDec}}, are collected in table~\ref{tab:twolooptopregionv}. The components of the region vector $\mathbf{u}^R$ follow the order, 
\begin{align}
\mathbf{u}^R=
\{u_{g_I}^R, u_{g_J}^R, u_{g_K}^R, u_{I}^R, u_{J}^R, u_{K}^R, 1\}\,.
\end{align}
This vector defines the scaling law or the Lee-Pomeransky parameters in region $R$ as follows,
\begin{align}
 \{x_{g_I}, x_{g_J}, x_{g_K}, x_{I}, x_{J}, x_{K}\}
 \xrightarrow{\;{R}\;}\{x_{g_I}\lambda^{u_{g_I}^R}, x_{g_J}\lambda^{u_{g_J}^R}, x_{g_K}\lambda^{u_{g_K}^R},x_{I}\lambda^{u_{I}^R}, x_{J}\lambda^{u_{J}^R}, x_{K}\lambda^{u_{K}^R}\}\,.
\end{align}
We note that these regions consist of five modes, which are summarized in table~\ref{tab:ModesSum}. Two of the modes appear first at two loops, while the remaining three have already been encountered at one loop (table~\ref{tab:oneloopregion}).  

In  appendices~\ref{app:2loopIR},~\ref{app:2loopNe} and~\ref{app:2loopUV}, we provide some details regarding the computation and the result of each region, after expansion to leading order in $\lambda$. The most complicated integral is the one corresponding to the hard region. To evaluate this integral we used the method of differential equations, which is setup in appendix~\ref{app:comp111webHH}. The remaining region integrals were performed directly in parameter space. 

\begin{table}
\centering
\begin{tabular}{|c|c|c|c|}
\hline
Integrand & Region vector $\mathbf{u}^R$ 
& $n_R$ & Region $R$\\
&
$\{u_{g_I}^R, u_{g_J}^R, u_{g_K}^R, u_{I}^R, u_{J}^R, u_{K}^R, 1\}$&&$\left[\text{mode}_{q_I},\text{mode}_{q_J},\text{mode}_{q_K}\right]$\\
\hline
\hline
\multirow{11}{*}{$\tilde{{\Y}}_{(IJK)}$} & $\{-1,-1,-1,0,0,0,1\}$ & $-1-1=-2$ & $\left[C_{\IR},C_{\IR},*\right]$\\
& $\{-1,0,0,0,1,0,1\}$ & $-1+0=-1$ & $\left[C_{\IR},C_{N},*\right]$\\
& $\{0,-1,0,1,0,0,1\}$ & $0-1=-1$ & $\left[C_{N},C_{\IR},*\right]$\\
& $\{0,0,-1,0,0,0,1\}$ & $0-1=-1$ & $\left[H,*,C_{\IR}\right]$\\
& $\{0,0,0,0,0,1,1\}$ & $0+0=0$ & $\left[H,H,*\right]$\\
& $\{0,0,0,1,1,0,1\}$ & $0+0=0$ & $\left[C_{N},C_{N},*\right]$\\
& $\{0,1,0,0,1,0,1\}$ & $0+0=0$ & $\left[H,*,C_{N}\right]$\\
& $\{1,0,0,1,0,0,1\}$ & $0+0=0$ & $\left[*,H,C_{N}\right]$\\
& $\{0,2,2,0,2,1,1\}$ & $0+2=2$ & $\left[H,*,C_{\UV}\right]$\\
& $\{2,0,2,2,0,1,1\}$ & $0+2=2$ & $\left[*,H,C_{\UV}\right]$\\
& $\{2,2,0,1,1,0,1\}$ & $2+0=2$ & $\left[H_{\UV},*,C_{N}\right]$\\
\hline
\end{tabular}
\caption{
Summary of the region analysis of 
the two-loop connected web  integral $\tilde{{\Y}}_{(IJK)}$. The second column shows the 
region vectors $\mathbf{u}^R$; the third displays  the coefficient $n_R$ identifying the overall scaling of each region, $\lambda^{n_R\epsilon}$ as a sum of the $n_R$ contributions of the two modes (see~eq.~(\ref{TransverseScaling_and_nR})); the fourth column displays the loop modes associated with two of the three gluon propagators in figure~\ref{fig:twolooptoppara}, defining the region in momentum space, as explained in the text. See Table~\ref{tab:ModesSum} for the definition of each mode. 
}
\label{tab:twolooptopregionv}
\end{table}

\begin{figure}
    \centering \includegraphics[width=0.4\linewidth]{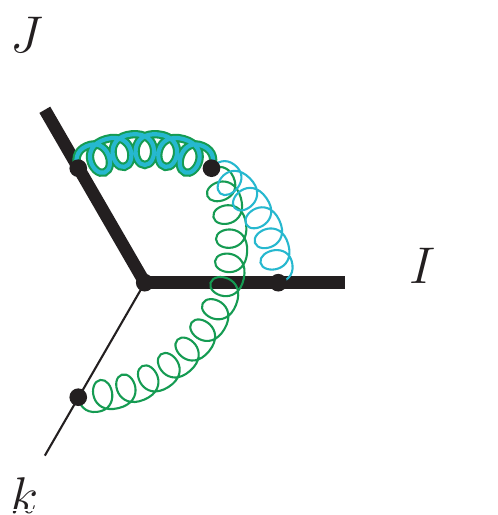}
    \caption{The diagram corresponding to region $[H,*,C_N]$. The two-fold line \hbox{\includegraphics[width=1.5cm]{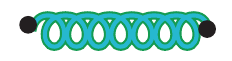}} represents simply the addition (owing to momentum conservation, $q_J=-q_I-q_K$) of the momentum components of the hard mode $H$ and the neutral collinear mode $C_{N}$. The scaling of such a gluon is $(\lambda^{-1},1,1)$ in lightcone coordinates.}
    \label{fig:RegionCNH}
\end{figure}

Before presenting the result, we would like to discuss the interpretation of the regions in momentum space, along the lines of the one-loop analysis in section~\ref{sec:one-loop_computation}. Considering the momentum representation in eq.~(\ref{Tripole_connected_web}) we note that the Dirac $\delta$ function in eq.~\eqref{twoloopmeasure} allows us to readily integrate any one of three momenta $\{q_I,q_J,q_K\}$, thus choosing the remaining two to serve as the two independent loop momenta, where the one we integrated over is fixed by  momentum conservation at the three-gluon vertex. However, in contrast to the integral before the region expansion, the choice of the two independent loop momenta in a given region matters. The principle in choosing the two independent momenta is that they should provide the complete  information regarding the modes. We therefore introduce suitable notation to encode which  two loop momenta are chosen to describe the modes in each region in the form of ordered brackets with three slots, corresponding to the three gluon propagators, following the order $\left[\text{mode}_{q_I},\text{mode}_{q_J},\text{mode}_{q_K}\right]$.
The entry with an asterisk   identifies  the propagator whose momentum is determined using momentum conservation.
This notation is used in 
table~\ref{tab:twolooptopregionv}, which lists all the regions of 
the (scalar) kinematic function is then $\tilde{\Y}_{(IJk)}^{\left[\text{mode}_{q_I},\text{mode}_{q_J},\text{mode}_{q_K}\right]}$, where one of the modes is replaced by an asterisk. 

As an example, consider the region vector $\{0,1,0,0,1,0,1\}$ in table~\ref{tab:twolooptopregionv}, which corresponds to the region $[H,*,C_N]$ shown in figure~\ref{fig:RegionCNH}. The scaling law of the two independent loop momenta in this region is
\begin{align}
\label{qIK_inCNH}
q_{I}\sim (1,1,1)\,
\,,
\qquad
q_{K}\sim (\lambda^{-1},\lambda,1)\,.
\end{align}
while 
\begin{align}
q_{J}=-q_{I}-q_{K} \sim (\lambda^{-1},1,1)
\end{align}
is determined from  by momentum conservation.
Note that if one attempt to describe this region instead by regarding $q_J$ and $q_K$ as the two independent loop momenta, then 
$q_I$ can be consistent with the hard mode scaling law, only if there is a finely-tuned cancellation between the positive components of the two momenta, 
\begin{align}
\label{extrareq}
    q_I^+=-q_J^+-q_K^+\sim O(\lambda^{-1})-O(\lambda^{-1})\sim O(\lambda^0).
\end{align}
Therefore, eq.~\eqref{extrareq} should be understood as an extra condition in addition the scaling law of $q_J$ and $q_K$.  A similar situation will happen if we keep $q_I$ and $q_J$. As a result, the only faithful description of the region purely in terms of a scaling law is according to eq.~(\ref{qIK_inCNH}).

Before we return to the asymptotic expansion of the original kinematic function ${\Y}_{(IJK)}$, we define, as an intermediate step, functions which sum the region integrals with the same type of modes, and thus the same $n_R$. Since $n_R$ corresponds to the sum of the respective $n_R$ values for each of the two independent loop momenta, as shown in general in eq.~(\ref{TransverseScaling_and_nR}) (and implemented   in the third column of table~\ref{tab:twolooptopregionv}), the total value of $n_R$ is the same independently of the association of modes with specific propagators, for instance, the two regions ${[H,*,C_{N}]}$ and ${[*,H,C_{N}]}$ have the same total $n_R$. We therefore introduce a notation for the unordered set of modes using curly brackets, for instance $\{H,C_{N}\}$ in the example above.

We now collect the regions into three classes according to eq.~(\ref{n_IR_classification}): 
IR regions, given by the first four rows in table~\ref{tab:twolooptopregionv},
\begin{align}
\label{111webRegions_IR}
\begin{split}
    {\Y}_{(IJk)}^{\{C_{\IR},C_{\IR}\}}\equiv\,\,&{\Y}_{(IJk)}^{[C_{\IR},C_{\IR},*]}=\vcenter{\hbox{\includegraphics[width=1.5cm]{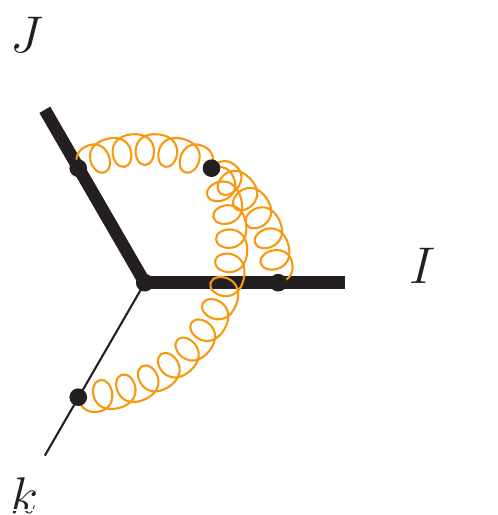}}}\,,
     \\
     {\Y}_{(IJk)}^{\{H,C_{\IR}\}}\equiv\,\,&{\Y}_{(IJk)}^{[H,*,C_{\IR}]}=\vcenter{\hbox{\includegraphics[width=1.5cm]{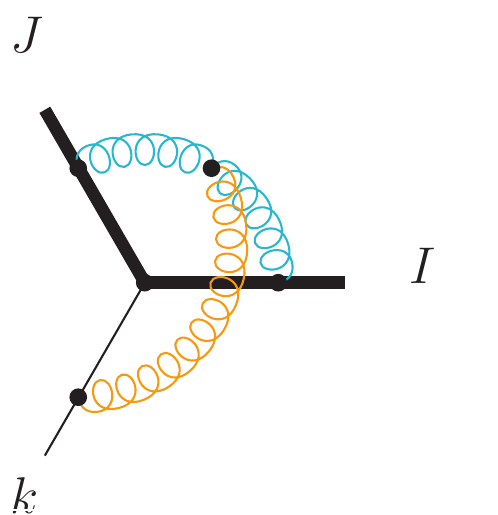}}}\,,
    \\
    {\Y}_{(IJk)}^{\{C_{N},C_{\IR}\}}\equiv\,\,&{\Y}_{(IJk)}^{[C_{\IR},C_{N},*]}+{\Y}_{(IJk)}^{[C_{N},C_{\IR},*]}=\vcenter{\hbox{\includegraphics[width=1.5cm]{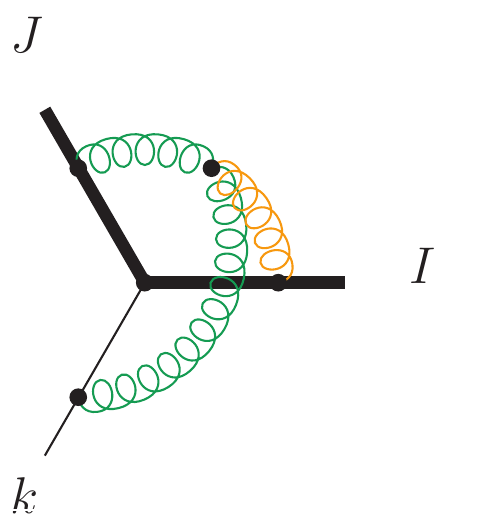}}}+\vcenter{\hbox{\includegraphics[width=1.5cm]{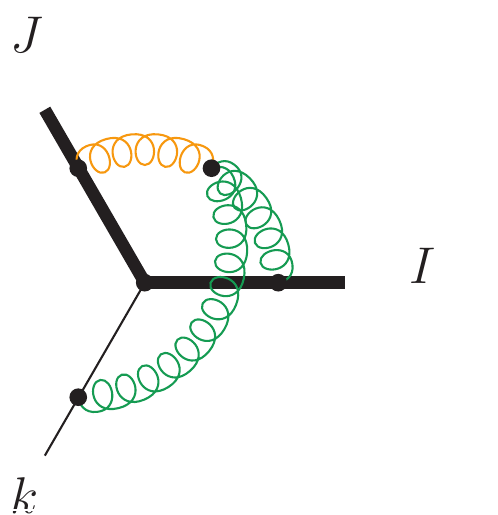}}}\,,
    \end{split}
\end{align}
neutral regions, given by the subsequent four rows (rows 5-8) in table~\ref{tab:twolooptopregionv},
\begin{align}
\label{111webRegions_Neutral}
\begin{split}
     {\Y}_{(IJk)}^{\{H,H\}}\equiv\,\,&{\Y}_{(IJk)}^{[H,H,*]}=\vcenter{\hbox{\includegraphics[width=1.5cm]{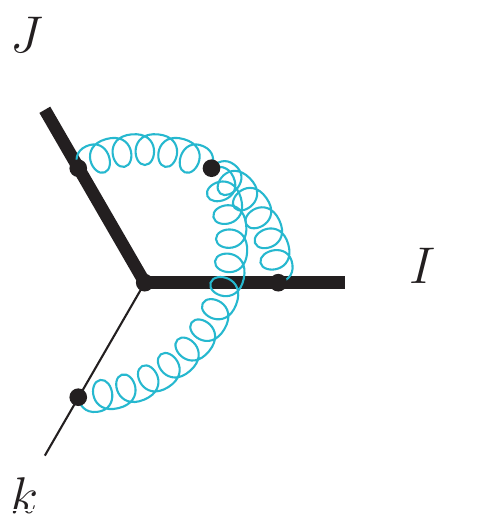}}}\,,
   \\
     {\Y}_{(IJk)}^{\{C_N,C_N\}}\equiv\,\,&{\Y}_{(IJk)}^{[C_N,C_N,*]}=\vcenter{\hbox{\includegraphics[width=1.5cm]{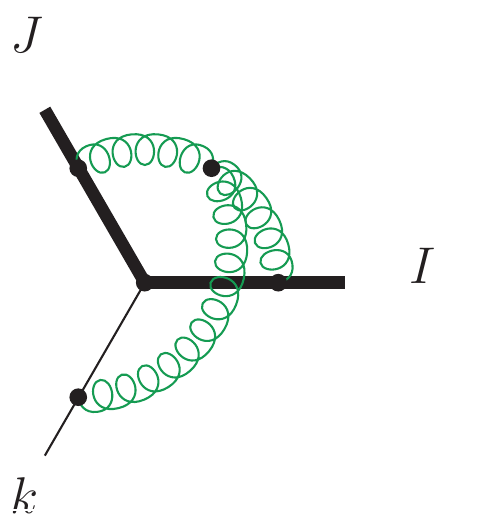}}}\,,
    \\
    {\Y}_{(IJk)}^{\{C_{N},H\}}\equiv\,\,&{\Y}_{(IJk)}^{[H,*,C_{N}]}+{\Y}_{(IJk)}^{[*,H,C_{N}]}=\vcenter{\hbox{\includegraphics[width=1.5cm]{fig/111HCNap.pdf}}}+\vcenter{\hbox{\includegraphics[width=1.5cm]{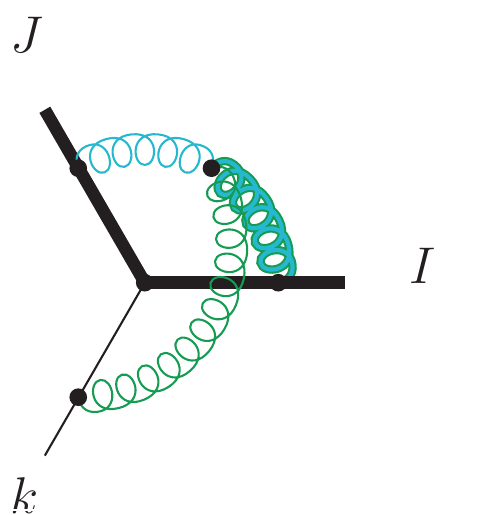}}}\,,
    \end{split}
\end{align}
and UV regions, given by the bottom three rows (rows 9-11) in table~\ref{tab:twolooptopregionv},
\begin{align}
\label{111webRegions_UV}
\begin{split}
     {\Y}_{(IJk)}^{\{C_N,H_{\UV}\}}\equiv\,\,&{\Y}_{(IJk)}^{[H_{\UV},*,C_N]}=\vcenter{\hbox{\includegraphics[width=1.5cm]{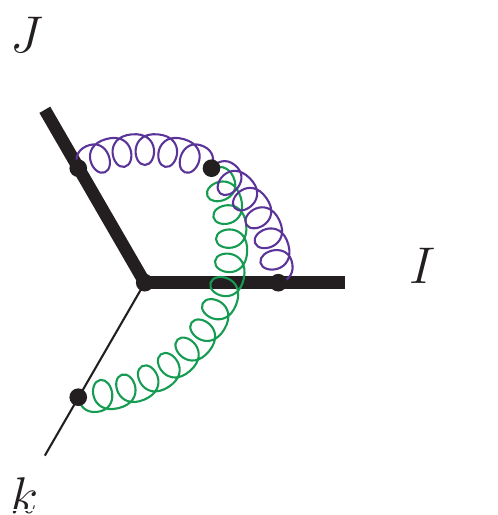}}}\,,
    \\
    {\Y}_{(IJk)}^{\{C_{\UV},H\}}\equiv\,\,&{\Y}_{(IJk)}^{[H,*,C_{\UV}]}+{\Y}_{(IJk)}^{[*,H,C_{\UV}]}=\vcenter{\hbox{\includegraphics[width=1.5cm]{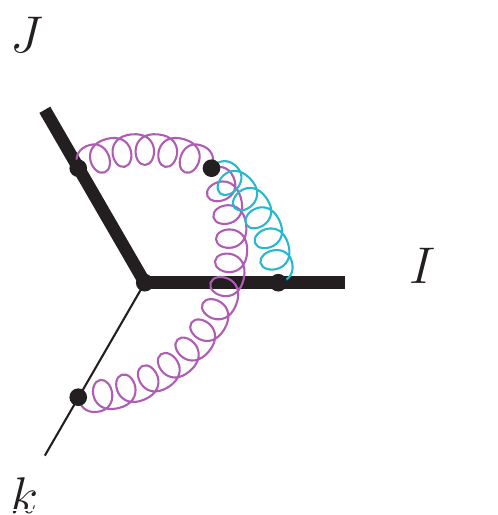}}}+\vcenter{\hbox{\includegraphics[width=1.5cm]{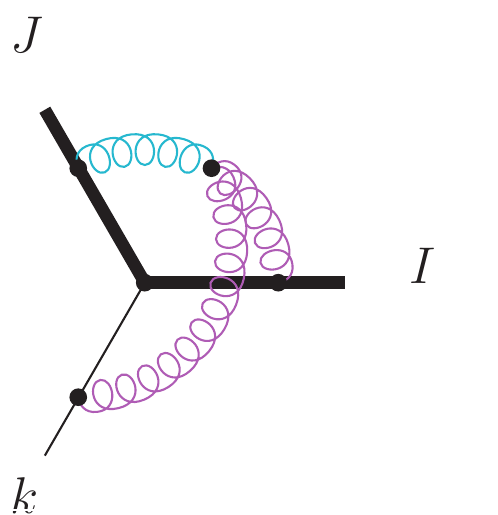}}}\,.
    \end{split}
\end{align}
Importantly, we observe that there exists no region which involves one loop in an IR mode with another loop in a UV mode. Only mixing within these two classes, and between each of them and neutral modes, appear as regions. This property will be useful in unambiguously associating the region contribution to the UV or IR renormalization in what follows.   

The results of the IR, neutral and UV regions, can be found in the appendices~\ref{app:2loopIR},
\ref{app:2loopNe} and \ref{app:2loopUV}, respectively. 
The IR regions, ${\Y}_{(IJk)}^{\{C_{\IR},C_{\IR}\}}$, ${\Y}_{(IJk)}^{\{H,C_{\IR}\}}$, and ${\Y}_{(IJk)}^{\{C_N,C_{\IR}\}}$ are given in eqs.~\eqref{111CIRCIR},~\eqref{111HCIR} and~\eqref{111CNCIR}, respectively. We note that these integrals yield triple and double poles in $\epsilon$ in addition to the single pole the correlator is expected to have.

The two neutral regions, ${\Y}_{(IJk)}^{\{C_N,C_N\}}$ and ${\Y}_{(IJk)}^{\{C_N,H\}}$, are zero,
\begin{align}
\label{111neutr}
    {\Y}_{(IJk)}^{\{C_N,C_N\}}= {\Y}_{(IJk)}^{\{C_N,H\}}=0\,.
\end{align}
The reason why they are trivial will be elucidated in section~\ref{sec:neutralregions}. 
The strict limit of ${\Y}_{(IJK)}$, which is the hard region ${\Y}_{(IJk)}^{\{H,H\}}$, gives
 \begin{align}
 \label{hard111}
 \begin{split}
       {\Y}_{(IJk)}^{\{H,H\}}=\,\,&\left(\frac{\alpha_s}{4\pi}\right)^2\left(\frac{\bar{m}^2}{\mu^2}\right)^{-2\epsilon}\bigg\{\frac{1}{\epsilon^2}\bigg[-\frac{1-y_{IJk}}{1+y_{IJk}}U_1(y_{IJk})-2\frac{1+\alpha_{IJ}^2}{1-\alpha_{IJ}^2}\log(y_{IJk})\log(\alpha_{IJ})\bigg]
       \\&+\frac{1}{\epsilon}\bigg[\frac{1+\alpha_{IJ}^2}{1-\alpha_{IJ}^2}\log(y_{IJk})\bigg(2V_1(\alpha_{IJ})+M_{100}(\alpha_{IJ})\bigg)-\frac{1}{2}\frac{1-y_{IJk}}{1+y_{IJk}}U_2(\alpha_{IJ},y_{IJk})
       \\&+2 \log^2\left(\alpha _{IJ}\right)\log \left(y_{IJk}\right)
       -\frac{2}{3}  \log ^3\left(y_{IJk}\right)-\frac{2}{3} \pi ^2 \log \left(y_{IJk}\right)\bigg]\bigg\}+{\cal O}(\epsilon^0)+{\cal O}(\lambda)\,,
         \end{split}
 \end{align}
 where the transcendental functions  $V_1$, $U_1$ and $U_2$ can be found in appendix~\ref{app:convfunc}. 
We note that the  leading order of the hard region features a double pole in $\epsilon$, which is a mix of UV and collinear (IR)  singularities.  

The UV regions ${\Y}_{(IJk)}^{\{C_N,H_{\UV}\}}$ and ${\Y}_{(IJk)}^{\{C_{\UV},H\}}$ are given in eq.~\eqref{111CNHUV} and eq.~\eqref{111CUVH}, respectively. These are new at two loops and their interpretation will be discussed in the next section. 
 Similarly to the IR regions, also the UV ones feature up to cubic poles in $\epsilon$.

According to the method of regions, and in line with eq.~(\ref{ExpansionCalY}), the expanded result for the web ${\Y}_{(IJK)}$ in the small $\beta_K^2$ limit is given by the sum of all the regions in eqs.~(\ref{111webRegions_IR}),  (\ref{111webRegions_Neutral}) and~(\ref{111webRegions_UV}):
\begin{align}
\label{111webRegionssum}
\begin{split}
    {\cal T}_{\lambda}\left[{\Y}_{(IJK)}\right]=\,\,&\lambda^{-2\epsilon}{\Y}_{(IJk)}^{\left\{C_{\IR},C_{\IR}\right\}}+\lambda^{-\epsilon}{\Y}_{(IJk)}^{\left\{C_{N},C_{\IR}\right\}}+\lambda^{-\epsilon}{\Y}_{(IJk)}^{\left\{C_{\IR},H\right\}}
    \\[.15cm]&+{\Y}_{(IJk)}^{\left\{H,H\right\}}+\textcolor{blue}{{\Y}_{(IJk)}^{\left\{C_{N},C_{N}\right\}}}+\textcolor{blue}{{\Y}_{(IJk)}^{\left\{C_{N},H\right\}}}
    \\[.15cm]&+
    \lambda^{2\epsilon}{\Y}_{(IJk)}^{\left\{H,C_{\UV}\right\}}+\lambda^{2\epsilon}{\Y}_{(IJk)}^{\left\{C_{N},H_{\UV}\right\}}\,,
    \end{split}
\end{align}
where the vanishing regions are marked in blue. The IR, neutral and UV regions, respectively, are arranged in the three lines. 
The first line  sums all IR regions, where $n_R=-1$ in the case of a single $C_{\text{IR}}$ loop-momentum mode 
or $-2$ for two such independent modes.  This sum features a triple pole in $\epsilon$ as the leading behavior. The second line in eq.~\eqref{111webRegionssum}, consists of the neutral regions with $n_R=0$. These involve the combination of two neutral loop-momentum modes, either hard or neutral-collinear, and are a natural generalization of the regions we have observed at one loop (see table~\ref{tab:oneloopregion}). However, since the latter two vanish, the sum of all neutral regions is exactly eq.~\eqref{hard111}. Finally, the third line in eq.~\eqref{111webRegionssum}, 
consists of the two UV regions, for which $n_R=2$. Similarly to the case of the IR regions, the sum of the UV regions also starts with a triple pole.

By summing up all the regions, the triple pole in $\epsilon$ from the UV and IR regions cancel each other out. In addition, the double poles present in the separate IR, neutral and UV components cancel entirely in the sum.  
Furthermore, the polylogarithmic  functions entering the coefficient of the single $\epsilon$ pole via $U_1$, $U_2$, $V_1$ and $M_{100}$, also 
cancel between the hard region and the IR and UV  regions. 
As a result, the leading order of the expansion ${\cal T}_{\lambda} \left[{\Y}_{(IJK)}\right]$ is -- as expected -- a single pole in $\epsilon$ times a sum of products of logarithms,
\begin{align}
\label{result111web}
     \begin{split}
        {\cal T}_{\lambda} \left[{\Y}_{(IJK)}\right]=\,\,&\left(\frac{\alpha_s}{4\pi}\right)^2\left(\frac{\bar{m}^2}{\mu^2}\right)^{-2\epsilon}\frac{1}{\epsilon}\log(y_{IJk})\bigg[-4\frac{1+\alpha_{IJ}^2}{1-\alpha_{IJ}^2}\log(\alpha_{IJ})\log(\lambda) 
       \\& +2 \log^2(\alpha_{IJ})
        +2\log^2(\lambda)-\frac{1}{2}\log^2(y_{IJk})\bigg]+{\cal O}(\epsilon^0)+{\cal O}(\lambda^1).
         \end{split}
 \end{align}
 Notice that eq.~\eqref{result111web} still contains $\log(\lambda)$, indicating that the limit $\beta_K^2\to 0$ of this web (on its own) is singular.
One can check that upon substituting $\lambda$ in terms of $\beta_K^2$ according to~eq.~(\ref{betaK2Lambda}), then eq.~\eqref{result111web} exactly matches the expansion of this web computed with three timelike Wilson lines~\cite{Mitov:2009sv,Ferroglia:2009ii,Ferroglia:2009ep}:
\begin{align}
     \begin{split}
        {\Y}_{(IJK)}=\,\,&\left(\frac{\alpha_s}{4\pi}\right)^2\left(\frac{\bar{m}^2}{\mu^2}\right)^{-2\epsilon}\frac{2}{\epsilon}\epsilon^{IJK}\frac{1+\alpha_{IJ}^2}{1-\alpha_{IJ}^2}\log(\alpha_{IJ})\log^2(\alpha_{IK})+{\cal O}(\epsilon^0)\,,
         \end{split}
 \end{align}
 where $\epsilon^{IJK}$ is the Levi-Civita tensor. Again we have verified that the MoR correctly reproduces the asymptotic expansion.

\subsubsection{Tripole contributions from multiple gluon exchange webs}
\label{sec:121web}

The remaining two-loop webs in eq.~(\ref{Tripole_webs}) are not fully connected, but consist of two separate gluon exchanges between the three Wilson lines. 
There are three different web configurations of this kind, $W_{(IJ)(IK)}$, $W_{(JK)(JI)}$ and $W_{(KI)(KJ)}$, which are related by permutations of external lines. 
Beyond computation of the soft~AD~\cite{Mitov:2009sv,Ferroglia:2009ii,Ferroglia:2009ep}, this type of web has been studied in detail in refs.~\cite{Gardi:2013saa,Falcioni:2014pka} for generic timelike Wilson lines, where it was used as the simplest example of multiple gluon exchange webs going beyond the cusp configuration. Here we consider its asymptotic expansion at small~$\beta_K^2$, pursing the same analysis as for the one-loop and the connected tripole webs above.

Each of the $W_{(IJ)(IK)}$, $W_{(JK)(JI)}$ and $W_{(KI)(KJ)}$ webs has a kinematic function which is a difference of two integrals, 
 \begin{subequations}
\label{Tripole_sub_webs}
\begin{align}
\begin{split}
\label{YIJIK}
 {\Y}_{(IJ)(IK)}=&\frac{1}{2}\left\{\vcenter{\hbox{\includegraphics[width=1.5cm]{fig/211IJKa.pdf}}} -\vcenter{\hbox{\includegraphics[width=1.5cm]{fig/211IJKb.pdf}}} \right\}
\\=\,\,&\left( \frac{\alpha_s}{4\pi}\right)^2
\frac{{\cal N}^2}{2}v_J\cdot v_Iv_K\cdot v_I \int\left[{\cal D}q_J\right]\left[{\cal D}q_K\right] \left[Y_{(IJ)(IK)}(q_J,q_K)-Y_{(IK)(IJ)}(q_K,q_J)\right],
  \end{split}
\\
\begin{split}
\label{YJKJI}
{\Y}_{(JK)(JI)}=&\frac{1}{2}\left\{\vcenter{\hbox{\includegraphics[width=1.5cm]{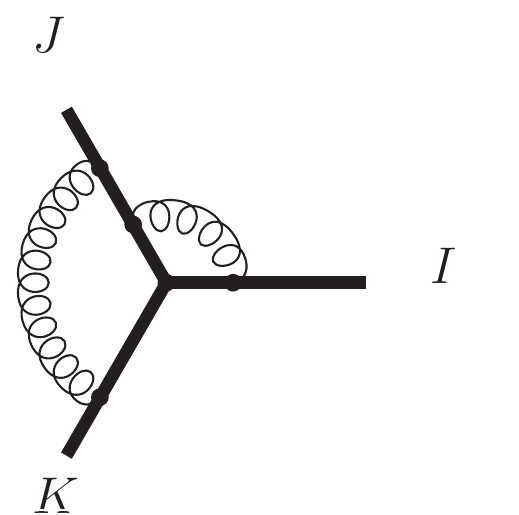}}}-\vcenter{\hbox{\includegraphics[width=1.5cm]{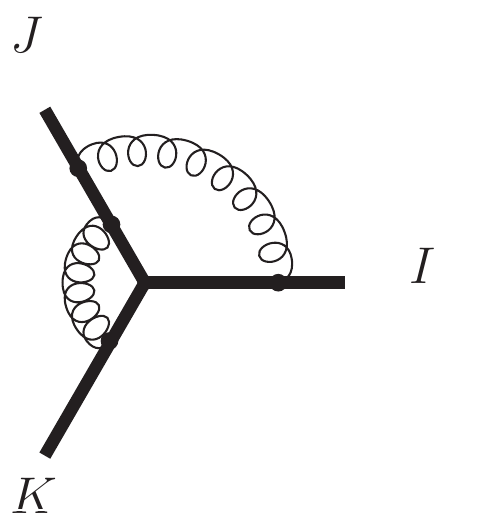}}}  \right\}
\\=&\,\,\left( \frac{\alpha_s}{4\pi}\right)^2
\frac{{\cal N}^2}{2}
 v_I\cdot v_Jv_K\cdot v_J\int\left[{\cal D}q_I\right]\left[{\cal D}q_K\right] \left[Y_{(JK)(JI)}(q_K,q_I)-Y_{(JI)(JK)}(q_I,q_K)\right],
  \end{split}
\\
\begin{split}
\label{YKIKJ}
{\Y}_{(KI)(KJ)}=&\frac{1}{2}\left\{\vcenter{\hbox{\includegraphics[width=1.5cm]{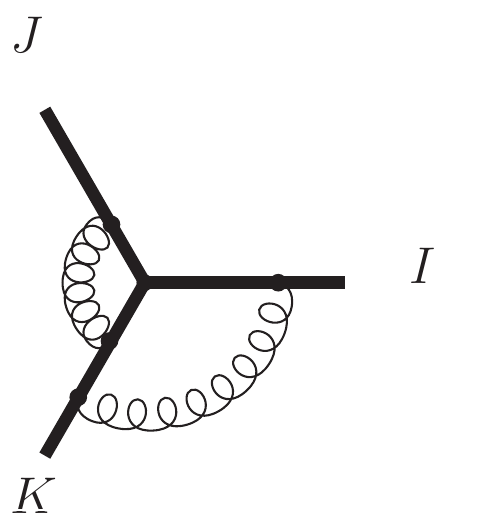}}} -\vcenter{\hbox{\includegraphics[width=1.5cm]{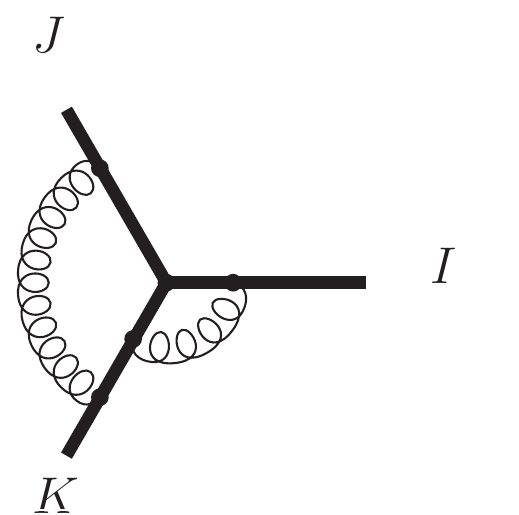}}}  \right\}
\\=&\,\,\left( \frac{\alpha_s}{4\pi}\right)^2
\frac{{\cal N}^2}{2} v_I\cdot v_Kv_J\cdot v_K\int\left[{\cal D}q_J\right]\left[{\cal D}q_I\right]
\left[Y_{(KI)(KJ)}(q_I,q_J)-Y_{(KJ)(KI)}(q_J,q_I)\right],
  \end{split}
 \end{align}
 \end{subequations}
where ${\cal N}$ and the measure $\left[ {\cal D}q\right]$ are defined in eq.~(\ref{calNandDq}), and the scalar integrand~$Y_{(IJ)(IK)}$ is defined by 
    \begin{align}
    \begin{split}
   Y_{(IJ)(IK)}(k_J,k_K)\equiv&\,\,\frac{1}{P_{g_J}}\frac{1}{P_{g_K}}\frac{1}{P_{J}}\frac{1}{P_{K}}\frac{1}{P_{I,J}}\frac{1}{P_{I,JK}}\,,
\end{split}
    \end{align}
with the six propagators given by
\begin{align}
\begin{split}
   P_{g_J}=\,\,&q_J^2+i\varepsilon\,,
   \qquad
   P_{J}=\frac{1}{\tilde{E}_{\beta_J}\left(1,q_I\right)}=-v_J\cdot q_J-1+i\varepsilon\,,
   \\
    P_{g_K}=\,\,&q_K^2+i\varepsilon\,,
    \qquad
P_{K}=\frac{1}{\tilde{E}_{\beta_K}\left(1,q_K\right)}=-v_K\cdot q_K-1+i\varepsilon
\\
  P_{I,J}=\,\,&\frac{1}{\tilde{E}_{\beta_I}\left(1,-q_J\right)}=v_I\cdot q_{J}-1+i\varepsilon\,,
  \\
 P_{I,JK}=\,\,&\frac{1}{\tilde{E}_{\beta_I}\left(2,-q_J-q_K\right)}=v_I\cdot (q_J+q_K)-2+i\varepsilon\,.
  \end{split}
\end{align}
With the definitions in eq.~\eqref{Tripole_sub_webs}, the kinematic function ${\Y}_{(IJ)(IK)}$ is by construction anti-symmetric on the interchange of $J$ and $K$,
\begin{align}
\label{antisymCalY}
  {\Y}_{(IJ)(IK)}=-{\Y}_{(IK)(IJ)}\,,  
\end{align}
in line with the antisymmetry of the colour factor $T_{IJK}$ in eq.~(\ref{Tripole_webs}). 

We define the scalar integral as our input of the region analysis, 
\begin{align}
    \tilde{{\Y}}_{(IJ)(IK)}=\left( \frac{\alpha_s}{4\pi}\right)^2
{\cal N}^2v_I\cdot v_Jv_I\cdot v_K\int\left[{\cal D}q_J\right]\left[{\cal D}q_K\right]Y_{(IJ)(IK)}(q_J,q_K),
\end{align}
and then
\begin{align}
  {\Y}_{(IJ)(IK)}= \frac{1}{2}\left[\tilde{{\Y}}_{(IJ)(IK)}-\tilde{{\Y}}_{(IK)(IJ)} \right]\,.
\end{align}
Note that in contrast to the web kinematic function ${\Y}_{(IJ)(IK)}$ which admits eq.~\eqref{antisymCalY}, here the subscript of $\tilde{{\Y}}_{(IJ)(IK)}$ identifies one of the two diagrams in the web.
The Lee-Pomeransky representation for $\tilde{\cal Y}_{(IJ)(IK)}$ is then 
\begin{align}
\label{ParametricIJIK}
\begin{split}
  \tilde{{\Y}}_{(IJ)(IK)}=\,\,&\left( \frac{\alpha_s}{4\pi}\right)^2
\frac{{\cal N}^2}{2}v_I\cdot v_Jv_I\cdot v_K\frac{\Gamma (2-\epsilon )}{\Gamma (-3 \epsilon )} \int_0^{\infty}dx_{g_J}\int_0^{\infty}dx_{g_K}
  \\&\int_0^{\infty}dx_{J}\int_0^{\infty}dx_{K}\int_0^{\infty}dx_{I,J}\int_0^{\infty}dx_{I,JK} \left[{\cal P}_{(IJ)(IK)}\right]^{\epsilon-2}, 
  \end{split}
\end{align}
and the polynomial ${\cal P}_{(IJ)(IK)}$ is given by
\begin{subequations}
\begin{align}
    \begin{split}
        {\cal P}_{(IJ)(IK)}=\,\,&\frac{x_{g_K}}{4} x_{J} (x_{I,J}+x_{I,JK}) \left(\alpha _{IJ}+\frac{1}{ \alpha _{IJ}}\right)+\frac{x_{g_J}}{4}  x_{K} x_{I,JK} \left(\alpha _{IK}+\frac{1}{\alpha _{IK}}\right)
        \\&+\frac{x_{g_J}}{4} \left(x_{K}^2+x_{I,JK}^2\right)+\frac{x_{g_K}}{4} \left(x_{J}^2+x_{I,J}^2+x_{I,JK}^2\right)
        \\&+\frac{x_{g_K}}{2}  x_{I,J} x_{I,JK}+x_{g_J} x_{g_K} \left(x_{J}+x_{K}+x_{I,J}+2 x_{I,JK}\right)+\textcolor{red}{x_{g_J}x_{g_K}}\,.
    \end{split}
\end{align}
\end{subequations}

\begin{figure}
\begin{subfigure}{0.5\linewidth}
    \centering \includegraphics[width=0.8\linewidth]{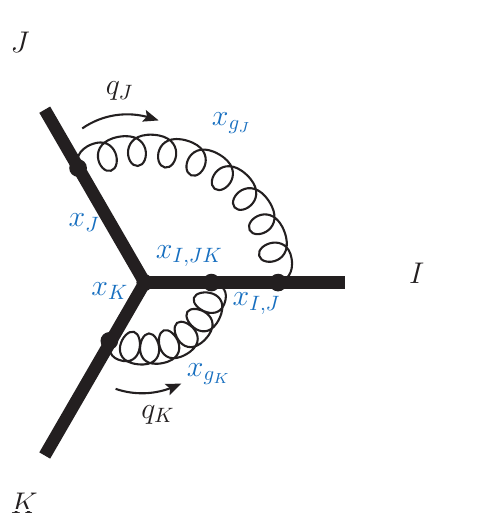}
    \caption{The integral  $\tilde{{\Y}}_{(IJ)(IK)}$}
    \label{fig:twoloopIJIK}
    \end{subfigure}
    \hfill
    \begin{subfigure}{0.5\linewidth}
    \centering \includegraphics[width=0.8\linewidth]{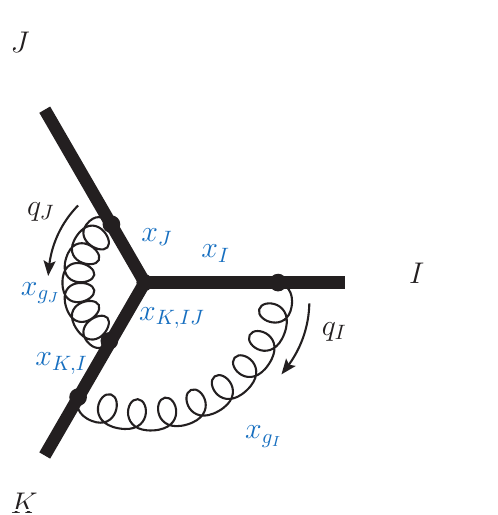}
    \caption{The integral  $\tilde{{\Y}}_{(KI)(KJ)}$}
    \label{fig:twoloopKIKJ}
    \end{subfigure}
     \caption{The propagators and the corresponding parameters for the integrals with two exchanged gluons.}
     \label{fig:twoloop112}
\end{figure}

\begin{table}
\centering
\begin{subtable}[]{0.9\textwidth}
\centering
\begin{tabular}{|c|c|c|c|}
\hline
Integrand & Region vector $\mathbf{u}^R$ & $n_R$ & Region $R$\\
&$\{u_{g_J},u_{g_K},u_{J},u_{K},u_{I,J},u_{I,JK},1\}$&&$\left[\text{mode}_{q_J},\text{mode}_{q_K}\right]$
\\
\hline
\hline
\multirow{4}{*}{$\tilde{{\Y}}_{(IJ)(IK)}$} & $\{0,-1,0,0,0,0,1\}$ & $0-1=-1$ & $\left[H,C_{\IR}\right]$ \\
& $\{0,0,0,0,0,1,1\}$ & $0+0=0$ & $\left[H,C_{N}\right]$ \\
& $\{0,0,0,1,0,0,1\}$ & $0+0=0$ & $\left[H,H\right]$ \\
& $\{2,0,1,0,1,1,1\}$ & $2+0=2$ & $\left[H_{\UV},C_N\right]$ \\
\hline
\end{tabular}
\caption{Region vectors for the integral $\tilde{{\Y}}_{(IJ)(IK)}$}
\label{tab:IJIKWebRegions}
\end{subtable}

\vspace{1em}

\begin{subtable}[]{0.9\textwidth}
\centering
\begin{tabular}{|c|c|c|c|}
\hline
Integrand & Region vector $\mathbf{u}^R$ & $n_R$ & Region $R$\\
&$\{u_{g_K},u_{g_J},u_{K},u_{J},u_{I,K},u_{I,KJ},1\}$&&$\left[\text{mode}_{q_K},\text{mode}_{q_J}\right]$
\\
\hline
\hline
\multirow{4}{*}{$\tilde{{\Y}}_{(IK)(IJ)}$} & $\{-1,0,0,0,0,0,1\}$ & $-1+0=-1$ & $\left[C_{\IR},H\right]$ \\
& $\{0,0,0,0,0,1,1\}$ & $0+0=0$ & $\left[C_{N},H\right]$ \\
& $\{0,0,1,0,0,0,1\}$ & $0+0=0$ & $\left[H,H\right]$ \\
& $\{0,2,0,1,1,1,1\}$ & $0+2=2$ & $\left[C_N,H_{\UV}\right]$ \\
\hline
\end{tabular}
\caption{Region vectors for the integral $\tilde{{\Y}}_{(IK)(IJ)}$}
\label{tab:IKIJWebRegions}
\end{subtable}

\vspace{1em}

\begin{subtable}[]{0.9\textwidth}
\centering
\begin{tabular}{|c|c|c|c|}
\hline
Integrand & Region vector $\mathbf{u}^R$ & $n_R$ & Region $R$\\
&$\{u_{g_I},u_{g_J},u_{I},u_{J},u_{K,I},u_{K,IJ},1\}$&&$\left[\text{mode}_{q_I},\text{mode}_{q_J}\right] $
\\
\hline
\hline
\multirow{9}{*}{$\tilde{{\Y}}_{(KI)(KJ)}$} & $\{-1,-1,0,0,0,0,1\}$ & $-1-1=-2$ & $\left[C_{\IR},C_{\IR}\right]$ \\
& $\{-1,0,0,0,0,1,1\}$ & $-1+0=-1$ & $\left[C_{\IR},H\right]$ \\
& $\{-1,0,0,1,0,0,1\}$ & $-1+0=-1$ & $\left[C_{\IR},C_{N}\right]$ \\
& $\{0,-1,1,0,0,0,1\}$ & $0-1=-1$ & $\left[C_{N},C_{\IR}\right]$ \\
& $\{0,0,0,0,1,1,1\}$ & $0+0=0$ & $\left[H,H\right]$ \\
& $\{0,0,1,0,0,1,1\}$ & $0+0=0$ & $\left[C_{N},H\right]$ \\
& $\{0,0,1,1,0,0,1\}$ & $0+0=0$ & $\left[C_{N},C_{N}\right]$ \\
& $\{0,2,0,2,1,1,1\}$ & $0+2=2$ & $\left[H,C_{\UV}\right]$ \\
& $\{2,0,2,0,1,1,1\}$ & $2+0=2$ & $\left[C_{\UV},H\right]$ \\
\hline
\end{tabular}
\caption{Region vectors for the integral $\tilde{{\Y}}_{(KI)(KJ)}$}
\label{tab:KIKJWebRegions}
\end{subtable}
\caption{
The region vectors $\mathbf{u}^R$ and the coefficient $n_R$ identifying the overall scaling of each region,~$\lambda^{n_R\epsilon}$, of the two-loop diagrams~$\tilde{{\Y}}_{(IJ)(IK)}$, $\tilde{{\Y}}_{(IK)(IJ)}$ and  
$\tilde{{\Y}}_{(KJ)(KI)}$. The rightmost column describes the region in momentum space in terms of the modes defined in table~\ref{tab:ModesSum}. In each case, the loop momentum mode is defined by the lightcone momentum components of the two gluons. }
\label{tab:121WebRegions}
\end{table}
Figure~\ref{fig:twoloop112} presents the interpretation of the Lee-Pomeransky parameters in relation with the diagrams: figure~\ref{fig:twoloopIJIK} for the web where only one of the gluons attaches to the (nearly) lightlike line $K$, and figure~\ref{fig:twoloopKIKJ} for the web where both gluons attach to this line.

We now perform the asymptotic expansion in $\lambda$ using the geometric method in parameter space with the help of {\tt{pySecDec}}, and immediately proceed to interpret the 
region vectors in momentum space, as we have done for the connected web.

The region vectors corresponding to the 
small $\beta_K^2$ expansion of the scalar integral $\tilde{\Y}_{(IJ)(IK)}$ of eq.~(\ref{ParametricIJIK}), and hence of the web  ${\Y}_{(IJ)(IK)}$ of eq.~(\ref{YIJIK}), computed using {\tt{pySecDec}}, are collected in table~\ref{tab:IJIKWebRegions}. The components of the region vector $\mathbf{u}^R$ follow the order, 
\begin{equation}
 {\bf u}^R=\{u_{g_J},u_{g_K},u_{J},u_{K},u_{I,J},u_{I,JK}\} 
\end{equation}
corresponding to the scaling of the propagators
in figure~\ref{fig:twoloopIJIK} as follows 
\begin{align}
\begin{split}
& \{x_{g_J},x_{g_K},x_{J},x_{K},x_{I,J},x_{I,JK}\}
 \xrightarrow{\;{R}\;}
 \\&\hspace*{50pt}
 \{x_{g_J}\lambda^{u_{g_J}^R}, x_{g_K}\lambda^{u_{g_K}^R}, x_{J}\lambda^{u_{J}^R},
 x_{K}\lambda^{u_{K}^R}, 
x_{I,J}\lambda^{u_{I,J}^R}, x_{I,JK}\lambda^{u_{I,JK}^R}\}\,.
\end{split}
\end{align}
Table~\ref{tab:IJIKWebRegions} indicates that there are only four regions in this case, one IR region, $[H,C_{\text{IR}}]$, with $n_R=-1$, two neutral regions, having $n_R=0$, and a single UV region, $[H_{\text{UV}},C_N]$, with $n_R=2$.
The reason for this simple region structure is clear: the gluon exchanged between the timelike lines is only indirectly affected by taking the $\beta_K^2\to 0$ limit, and it remains hard, or UV hard, throughout. 
A similar expansion is obtained for the 
$\tilde{\Y}_{(JK)(JI)}$ web in eq.~(\ref{YJKJI}), which can be obtained from ${\Y}_{(IJ)(IK)}$ via the cyclic permutation, $I\to J$, $J\to K$ and $K\to I$.

A rather different situation is encountered in the case of the region expansion of  the scalar integral $\tilde{\Y}_{(KI)(KJ)}$ and hence of the web  ${\Y}_{(KI)(KJ)}$ of eq.~(\ref{YKIKJ}). Here the region vectors computed using {\tt{pySecDec}} are collected in table~\ref{tab:KIKJWebRegions}. 
The components of the region vector $\mathbf{u}^R$ follow the order, 
\begin{equation}
 {\bf u}^R=\{u_{g_I},u_{g_J},u_{I},u_{J},u_{K,I},u_{K,IJ}\} 
\end{equation}
corresponding to the scaling of the propagators
in figure~\ref{fig:twoloopIJIK} as follows 
\begin{align}
\begin{split}
& \{x_{g_I},x_{g_J},x_{I},x_{J},x_{K,I},x_{K,IJ}\}
 \xrightarrow{\;{R}\;}
 \\&\hspace*{50pt}\{x_{g_I}\lambda^{u_{g_I}^R}, x_{g_J}\lambda^{u_{g_J}^R}, x_{I}\lambda^{u_{I}^R},
 x_{J}\lambda^{u_{J}^R}, 
x_{K,I}\lambda^{u_{K,I}^R}, x_{K,IJ}\lambda^{u_{K,IJ}^R}\}\,.
\end{split}
\end{align}
Compared to~\ref{tab:IJIKWebRegions},  table~\ref{tab:KIKJWebRegions} displays a much richer region structure, reminiscent of that of the connected web in table~\ref{tab:twolooptopregionv}.   The first four rows of table~\ref{tab:KIKJWebRegions} correspond to IR regions, with either both loops in a IR-collinear mode $[C_{\text{IR}},C_{\text{IR}}]$, with $n_R=-2$, or just one of them, with $n_R=-1$, the second loop remaining hard or collinear neutral.
The next three rows in table~\ref{tab:KIKJWebRegions} correspond to neutral regions with $n_R=0$, while the last two rows correspond to UV ones. 

Following the same steps we have taken in 
section~\ref{sec:connectedweb} in the case of the connected web, we now proceed to  
build up the asymptotic expansion for the three web kinematic functions ${\Y}_{(IJ)(IK)}$, ${\Y}_{(JK)(JI)}$ and ${\Y}_{(KI)(KJ)}$. 
To begin, in analogy with eqs.~(\ref{111webRegions_IR})~to~(\ref{111webRegions_UV}), we classify the regions contributing to each web using curly brackets to denote an unordered set of modes. 
Based on tables~\ref{tab:IJIKWebRegions}, \ref{tab:IKIJWebRegions} and~\ref{tab:KIKJWebRegions}, the IR regions kinematic functions for the three webs from are defined as follows. For the $(IJ)(IK)$ web:
\begin{align}
    \begin{split}
        {\Y}_{(IJ)(IK)}^{\left\{C_{\IR},H\right\}}=\,\,&\frac{1}{2}\left[\tilde{{\Y}}_{(IJ)(IK)}^{[H,C_{\IR}]}-\tilde{{\Y}}_{(IK)(IJ)}^{[C_{\IR},H]}\right]=\frac{1}{2}\left(\vcenter{\hbox{\includegraphics[width=1.5cm]{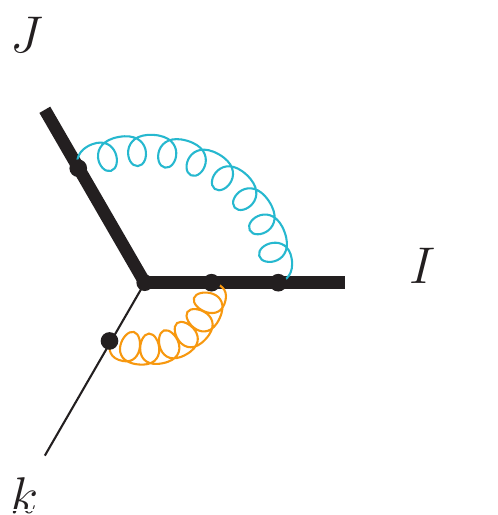}}} -\vcenter{\hbox{\includegraphics[width=1.5cm]{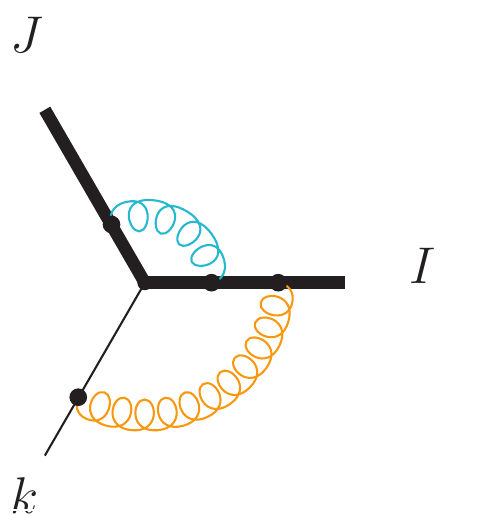}}}  \right)\,,
\end{split}
\end{align}
for the ${(JK)(JI)}$ web:
\begin{align}
        \begin{split}
        {\Y}_{(JK)(JI)}^{\left\{C_{\IR},H\right\}}=\,\,&\frac{1}{2}\left[\tilde{{\Y}}_{(JK)(JI)}^{[C_{\IR},H]}-\tilde{{\Y}}_{(JI)(JK)}^{[H,C_{\IR}]}\right]=\frac{1}{2}\left(\vcenter{\hbox{\includegraphics[width=1.5cm]{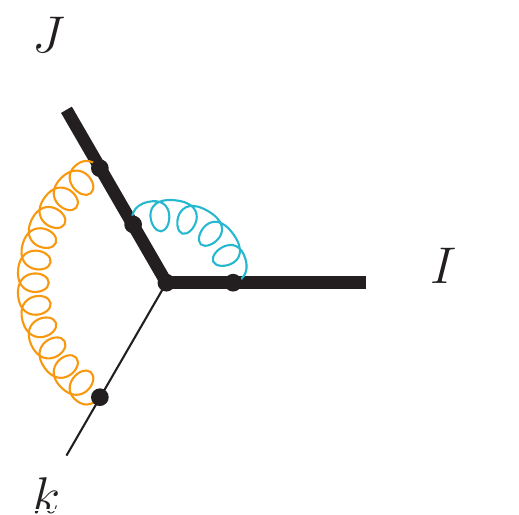}}} -\vcenter{\hbox{\includegraphics[width=1.5cm]{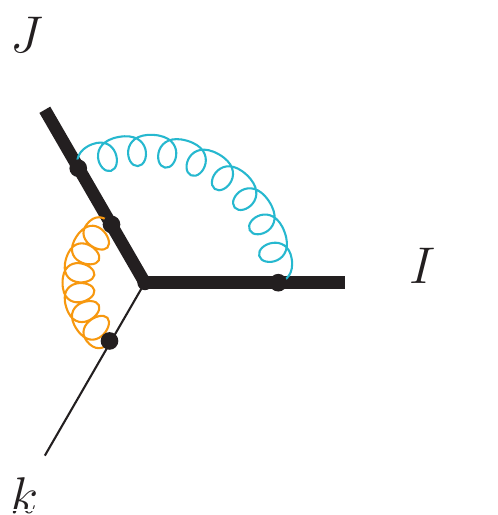}}}  \right)\,,
    \end{split}
\end{align}
and, finally, for the ${(KI)(KJ)}$ web:
\begin{align}
        \begin{split}
        {\Y}_{(KI)(KJ)}^{\left\{C_{\IR},H\right\}}=\,\,&\frac{1}{2}\left[\tilde{{\Y}}_{(KI)(KJ)}^{[C_{\IR},H]}-\tilde{{\Y}}_{(KJ)(KI)}^{[H,C_{\IR}]}\right]
        =\frac{1}{2}\left(\vcenter{\hbox{\includegraphics[width=1.5cm]{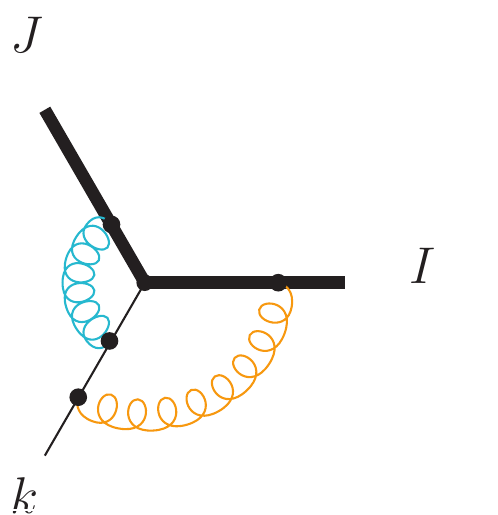}}} -\vcenter{\hbox{\includegraphics[width=1.5cm]{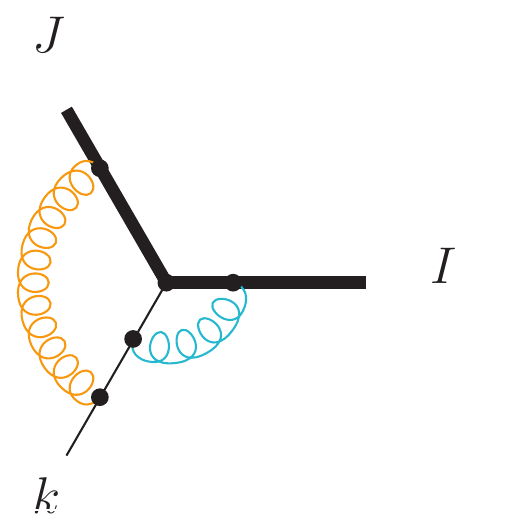}}}\right)  \,,
        \\[.15cm]
        {\Y}_{(KI)(KJ)}^{\left\{C_{\IR},C_{\IR}\right\}}=\,\,&\frac{1}{2}\left[\tilde{{\Y}}_{(KI)(KJ)}^{[C_{\IR},C_{\IR}]}-\tilde{{\Y}}_{(KJ)(KI)}^{[C_{\IR},C_{\IR}]}\right]=\frac{1}{2}\left(\vcenter{\hbox{\includegraphics[width=1.5cm]{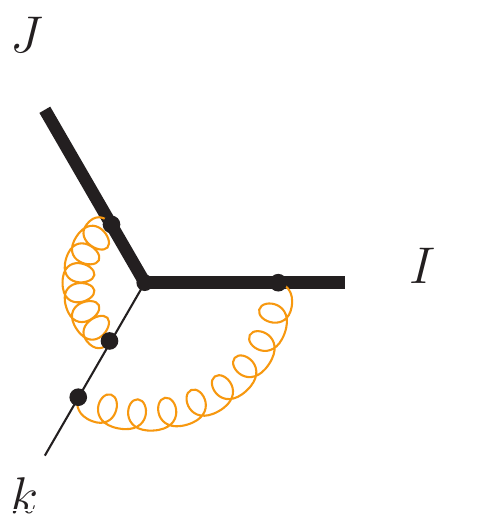}}} -\vcenter{\hbox{\includegraphics[width=1.5cm]{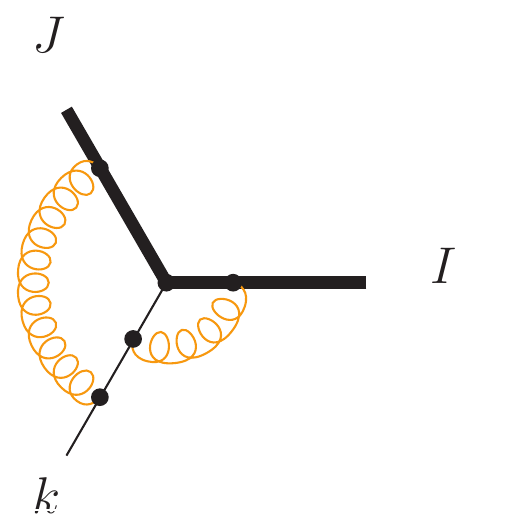}}}  \right)\,,
       \\[.15cm]
       {\Y}_{(KI)(KJ)}^{\left\{C_{N},C_{\IR}\right\}}=\,\,&\frac{1}{2}\left[\tilde{{\Y}}_{(KI)(KJ)}^{[C_{N},C_{\IR}]}-\tilde{{\Y}}_{(KJ)(KI)}^{[C_{\IR},C_{N}]}\right]+\frac{1}{2}\left[\tilde{{\Y}}_{(KI)(KJ)}^{[C_{\IR},C_{N}]}-\tilde{{\Y}}_{(KJ)(KI)}^{[C_{N},C_{\IR}]}\right]
       \\[.15cm]
       =\,\,&\frac{1}{2}\left(\vcenter{\hbox{\includegraphics[width=1.5cm]{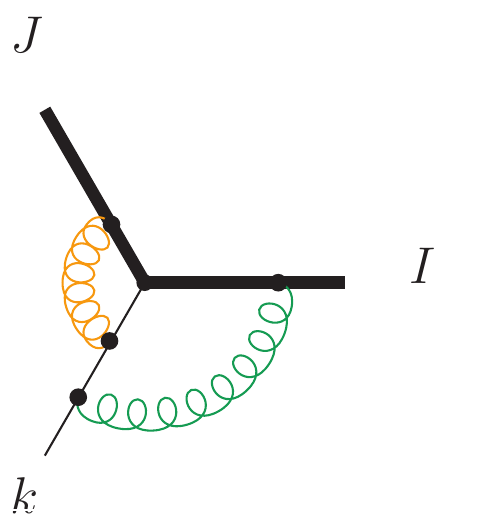}}} -\vcenter{\hbox{\includegraphics[width=1.5cm]{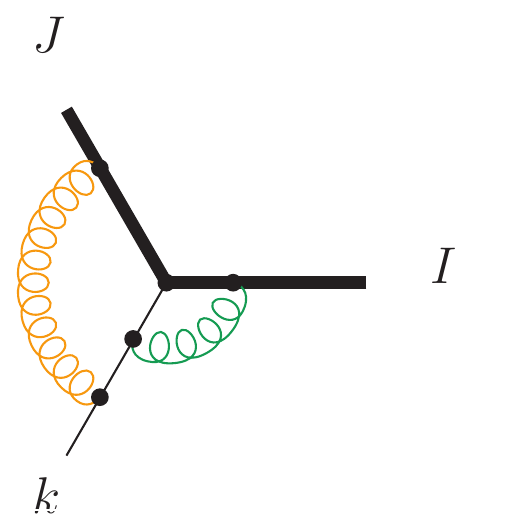}}}  \right)+\frac{1}{2}\left(\vcenter{\hbox{\includegraphics[width=1.5cm]{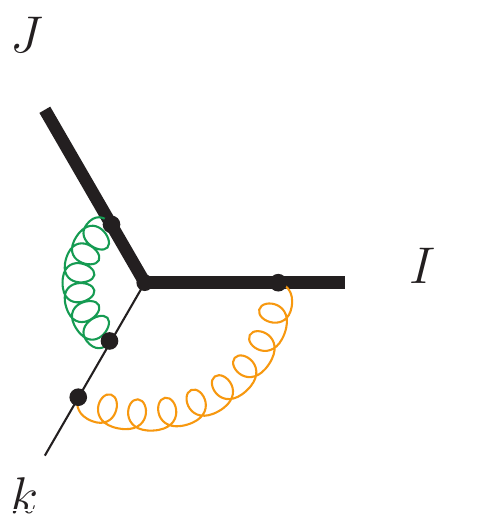}}} -\vcenter{\hbox{\includegraphics[width=1.5cm]{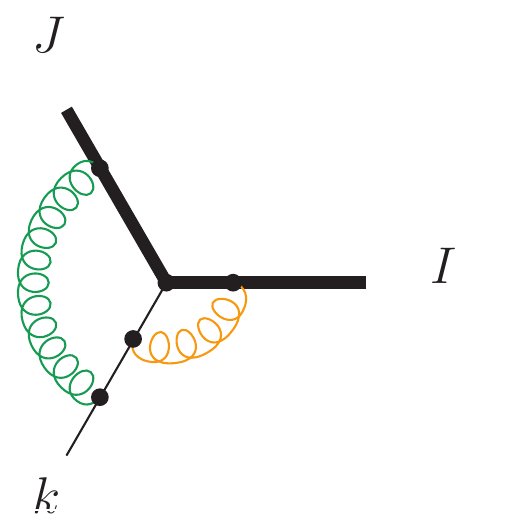}}}  \right) \,. 
    \end{split}
\end{align}
Similarly, the neutral regions for the three webs, respectively, defined as follows. For the $(IJ)(IK)$ web:
\begin{align}
\begin{split}
    {\Y}_{(IJ)(IK)}^{\left\{H,H\right\}}=\,\,&\frac{1}{2}\left[\tilde{{\Y}}_{(IJ)(IK)}^{[H,H]}-\tilde{{\Y}}_{(IK)(IJ)}^{[H,H]}\right]=\frac{1}{2}\left(\vcenter{\hbox{\includegraphics[width=1.5cm]{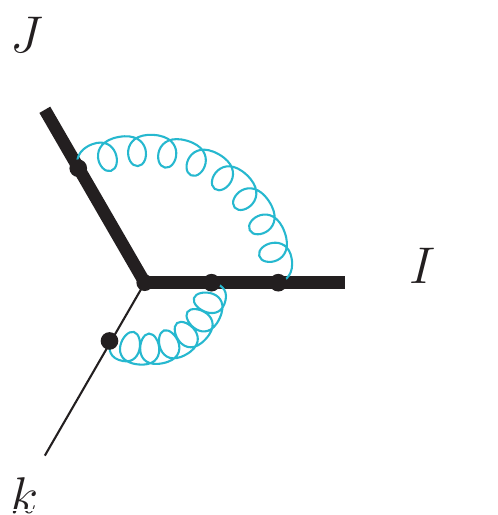}}} -\vcenter{\hbox{\includegraphics[width=1.5cm]{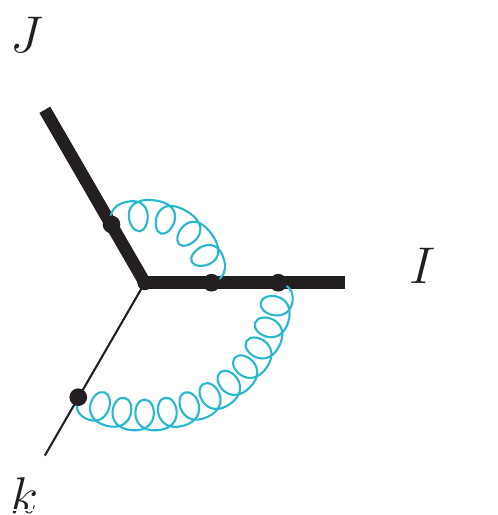}}}  \right)\,,
    \\[.15cm]
     {\Y}_{(IJ)(IK)}^{\left\{C_N,H\right\}}=\,\,&\frac{1}{2}\left[\tilde{{\Y}}_{(IJ)(IK)}^{[H,C_N]}-\tilde{{\Y}}_{(IK)(IJ)}^{[C_N,H]}\right]=\frac{1}{2}\left(\vcenter{\hbox{\includegraphics[width=1.5cm]{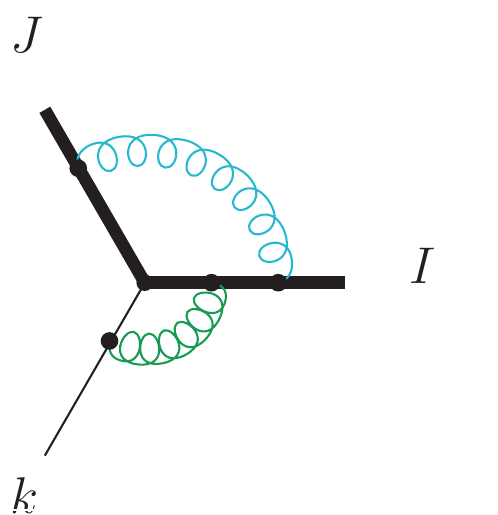}}} -\vcenter{\hbox{\includegraphics[width=1.5cm]{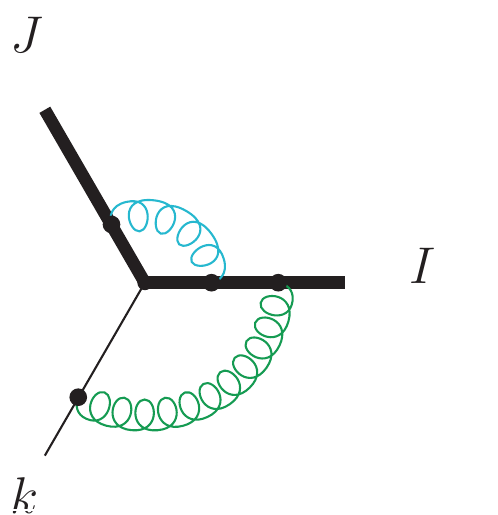}}}  \right)\,,
 \end{split}
\end{align}
For the ${(JK)(JI)}$ web:
\begin{align}
\begin{split}
    {\Y}_{(JK)(JI)}^{\left\{H,H\right\}}=\,\,&\frac{1}{2}\left[\tilde{{\Y}}_{(JK)(JI)}^{[H,H]}-\tilde{{\Y}}_{(JI)(JK)}^{[H,H]}\right]=\frac{1}{2}\left(\vcenter{\hbox{\includegraphics[width=1.5cm]{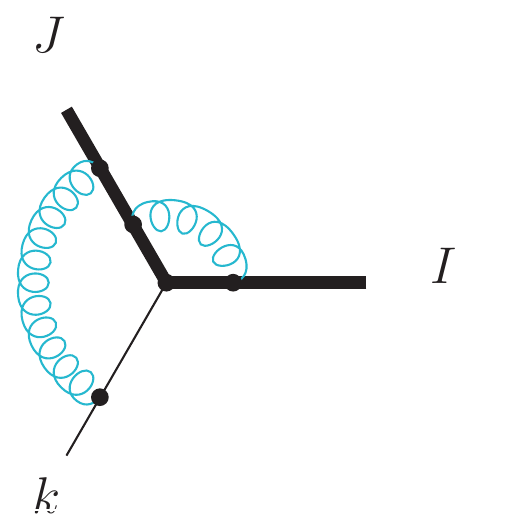}}} -\vcenter{\hbox{\includegraphics[width=1.5cm]{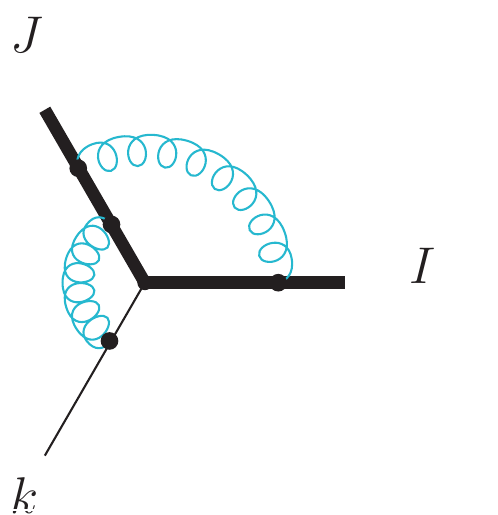}}}  \right)\,,
      \\[.15cm]
      {\Y}_{(JK)(JI)}^{\left\{C_N,H\right\}}=\,\,&\frac{1}{2}\left[\tilde{{\Y}}_{(JK)(JI)}^{\left\{C_N,H\right\}}-\tilde{{\Y}}_{(JI)(JK)}^{[H,C_N]}\right]=\frac{1}{2}\left(\vcenter{\hbox{\includegraphics[width=1.5cm]{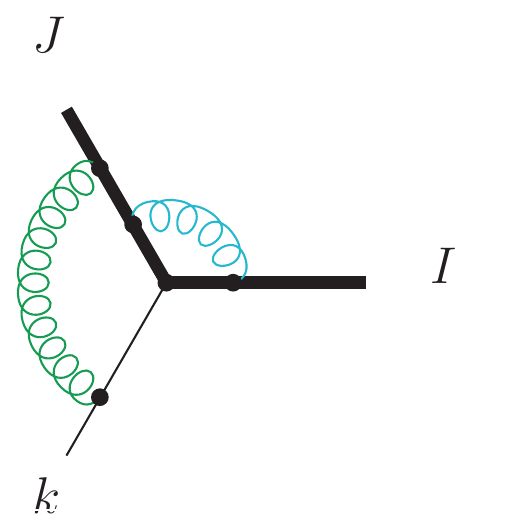}}} -\vcenter{\hbox{\includegraphics[width=1.5cm]{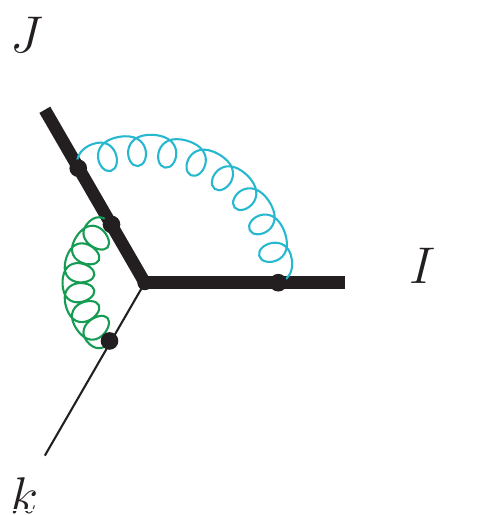}}}  \right)\,,
 \end{split}
\end{align}
and for the ${(KI)(KJ)}$ web:
\begin{align}
\begin{split}
     {\Y}_{(KI)(KJ)}^{\left\{H,H\right\}}=\,\,&\frac{1}{2}\left[\tilde{{\Y}}_{(KI)(KJ)}^{[H,H]}-\tilde{{\Y}}_{(KJ)(KI)}^{[H,H]}\right]=\frac{1}{2}\left(\vcenter{\hbox{\includegraphics[width=1.5cm]{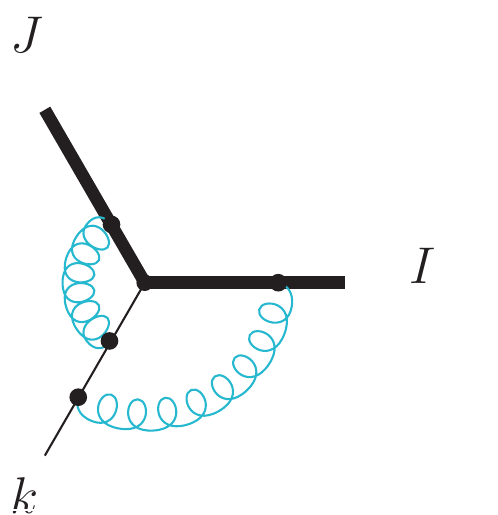}}} -\vcenter{\hbox{\includegraphics[width=1.5cm]{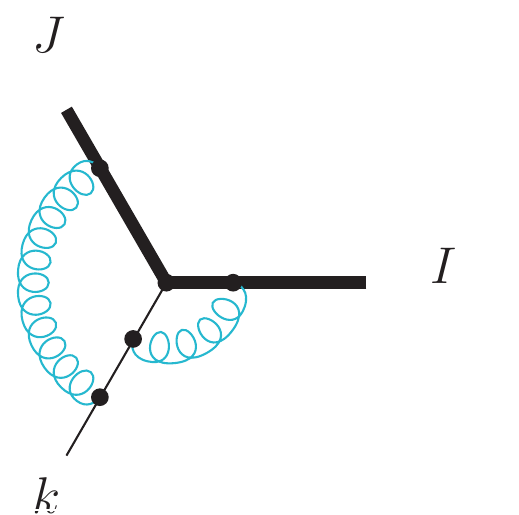}}}  \right)\,,
       \\[.15cm]
       {\Y}_{(KI)(KJ)}^{\left\{C_N,H\right\}}=\,\,&\frac{1}{2}\tilde{{\Y}}_{(KI)(KJ)}^{\left\{C_N,H\right\}}-\frac{1}{2}\tilde{{\Y}}_{(KJ)(KI)}^{\left\{C_N,H\right\}}=\frac{1}{2}\vcenter{\hbox{\includegraphics[width=1.5cm]{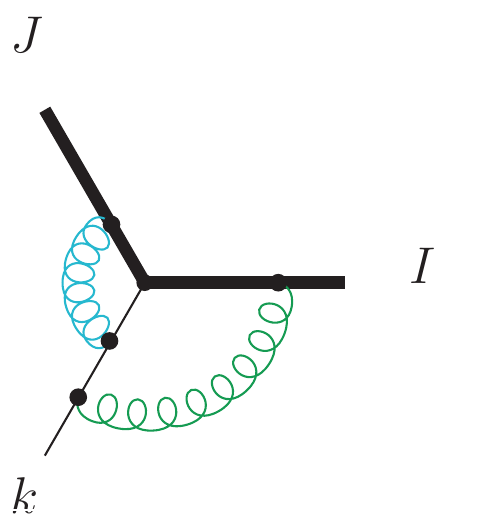}}} -\frac{1}{2}\vcenter{\hbox{\includegraphics[width=1.5cm]{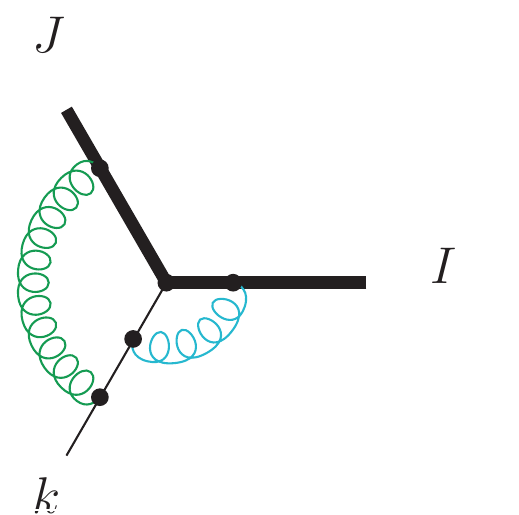}}}\,,
        \\[.15cm]
    {\Y}_{(KI)(KJ)}^{\left\{C_N,C_N\right\}}=\,\,&\frac{1}{2}\left[\tilde{{\Y}}_{(KI)(KJ)}^{[C_N,C_N]}-\tilde{{\Y}}_{(KJ)(KI)}^{[C_N,C_N]}\right]=\frac{1}{2}\left(\vcenter{\hbox{\includegraphics[width=1.5cm]{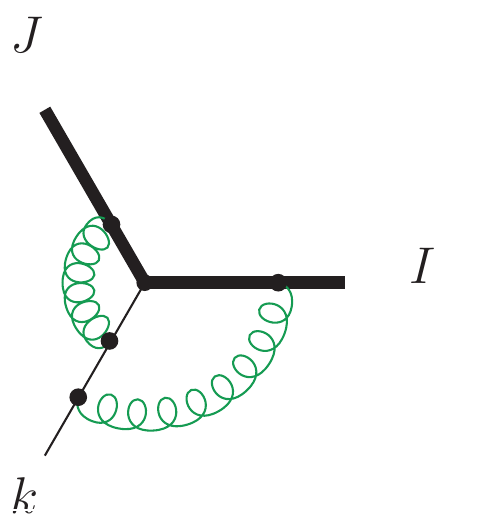}}} -\vcenter{\hbox{\includegraphics[width=1.5cm]{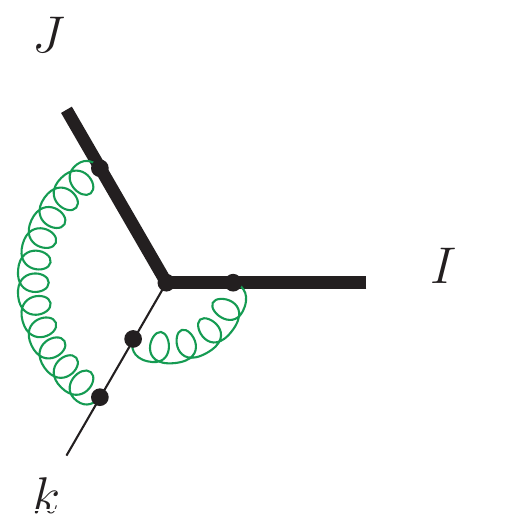}}}  \right)\,\,.
    \end{split}
\end{align}
Lastly, the UV regions, respectively, are defined as follows. For the $(IJ)(IK)$ web:
\begin{align}
    \begin{split}
      {\Y}_{(IJ)(IK)}^{\left\{C_{N},H_{\UV}\right\}}=\,\,&\frac{1}{2}\left[\tilde{{\Y}}_{(IJ)(IK)}^{[H_{\UV},C_{N}]}-\tilde{{\Y}}_{(IK)(IJ)}^{[C_{N},H_{\UV}]}\right]= \frac{1} {2}\left(\vcenter{\hbox{\includegraphics[width=1.5cm]{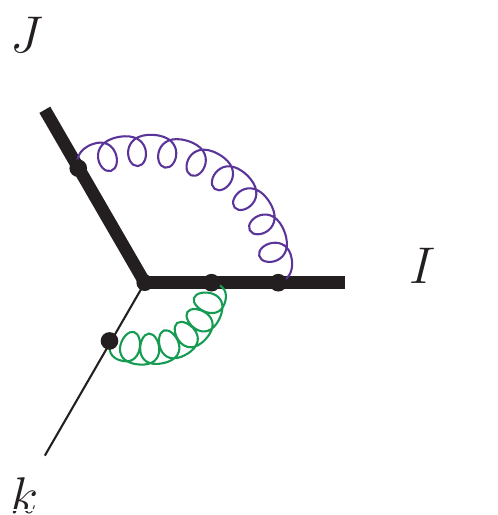}}} -\vcenter{\hbox{\includegraphics[width=1.5cm]{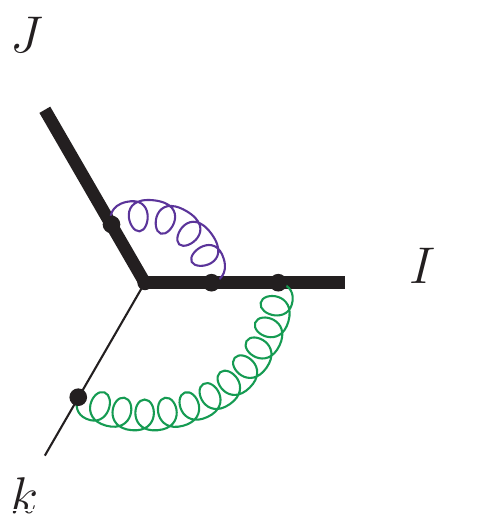}}}\right) \,,
 \end{split}
\end{align}
For the $(JK)(JI)$ web:
\begin{align}
    \begin{split}
        {\Y}_{(JK)(JI)}^{\left\{C_{N},H_{\UV}\right\}}=\,\,&\frac{1}{2}\left[\tilde{{\Y}}_{(JK)(JI)}^{[C_{N},H_{\UV}]}-\tilde{{\Y}}_{(JI)(JK)}^{[H_{\UV},C_{N}]}\right]= \frac{1} {2}\left(\vcenter{\hbox{\includegraphics[width=1.5cm]{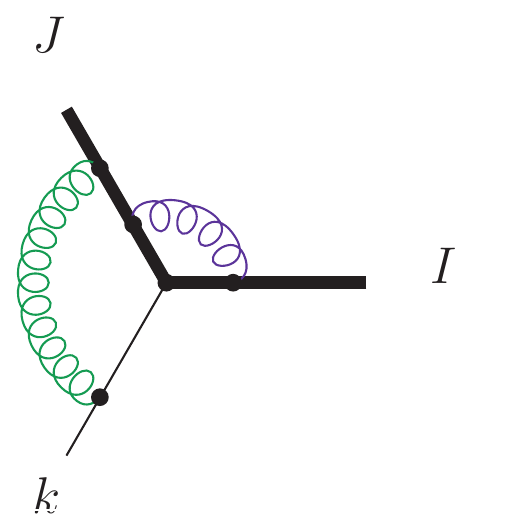}}} -\vcenter{\hbox{\includegraphics[width=1.5cm]{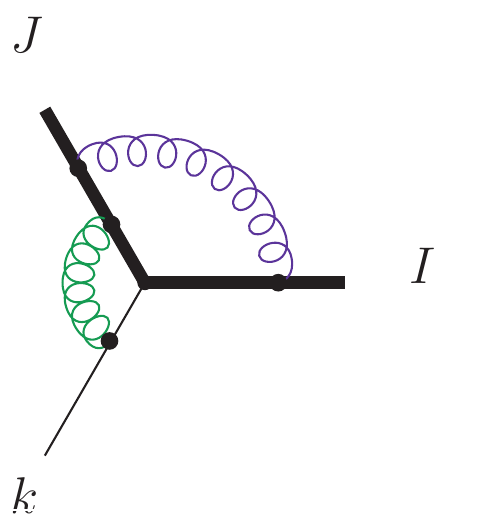}}}\right) \,,  
         \end{split}
\end{align}
and, finally, for the $(KI)(KJ)$ web:
\begin{align}
    \begin{split}
        {\Y}_{(KI)(KJ)}^{\left\{C_{\UV},H\right\}}=\,\,&\frac{1}{2}\left[\tilde{{\Y}}_{(KI)(KJ)}^{[C_{\UV},H]}-\tilde{{\Y}}_{(KJ)(KI)}^{[H,C_{\UV}]}\right]+\frac{1}{2}\left[\tilde{{\Y}}_{(KI)(KJ)}^{[H,C_{\UV}]}-\tilde{{\Y}}_{(KJ)(KI)}^{[C_{\UV},H]}\right]
       \\[.15cm]
       =\,\,&\frac{1} {2}\left(\vcenter{\hbox{\includegraphics[width=1.5cm]{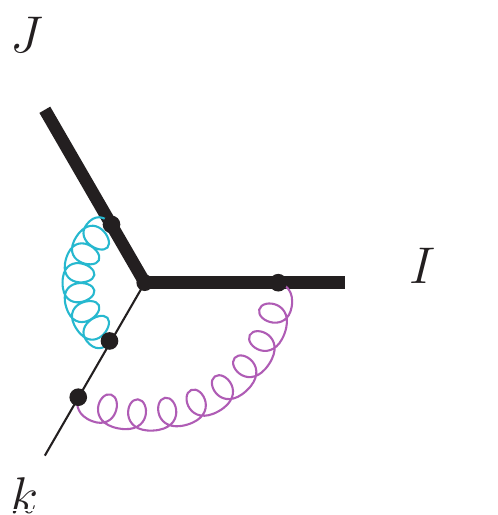}}} -\vcenter{\hbox{\includegraphics[width=1.5cm]{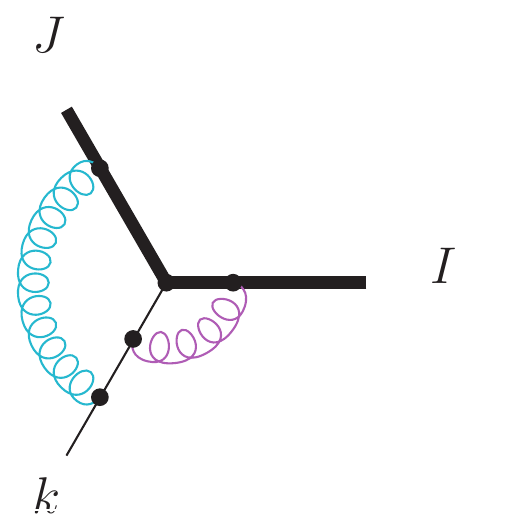}}}\right) + \frac{1} {2}\left(\vcenter{\hbox{\includegraphics[width=1.5cm]{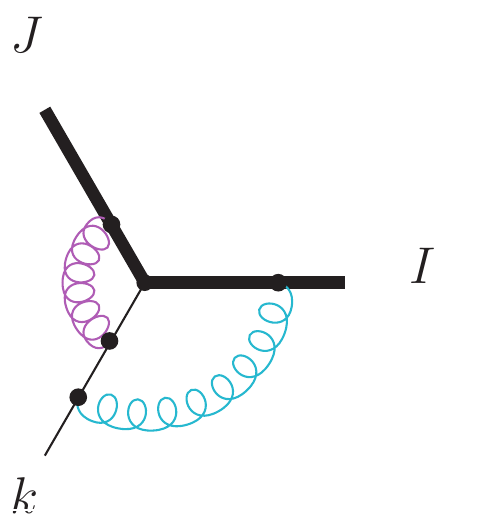}}} -\vcenter{\hbox{\includegraphics[width=1.5cm]{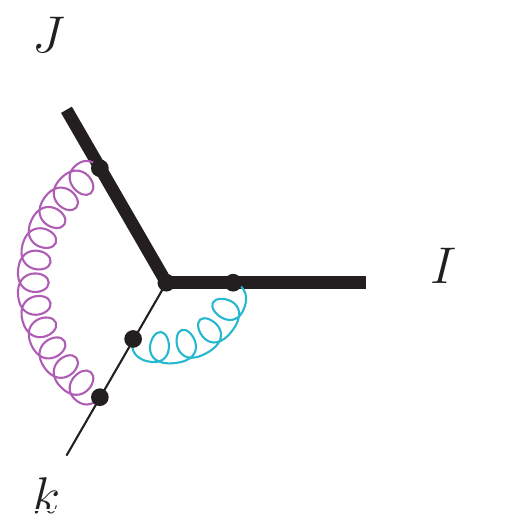}}}\right)\,.
    \end{split}
\end{align}

With the above definitions in place, according to the MoR, the asymptotic expansions of the three webs are, respectively,  
\begin{subequations}
\label{exp211}
\begin{align}
\begin{split}
   {\cal T}_{\lambda} \left[{\Y}_{(IJ)(IK)}\right]=\,\,&\lambda^{-\epsilon}{\Y}_{(IJ)(IK)}^{\left\{C_{\IR},H\right\}}+{\Y}_{(IJ)(IK)}^{\left\{H,H\right\}}+{\Y}_{(IJ)(IK)}^{\left\{C_N,H\right\}}+\lambda^{2\epsilon}{\Y}_{(IJ)(IK)}^{\left\{C_{N},H_{\UV}\right\}}
   \end{split}
   \\[.25cm]
   \begin{split}
      {\cal T}_{\lambda} \left[{\Y}_{(JK)(JI)}\right]=\,\,&\lambda^{-\epsilon}{\Y}_{(JK)(JI)}^{\left\{C_{\IR},H\right\}}+{\Y}_{(JK)(JI)}^{\left\{H,H\right\}}+{\Y}_{(JK)(JI)}^{\left\{C_N,H\right\}}+\lambda^{2\epsilon}{\Y}_{(JK)(JI)}^{\left\{C_{N},H_{\UV}\right\}}  
   \end{split}
   \\[.25cm]
\begin{split}
     {\cal T}_{\lambda} \left[{\Y}_{(KI)(KJ)}\right]=\,\,&\lambda^{-2\epsilon}{\Y}_{(KI)(KJ)}^{\left\{C_{\IR},C_{\IR}\right\}}+\lambda^{-\epsilon}{\Y}_{(KI)(KJ)}^{\left\{C_{N},C_{\IR}\right\}}+\lambda^{-\epsilon}{\Y}_{(KI)(KJ)}^{\left\{C_{\IR},H\right\}}
     \\[.25cm]
     &+{\Y}_{(KI)(KJ)}^{\left\{H,H\right\}}+{\Y}_{(KI)(KJ)}^{\left\{C_N,H\right\}}+{\Y}_{(KI)(KJ)}^{\left\{C_N,C_N\right\}}+\lambda^{2\epsilon}{\Y}_{(KI)(KJ)}^{\left\{C_{\UV},H\right\}}
\end{split}
\end{align}
\end{subequations}

Let us now turn to discuss the results for the asymptotic expansion of each of the three webs as computed region by region in appendices~\ref{app:2loopIR},~\ref{app:2loopNe} and~\ref{app:2loopUV},
and then assembled according to eq.~(\ref{exp211}).

Starting with the $(IJ)(IK)$ web,
the hard region $ {\Y}_{(IJ)(Ik)}^{\{H,H\}}$ can be found in eq.~\eqref{211Hard}. The leading order is a double pole in $\epsilon$, which will be cancelled by other regions. The IR region,  ${\Y}_{(IJ)(IK)}^{\left\{C_{\IR},H\right\}}$, the neutral region, ${\Y}_{(IJ)(IK)}^{\left\{C_N,H\right\}}$, and the UV region, ${\Y}_{(IJ)(IK)}^{\left\{C_{N},H_{\UV}\right\}}$, can be found in eqs.~\eqref{HCIR211},~\eqref{211CNH} and~\eqref{211CNHUV}, respectively. By summing up all the regions contributing to the right-hand side of eq.~\eqref{exp211}, we get the following result:
\begin{align}
\label{211res}
\begin{split}
    {\cal T}_{\lambda} \left[{\Y}_{(IJ)(IK)}\right]=\,\,&\left(\frac{\alpha_s}{4\pi}\right)^2\left(\frac{\bar{m}^2}{\mu^2}\right)^{-2\epsilon}\frac{1+\alpha_{IJ}^2}{1-\alpha_{IJ}^2}\frac{1}{\epsilon}\bigg\{-\log \left(\alpha _{IJ}\right) \left[\frac{1}{4} \log ^2\left(\frac{\lambda ^2}{y_{IJk}}\right)+\frac{2}{3} \pi ^2\right]
    \\&-\frac{1}{2}\log \left(\frac{\lambda ^2}{y_{IJk}}\right) \bigg[M_{100}\left(\alpha _{IJ}\right)+V_1\left(\alpha _{IJ}\right)\bigg]\bigg\}+{\cal O}(\epsilon^0)+{\cal O}(\lambda)\,,
    \end{split}
\end{align}
where the functions $M_{100}$ and $V_1$ can be found in appendix~\ref{app:convfunc}.
Notice that $\log\left(\lambda\right)$ is not completely cancelled in the sum of regions. The result in eq.~\eqref{211res} is in agreement with the expansion performed on the original web ${\Y}_{(IJ)(IK)}$ defined with three timelike lines. The latter was computed in refs.~\cite{Mitov:2009sv,Ferroglia:2009ii,Mitov:2010xw,Gardi:2013saa}, and takes the form\footnote{In ref.~\cite{Gardi:2013saa}, eq.~\eqref{211res} is written in terms of a function $S_1(\alpha)$ as well as ordinary logarithms. The function $S_1(\alpha)$ defined there is exactly $M_{100}\left(\alpha \right)+V_1\left(\alpha \right)$ in the present paper.  }
\begin{align}
\label{allTimelikeIJIK}
\begin{split}
  {\Y}_{(IJ)(IK)}=\,\,& \frac{1}{2}\left\{\vcenter{\hbox{\includegraphics[width=1.5cm]{fig/211IJKa.pdf}}} -\vcenter{\hbox{\includegraphics[width=1.5cm]{fig/211IJKb.pdf}}} \right\}
  \\=\,\,&\left(\frac{\alpha_s}{4\pi}\right)^2\left(\frac{\bar{m}^2}{\mu^2}\right)^{-2\epsilon}\frac{1+\alpha_{IJ}^2}{1-\alpha_{IJ}^2}\frac{1+\alpha_{IK}^2}{1-\alpha_{IK}^2}\frac{1}{\epsilon}\bigg\{\log\left(\alpha_{IJ}\right)\bigg[M_{100}\left(\alpha_{IK}\right)+V_1\left(\alpha_{IK}\right)\bigg]
  \\&-\log\left(\alpha_{IK}\right)\bigg[M_{100}\left(\alpha_{IJ}\right)+V_1\left(\alpha_{IJ}\right)\bigg]\bigg\}+{\cal O}(\epsilon^0)\,.
  \end{split}
\end{align}

The expression for the asymptotic expansion of the second web, ${\Y}_{(JK)(JI)}$, can be obtained from the above by an $(I,J)$ permutation. Note that the neutral region ${\Y}_{(IJ)(IK)}^{\left\{C_N,H\right\}}$ is $(I,J)$ symmetric, and therefore it will cancel upon summing up the two webs.

Finally, consider the $(KI)(KJ)$ web. We first observe that the hard region $ {\Y}_{(kI)(kJ)}^{\{H,H\}}$ is even more singular than in the previous cases considered, however, the expression is simpler (see eq.~\eqref{112Hard}),
\begin{align}
\label{hard121}
     \begin{split}
         {\Y}_{(kI)(kJ)}^{\{H,H\}}=\,\,&\left(\frac{\alpha_s}{4\pi}\right)^2\left(\frac{\bar{m}^2}{\mu^2}\right)^{-2\epsilon}\!\log(y_{IJk})\bigg\{\frac{1}{\epsilon^3}+\frac{1}{\epsilon}\left[\frac{1}{3}\log^2(y_{IJk})+\frac{11}{6}\pi^2\right]\bigg\}+{\cal O}(\epsilon^0)
         \\
         &
         +{\cal O}(\lambda)\,.
     \end{split}
 \end{align}
The strict limit of this web contains a triple pole in  $\epsilon$. Similarly to the connected web, $W_{(IJK)}$, the other neutral regions are vanishing, 
\begin{align}
\label{121neutr}
   {\Y}_{(kI)(kJ)}^{\{C_N,C_N\}}={\Y}_{(kI)(kJ)}^{\{C_N,H\}}=0\,. 
\end{align}
In addition to these neutral regions, the purely IR collinear region is also trivial due to the antisymmetry,
\begin{align}
\label{CNCN_MGEW}
   {\Y}_{(kI)(kJ)}^{\{C_{\IR},C_{\IR}\}}=0\,. 
\end{align}
The remaining IR regions of this web, ${\Y}_{(kI)(kJ)}^{\{C_N,C_{\IR}\}}$ and ${\Y}_{(kI)(kJ)}^{\{C_{\IR},H\}}$ 
are given in eqs.~\eqref{112CNCIR} and~\eqref{112CIRH}, respectively, while the UV region  ${\Y}_{(kI)(kJ)}^{\{C_{\UV},H\}}$ can be found in eq.~\eqref{112CUVH}.
 The sum of all the regions of this web is then,
 \begin{align}
 \begin{split}
     {\cal T}_{\lambda} \left[{\Y}_{(KI)(KJ)}\right]=\,\,&\left(\frac{\alpha_s}{4\pi}\right)^2\left(\frac{\bar{m}^2}{\mu^2}\right)^{-2\epsilon}\frac{1}{\epsilon}\log(y_{IJk})\bigg[\frac{1}{4}\log^2(y_{IJk})+\frac{2}{3}\pi^2-\log^2(\lambda)\bigg]
     \\&+{\cal O}(\epsilon^0)+{\cal O}(\lambda)\,,
     \end{split}
 \end{align}
 which also matches the small $\beta_K^2$ expansion performed on the expression for the web ${\Y}_{(KI)(KJ)}$ with three timelike lines (a permutation of the expression quoted in eq.~(\ref{allTimelikeIJIK}) above).
 
Finally, summing up the asymptotic expansions for the three multiple-gluon-exchange webs we obtain
\begin{align}
\begin{split}
    {\cal T}_{\lambda}\left[{\Y}_{(IJ)(IK)}+{\Y}_{(JK)(JI)}+{\Y}_{(KI)(KJ)}\right]=&
    \\[.15cm]&\hspace{-60mm}\left(\frac{\alpha_s}{4\pi}\right)^2\left(\frac{\bar{m}^2}{\mu^2}\right)^{-2\epsilon}\frac{1}{\epsilon}\log(y_{IJk})\bigg\{\frac{1+\alpha_{IJ}^2}{1-\alpha_{IJ}^2}\bigg[M_{100}(\alpha_{IJ})+V_1(\alpha_{IJ})+2\log(\lambda)\log(\alpha_{IJ})\bigg]
    \\[.15cm]
    &\hspace{-10mm} +\frac{1}{4}\log^2(y_{IJk})+\frac{2}{3}\pi^2-\log^2(\lambda)\bigg\}+{\cal O}(\epsilon^0)+{\cal O}(\lambda)\,,
    \end{split}
\end{align}
where the functions $M_{100}$ and $V_1$ can be found in appendix~\ref{app:convfunc}. 

To summarize, we have confirmed for each separate web, that the MoR successfully reproduces the 
small-$\beta_K^2$ expansion of the original expression obtained with timelike lines. In particular, all higher-order poles in $\epsilon$ cancel between the regions, leaving behind the correct single pole, which eventually contributes to the soft anomalous dimension.

Comparing tables~\ref{tab:IJIKWebRegions}, \ref{tab:IKIJWebRegions} and~\ref{tab:KIKJWebRegions} to table~\ref{tab:twolooptopregionv}, we observe that all the loop-momentum modes, and also all their specific pairings into regions for the three web functions ${\Y}_{(IJ)(IK)}$, ${\Y}_{(JK)(JI)}$ and ${\Y}_{(KI)(KJ)}$ have already been encountered in the case of the connected web ${\Y}_{(IJK)}$. Because these four webs also share the same colour structure,  $\mathbf{T}_{(IJK)}$, this means, in particular, that their contributions from separate regions can be combined into \emph{Region functions} ${\cal R}^R_{(IJK)}$ with the dependence on $\lambda^{n_R\epsilon}$ factored out, as in eq.~(\ref{RegionFunction}). With these one can further form invariant region functions~$\mathbb{R}^{(n_R)}_{(IJK)}$, as we do in the next section.

\subsubsection{Tripole region functions at two loops}
\label{sec:TripoleRegionFunctions}

Let us now use the results computed in the previous sections to construct the complete asymptotic expansion of the exponent of the correlator in eq.~(\ref{nloopexp}). 
Rather than considering the connected and multiple-gluon-exchange webs separately, we will combine them into ordinary region functions, ${\cal R}^R$, and invariant ones, ${\mathbb R}^{(n_R)}$, according to eq.~(\ref{RegionFunction}).

The gauge-invariant region functions for the two-loop colour tripole $\mathbf{T}_{(IJK)}$ are given by the asymptotic expansion of all four contributing webs,  
\begin{align}
\label{regionfunctionsumMathbb}
\begin{split}
    {\cal T}_{\lambda}\left[{\Y}_{(IJK)}+{\Y}_{(IJ)(IK)}+{\Y}_{(Jk)(JI)}+{\Y}_{(kI)(kJ)}\right]&
    \\[.15cm]
    &\hspace{-60mm}=\sum_{R}\lambda^{n_{R}\epsilon}{\cal R}_{(IJk)}^{R}
    =\sum_{n_R}\lambda^{n_{R}\epsilon}{\mathbb R}_{(IJk)}^{(n_R)}
    \\[.15cm]&\hspace{-60mm}
    =
    \lambda^{-2\epsilon} {\mathbb R}_{(IJk)}^{(-2)}
    +\lambda^{-\epsilon} {\mathbb R}_{(IJk)}^{(-1)}
    +  {\mathbb R}_{(IJk)}^{(0)}
    +\lambda^{2\epsilon}{\mathbb R}_{(IJk)}^{(2)}\,,
    \end{split}
\end{align}
with
\begin{subequations}
\label{mathbbRdef}
    \begin{align}
    {\mathbb R}_{(IJk)}^{(-2)} &= {\cal R}_{(IJk)}^{\{C_{\IR},C_{\IR}\}}\label{mathbbRminus2}\\
    {\mathbb R}_{(IJk)}^{(-1)} &= {\cal R}_{(IJk)}^{\{C_{\IR},H\}}
    +{\cal R}_{(IJk)}^{\{C_{N},C_{\IR}\}} \label{mathbbRminus1}\\
     {\mathbb R}_{(IJk)}^{(0)} &=  {\cal R}_{(IJk)}^{\{H,H\}}+{\cal R}_{(IJk)}^{\{C_{N},H\}}+{\cal R}_{(IJk)}^{\{C_{N},C_{N}\}} \label{mathbbR0}\\
     {\mathbb R}_{(IJk)}^{(2)} &= {\cal R}_{(IJk)}^{\{C_{N},H_{\UV}\}}+{\cal R}_{(IJk)}^{\{C_\UV,H\}} 
     \label{mathbbR2}
\end{align}
\end{subequations}
where the negative, zero and positive superscripts of the invariant region function $\mathbb{R}_{(IJk)}^{(n_R)}$ correspond respectively to IR regions, neutral regions and UV regions. 
Note that the subscripts of all region functions are the same: they indicate the colour structure they are associated with, $\mathbf{T}_{(IJK)}$, in contrast to the subscript of ${\Y}$, which correspond to specific webs.

The IR region functions appearing in the first line of eqs.~\eqref{mathbbRminus2} and~\eqref{mathbbRminus1} are given by 
 \begin{subequations}
   \begin{align}
   \begin{split}
   {\cal R}_{(IJk)}^{\left\{C_{\IR},C_{\IR}\right\}}=\,\,&{\Y}_{(IJk)}^{\left\{C_{\IR},C_{\IR}\right\}}+{\Y}_{(kI)(kJ)}^{\left\{C_{\IR},C_{\IR}\right\}}
    \end{split}
   \\[.2cm]
   \begin{split}
   \label{eq:C_IRH}
    {\cal R}_{(IJk)}^{\left\{C_{\IR},H\right\}}=\,\,& {\Y}_{(IJk)}^{\left\{C_{\IR},H\right\}}+{\Y}_{(IJ)(Ik)}^{\left\{C_{\IR},H\right\}}+{\Y}_{(Jk)(JI)}^{\left\{C_{\IR},H\right\}}+{\Y}_{(kI)(kJ)}^{\left\{C_{\IR},H\right\}}
     \end{split}
   \\[.2cm]
   \begin{split}
   {\cal R}_{(IJk)}^{\left\{C_N,C_{\IR}\right\}}=\,\,& {\Y}_{(IJk)}^{\left\{C_N,C_{\IR}\right\}}+{\Y}_{(kI)(kJ)}^{\left\{C_N,C_{\IR}\right\}}
     \end{split}
     \end{align}
     \end{subequations}
     All of these IR regions are non-trivial and contribute to the correlator. The results of the region functions 
     ${\cal R}_{(IJk)}^{\left\{C_{\IR},C_{\IR}\right\}}$,
     ${\cal R}_{(IJk)}^{\left\{C_{\IR},H\right\}}$ and 
     ${\cal R}_{(IJk)}^{\left\{C_N,C_{\IR}\right\}}$ 
     can be found in eqs.~\eqref{RegionCIRCIR},~\eqref{RegionCIRH} and~\eqref{RegionCNCIR}, respectively.

In eq.~\eqref{mathbbR0}, the neutral region functions are 
\begin{subequations}
\label{twoloopallregions}
\begin{align}
\begin{split}
    {\cal R}_{(IJk)}^{\left\{H,H\right\}}=\,\,& {\Y}_{(IJk)}^{\left\{H,H\right\}}+{\Y}_{(IJ)(Ik)}^{\left\{H,H\right\}}+{\Y}_{(Jk)(JI)}^{\left\{H,H\right\}}+{\Y}_{(kI)(kJ)}^{\left\{H,H\right\}}
    \end{split}
   \\[.2cm]
   \begin{split}
\label{CNH}
   {\cal R}_{(IJk)}^{\left\{C_N,H\right\}}=\,\,& {\Y}_{(IJk)}^{\left\{C_N,H\right\}}+{\Y}_{(IJ)(Ik)}^{\left\{C_N,H\right\}}+{\Y}_{(Jk)(JI)}^{\left\{C_N,H\right\}}+{\Y}_{(kI)(kJ)}^{\left\{C_N,H\right\}}
     \end{split}
   \\[.2cm]
   \begin{split}
   {\cal R}_{(IJk)}^{\left\{C_N,C_N\right\}}=\,\,&{\Y}_{(IJk)}^{\left\{C_N,C_N\right\}}+{\Y}_{(kI)(kJ)}^{\left\{C_N,C_N\right\}}\,.
    \end{split}
    \end{align}
    \end{subequations}
As discussed in previous sections, the neutral regions except for the hard are vanishing, see eqs.~\eqref{111neutr} and~\eqref{121neutr}, so 
\begin{align}
     {\mathbb R}_{(IJk)}^{(0)}={\cal R}_{(IJk)}^{\{H,H\}}.
\end{align}
The result of the hard region $ {\cal R}_{(IJk)}^{\{H,H\}}$ can be found in eq.~\eqref{RegionHH}. 

The two UV region functions in  eq.~\eqref{mathbbR2} are 
\begin{subequations}
    \begin{align}
   \begin{split}
 {\cal R}_{(IJk)}^{\left\{C_N,H_{\UV}\right\}}=\,\,&{\Y}_{(IJk)}^{\left\{C_N,H_{\UV}\right\}}+{\Y}_{(IJ)(Ik)}^{\left\{C_N,H_{\UV}\right\}}+{\Y}_{(Jk)(JI)}^{\left\{C_N,H_{\UV}\right\}}
   \end{split}
   \\[.2cm]
   \begin{split}
  {\cal R}_{(IJk)}^{\{C_{\UV},H\}}=\,\,& {\Y}_{(IJk)}^{\left\{C_{\UV},H\right\}}+{\Y}_{(kI)(kJ)}^{\left\{C_{\UV},H\right\}}\,.
   \end{split}
 \end{align}
 \end{subequations}
The results for ${\cal R}_{(IJk)}^{\left\{C_N,H_{\UV}\right\}}$ 
and ${\cal R}_{(IJk)}^{\left\{C_{\UV},H\right\}}$ 
can be found in eq.~\eqref{RegionCNHUV} and~\eqref{RegionCUVH}, respectively. 
One can check that the two UV regions cancel each other on the correlator level,
\begin{align}
\label{UVcancel}
    {\mathbb R}_{(IJk)}^{(2)}={\cal R}_{(IJk)}^{\{C_{\UV},H\}}+ {\cal R}_{(IJk)}^{\{C_{N},H_{\UV}\}}=0\,.
\end{align}
The UV modes completely disappear as this invariant function ${\mathbb R}_{(IJk)}^{(2)}$ is vanishing. Note that this is not true on the web-by-web level.

Finally, at the correlator level, accounting for the vanishing UV contribution and all of the trivial neutral regions with the exception of the hard one, the asymptotic expansion reads
\begin{align}
\label{expanstwoloop}
\begin{split}
    {\cal T}_{\lambda}\left[{\Y}_{(IJK)}+{\Y}_{(IJ)(IK)}+{\Y}_{(Jk)(JI)}+{\Y}_{(kI)(kJ)}\right]&
    \\[.15cm]&\hspace{-60mm}
    =
    \lambda^{-2\epsilon} {\cal R}_{(IJk)}^{\{C_{\IR},C_{\IR}\}}
    +\lambda^{-\epsilon}\left({\cal R}_{(IJk)}^{\{C_{\IR},H\}}
    +{\cal R}_{(IJk)}^{\{C_{N},C_{\IR}\}}\right)+ {\cal R}_{(IJk)}^{\{H,H\}}
    \\  =
    \lambda^{-2\epsilon}  {\mathbb R}_{(IJk)}^{(-2)}
    +\lambda^{-\epsilon} {\mathbb R}_{(IJk)}^{(-1)}+  {\mathbb R}_{(IJk)}^{(0)}\,.
   \end{split}
\end{align}
The surviving modes at the correlator level are $H$, $C_N$ and $C_{\IR}$ which have been all observed at one loop. The remaining region functions all contain triple poles in $\epsilon$. The results of the hard region ${\cal R}_{(IJk)}^{\{H,H\}}$ and the sum of the IR regions at the leading order in power expansion are
\begin{align}
\label{allHard}
\begin{split}
 {\mathbb R}_{(IJk)}^{(0)}=\,\,&\left(\frac{\alpha_s}{4\pi}\right)^2\left(\frac{\bar{m}^2}{\mu^2}\right)^{-2\epsilon}\bigg\{\frac{1}{\epsilon^3}\log(y_{IJk})
 \\&-\frac{1}{\epsilon^2}\bigg[\frac{1-y_{IJk}}{1+y_{IJk}}U_1(y_{IJk})+2\frac{1+\alpha_{IJ}^2}{1-\alpha_{IJ}^2}\log(y_{IJk})\log(\alpha_{IJ})\bigg]
       \\&+\frac{1}{\epsilon}\bigg[\frac{1+\alpha_{IJ}^2}{1-\alpha_{IJ}^2}\log(y_{IJk})\bigg(2V_1(\alpha_{IJ})+M_{100}(\alpha_{IJ})\bigg)
       \\&\hspace{10mm}-\frac{1}{2}\frac{1-y_{IJk}}{1+y_{IJk}}U_2(\alpha_{IJ},y_{IJk})+2 \log \left(y_{IJk}\right)\log^2\left(\alpha _{IJ}\right) 
       \\&\hspace{10mm}-\frac{1}{3}  \log ^3\left(y_{IJk}\right)+\frac{7}{6} \pi ^2 \log \left(y_{IJk}\right)\bigg]\bigg\}+{\cal O}(\epsilon^0)+{\cal O}(\lambda),
       \end{split}
       \end{align}
and       
       \begin{align}
       \label{allIR}
    \begin{split}
        \lambda^{-2\epsilon} {\mathbb R}_{(IJk)}^{(-2)}
    +\lambda^{-\epsilon}{\mathbb R}_{(IJk)}^{(-1)}=&\left(\frac{\alpha_s}{4\pi}\right)^2\left(\frac{\bar{m}^2}{\mu^2}\right)^{-2\epsilon}\bigg\{-\frac{1}{\epsilon^3}\log(y_{IJk})
   \\&+\frac{1}{\epsilon^2}\bigg[\frac{1-y_{IJk}}{1+y_{IJk}}U_1(y_{IJk})+2\frac{1+\alpha_{IJ}^2}{1-\alpha_{IJ}^2}\log(y_{IJk})\log(\alpha_{IJ})\bigg]
       \\&+\frac{1}{\epsilon}\bigg[-\frac{1+\alpha_{IJ}^2}{1-\alpha_{IJ}^2}\log(y_{IJk})\bigg(V_1(\alpha_{IJ})+2\log(\lambda)\log(\alpha_{IJ})\bigg)
       \\&+\frac{1}{2}\frac{1-y_{IJk}}{1+y_{IJk}}U_2(\alpha_{IJ},y_{IJk})+ \log ^2(\lambda ) \log \left(y_{IJk}\right)
       \\&+\frac{1}{12} \log ^3\left(y_{IJk}\right)-\frac{1}{2}\pi ^2 \log \left(y_{IJk}\right)\bigg]\bigg\}+{\cal O}(\epsilon^0)+{\cal O}(\lambda).
    \end{split}   
\end{align}
The expansion of the correlator is then 
\begin{align}
    \begin{split}
    \label{w2ex}
  \left(\frac{\alpha_s}{4\pi}\right)^2 {\cal T}_{\lambda}\left[w^{(2)}\right]=\,\,&\sum_{I<J<K}\mathbf{T}_{IJK}{\cal T}_{\lambda}\left[{\Y}_{(IJk)}^{(2)}+{\Y}_{(IJ)(IK)}^{(2)}+{\Y}_{(JK)(JI)}^{(2)}+{\Y}_{(KI)(KJ)}^{(2)}\right]+\cdots
  \\=\,\,&\sum_{I<J}\mathbf{T}_{IJk}\left(\frac{\alpha_s}{4\pi}\right)^2\left(\frac{\bar{m}^2}{\mu^2}\right)^{-2\epsilon}\frac{1}{\epsilon}\log(y_{IJk})
   \\&\bigg\{\frac{1+\alpha_{IJ}^2}{1-\alpha_{IJ}^2}\bigg[V_1(\alpha_{IJ})+M_{100}(\alpha_{IJ})-2\log(\lambda)\log(\alpha_{IJ})\bigg]
       \\&+ \log ^2(\lambda ) +2 \log^2\left(\alpha _{IJ}\right)-\frac{1}{4} \log ^2\left(y_{IJk}\right)+\frac{2}{3}\pi ^2 \bigg\}+{\cal O}(\epsilon^0)+{\cal O}(\lambda).
       \\&+\cdots.
    \end{split}
 \end{align}
As expected, all triple and double poles in $\epsilon$ cancel in  the sum of  eqs.~\eqref{allHard} and~\eqref{allIR}. Notice that there is still $\lambda$ dependence even at the correlator level. We shall verify below that it is eliminated upon extracting the soft~AD.

We comment that ${\cal T}_{\lambda}\left[w^{(2)}\right]$ in eq.~\eqref{w2ex} does have double poles in $\epsilon$ in colour dipole terms, which we have not written explicitly. These double poles are related exclusively to the running  coupling~\cite{Gardi:2011yz}, and are proportional to $b_0$ of eq.~\eqref{betab0},
\begin{align}
    {\cal T}_{\lambda}\left[w^{(2,-2)}\right]=-\frac{b_0}{2}{\cal T}_{\lambda}\left[w^{(1,-1)}\right].
\end{align}
Focusing here on the tripole colour structure which is free of $b_0$, it will be convenient to simply set $b_0=0$ in the remainder of this section, so the leading order of $ {\cal T}_{\lambda}\left[w^{(2)}\right]$ is single pole in $\epsilon$, as obtained eq.~(\ref{w2ex}). 

We now split the two-loop correlator $w^{(2)}$ into the separate hard and IR region contributions, following  eq.~\eqref{correlatorsplit},
\begin{equation}
   {\cal T}_{\lambda}\left[w^{(2)}\right]=h^{(2)}+r^{(2)}\,,
 \end{equation}
Both $h^{(2)}$ and $r^{(2)}$ contain $\mathbf{T}_{IJk}$, as well as other colour structures which we represent by ellipsis:
 \begin{subequations}
\begin{align}
\begin{split}
\left( \frac{\alpha_s}{4\pi}\right)^2h^{(2)}=\,\,&\sum_{I<J}\left( \frac{\alpha_s}{4\pi}\right)^2h^{(2)}_{(IJk)}+\cdots\,,
\end{split}
\\
\begin{split}
\left( \frac{\alpha_s}{4\pi}\right)^2r^{(2)}=\,\,&\sum_{I<J}\left( \frac{\alpha_s}{4\pi}\right)^2r^{(2)}_{(IJk)}+\cdots\,,
 \end{split}
 \end{align}
\end{subequations}
The tripole-term components may be written in terms of region functions as follows:
\begin{subequations}
\label{twolooprs}
\begin{align}
\begin{split}
   h^{(2)}_{(IJk)}=\textbf{T}_{IJk}{\mathbb R}_{(IJk)}^{(0)}
    \end{split}
    \\
    \begin{split}
      r^{(2)}_{(IJk)}=\textbf{T}_{IJk}\left[\lambda^{-2\epsilon} {\mathbb R}_{(IJk)}^{(-2)}
    +\lambda^{-\epsilon}{\mathbb R}_{(IJk)}^{(-1)}\right].
    \end{split}
\end{align}
\end{subequations}
The kinematic functions in $h_{(IJk)}^{(2)}$ and $r_{(IJk)}^{(2)}$ are exactly given in eq.~\eqref{allHard} and~\eqref{allIR},  respectively. 

Using eq.~\eqref{twoloopGWeb}, the soft~AD is given by 
\begin{align}
\label{Gamma2sr}
\begin{split}
\frac{1}{\epsilon}\Gamma_{\UV}^{(2)}=\,\,&\frac{1}{\epsilon}{\cal T}_{\lambda}\left[-4w^{(2,-1)}-2\left[w^{(1,-1)},w^{(1,0)}\right]\right]
\\=\,\,&-\frac{4}{\epsilon}\left(h^{(2,-1)}+r^{(2,-1)}\right)-\frac{2}{\epsilon}\left[h^{(1,-1)}+r^{(1,-1)},h^{(1,0)}+r^{(1,0)}\right]
\\=\,\,&
\sing
\left(\,\,-4
\left(h^{(2)}+r^{(2)}\right)
-2\left[
\sing \left(
h^{(1)}+r^{(1)}
\right)
, \, \reg \left(
h^{(1)}+r^{(1)}
\right)\right] 
\right)\,,
\end{split}
\end{align}
where we have omitted the trivial combinations $h^{(2,-3)}+r^{(2,-3)}$, $h^{(2,-2)}+r^{(2,-2)}$, and $h^{(1,-2)}+r^{(1,-2)}$ in the second line.
The operators $\sing$ and $\reg$ are defined in eq.~\eqref{defineSingReg}, and the one-loop result can be found in eqs.~\eqref{onelooph} 
and~\eqref{oneloopr}.  Upon substituting these one-loop expressions into eq.~(\ref{Gamma2sr}), and using 
eq.~(\ref{Commutator_of_Dipoles}) to obtain a tripole colour structure from the commutator, we finds that the $\log(\lambda)$ terms appearing in the two-loop expressions cancel by the commutator term, and the soft~AD $\Omega_{(IJk)}$ of eq.~\eqref{mixtwoloop} is exactly reproduced.

\subsection{Neutral modes}
\label{sec:neutralregions}

Among the modes collected in table~\ref{tab:ModesSum}, only three survive at the correlator level, these are the hard mode $H$, the IR-collinear mode $C_{\IR}$ and the neutral collinear mode $C_N$; these are the relevant modes at both one and two loops, see eqs.~\eqref{expansoneloop} and~\eqref{expanstwoloop}. The first two are also ubiquitous in the on-shell expansion of wide-angle scattering amplitudes. However, the neutral collinear mode $C_N$, which features both large and small momentum components,  is special to the expansion of correlators of Wilson lines. 
In this section, we claim that the presence of this second neutral mode, alongside the hard mode, is a consequence of the rescaling symmetry of semi-infinite Wilson lines. 

\subsubsection{Rescaling symmetry and the complementary lightcone expansion}

The momentum-space analysis in section~\ref{sec:one-loop_computation} at one loop, and in sections~\ref{sec:connectedweb} and~\ref{sec:121web} at two loops, was performed in a special frame $O$ where $\beta_k=(\beta_k^+,0,0)$,  with the opposite lightlike direction being $\bar{\beta}_k=(0,\bar{\beta}_k^-,0)$. In this frame, the timelike velocity $\beta_K$ behave as $\beta_K\sim(1,\lambda^2,\lambda)$; it has a large ``plus'' momentum component, and thus is close to the lightlike direction defined by $\beta_k$. The other two timelike velocities are hard, i.e., $\beta_I\sim \beta_J\sim(1,1,1)$. The scalar products behave as
\begin{align}
\label{limi1beta}
    \beta_K^2\sim {\cal O}(\lambda^2)\,,
    \qquad
    \beta_K\cdot\beta_I\sim\beta_K\cdot\beta_J\sim\beta_I\cdot\beta_J\sim \beta_I^2\sim\beta_J^2\sim{\cal O}(\lambda^0)\,.
\end{align}
Because of the rescaling symmetry, the kinematic variables entering the correlator are the three scalar products of the normalized velocities, 
\begin{align}
\label{limi1v}
    v_K\cdot v_I\sim v_K\cdot v_J\sim {\cal O}(\lambda^{-1})\,,
    \qquad
    v_I\cdot v_J\sim{\cal O}(\lambda^0)\,,
\end{align}
where $v_I=\beta_I/\sqrt{\beta_I^2}$.
The scaling law of the normalized velocities in frame $O$ is given by 
\begin{align}
\label{limi1vscale}
    v_K\sim(\lambda^{-1},\lambda,1)
    \qquad
    v_I\sim v_J\sim (1,1,1).
\end{align}
One can of course consider other Lorentz frames, in which the scaling law of the velocities would differ from~(\ref{limi1vscale}), without modifying the invariants (\ref{limi1v}). 

Consider specifically a Lorentz transformation to 
the frame $O'$ which is related to the frame $O$ through the boost $B_3\left(\eta\right)$ along the third\footnote{The third spatial  direction is chosen as the same as that of the spacial momentum of $\beta_k=(\beta_k^+,0,0)$. The lightcone coordinates are defined as 
$$\beta_k^+=\frac{\beta_k^0+\beta_k^3}{\sqrt{2}}\,,
\qquad
\beta_k^-=\frac{\beta_k^0-\beta_k^3}{\sqrt{2}}=0\,.$$
} spatial direction  with rapidity $\eta=\log(\lambda)$. The normalized velocities then become 
\begin{subequations}
\label{vframe}
\begin{align}
    \begin{split}
v_K\sim(\lambda^{-1},\lambda,1)&\xrightarrow{\;B_3\left(\log(\lambda)\right)\;} v_K\sim(1,1,1)\,,
\end{split}
    \\
    \begin{split}
v_I\sim(1,1,1)&\xrightarrow{\;B_3\left(\log(\lambda)\right)\;}v_I\sim(\lambda,\lambda^{-1},1)\,,
    \end{split}
    \\
    \begin{split}
    \label{vJ_Bossted_kbar}
v_J\sim(1,1,1)&\xrightarrow{\;B_3\left(\log(\lambda)\right)\;}v_J\sim(\lambda,\lambda^{-1},1)\,.
    \end{split}
\end{align}
\end{subequations}

We observe that the behaviour of the velocities in the frame $O'$ can also be obtained from a \emph{different physical limit} where $\beta_K$ remains generic, $\beta_K\sim(1,1,1)$, while  
the other two external velocities, $\beta_I$ and $\beta_J$, become collinear to the opposite lightlike direction defined by $\bar{\beta}_k$, i.e., $\beta_I\sim\beta_J\sim(\lambda^2,1,\lambda)$. The Lorentz invariants constructed by ordinary velocities behave as 
\begin{align}
\label{limi2beta}
  \beta_K^2\sim\beta_K\cdot \beta_J\sim\beta_K\cdot\beta_I\sim{\cal O}(\lambda^0)\,,
  \qquad 
\beta_I^2\sim\beta_J^2\sim\beta_I\cdot\beta_J\sim{\cal O}(\lambda^2)\,,
\end{align}
which is a completely different expansion compared with that defined in eq.~\eqref{limi1beta}. In the extreme limit eq.~(\ref{limi2beta}) describes two collinear massless particles $I$ and $J$ interacting with one massive particle $K$ (plus any number of non-coloured particles, sharing the recoil momentum). This can be contrasted to the original limit (\ref{limi1beta}), where a single massless particle $K$ is interacting with two massive particles, $I$ and $J$, scattering at generic angles (plus any number of non-coloured particles). 

This puzzling situation is explained by the fact the behaviour of the scalar products of the normalized velocities (\ref{limi1v}), is consistent with both the original limit we considered, eq.~(\ref{limi1beta}), and the one we formulated in eq.~(\ref{limi2beta}). In other words, knowing only (\ref{limi1v}) one cannot distinguish between the two physical situations.  
This can be most easily understood by re-expressing (\ref{limi1v}) back in terms non-normalized velocities, namely
\begin{align}
\label{limi1v_again}
    \frac{\beta_K\cdot \beta_I}{\sqrt{\beta_K^2\,\beta_I^2}}
    \sim
    \frac{\beta_K\cdot \beta_J}{\sqrt{\beta_K^2\,\beta_J^2}} \sim {\cal O}(\lambda^{-1})
   \,,
    \qquad
    \frac{\beta_I\cdot \beta_J}{\sqrt{\beta_I^2\,\beta_J^2}} \sim{\cal O}(\lambda^0)\,.
\end{align}
It straightforward to check that either the limit of eq.~(\ref{limi1beta}) or that in eq.~(\ref{limi2beta}) can be consistent with eq.~(\ref{limi1v_again}).

Referring to the original expansion in eq.~\eqref{limi1beta} as the \emph{lightcone expansion}, the new expansion defined in eq.~\eqref{limi2beta},  may be called the \emph{complementary lightcone expansion}, referring to the fact that the complementary set of Wilson lines (or particles) to line $K$, are now highly boosted in the ``minus'' direction.
Although these are two distinct limits, because of the rescaling symmetry, the integrals before the expansion really depend only on the scalar products of normalized velocities. Therefore, the two expansions, while having different physical interpretations, are described by the very same result.

Let us point out another instance where the same phenomenon was observed~\cite{Duhr:2025cye,GZ-PRL,GZ-TBP}.
This is the case of the soft~AD of four  coloured particles, one of which ($Q$) is heavy and the rest ($i$, $j$ and $k$) are massless. This object depends on only three rescaling invariant cross ratios~\cite{Liu:2022elt}:
\begin{align}
\label{rIJKQ}
\begin{split}
r_{ijQ}&=\frac{(\beta_i\cdot \beta_j) \beta_Q^2}{(\beta_i\cdot\beta_Q)\,\,(\beta_j\cdot\beta_Q)}\,, 
\\
r_{ikQ}&=\frac{(\beta_i\cdot \beta_k) \beta_Q^2}{(\beta_i\cdot\beta_Q)\,\,(\beta_k\cdot\beta_Q)} \,,
\\
r_{jkQ}&=\frac{(\beta_j\cdot \beta_k) \beta_Q^2}{(\beta_j\cdot\beta_Q)\,\,(\beta_k\cdot\beta_Q)} \,.
\end{split}
\end{align}
As a consequence of the  rescaling symmetry with respect to $\beta_Q$, there are two physically-distinct limits of the velocities,  which correspond to the very same limit of the cross ratios in eq.~(\ref{rIJKQ}), namely \hbox{$r_{ijQ}\sim r_{ikQ}\sim r_{jkQ}\sim\lambda$}, with the ratios between them remaining finite for $\lambda\to 0$. The two limits are the massless limit:
\begin{align}
\beta_Q^2\sim \lambda,
\qquad 
\beta_i\cdot \beta_j\sim 
\beta_i\cdot \beta_k\sim 
\beta_j\cdot \beta_k\sim {\cal O}(\lambda^0)\,,
\end{align}
and the triple collinear limit:
\begin{align}
    \beta_Q^2\sim {\cal O}(\lambda^0),
\qquad 
\beta_i\cdot \beta_j\sim 
\beta_i\cdot \beta_k\sim 
\beta_j\cdot \beta_k\sim \lambda\,,
\end{align} 
where in either of these cases, one considers 
\begin{align}
\beta_i\cdot \beta_Q\sim 
\beta_j\cdot \beta_Q\sim 
\beta_k\cdot \beta_Q\sim 
{\cal O}(\lambda^0)\,.
\end{align}
These two limits represent, respectively, the lightcone expansion of particle $Q$, and the complementary lightcone expansion, where all other particles become collinear. We stress that this is true in the presence of an arbitrary momentum recoil, so this is a non-trivial relation two between physically distinct limits of the anomalous dimension,  which coincide because of the rescaling symmetry. 
The considerations above indicate that this is a general feature.  

\subsubsection{Degeneracy of neutral modes}

Let us return to our original set up of three Wilson lines, one of which is becoming lightlike, and turn to discuss the workings of the MoR. We observe that although the relevant kinematic invariants in eq.~(\ref{limi2beta}) are the same in the two expansions, \emph{the loop-momentum modes are different}. To see this, recall that the two expansions are naturally described in different frames, as we have seen for the velocity components in eq.~(\ref{vframe}). Consider now the three loop-momentum modes appearing in the expansion of the correlator, $H$, $C_{\IR}$ and~$C_N$. In the frame~$O'$, they become,  respectively, 
\begin{subequations}
\label{comlimit}
\begin{align}
    \begin{split}
H\sim(1,1,1)&\xrightarrow{\;B_3\left(\log(\lambda)\right)\;}\bar{C}_N\sim(\lambda,\lambda^{-1},1)\,,
\end{split}
    \\
    \begin{split}
    C_{\IR}\sim (1,\lambda,\sqrt{\lambda})&\xrightarrow{\;B_3\left(\log(\lambda)\right)\;} \bar{C}_{\IR}\sim (\lambda,1,\sqrt{\lambda})\,,
    \end{split}
    \\
    \label{C_N_new frame}C_N\sim(\lambda^{-1},\lambda,1)&\xrightarrow{\;B_3\left(\log(\lambda)\right)\;} H\sim(1,1,1)\,.
\end{align}
\end{subequations}
Notice that the special neutral mode $C_N$ in frame $O$ has become the hard mode in $O'$, which is associated with the strict limit of the complementary lightcone expansion. The two types of neutral modes behave as the hard mode in the two different frames,  respectively. 

With the modes identified in frame $O'$, the complementary lightlike expansion is, for example, at one loop,
\begin{align}
    {\cal T}_{\lambda}\left[{\Y}_{JK}\right]={\cal R}_{(j_{\bar{k}}K)}^{\{\bar{C}_N\}}+{\cal R}_{(j_{\bar{k}}K)}^{\{H\}}+\lambda^{-\epsilon}{\cal R}_{(j_{\bar{k}}K)}^{\{\bar{C}_{\IR}\}}\,,
\end{align}
where the subscript  $J$ has been replaced by $j_{\bar{k}}$, identifying the limit where $v_J$ is highly boosted in the direction collinear to $\bar{\beta}_k$, as in (\ref{vJ_Bossted_kbar}) with $\lambda\to 0$.  
Because the region functions are Lorentz invariant, eq.~(\ref{comlimit}) implies a direct relation between the two expansions:
\begin{subequations}
\begin{align}
    \begin{split}
     {\cal R}_{(j_{\bar{k}}K)}^{\{\bar{C}_N\}}={\cal R}_{(Jk)}^{\{H\}}\,,   
    \end{split}
    \\
    \begin{split}
     {\cal R}_{(j_{\bar{k}}K)}^{\{\bar{C}_{\IR}\}}={\cal R}_{(Jk)}^{\{C_{\IR}\}}\,, 
    \end{split}
    \\
    \begin{split}
     {\cal R}_{(j_{\bar{k}}K)}^{\{H\}}={\cal R}_{(Jk)}^{\{C_N\}}\,.  
    \end{split}
\end{align}
\end{subequations}
In the remaining of this section, we will always use the notations in frame $O$.

We stress that the strict limits of the two expansions are different. At one loop, they are accidentally the same, see the result of ${\cal R}_{(Jk)}^{\{H\}}$ and ${\cal R}_{(Jk)}^{\{C_N\}}$ in eqs.~\eqref{oneloophard} and~\eqref{oneloopNeutC}, respectively.  
At two loops the situation is entirely different. While hard region of the original lightcone expansion, ${\cal R}_{(IJk)}^{\{H,H\}}$, is a non-trivial function given in eq.~\eqref{RegionHH}, the hard region of the complementary lightcone expansion, ${\cal R}_{(IJk)}^{\{C_N,C_N\}}$, is vanishing. 

In fact, the $\{C_N,C_N\}$ region vanishes for each web separately, as reported in eqs.~(\ref{111neutr}) and~(\ref{CNCN_MGEW}) above.
Let us now explain, as an example, the reason why the contribution of this region is vanishing for the 
connected web, i.e., why
\[
{\Y}_{(IJk)}^{\{C_N,C_N\}} =0\,.
\]
In frame $O$, the $C_N$ mode, characterized by a large $q^+$, extracts only the ``minus'' lightcone component of the timelike velocity in the denominator of the eikonal propagator,
\begin{align}
\label{CNsimp}
    \frac{1}{-v\cdot q-1+i\varepsilon}\xrightarrow{\;\text{mode}_q=C_N\;}\frac{\lambda}{-v^-q^++i\varepsilon}+O(\lambda^2)\equiv\frac{\lambda}{-\kappa\bar{\beta}_k\cdot q+i\varepsilon}+O(\lambda^2),
\end{align}
where $\kappa$ is the ratio of $v^-$ and $\bar{\beta}_k^-$. Here the timelike velocity $v\sim(1,1,1)$ can be either~$v_I$ or~$v_J$. 
Equation~(\ref{CNsimp}) leads to a simplification of the integrand.  In particular, the integrand $Y_{(IJK)}$ defined in eq.~\eqref{connIntegrand} becomes
\begin{align}
\label{ComLInt}
\begin{split}
    Y_{(IJK)}\xrightarrow{\;\left\{C_N,C_N\right\}\;}\,\,&
\frac{\lambda^2}{\kappa_I\kappa_J}\,
    \frac{1}{q_I^2+i\varepsilon}\frac{1}{q_J^2+i\varepsilon}\frac{1}{q_K^2+i\varepsilon}
    \\
    &\times
    \left(\frac{1}{-\bar{\beta}_k\cdot q_I+i\varepsilon}\right)
    \left(\frac{1}{-\bar{\beta}_k\cdot q_J+i\varepsilon}\right)
    \left(
    \frac{1}{-v_K\cdot q_K-1+i\varepsilon}
    \right)
    +O(\lambda^3)\,.
    \end{split}
\end{align}
Having factorized $\kappa_I\kappa_J$, the leading term integrates to a constant, and vanishes upon including the $(I,J)$ antisymmetric numerator. 
Finally, we note that eq.~\eqref{ComLInt} is consistent with directly applying the complementary lightcone expansion on $Y_{(IJK)}$.

To conclude, we have shown that the two neutral modes can be interpreted, respectively, as the hard modes of two physically-distinct expansions. We have learnt that the reason these two expansions are linked, is that the integrals only depends on a subset of the kinematic invariants, namely the rescaling-invariant subset.
Rescaling symmetry therefore provides the fundamental reason for the degeneracy of the neutral modes, which was observed already in the Lorentz-invariant parameters-space analysis.   

\section{Renormalization of  correlators with timelike and lightlike Wilson lines}
\label{sec:RenMixedCorr}

In the MoR, the leading-power term of the hard region has a special status as the strict limit, taken at integrand level. In the context of our lightcone expansion of Wilson-line correlators one can therefore view the (leading power term of the) hard region as the computation of a mixed correlator, involving both timelike and lightlike lines. 

For correlators of timelike Wilson lines in a complete regularization scheme, multiplicative renormalizability enables us to extract the soft~AD from the computation of 
the $1/\epsilon$ pole. However, as we have seen in section~\ref{sec:twoloopcomp}, the hard region of the mixed tripole term $\mathbb{R}_{(IJk)}^{(0)}$, where the $k$ line is strictly lightlike, contains a third-order pole in $\epsilon$, see eq.~\eqref{allHard}. Since the tripole is a new colour structure appearing first at two loops, such higher-order poles cannot be attributed to the short-distance renormalization of the strong coupling -- nor any operator -- and hence they immediately violate the multiplicative renormalizability of the correlator.

Fortunately, using MoR, the origin of these extra poles becomes clear, with the modes as well as the regions being classified as IR, neutral, and UV according to their scaling exponent $n_R$ of eq.~(\ref{TransverseScaling_and_nR}). This classification tells us the characteristic loop-momentum virtuality giving rise to the various singularities.  
We will now use this information in order to disentangle between short- and long-distance effects in a correltor containing strictly lightlike Wilson lines. 

\begin{figure}
    \centering \includegraphics[width=0.5\linewidth]{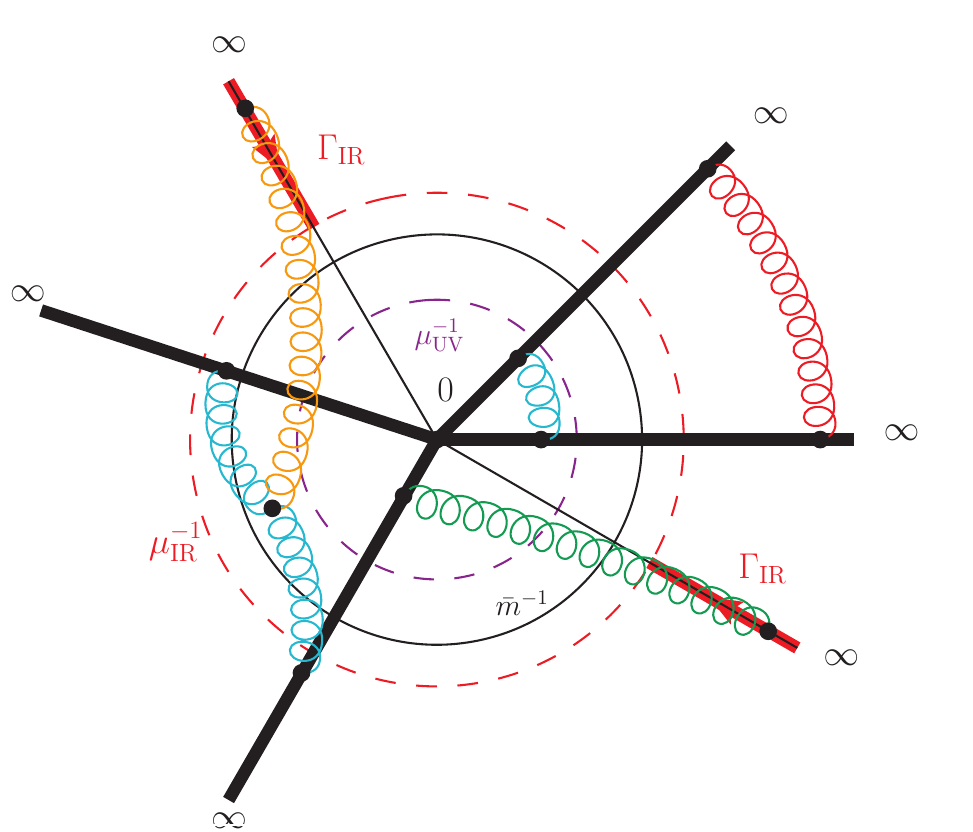}
    \caption{Configuration-space sketch of different modes and the corresponding renormalization of the mixed correlator involving timelike (thick) and lightlike (thin)  Wilson lines. The modes are colour-coded following the conventions of table~\ref{tab:ModesSum}. In particular, the connected tripole is in the $\{H,C_{\IR}\}$ region (one loop is hard and the second is IR-collinear), while the single gluon exchange in green is in a neutral collinear ($C_N$) region.  In turn, the red gluon represents a long-distance (soft) exchange that is removed by the IR regulator, while the blue gluon represents a genuine contribution of short-distance origin to the soft~AD. The two renormalization scales $\muIR$ and $\muUV$ are represented by dashed lines, while the regulator $\bar{m}$ is represented by the full line in between the two.   }
    \label{mixedcorrelator}
\end{figure}

Let us recall the region structure we obtained at one and two loops in section~\ref{sec:twoloopcomp}. At the correlator level UV regions do not contribute (see eq.~\eqref{UVcancel}), leaving behind only neutral
and IR regions. Furthermore, neutral regions other than the hard only contribute to the dipole colour structure, not to the tripole.\footnote{The neutral regions correspond contributions that may not be uniquely identified as either of short or long-distance origin; we will return to discuss non-hard neutral regions later on in this section. }
As a result, a simple picture emerges: tripole singularities arise from just two sources, the hard region, (i.e. the strict limit) of eq.~(\ref{allHard}), and IR regions, eq.~(\ref{allIR}), representing contributions of long-distance origin. It is natural to assume that this broad classification of the origin of singularities applies in general in the lightcone expansion, although here we will only study it at two loops. 

Based on the region analysis, to ``renormalize'' the mixed timelike-lightlike correlator, it is necessary to consider an additional ``renormalization'' procedure at infinity in configuration space, where IR singularities are generated; this is illustrated in figure~\ref{mixedcorrelator} along with some of the relevant regions.
We put the word renormalization in quotation marks to emphasise that this operation does not correspond to short-distance renormalization of any operator. Rather, it is similar to removing long-distance singularities from on-shell scattering amplitudes, see e.g.~\cite{Catani:1998bh,Aybat:2006wq,Becher:2009qa,Gardi:2009qi,Gardi:2009zv}.

We thus propose the following conjecture for the singularity structure of the mixed correlator:
\begin{align}
\label{addren}
\begin{split}
&\left<\phi_{\beta_{1}}\cdots\phi_{\beta_{N}}\Phi^{(m)}_{\beta_{N+1}}\!\cdots\Phi^{(m)}_{\beta_{N+M}}\right>_{\text{ren.}(\muIR, \muUV)}\\
&\hspace*{100pt}=\,\,Z_{\IR}(\muIR)\left<\phi_{\beta_{1}}\cdots\phi_{\beta_{N}}\Phi^{(m)}_{\beta_{N+1}}\!\cdots\Phi^{(m)}_{\beta_{N+M}}\right>Z_{\UV}(\muUV)\,,
\end{split}
\end{align}
with the renormalization factors taking the form

\begin{subequations}
\label{GammaUVIR}
\begin{align}
\begin{split}
\label{GammaUVdef}
 Z_{\UV}(\muUV)&\equiv \text{P}\exp\left[\int_{\muUV}^{\infty}\frac{d\tau}{\tau}\Gamma_{\UV}\right]\,,
\end{split}
\\
\begin{split}
\label{GammaIRdef}
   Z_{\IR}(\muIR)&\equiv\text{P}\exp\left[\int_{0}^{\muIR}\frac{d\tau}{\tau}\Gamma_{\IR}\right]\,,
\end{split}
 \end{align}
 \end{subequations}
 where $\muUV$ and~$\muIR$ serve as two separate factorization scales, such that~\hbox{$\muUV\ge \bar{m}\ge\muIR$}.

\begin{figure}
    \centering \includegraphics[width=0.5\linewidth]{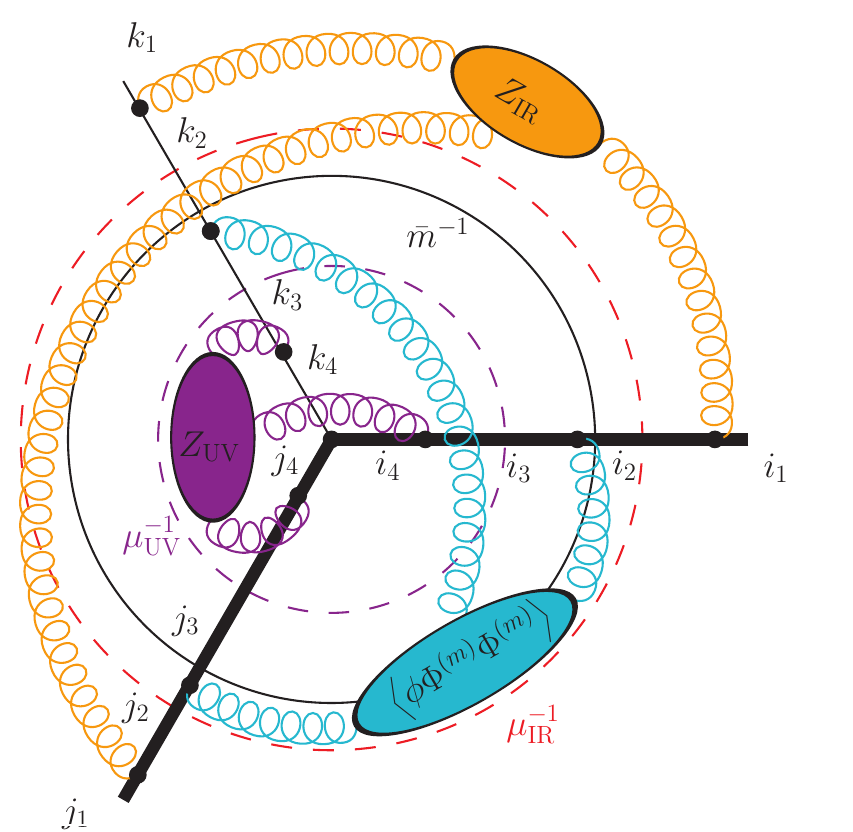}
    \caption{The factorization structure of eq.~(\ref{addren}) for a correlator consisting of one lightlike Wilson line and two timelike ones. The attachments along the Wilson lines are consistent with the colour ordering prescribed there. The indices $i_n$, $j_n$ and $k_n$ represent colour indices in the representation of each of the corresponding Wilson lines; these indices are summed over when the matrices are multiplied. Note that $Z_{\IR}$ would be trivial if the line $k$ were taken to be massive, leading to the one-sided factorization of figure~\ref{fig:FactorizationT}.   }
\label{fig:ZIRUVmixedcorrelator}
\end{figure}

 The general structure of the renormalization in eq.~(\ref{addren}) is sketched in figure~\ref{fig:ZIRUVmixedcorrelator} in configuration space. 
 In analogy with the treatment of singularities in on-shell amplitudes, the renormalization process of the mixed correlator consists of two separate steps, one in which short-distance singularities are compensated for by $Z_{\UV}$ and a second where the remaining, long-distance ones are removed by $Z_{\IR}$.\footnote{The difference to the amplitude case is of course that here both operations concern the correlator itself. Of course, additional renormalization of the coupling is assumed here, but will not be discussed explicitly.} 
Conceptually, these operate at two different cutoff scales $\muUV$ and $\muIR$. Of course, in practice these two scales can identified as a single renormalization point $\mu$. We keep them distinct in order to clearly demonstrate that the two renormalization procedures are independent.
 Note that the (path) ordering in eq.~(\ref{GammaUVIR}) is such that the contributions of high virtulaity are always placed to the right of those of lower virtuality. This is in line with the corresponding renormalization-group equations:
\begin{subequations}
\begin{align}
 \begin{split}
 \label{UV_RGeq}
\frac{d}{d\log\muUV}
Z_{\mathrm{UV}}(\muUV)=\,\,&-\Gamma_{\mathrm{UV}}(\alpha_s(\muUV))
\, Z_{\mathrm{UV}}(\muUV),
\end{split}
\\
    \begin{split}
    \label{IR_RGeq}
\frac{d}{d\log\muIR}
Z_{\mathrm{IR}}(\muIR) =\,\,&Z_{\mathrm{IR}}(\muIR)\, 
\Gamma_{\mathrm{IR}}(\muIR, \alpha_s(\muIR))\,.
\end{split}
\end{align}
\end{subequations}
On the right-hand side of eq.~(\ref{addren}), the mixed correlator (involving both timelike and lightlike lines) is just the hard region. To remove all its singularities we multiply it from the right by~$Z_{\UV}$, and from the left by $Z_{\IR}$. 
Because the UV regions cancel (at the correlator level), the  UV singularities of the mixed correlator, which are compensated by $Z_{\UV}$, directly correspond to the soft~AD, and $\Gamma_{\UV}$ is thus the same as that given in eq.~\eqref{GammaUV}. $Z_{\UV}$ is the proper short-distance renormalization factor which can be computed from the correlator of timelike Wilson lines, as in section~\ref{sec:correlators}. Note that $\Gamma_{\UV}$ therefore depends on the scale only via the argument of the coupling, $\alpha_s$, as shown in eq.~(\ref{UV_RGeq}). 
In turn, the factor~$Z_{\IR}$ in eq.~(\ref{addren}) encodes IR singularities; its role is to remove all the extra long-distance singularities that have been generated in the hard region by taking the Wilson lines to be strictly lightlike at integrand level.
In line with this, its dependence on the scale is not only via the argument of the coupling, but also explicit, as in eq.~(\ref{IR_RGeq}), reflecting the presence of overlapping  singularities.

With the region functions computed, we have enough information to determine $\Gamma_{\IR}$ for the tripole colour structure. Let us proceed by rewriting the right-hand side of eq.~\eqref{addren} as exponentiated perturbative expansions of each factor. Firstly, we construct the expansion of the yet unknown $Z_{\IR}$ as
\begin{align}
\label{ZIR_zeta}
\begin{split}
Z_{\mathrm{IR}}(\muIR)=\,\,&\exp\left\{\sum_{n}\left[\frac{\alpha_s(\bar{m})}{4\pi}\right]^n\left(\frac{\bar{m}^2}{\muIRsq}\right)^{n\epsilon}\zeta^{(n)}\right\}
\\=\,\,&\exp\left\{\sum_{n}\left[\frac{\alpha_s(\bar{m})}{4\pi}\right]^n\left(\frac{\bar{m}^2}{\muIRsq}\right)^{n\epsilon}\sum_{l<0}\zeta^{(n,l)}\epsilon^l\right\}\,
.
\end{split}
\end{align}
Assuming minimal subtraction, $\zeta^{(n)}$ only contains negative powers of $\epsilon$, i.e., $l<0$. Notice that the strong coupling $\alpha_s$ is evaluated at the scale $\bar{m}$ instead of $\muIR$, and the evolution of $\alpha_s$, controlled by  eq.~\eqref{RunningCoupling}, generates the additional factor $\left(\frac{\bar{m}^2}{\muIRsq}\right)^{n\epsilon}$. To simplify the derivation, we have set $b_0=0$. The same setup for the strong coupling will be used for  $\Gamma_{\UV}$ below.

Next, the mixed correlator is the strict lightlike limit of the correlator with respect to~$\beta_i^2$ for $i=1,\ldots, N$, the so-called hard region,
\begin{align}
\label{HardCorre}
\begin{split}
\left<\phi_{\beta_{1}}\cdots\phi_{\beta_{N}}\Phi^{(m)}_{\beta_{N+1}}\cdots\Phi^{(m)}_{\beta_{N+M}}\right>=\,\,&\exp\left\{\sum_{n}\left[\frac{\alpha_s(\bar{m})}{4\pi}\right]^nh^{(n)}\right\}
\\=\,\,&\exp\left\{\sum_{n}\left[\frac{\alpha_s(\bar{m})}{4\pi}\right]^n\sum_{l}h^{(n,l)}\epsilon^l\right\}\,,
\end{split}
\end{align}
Here of course both negative and positive powers of $\epsilon$ appear.
Finally, the UV renormalization takes a similar form of that in eq.~\eqref{PertExpGammaT}, 
\begin{align}
    \begin{split}
        Z_{\UV}(\muUV)
        =\,\,&\exp\left\{\frac{\alpha_s(\bar{m})}{4\pi}\left(\frac{\bar{m}^2}{\muUVsq}\right)^{\epsilon}\frac{1}{2\epsilon}\Gamma_{\UV}^{(1)}+\left[\frac{\alpha_s(\bar{m})}{4\pi}\right]^2\left(\frac{\bar{m}^2}{\muUVsq}\right)^{2\epsilon}\frac{1}{4\epsilon}\Gamma_{\UV}^{(2)}
      +{\cal O}(\alpha_s^3)\right\}\,,
    \end{split}
\end{align}
except that we set $b_0=0$ for simplicity.
The first two orders, $\Gamma_{\UV}^{(1)}$ and $\Gamma_{\UV}^{(2)}$, written in terms of $h^{(n)}$ and $r^{(n)}$, have been presented in eqs.~\eqref{Gamma1sr} and~\eqref{Gamma2sr}, respectively.

Using the expressions given in eqs.~\eqref{HardCorre},~\eqref{Gamma1sr} and~\eqref{Gamma2sr}, $\zeta^{(1)}$ and $\zeta^{(2)}$ of eq.~(\ref{ZIR_zeta}) can be obtained by requiring the left-hand-side of eq.~\eqref{addren} is finite.
More specifically, by expanding the product of the three non-Abelian exponentials on the right-hand side of eq.~\eqref{addren}, and then extracting the singularities, we have, at one loop, 
\begin{align}
\label{oneloopmatching}
    \begin{split}
      0=\sing\left[ \left(\frac{\bar{m}^2}{\muIRsq}\right)^{\epsilon} \zeta^{(1)}+h^{(1)}+\left(\frac{\bar{m}^2}{\muUVsq}\right)^{\epsilon}\frac{1}{2\epsilon}\Gamma_{\UV}^{(1)}\right].
    \end{split}
\end{align}
The leading order of $h^{(1)}$ is a double pole in $\epsilon$. In order to satisfy eq.~\eqref{oneloopmatching}, the leading order of $\zeta^{(2)}$ should also be the second order pole in $\epsilon$. With the powers in $\epsilon$ made manifest, we have, at the leading order,
\begin{align}
\label{oneloopleading}
\begin{split}
    0=\zeta^{(1,-2)}+h^{(1,-2)}-\left(h^{(1,-2)}+r^{(1,-2)}\right)\,
    \quad
    \Rightarrow
    \quad
    \zeta^{(1,-2)}=r^{(1,-2)}\,.
    \end{split}
\end{align}
Note that $\frac{1}{2\epsilon}\Gamma_{\UV}^{(1)}$ has been replaced by 
$-\left(h^{(1)}+r^{(1)}\right)$ according to eq.~\eqref{Gamma1sr}. At ${\cal O}(\epsilon^{-1})$, eq.~\eqref{oneloopmatching} requires
\begin{align}
\begin{split}
    0=\,\,&\zeta^{(1,-1)}+\log\left(\frac{\bar{m}^2}{\muIRsq}\right)\zeta^{(1,-2)}+h^{(1,-1)}
    \\&-\left(h^{(1,-1)}+r^{(1,-1)}\right)+\log\left(\frac{\bar{m}^2}{\muUVsq}\right)\left(h^{(1,-2)}+r^{(1,-2)}\right)\,.
    \end{split}
\end{align}
According to the definition of $h^{(n)}$ and $r^{(n)}$ (eq.~\eqref{correlatorsplit}), the coefficient of the UV logarithm $\log\left(\frac{\bar{m}^2}{\muUVsq}\right)$ is vanishing. This is because the one-loop correlator $w^{(1)}$ in the asymptotic expansion is dominant by a single pole in $\epsilon$, such that $ {\cal T}_{\lambda}\left[w^{(2,-2)}\right]=h^{(1,-2)}+r^{(1,-2)}=0$. Therefore, eq.~(\ref{NonTrivilZeta2}) implies that 
\begin{align}
\label{oneloopNTleading}
   \zeta^{(1,-1)}= r^{(1,-1)}-\log\left(\frac{\bar{m}^2}{\muIRsq}\right)r^{(1,-2)}\,,
\end{align}
with IR logarithm $\log\left(\frac{\bar{m}^2}{\muIRsq}\right)$ surviving. 

At two loops, the finiteness of eq.~\eqref{addren} gives the following constraint,
\begin{align}
\label{nonAb2loop}
\begin{split}
  0=\,\,&\sing\bigg[ \left(\frac{\bar{m}^2}{\muIRsq}\right)^{2\epsilon} \zeta^{(2)}+h^{(2)}+\left(\frac{\bar{m}^2}{\muUVsq}\right)^{2\epsilon}\frac{1}{4\epsilon}\Gamma_{\UV}^{(2)}
  \\&\hspace*{20pt}+\left(\frac{\bar{m}^2}{\muIRsq}\right)^{2\epsilon} \frac{1}{2} \left(\zeta^{(1)}\right)^2+\frac{1}{2} \left(h^{(1)}\right)^2+\left(\frac{\bar{m}^2}{\muUVsq}\right)^{2\epsilon}\frac{1}{2} \left(\frac{1}{2\epsilon}\Gamma_{\UV}^{(1)}\right)^2
  \\&\hspace*{20pt}+\, \left(\frac{\bar{m}^2}{\muIRsq}\right)^{\epsilon} \zeta^{(1)}h^{(1)}+\left(\frac{\bar{m}^2}{\muUVsq}\right)^{\epsilon}h^{(1)}\frac{1}{2\epsilon}\Gamma_{\UV}^{(1)}+\left(\frac{\bar{m}^2}{\muIRsq}\right)^{\epsilon} \left(\frac{\bar{m}^2}{\muUVsq}\right)^{\epsilon}\zeta^{(1)}\frac{1}{2\epsilon}\Gamma_{\UV}^{(1)}\bigg]\,.
  \end{split}
\end{align}
The leading order of terms such as $\left(h^{(1)}\right)^2$ or $\left(\zeta^{(1)}\right)^2$, have a quadruple pole in $\epsilon$. The resulting constraint on the quadrupole pole in $\epsilon$ is
\begin{align}
\label{twoloopqudru}
    0=\zeta^{(2,-4)}+\frac{1}{2}\left[r^{(1,-2)},h^{(1,-2)}\right]+\left(r^{(1,-2)}\right)^2-2\zeta^{(1,-2)}r^{(1,-2)}+\left(\zeta^{(1,-2)}\right)^2\,.
\end{align}
However, by plugging in the results given in eqs.~\eqref{oneloopleading},~\eqref{oneloopNTleading},~\eqref{HardCorre} and~\eqref{Gamma1sr}, one finds that the leading order $\zeta^{(2,-4)}$ is vanishing. At the order of the triple pole in $\epsilon$, $r^{(2)}$ starts to contribute. Then, eq.~\eqref{nonAb2loop} requires 
\begin{align}
    \zeta^{(2,-3)}=r^{(2,-3)}+\frac{1}{2}\left[h^{(1,-2)},r^{(1,-1)}\right]+\frac{1}{2}\left[h^{(1,-1)},r^{(1,-2)}\right]\,,
\end{align}
where the one-loop $\zeta^{(1,l)}$ has been substituted into $r^{(1,l)}$. 
Similarly to the quadrupole case, upon substituting the explicit results we find that $\zeta^{(2,-3)}$ is trivial. However, this conclusion relies on cancellation between the two-loop region $r^{(2,-3)}$ and the commutators constructed by one-loop functions. This fact
is significant and reflects the effect of the renormalization procedure, not merely a cancellation between one-loop functions as in eq.~\eqref{twoloopqudru}. 

Starting from the second order pole in $\epsilon$, $\zeta^{(2)}$ is non-trivial. We directly give the results of $ \zeta^{(2,-2)}$ and $\zeta^{(2,-1)}$, skipping some straightforward algebra. We find: 
\begin{subequations}
\label{NonTrivilZeta2}
\begin{align}
\begin{split}
    \zeta^{(2,-2)}=\,\,&r^{(2,-2)}+\frac{1}{2}\left[h^{(1,-1)},r^{(1,-1)}\right]+\log\left(\frac{\bar{m}^2}{\muIRsq}\right)\frac{1}{2}\left[r^{(1,-1)},r^{(1,-2)}\right]\,,
\end{split}
\\
\begin{split}
    \zeta^{(2,-1)}=\,\,&r^{(2,-1)}+\frac{1}{2}\left[h^{(1,-1)},r^{(1,0)}\right]+\frac{1}{2}\left[h^{(1,0)},r^{(1,-1)}\right]+\frac{1}{2}\left[r^{(1,-1)},r^{(1,0)}\right]
    \\&-\log\left(\frac{\bar{m}^2}{\muIRsq}\right)\left\{2r^{(2,-2)}+\left[h^{(1,-1)},r^{(1,-1)}\right]\right\}+\log^2\left(\frac{\bar{m}^2}{\muIRsq}\right)\frac{1}{2}\left[r^{(1,-2)},r^{(1,-1)}\right]\,,
    \end{split}
\end{align}
\end{subequations}
where the commutators that vanish upon substituting the corresponding $h^{(1,l)}$ and $r^{(1,l)}$ have been omitted. An important feature of eq.~\eqref{NonTrivilZeta2} is that only the IR logarithm appears in $\zeta^{(2)}$. This demonstrates that IR and UV singularities are disentangled. As a result, $\zeta^{(1)}$ and $\zeta^{(2)}$ are given below,
\begin{subequations}
\label{zeta12}
    \begin{align}
\begin{split}
\label{gamma1}
\zeta^{(1)}=\,\,&\sum_{I}\textbf{T}_I\cdot\textbf{T}_k\left\{\frac{1}{\epsilon^2}+\frac{1}{\epsilon}\left[\log\left(\frac{(2\beta_I\cdot\beta_k)^2}{t_k\lambda^2\beta_I^2}\right)-\log\left(\frac{\bar{m}^2}{\muIRsq}\right)\right]\right\}+\cdots,
\end{split}
      \\
      \begin{split}
      \label{gamma2}
 \zeta^{(2)}=\,\,&\sum_{I<J}\mathbf{T}_{IJk}\bigg\{\frac{1}{\epsilon^2}\left[\frac{1-y_{IJk}}{1+y_{IJk}}U_{1}(y_{IJk})+\log\left(y_{IJk}\right)\log\left(\frac{\bar{m}^2}{\muIRsq}\right)\right]
 \\&+\frac{1}{\epsilon}\bigg[\frac{1}{3}\log^3(y_{IJk})
 -\log\left(y_{IJk}\right)\log^2\left(\frac{\bar{m}^2}{\muIRsq}\right)
 \\&+\frac{1-y_{IJk}}{1+y_{IJk}}\left(\frac{1}{2}U_2(\alpha_{IJ},y_{IJk})-2U_{1}(y_{IJk})\log\left(\frac{\bar{m}^2}{\muIRsq}\right)\right)\bigg]\bigg\}+\cdots\,,
   \end{split}
 \end{align}
\end{subequations}
where, as above, the ellipsis stand for additional colour structures, such as colour singlet terms at one loop and dipole terms at two loops, which we have not computed.

Next we would like to use the results for $\zeta^{(n)}$ for $n=1$ and $2$ to infer the corresponding mixed-correlator IR anomalous dimension $\Gamma_{\IR}$ in eq.~(\ref{ZIR_zeta}). 
We note that both $\zeta^{(1)}$ and $\zeta^{(2)}$ in eq.~(\ref{zeta12}) are dominated by double poles in~$\epsilon$. This suggests that there is a single factor of $\log(\tau)$ appearing explicitly in $\Gamma_{\IR}$, in addition to the dependence on $\tau$ via the running coupling.
Based on this we find that the following simple ansatz for $\Gamma_{\IR}$ in eq.~(\ref{ZIR_zeta}) generates exactly $\zeta^{(1)}$ and $\zeta^{(2)}$ given in eq.~\eqref{zeta12},
\begin{align}
\label{GammaIR}
\begin{split}
    \Gamma_{\IR}(\tau,\alpha_s(\tau^2))=\,\,&-\sum_{I}\frac{1}{2}\textbf{T}_{I}\cdot \textbf{T}_{k}\gamma_{\text{cusp}}(\alpha_s)\log\left(\frac{\tau^2}{\bar{m}^2}\frac{(2\beta_I\cdot\beta_k)^2}{t_k\lambda^2\beta_I^2}\right)
    \\&+\sum_{I<J}\textbf{T}_{IJk}\bigg[\psi_{(IJk)}\left(y_{IJk},\alpha_s,\frac{\bar{m}^2}{\muIRsq}\right)\log\left(\frac{\tau^2}{\bar{m}^2}\right)
    \\&+\Psi_{(IJk)}\left(\left\{\alpha_{IJ},y_{IJk}\right\},\alpha_s,\frac{\bar{m}^2}{\muIRsq}\right)\bigg]+\cdots\,,
     \end{split}
 \end{align}
where the leading-order (mixed-correlator) IR ADs for the colour-tripole terms are 
\begin{subequations}
\label{GgIJk}
\begin{align}
\begin{split}
   \psi_{(IJk)}^{(2)}=\,\,&-8\left[\frac{1-y_{IJk}}{1+y_{IJk}}U_{1}(y_{IJk})+\log\left(y_{IJk}\right)\log\left(\frac{\bar{m}^2}{\muIRsq}\right)\right],
   \end{split}
\\
\begin{split}
   \Psi_{(IJk)}^{(2)}=\,\,&-4\left[\frac{1-y_{IJk}}{1+y_{IJk}}\frac{1}{2}U_2(\alpha_{IJ},y_{IJk})+\frac{1}{3}\log^3(y_{IJk})-\log\left(y_{IJk}\right)\log^2\left(\frac{\bar{m}^2}{\muIRsq}\right)\right].
   \end{split}
     \end{align}
\end{subequations}
where, of course, at one loop, tripole terms do not contribute,
\begin{align}
    \psi_{(IJk)}^{(1)}= \Psi_{(IJk)}^{(1)}=0\,.
\end{align}
To avoid confusion we stress again that the mixed-correlator IR anomalous dimension, which controls long-distance singularities of a correlator involving an non-regularized lightlike Wilson line,  is a very different object to the usual soft anomalous dimension of eq.~\eqref{mixtwoloop}, which instead controls long-distance singularities in on-shell amplitudes. The latter corresponds to the UV anomalous dimension of the mixed correlator. 
The two anomalous dimensions we determined here control singularities of different origin -- IR versus UV -- in the mixed correlator itself, and appear respectively on the left and right hand side of the correlator in eq.~(\ref{addren}). 

We also note that in contrast to the UV anomalous dimension, which is independent of the IR regulator $m$, as it must be, the IR anomalous dimension does depend on $m$. This is of course expected, as $Z_{\IR}$ is designed to cancel the poles generated by the lightlike line(s), which are not regularized. In particular, $\Gamma_{\IR}$ should vanish entirely for a correlator of timelike lines (in a complete regularization scheme) and must be sensitive to changing or removing the regulator where it is active.

Having determined both of the UV and IR anomalous dimensions explicitly in the case of the tripole, it is interesting to compare their analytic structure, which are very different indeed. In particular we find that while the UV anomalous dimension, $\Gamma_{\UV}$, given in eq.~\eqref{mixtwoloop}, only contains one rational function,  \begin{equation}\label{ratFac1}\frac{1+\alpha_{IJ}^2}{1-\alpha_{IJ}^2},
\end{equation}
the IR anomalous dimension $\Gamma_{\IR}$  contains another,  
\begin{equation}\label{ratFac2}
    \frac{1-y_{IJk}}{1+y_{IJk}}\,.
\end{equation}
By Bose symmetry, the kinematic function  multiplying the tripole colour structure in eq.~(\ref{GammaIR}) must be antisymemtric in $(I,J)$ and hence, antisymmetric in \hbox{$y_{IJk}\to 1/y_{IJk}$}. In eq.~(\ref{GgIJk}) this is realised either through the odd powers of the logarithms, or, through the antisymmetry of the rational factor (\ref{ratFac2}) in conjunction with the symmetry of the corresponding transcendental functions $U_i$.
Further note that the denominators of the two rational factors in eqs.~(\ref{ratFac1}) and~(\ref{ratFac2}), also appear as the symbol letters in the transcendental functions which multiply them, such as $U_{2}(\alpha_{IJ},y_{IJk})$
and $M_{100}(\alpha_{IJ})$ of eqs.~(\ref{U2def})
and (\ref{M100def}), respectively. 
Note that, as often happens, the transcendental functions multiplying the rational factor (\ref{ratFac2}) are power suppressed near $y_{IJk}=-1$,
\begin{align}
    U_1(y_{IJk})\sim U_2(\alpha_{IJ},y_{IJk})\sim {\cal O}(1+y_{IJk})\,,
\end{align} so the pole at $y_{IJk}=-1$ is spurious and the function $\Gamma_{\IR}$ is regular in the limit \hbox{$y_{IJk}\to -1$}.  This is similar to what happens in the $\alpha_{IJ}=1$  limit of eqs.~(\ref{OneLoopCusp}) and~\eqref{mixtwoloop}. Both situations correspond to the straight-line limit, where lines $I$ and $J$ are in the same direction, but one is in the initial state and the other in the final state. This special limit was used to set the boundary condition for the connected tripole web 
in appendix~\ref{app:comp111webHH}, following  eq.~(\ref{StraightLine_condition}) there.
 
To write down the full expression of $\Gamma_{\IR}$, we also have to consider  configurations involving only lightlike Wilson lines, 
such as lightlike dipoles and singlets which are hidden in the ellipsis in eq.~\eqref{zeta12}. Fortunately, with the exponential regulator defined in eq.~\eqref{Phim}, purely lightlike webs of the mixed correlator are scaleless, where IR and UV singularities exactly cancel each other out. Therefore, the corresponding terms appearing in $\Gamma_{\IR}$ should be exactly the same as those in $\Gamma_{\UV}$, given in eq.~\eqref{GammaUV}. Therefore, the IR anomalous dimension $\Gamma_{\IR}$ is given by 
\begin{align}
\label{GammaIR_Complete}
\begin{split}
\Gamma_{\IR}(\tau,\alpha_s(\tau^2))=\,\,&\sum_{i}\tilde{\gamma}_{i}(\alpha_s)
-\sum_{i}\sum_{I}\frac{1}{2}\mathbf{T}_{i}\cdot\mathbf{T}_{I}\gamma_{\text{cusp}}\left(\alpha_s\right)\log\left(\frac{\tau^2}{\bar{m}^2}\frac{(2\beta_I\cdot\beta_i)^2}{t_i\lambda^2\beta_I^2}\right)
\\&-\sum_{i<j}\frac{1}{2}\mathbf{T}_{i}\cdot\mathbf{T}_{j}\gamma_{\text{cusp}}\left(\alpha_s\right)\log\left(\frac{(2\beta_{i}\cdot\beta_{j})^2}{t_it_j\lambda^4}\right)
\\&
+\sum_{I<J}\sum_{k}\textbf{T}_{IJk}\bigg[\psi_{(IJk)}\left(y_{IJk},\alpha_s,\frac{\bar{m}^2}{\muIRsq}\right)\log\left(\frac{\tau^2}{\bar{m}^2}\right)
\\&+\Psi_{(IJk)}\left(\left\{\alpha_{IJ},y_{IJk}\right\},\alpha_s,\frac{\bar{m}^2}{\muIRsq}\right)\bigg].
\end{split}
 \end{align}
 Note that the colour structures free a lightlike index, for example $\mathbf{T}_{IJK}$, do not appear in~$\Gamma_{\IR}$, since webs involving only timelike lines 
 are fully regularized in the IR and thus only generate UV singularities.

We managed to obtain the renormalization factors $Z_{\IR}$ and $Z_{\UV}$ for the mixed correlator. However, one may not be entirely satisfied with the $Z_{\IR}$ and $Z_{\UV}$ presented in eqs.~\eqref{GammaIR_Complete} and~\eqref{GammaUV}, because the mixed correlator itself has no $\lambda$ dependence: the strict limit is well-defined, having no memory of the expansion parameter.  
Apparently, our procedure of determining the separate renormalization factors  $\Gamma_{\IR}$ and $\Gamma_{\UV}$, has led to both being dependent on $t_k\lambda^2$. One should therefore expect that this dependence can be removed. 
In fact, we will see that the $\lambda$ dependence in eqs.~\eqref{GammaIR} and~\eqref{GammaUV} can be regarded as the ambiguity in separating the IR and UV singularities for colour dipole terms.  

To demonstrate this and completely remove the $\lambda$ dependence from both the UV and the IR anomalous dimensions, we now identify both renormalization scales as the regulator scale itself,~that is  $\muIR=\muUV=\bar{m}$. In this case the logarithms depending on the ratio  scales in eq.~\eqref{zeta12} vanish\footnote{The dependence on the regulator $\bar{m}$  in the IR AD persists via the terms proportional to  $\psi_{(IJk)}$ and~$\gamma_{\text{cusp}}$, see eq.~\eqref{GammaIR}), which generate higher-order poles through the integration in eq.~\eqref{GammaIRdef}.} and these scales are eliminated from $\zeta^{(i)}$ and thus from the factor~$Z_{\IR}$.
With this, let us now prove that the $\lambda$ dependence can be eliminated, while recovering the correct dependence on the  dimensionful momenta. 
To this end we introduce the following exponential, 
\begin{align}
\label{Zlambda}
    Z_{0}(a,b)\equiv\text{P}\exp\left[\int_{a}^{b}\frac{d\tau}{\tau}\Gamma_{0}(\tau,\alpha_{s}(\tau))\right],
\end{align}
where $\Gamma_{0}$ contains only dipole terms and takes the general form 
\begin{align}
\label{Gamma0}
\begin{split}
\Gamma_{0}(\tau,\alpha_{s}(\tau))\equiv\,\,&\sum_{i}\sum_{I}\mathbf{T}_{i}\cdot\mathbf{T}_{I}\Gamma_{i}+\sum_{i<j}\mathbf{T}_{i}\cdot\mathbf{T}_{j}\left(\Gamma_{i}+\Gamma_{j}\right).
\end{split}
 \end{align}
Notice that in the first term on the right-hand side of eq.~\eqref{Gamma0}, $\Gamma_i$ is independent of $I$, that is, it is the same common factor for all timelike Wilson lines~$I$. 
 
 We now show  that upon taking colour conservation into account,  the dipole terms taking the form of eq.~\eqref{Gamma0} can be inserted into $\Gamma_{\IR}$ and $\Gamma_{\UV}$ simultaneously without changing the result. Therefore, $Z_0$ encapsulates the freedom on shifting terms of the form of $\Gamma_0$ between the IR and the UV factors in eq.~(\ref{addren}). 
 Note that the $\lambda$ dependence of $\Gamma_{\IR}$ and $\Gamma_{\UV}$ in eqs.~\eqref{GammaIR_Complete} and~\eqref{GammaUV} exactly satisfy the form of $\Gamma_0$. Importantly they are all proportional to $\gamma_{\text{cusp}}$.
 Additionally, recall that the neutral regions at one loop, ${\cal R}_{(Jk)}^{\{C_N\}}$ and ${\cal R}_{(Ik)}^{\{C_N\}}$, are constants given in eq.~\eqref{oneloopNeutC}. 
 They contribute to $\zeta^{(1)}$ and hence to the dipole terms in $\Gamma_{\IR}$. These neutral collinear contributions may also be understood as representing the ambiguity in separating between UV and IR singularities.\footnote{In section~\ref{sec:twoloopcomp}, we assumed that there is only one lightlike line $k$ with an arbitrary number of timelike lines (at least two). In that case only the first term on the right-hand side of eq.~\eqref{Gamma0} survives. Then, the one-loop neutral regions, ${\cal R}_{(Jk)}^{\{C_N\}}$ and ${\cal R}_{(Ik)}^{\{C_N\}}$, match exactly the form of $\Gamma_0$. We can interpret these regions as either IR or UV singularities. In the present section, we generalize the setup to any number of lightlike lines, such that the second term in eq.~\eqref{Gamma0} also contributes. 
 } 
  
To proceed with the proof, we collect the coefficient of $\Gamma_i$ on the right-hand side of eq.~\eqref{Gamma0}, 
\begin{align}
\label{colouId}
\begin{split}
 \Gamma_{0}=\,\,&\sum_{i}\mathbf{T}_{i}\cdot\left(\sum_{j\neq i}\mathbf{T}_{j}+\sum_{I}\mathbf{T}_{I}\right)\Gamma_{i}=-\sum_{i}C_i\mathbf{1}\Gamma_i\,.
    \end{split}
\end{align}
At the second equal sign, we use colour conservation and $\Gamma_{0}$ becomes proportional to the identity $\mathbf{1}$ in colour space,
times $C_i$, the quadratic Casimir of line $i$. 
Additionally, by setting $a$ and $b$ in eq.~(\ref{Zlambda}) to be $0$ and $\infty$ respectively, the exponent of $Z_{0}(a,b)$ is scaleless,
\begin{align}
Z_0(0,\infty)=\text{P}\exp\left[\int_{0}^{\infty}\frac{d\tau}{\tau}\Gamma_{0}\right]=\mathbf{1}.
\end{align}
 Therefore, we have the freedom to insert $Z_0$ in the right-hand side of eq.~\eqref{addren} at no cost. We then split $Z_0$ into IR and UV parts,
\begin{align}
\label{Z0IRUV}
Z_0(0,\infty)=Z_0(0,\bar{m})Z_0(\bar{m},\infty)=\mathbf{1}.
\end{align}
Note that the order of $Z_0(0,\bar{m})$ and $Z_0(\bar{m},\infty)$ is arbitrary because $\Gamma_0$ is proportional to colour identity; see eq.~\eqref{colouId}. We then absorb $Z_0(0,\bar{m})$ and $Z_0(\bar{m},\infty)$ into $\Gamma_{\IR}$ and $\Gamma_{\UV}$, respectively,
\begin{subequations}
\label{Gammabar}
\begin{align}
\begin{split}
\label{ZUVbar}
Z_{\UV}\rightarrow\bar{Z}_{\UV}\equiv Z_0(0,\bar{m})Z_{\UV},
\end{split}
\\
\begin{split}
\label{ZIRbar}
Z_{\IR}\rightarrow\bar{Z}_{\IR}\equiv Z_0(\bar{m},\infty)Z_{\IR},
\end{split}
\end{align}
\end{subequations}
while the ADs become,
\begin{subequations}
\begin{align}
\begin{split}
\label{GammaUVbardef}
\Gamma_{\UV}\rightarrow\bar{\Gamma}_{\UV}\equiv \Gamma_{\UV}+\Gamma_0,
\end{split}
\\
\begin{split}
\label{GammaIRbardef}
\Gamma_{\IR}\rightarrow\bar{\Gamma}_{\IR}\equiv \Gamma_{\IR}+\Gamma_0\,.
\end{split}
\end{align}
\end{subequations}

Because~$\Gamma_0$ does not contain any terms that  involve only timelike lines, we can 
interpret~$\Gamma_0$ as the ambiguity  in  distinguishing between IR and UV singularities for massless lines. 
In particular, individual terms in $\Gamma_0$ involving  $\Gamma_i$, carry the information on how the UV versus IR separation is made for line $i$. 
By making a special choice of $\Gamma_i$ such that 
\begin{align}
\label{Gammai}
    \Gamma_i=-\frac{1}{2}\gamma_{\text{cusp}}(\alpha_s)\log\left(\frac{\bar{m}^2}{\tau^2}t_i\lambda^2\right),
\end{align}
the $\lambda$ dependence in both eq.~\eqref{GammaUV} and eq.~\eqref{GammaIR} are cancelled.
The $\lambda$-independent ADs are then
\begin{subequations}
\label{GammaBarIRUV}
\begin{align}
\begin{split}
\label{GammaUVbar}
\\\bar{\Gamma}_{\UV}(\tau,\alpha_s(\tau^2))=\,\,&\sum_{i}\tilde{\gamma}_{i}(\alpha_s)+\sum_{I}\Omega_{(I)}(\alpha_s)+\sum_{I<J}\mathbf{T}_{I}\cdot\mathbf{T}_{J}\Omega_{(IJ)}(\alpha_{IJ},\alpha_s)
\\&+\sum_{i<j}\mathbf{T}_i\cdot\mathbf{T}_j\gamma_{\text{cusp}}(\alpha_s)\log
\frac{\tau^2}{-2p_i\cdot p_j}
\\&
+\sum_{i}\sum_{I}\mathbf{T}_i\cdot\mathbf{T}_I\gamma_{\text{cusp}}(\alpha_s)\log\frac{\tau\sqrt{p_I^2}}{-2p_i\cdot p_{I}}
\\&+\sum_{I<J<K}\textbf{T}_{IJK}\Omega_{(IJK)}(\{\alpha_{IJ},\alpha_{JK},\alpha_{IK}\},\alpha_s)
\\&
+\sum_{I<J}\sum_{k}\textbf{T}_{IJk}\Omega_{(IJk)}\left(\left\{\alpha_{IJ},y_{IJk}\right\},\alpha_s\right)
\\=\,\,&\sum_{i}\tilde{\gamma}_{i}(\alpha_s)+\Gamma^T+\Gamma^L,
\end{split}
\\
\begin{split}
\label{GammaIRbar}
   \bar{\Gamma}_{\IR}(\tau,\alpha_s(\tau^2))=\,\,&\sum_{i}\tilde{\gamma}_{i}(\alpha_s)-\sum_{i}\sum_{I}\frac{1}{2}\mathbf{T}_{i}\cdot\mathbf{T}_{I}\gamma_{\text{cusp}}\left(\alpha_s\right)\log\left(\frac{(2\beta_I\cdot\beta_i)^2}{\beta_I^2}\right)
\\&+\sum_{i<j}\mathbf{T}_i\cdot\mathbf{T}_j\gamma_{\text{cusp}}(\alpha_s)\log
\frac{\tau^2}{-2p_i\cdot p_j}
\\&
+\sum_{I<J}\sum_{k}\textbf{T}_{IJk}\bigg[\psi_{(IJk)}(y_{IJk},\alpha_s)\log\left(\frac{\tau^2}{\bar{m}^2}\right)
\\&+\Psi_{(IJk)}\left(\left\{\alpha_{IJ},y_{IJk}\right\},\alpha_s\right)\bigg],
\end{split}
 \end{align}
 \end{subequations}
 where we have recovered the dimensionful momentum $p_i$ defined by
\begin{align}
\label{dimfulp}
p_i^{\nu}\equiv\bar{m}\beta_i^{\nu}\,,
\end{align}
defining the normalization of the dimensionless velocity $\beta_i$ for lightlike lines. Notice that, in contrast to eq.~\eqref{GammaIR_Complete},  the tripole terms in $\bar{\Gamma}_{\IR}$ do not depend on $\bar{m}/\muIR$, since $\muIR$ has been chosen to be $\bar{m}$, and we have, 
\begin{subequations}
\label{GgIJkId}
\begin{align}
\begin{split}
   \psi_{(IJk)}^{(2)}\left(y_{IJk},\alpha_s\right)=\,\,&-8\frac{1-y_{IJk}}{1+y_{IJk}}U_{1}(y_{IJk}),
   \end{split}
\\
\begin{split}
   \Psi_{(IJk)}^{(2)}\left(\left\{\alpha_{IJ},y_{IJk}\right\},\alpha_s\right)=\,\,&-4\left[\frac{1-y_{IJk}}{1+y_{IJk}}\frac{1}{2}U_2(\alpha_{IJ},y_{IJk})+\frac{1}{3}\log^3(y_{IJk})\right].
   \end{split}
     \end{align}
\end{subequations} 

The ADs, $\Gamma^T$ and $\Gamma^L$, in the second line of eq.~\eqref{GammaUVbar}, are defined in eqs.~\eqref{GammaT} and~\eqref{GammaL}, respectively. The colour singlet term $\tilde{\gamma_i}$ is the same as that in $\Gamma_{\UV}$ in eq.~\eqref{GammaUV}. 
Notice that with the special choice of $\Gamma_i$ made in eq.~\eqref{Gammai}, the UV AD $\bar{\Gamma}_{\UV}$ in eq.~\eqref{GammaUVbar} is exactly the soft~AD of on-shell amplitudes, up to the colour singlet term.\footnote{In the soft~AD of the amplitudes, the colour singlet term $\gamma_{i}$ is known as the collinear AD. To obtain the correct collinear AD, partonic hard collinear singularities should also be considered, which is not discussed in this paper.}

Finally, the renormalization of the mixed correlator becomes
\begin{align}
\label{addren2}
\begin{split}
\left<\phi_{\beta_{1}}\cdots\phi_{\beta_{N}}\Phi^{(m)}_{\beta_{N+1}}\cdots\Phi^{(m)}_{\beta_{N+M}}\right>_{\text{ren.}}=\bar{Z}_{\IR}(\bar{m})\left<\phi_{\beta_{1}}\cdots\phi_{\beta_{N}}\Phi^{(m)}_{\beta_{N+1}}\cdots\Phi^{(m)}_{\beta_{N+M}}\right>\bar{Z}_{\UV}(\bar{m}).
\end{split}
 \end{align}
The structure we have here is invariant with respect to the  way in which the lightlike limit is taken, i.e., there is no dependence on $\lambda$ in either factor. 
At the same time, we lose the freedom to choose a $t_k$ as we did in eq.~\eqref{tk}, such that the limit preserves rescaling symmetry. We emphasize that $\bar{\Gamma}_{\UV}$ captures soft singularities of on-shell amplitudes, while~$\bar{\Gamma}_{\IR}$ describes special IR singularities of the mixed correlator and does not directly appear in physical processes. Understanding the renormalization structure of the mixed correlator itself, is an important step forward, as this object provides a new route towards efficient computation of $\bar{\Gamma}_{\UV}$ at higher loop orders.

\section{Conclusions}
\label{sec:conc}

In this paper we presented a new method to compute the soft anomalous dimension controlling long-distance singularities of amplitudes with both massive and massless particles. As reviewed in section~\ref{sec:correlators}, the standard approach to this problem is based on the computation of correlators of fully IR-regularized,  \emph{timelike} semi-infinite Wilson lines, which are subsequently expanded  near the lightcone, so as to place the massless particles on shell. 
A key example of such a computation is the determination of the three-loop soft anomalous dimension for massless scattering~\cite{Almelid:2015jia}. In this case configuration-space integrals contributing to the correlator of four timelike lines were performed to derive multifold Mellin-Barnes representations, which were subsequently used to take the simultaneous lightlike limit of all Wilson lines, at which point major simplifications took place, and the remaining integrals could be performed. Such a computation may be hard to complete if the limit of interest is more complicated, e.g. if only a subset of the Wilson lines become lightlike.
The reason for taking all Wilson lines to be non-lightlike in the first place is rather fundamental: it guarantees multiplicative renormalizability of the correlator, making it possible to deduce the soft anomalous dimension from the single UV $1/\epsilon$ pole of the correlator.
Of course, this advantage comes at the a high price in terms of the increased complexity of the integrals. One may wonder whether there is a strategy that makes better use of the simpler nature of the lightlike limit at the outset. This our starting point in this paper.

Our approach is based on replacing said correlator by its asymptotic expansion near the lightcone, using the method of regions. This gives rise, in particular, to a ``hard region'' integral, where the strict lightlike limit is taken prior to the loop-momentum integration. 
These hard-region integrals are precisely those defined by a correlator consisting of both timelike and lightlike lines, so that each external particle is represented by a Wilson line along its exact classical trajectory. The computation of this object has been avoided in the past because its singularity structure is complex and is hard to interpret. In particular, it is not multiplicatively renormalizable in the usual sense. Moreover, in contrast the case of  timelike Wilson lines, such a correlator cannot be fully regularized in the IR without breaking gauge invariance and rescaling symmetry. In the absence of complete regularization of long-distance and collinear singularities, higher-order poles in $\epsilon$ are generated, obscuring the connection between the correlator and the sought-after anomalous dimension. 

In this paper we solved this problem, outlining a  general strategy for determining the soft~AD from the sum of all region integrals, see eqs.~(\ref{Gamma1sr}) and (\ref{Gamma2sr}) at one and two loops, respectively. 
Moreover, we have shown how the UV and IR singularities can be systematically disentangled. 
As shown in eq.~(\ref{addren}), from the perspective of the mixed correlator, the sought-after soft~AD still corresponds to UV renormalization~(\ref{GammaUVdef}), except that it is now recovered through an asymptotic expansion. Of course, the UV renormalization of the mixed correlator 
does not remove all its  singularities  -- the remaining ones can be identified as of IR origin, defining $\Gamma_\IR$
in eq.~(\ref{GammaIRdef}).
The latter is a very different object, which depends on the way the correlator is regularized in the IR. 
We have seen that the classification of singularities based on their non-analytic scaling, $\lambda^{n_R\epsilon}$, with positive versus negative $n_R$ within the MoR, nicely translates into the separate $Z_\UV$ and $Z_\IR$ factors in eq.~(\ref{addren}),  respectively. Interestingly, neutral regions other than the hard, have a unique role as, owing to colour conservation, they describe the inherent ambiguity in assigning dipole contributions obtained using the MoR to either of the two anomalous dimensions. This was used in eq.~(\ref{GammaBarIRUV}) to eliminate the freedom in choosing the MoR expansion parameter, and restore the proper dependence of the ADs on dimensionful momenta. 

In this paper we demonstrated our strategy at two loops, where the key example has been the correlator of three Wilson lines, one of which becomes lightlike. Specifically, the fully-antisymmetric contribution to the soft~AD proportional to the tripole  
$\textbf{T}_{IJK}$ (see eq.~(\ref{fTTT})), served us as case study.
In contrast to dipole terms, the lightlike limit of the tripole soft~AD is smooth, tending to a finite limit, see eq.~\eqref{mixtwoloop}. Such a finite lightlike limit is expected to arise also in more complicated higher-order contributions to the soft~AD, in fact, in any contribution depending on rescaling-invariant kinematic variables.  
It is useful to note that the finite limit of the soft~ADs does not simply translate into a finite limit of the corresponding correlator term. Indeed, the two-loops correlator in eq.~\eqref{w2ex} is still logarithmically-divergent in the expansion parameter $\lambda$. These logarithms are associated with UV subdivergences generated by sequentially contracting individual gluon exchanges to the multi-Wilson-line vertex. They are therefore removed by commutators of one-loop webs when computing the soft~AD, see eq.~(\ref{Gamma2sr}). This subtraction has the very same effect to the definition of subtracted webs~\cite{Gardi:2013saa}. 
  
In section~\ref{sec:twoloopcomp} we performed a detailed MoR-based analysis of one-loop and two-loop webs, with a special focus on the contributions to the tripole colour structure, originally computed in refs.~\cite{Mitov:2009sv,Ferroglia:2009ii,Ferroglia:2009ep}. 
For each web we verified that the asymptotic expansion obtained through a sum of all region integrals correctly reproduces the leading term in the lightcone expansion, $\beta_K^2\to 0$, of the corresponding web computed with all timelike lines. Beyond providing assurance for the validity of the MoR in this, somewhat unorthodox application, our central aim in this analysis was to understand the modes and the regions which arise, and then further investigate how these contribute to the correlator, and hence to the anomalous dimension. Working in Euclidean kinematics, where all $\alpha_{IJ}>0$, there are only endpoint singularities in parameters space, and then the geometric MoR construction is guaranteed to provide the complete set of regions. Starting with the set of region vectors computed by {\tt{pySecDec}} (defined in the Lee-Pomeransky representation) we determined the set of loop-momentum modes and regions. 
All the regions are summarized in tables~\ref{tab:oneloopregion},~\ref{tab:121WebRegions}, and~\ref{tab:twolooptopregionv}, where they are expressed in terms of the five modes defined in table~\ref{tab:ModesSum}. These include a single IR-collinear mode, $C_{\IR}$, two neutral modes, the hard, $H$, and the neutral-collinear mode, $C_N$, and two UV modes, UV-collinear, $C_{\UV}$, and 
UV-hard, $H_{\UV}$.  The latter two appear starting from two loops.

We observe that the set of modes and regions is similar in different webs.
It is therefore insightful to define
\emph{region functions}, (see eq.~\eqref{ExpansionCalY}) which sum up the contributions to the correlator from different webs, which share a common region $R$ as well as a common colour factor. 
Furthermore, we define 
invariant region functions in the final line of eq.~\eqref{ExpansionCalY}, which sum up all region functions with a common overall asymptotic behaviour, $\lambda^{n_R\epsilon}$, but different set of modes. 
In terms of these invariant region functions we find a remarkably simple picture. First, all UV regions completely cancel out at correlator level. Second, neutral regions other than the hard only contribute a constant to the dipole structure, as mentioned above. 
Thus, the tripole structure of the correlator receives only two types of contributions, the hard region and the IR regions, summarized in eqs.~(\ref{allHard}) and~(\ref{allIR}), respectively. The UV singularities of the correlator, $Z_{\UV}$ in (\ref{addren}), from which the soft~AD is determined, are then recovered in the sum of all regions, while the remaining singularities, corresponding to~$Z_{\IR}$, are driven by the IR regions according to eqs.~(\ref{ZIR_zeta}) and~(\ref{NonTrivilZeta2}).

In contrast to the IR modes, which are common in on-shell expansions, the presence of multiple neutral modes with hard virtuality, which arise in our lightcone expansion, is unique to correlators of semi-infinite Wilson lines. 
In section~\ref{sec:neutralregions} showed that under a certain Lorentz boost, the two neutral modes, $H$ and $C_N$, behave respectively as the hard mode in two different Lorentz frames; see eq.~\eqref{comlimit}.
Moreover, in terms of the Wilson-line velocities, these two hard modes correspond to two physically-distinct limits, the lightcone limit of eq.~(\ref{limi1beta}) and the complementary lightcone limit of eq.~(\ref{limi2beta}).
While in the lightcone expansion one particle is becoming massless, in the complementary one, all other particles become massless and collinear to each other.  
A key observation is that the two limits  share the same mathematical result due to the rescaling symmetry of the Wilson lines. As explained following eq.~(\ref{comlimit}), this is a generalization of the identification between the massless limit and the triple collinear limit discussed in refs.~\cite{Duhr:2025cye,GZ-PRL,GZ-TBP}.
Another conclusion from this analysis is  that the observed degeneracy of the neutral modes stems from the rescaling symmetry of the correlator of semi-infinite Wilson lines. 

While in this paper we developed and tested the approach using well-known one- and two-loop examples, its generalization 
to higher loop orders with arbitrary numbers of lightlike and timelike Wilson lines is in principle straightforward.
Indeed, we have been able to apply this method to evaluate the correlator of three lightlike lines and one timelike line at three-loop order, and determine the function multiplying the quadrupole colour structure. This result, along with earlier work by Liu and Schalch~\cite{Liu:2022elt}, completes the knowledge of the soft~AD for the scattering of a single massive particle and any number of massless ones at three loops. Details of this three-loop calculation and the results for the soft AD will soon be published~\cite{GZ-PRL,GZ-TBP}.

\acknowledgments
We are grateful to Wen Chen, Claude Duhr, Stephen Jones, Gregory Korchemsky, Yao~Ma, Andrew McLeod and Kai~Yan for helpful discussions. 
ZZ is supported by the China Scholarship Council PhD programme. 
EG is supported by the STFC Consolidated Grant \emph{Particle Physics at the Higgs Centre}. 
For the purpose of open access, the authors have applied a Creative Commons Attribution (CC BY) licence to any Author Accepted Manuscript version arising from this submission.


 \appendix

\section{Basis of transcendental functions}
\label{app:convfunc}

In this appendix we define the transcendental functions that are used to express the results for the webs we compute. Since we restrict the computation to ${\cal O}(\epsilon^{-1})$ at two loops, we only need polylogarithmic functions up to weight 3. We work with a basis of uniform weight functions, and express them using Goncharov   polylogarithms~\cite{Goncharov:2009lql}, adopting the notation of {\tt{PolyLogTools}}~\cite{Duhr:2019tlz}. These generalize both harmonic~\cite{Gehrmann:2000zt} and classical polylogarithms, but the low-weight functions used here can in fact all be expressed as classical polylogarithms.

We classify the functions into three groups, depending on where they appear. First, we require a function from the basis transcendental functions appearing in  multiple-gluon-exchange webs, $M_{abc}(\alpha)$, defined in  refs.~\cite{Gardi:2013saa,Falcioni:2014pka}:
\begin{align}
\label{M100def}
\begin{split}
    M_{100}(\alpha)&=
  -\frac{\pi^2}{3} + 4 G(-1, 0, \alpha) - 4 G(0, 0, \alpha) + 4 G(1, 0, \alpha)  \\
&=-\frac{\pi^2}{3}
+ 2\text{Li}_{2}\left(\alpha^2\right)+4\log\left(\alpha\right)\log\left(1-\alpha^2\right)-2\log^2\left(\alpha\right)\,.
    \end{split}
\end{align}

Next, for the one-loop web computation~in~eq.~(\ref{Oneloop2F1result}) at ${\cal O}(\epsilon^0)$ and ${\cal O}(\epsilon^1)$ we define:
\begin{subequations}
\begin{align}
\label{Vdef}
    \begin{split}
        V_1(\alpha)&\,\equiv-\frac{1}{3}  \left[12 G\left(-1,0,\alpha \right)-6 G\left(0,0,\alpha \right)+\pi ^2\right]
        \\
        &\,=-4 \text{Li}_2(-\alpha)+\log^2(\alpha)-4 \log (\alpha+1) \log (\alpha)-\frac{\pi^2}{3}
        \,,
    \end{split}
\end{align}
and
\begin{align}
    \begin{split}
      V_2(\alpha)\equiv \,\,&-\frac{1}{6} \pi ^2 \left[4 G(-1,\alpha)+3 G(0,\alpha)\right]-8 G(-1,-1,0,\alpha)
      \\&+4 G(-1,0,0,\alpha)+4 G(0,-1,0,\alpha)-2 G(0,0,0,\alpha)+4 \zeta (3)\,.
    \end{split}
\end{align}
\end{subequations}

Finally, for the two-loop web computation involving two timelike Wilson lines and one approaching the lightlike limit, we also need functions depending also on the kinematic variable $y_{IJk}$ of eq.~(\ref{yIJk}), which we defined as follows:
\begin{subequations}
\begin{align}
\begin{split}
        U_1(y_{IJk})\equiv -G\left(0,0,y_{IJk}\right)-\frac{\pi ^2}{2}\,,
    \end{split}
\end{align}
\begin{align}
\label{U2def}
    \begin{split}
        U_2(\alpha_{IJ},y_{IJk})\equiv \,\,&-4 G\left(0,0,0,\alpha _{IJ}\right)+8 G\left(1,0,0,\alpha _{IJ}\right)+4 G\left(0,0,\alpha _{IJ}\right) G\left(0,y_{IJk}\right)
        \\&+6 \pi ^2 G\left(-1,y_{IJk}\right)-\pi ^2 G\left(0,y_{IJk}\right)+12 G\left(-1,0,0,y_{IJk}\right)
        \\&-2 G\left(0,0,0,y_{IJk}\right)-4 {\mathbb{U}}\left(\frac{y_{IJk}}{\alpha _{IJ}}\right)-4 {\mathbb{U}}\left(\alpha _{IJ} y_{IJk}\right)+12 \zeta (3)\,,
    \end{split}
\end{align}
\end{subequations}
where
\begin{equation}
    {\mathbb{U}}(x)=G(-1,0,0,x)+\frac{\pi^2}{2}G(-1,x).
\end{equation}

\section{Computation and result for regions at one loop}
\label{app:oneloop}

In this appendix we  provide the details on the Feynman integral computation of the region functions for one-loop webs in the limit $\beta_K^2\to 0$ using the MoR. 
We perform the asymptotic expansion of the kinematic functions ${\cal Y}$ given in section~\ref{sec:one-loop_computation} using the package {\tt{AmpRed}}~\cite{Chen:2024xwt}. Then, each region ${\cal Y}^R$ can be reduced to a finite set of independent master integrals using integration-by-parts (IBP) identities~\cite{Chetyrkin:1981qh}. In practice, we use a combination of the packages {\tt{AmpRed}}~\cite{Chen:2024xwt} and {\tt{Kira}}~\cite{Maierhofer:2017gsa,Klappert:2020nbg,Lange:2025fba} to do the reduction. For each master integral $I^R$ that contributes to the region $R$ of a particular web, ${\cal Y}^R$, we directly perform the integrations in parametric space. More specifically, we use Feynman parametrization,
\begin{align}
\label{FeynmanPara}
 I^R=\frac{\Gamma\left(\nu-\frac{LD}{2}\right)}{\prod_{e=1}^{E}\Gamma\left(\nu_e\right)} \int_{a>0}\left[da\right]\prod_{e=1}^{E}a_e^{\nu_e-1}\frac{\left({\cal U}\right)^{\nu-\frac{(L+1)D}{2}}}{\left({\cal F}\right)^{\nu-\frac{LD}{2}}} \,, 
\end{align}
where $L$ is the number of loops, $E$ the number of (internal) edges, $\nu_e$ is the power of a given propagator and $\nu$ is the sum of $\nu_e$. The spacetime dimension $D$ is set to be $D=4-2\epsilon$ through all the calculations. The notation in eq.~\eqref{FeynmanPara} is consistent with that of eq.~\eqref{LeePom} where the integral is expressed using  the Lee-Pomeransky representation. The integral~(\ref{FeynmanPara}) is defined in projective space, where the measure is 
\begin{align}
\begin{split}
[da]=\prod_{e=1}^E da_e \delta\left(1-\sum_{e=1}^Ea_e\right)\,.
\end{split}
 \end{align}
Owing to the projective space redundancy (the 
 Cheng-Wu theorem) there is the freedom to set the sum of a subset of parameters to be one, and all the remaining parameters are integrated over all positive values.

 The following sections contain the IBP reductions for regions ${\cal Y}^R$, the Feynman parameter representations of the master integrals $I^R$, as well as the results of ${\cal Y}^R$. To facilitate straightforward check of the computation, we will also   specify the order of integration we used to evaluate the parametric master integrals which do not simply evaluate  into Euler gamma functions.
 
\subsection{\texorpdfstring{Region ${\cal R}_{(IJ)}^{\{H\}}$}{R(IJ){H}}}
\label{IJ_Hard}

The hard region of the $(IJ)$ web is clearly not affected by the limit $\beta_K^2\to 0$. This web is therefore given by
\begin{align}
 \begin{split}
    {\Y}_{(IJ)}^{\{H\}}=\,\,&\vcenter{\hbox{\includegraphics[width=1.5cm]{fig/11IJH.pdf}}}=\frac{\alpha_s}{4\pi}{\cal N}\frac{\alpha_{IJ}^2+1}{2\alpha_{IJ}}I_{(IJ)}^{\{H\}}(\alpha_{IJ})+{\cal O}(\lambda),
 \end{split} 
 \end{align}
 where ${\cal N}$ is defined in eq.~(\ref{calYIJ}) and the integral $I_{(IJ)}^{\{H\}}$ is defined by
 \begin{align}
     \begin{split}
         I_{(IJ)}^{\{H\}}=4^{\epsilon +1}\Gamma(1+\epsilon)\int_{a>0}[da] a_1^{2 \epsilon -1} \left[a_3 a_2 \left(\alpha_{IJ}+\frac{1}{\alpha_{IJ}}\right)+a_2^2+a_3^2+4 a_1 \left(a_2+a_3\right)\right]^{-\epsilon -1}.
     \end{split}
 \end{align} 
  By setting $a_3=1$ (Cheng-Wu theorem) and integrating $a_1$ and $a_2$ sequentially, we get the following result
 \begin{align}
 \label{Oneloop2F1result}
     \begin{split}
        {\cal R}_{(IJ)}^{\{H\}}= {\Y}_{(IJ)}^{\{H\}}=\,\,&\left(\frac{m^2}{\mu^2\pi}\right)^{-\epsilon}2\frac{1+ \alpha_{IJ}^2 }{1-\alpha_{IJ}^2}\Gamma (-\epsilon ) \Gamma (2 \epsilon )\bigg[(\alpha_{IJ}+1)^{\epsilon}\, _2F_1\left(1-\epsilon ,\epsilon ;\epsilon +1;\frac{\alpha_{IJ}}{\alpha_{IJ}+1}\right)
          \\&-\left(\frac{1}{\alpha_{IJ}}+1\right)^{\epsilon }\, _2F_1\left(1-\epsilon ,\epsilon ;\epsilon +1;\frac{1}{\alpha_{IJ}+1}\right)\bigg]+{\cal O}(\lambda)
          \\=\,\,&\left(\frac{\bar{m}^2}{\mu^2}\right)^{-\epsilon}\frac{1+ \alpha_{IJ}^2 }{1-\alpha_{IJ}^2}\bigg[-\frac{2}{\epsilon}\log(\alpha_{IJ})+V_1(\alpha_{IJ})+\epsilon V_2(\alpha_{IJ})\bigg]+{\cal O}(\epsilon^2)+{\cal O}(\lambda)\,,
     \end{split}
 \end{align}
where we expanded to ${\cal O}(\epsilon^1)$, since these terms enter the renormalization of the correlator at two loops.

\subsection{\texorpdfstring{Region ${\cal R}_{(Jk)}^{\{H\}}$ and Region ${\cal R}_{(Ik)}^{\{H\}}$}{R(Jk){H} and R(Ik){H}}}

In contrast to ${\cal R}_{(IJ)}^{\{H\}}$ of section~\ref{IJ_Hard}, the integrals involving the line $K$, ${\cal R}_{(Jk)}^{\{H\}}$ and ${\cal R}_{(Ik)}^{\{H\}}$ are greatly simplified by taking the limit.
The read
\begin{subequations}
\begin{align}
 \begin{split}
  {\Y}_{(Jk)}^{\{H\}}=\vcenter{\hbox{\includegraphics[width=1.5cm]{fig/11JkH.pdf}}}=\frac{\alpha_s}{4\pi}{\cal N}\frac{1-2\epsilon}{2\epsilon}I_{(Jk)}^{\{H\}}+{\cal O}(\lambda),
\end{split} 
 \\
 \begin{split}
     {\Y}_{(Ik)}^{\{H\}}=\vcenter{\hbox{\includegraphics[width=1.5cm]{fig/11IKH.pdf}}}=\frac{\alpha_s}{4\pi}{\cal N}\frac{1-2\epsilon}{2\epsilon}I_{(Jk)}^{\{H\}}+{\cal O}(\lambda).
 \end{split} 
 \end{align}
 \end{subequations}
 The integral $I_{(Jk)}^{\{H\}}$ in parametric space is defined in
 \begin{align}
     \begin{split}
       I_{(Jk)}^{\{H\}}= 4^{\epsilon }\Gamma (\epsilon ) \int_{a>0}[da] a_1^{2 \epsilon -2} \left( 4 a_1a_2+a_2^2\right)^{-\epsilon }.
     \end{split}
 \end{align}
 The integral is easy to perform and the result is 
 \begin{align}
 \label{oneloophard}
     \begin{split}
      {\Y}_{(Jk)}^{\{H\}}= {\Y}_{(Ik)}^{\{H\}}=\,\,&\frac{\alpha_s}{4\pi}\left(\frac{m^2}{\mu^2\pi}\right)^{-\epsilon}2 \Gamma (-\epsilon ) \Gamma (2 \epsilon )+{\cal O}(\lambda)
      \\=\,\,&\frac{\alpha_s}{4\pi}\left(\frac{\bar{m}^2}{\mu^2}\right)^{-\epsilon}\left[-\frac{1}{\epsilon ^2}-\frac{5 \pi ^2}{12}+\epsilon\frac{7  }{3} \zeta (3)\right]+{\cal O}\left(\epsilon ^2\right)+{\cal O}(\lambda)
       \end{split}
 \end{align}
 \subsection{\texorpdfstring{Region ${\cal R}_{(Jk)}^{\{C_N\}}$ and Region ${\cal R}_{(Ik)}^{\{C_N\}}$}{R(Jk){CN} and R(Ik){CN}}}
\begin{subequations}
\begin{align}
 \begin{split}
  {\Y}_{(Jk)}^{\{C_N\}}=\vcenter{\hbox{\includegraphics[width=1.5cm]{fig/11JKHCN.pdf}}}=\frac{\alpha_s}{4\pi}{\cal N}\frac{1-2\epsilon}{2\epsilon}I_{(Jk)}^{\{H\}}+{\cal O}(\lambda),
\end{split} 
 \\
 \begin{split}
    {\Y}_{(Ik)}^{\{C_N\}}=\vcenter{\hbox{\includegraphics[width=1.5cm]{fig/11IKCN.pdf}}}=\frac{\alpha_s}{4\pi}{\cal N}\frac{1-2\epsilon}{2\epsilon}I_{(Jk)}^{\{H\}}+{\cal O}(\lambda).
 \end{split} 
 \end{align}
 \end{subequations}
 The result is exactly the same as the hard regions,
 \begin{align}
 \label{oneloopNeutC}
     \begin{split}
      {\Y}_{(Jk)}^{\{C_N\}}= {\Y}_{(Ik)}^{\{C_N\}}=\,\,& \frac{\alpha_s}{4\pi}\left(\frac{m^2}{\mu^2\pi}\right)^{-\epsilon}2 \Gamma (-\epsilon ) \Gamma (2 \epsilon )+{\cal O}(\lambda)
      \\=\,\,&\frac{\alpha_s}{4\pi}\left(\frac{\bar{m}^2}{\mu^2}\right)^{-\epsilon}\left[-\frac{1}{\epsilon ^2}-\frac{5 \pi ^2}{12}+\epsilon\frac{7  }{3} \zeta (3)\right]+{\cal O}\left(\epsilon ^2\right)+{\cal O}(\lambda)
       \end{split}
 \end{align}
 \subsection{\texorpdfstring{Region ${\cal R}_{(Jk)}^{\{C_{\IR}\}}$ and Region ${\cal R}_{(Ik)}^{\{C_{\IR}\}}$}{R(Jk){CIR} and R(Ik){CIR}}}
\begin{subequations}
\begin{align}
 \begin{split}
  {\Y}_{(Jk)}^{\{C_{\IR}\}}=\vcenter{\hbox{\includegraphics[width=1.5cm]{fig/11JKHCIR.pdf}}}=\frac{\alpha_s}{4\pi}{\cal N}\frac{1}{2\sqrt{y_{IJk}}}I_{(Jk)}^{\{C_{\IR}\}}(y_{IJk})+{\cal O}(\lambda),
\end{split} 
 \\
 \begin{split}
     {\Y}_{(Ik)}^{\{C_{\IR}\}}=\vcenter{\hbox{\includegraphics[width=1.5cm]{fig/11IKCIR.pdf}}}=\frac{\alpha_s}{4\pi}{\cal N}\frac{\sqrt{y_{IJk}}}{2}I_{(Jk)}^{\{C_{\IR}\}}(y_{IJk}^{-1})+{\cal O}(\lambda),
 \end{split} 
 \end{align}
 \end{subequations}
where the integral $I_{(Jk)}^{\{C_{\IR}\}}$ is defined in 
  \begin{align}
     \begin{split}
         I_{(Jk)}^{\{C_{\IR}\}}=4^{\epsilon+1} \Gamma (\epsilon +1)\int_{a>0}[da]a_1^{2 \epsilon -1} \left[\frac{1}{\sqrt{y_{IJk}}}a_2 a_3+4 a_1 \left(a_2+a_3\right)\right]^{-\epsilon -1},
     \end{split}
\end{align}
 The integral is also easy to perform. The result for this region is 
 \begin{subequations}
 \begin{align}
     \begin{split}
        {\Y}_{(Jk)}^{\{C_{\IR}\}}=\,\,&\frac{\alpha_s}{4\pi}\left(\frac{m^2}{\mu^2\pi}\right)^{-\epsilon}2 \pi \csc(\pi  \epsilon )   \Gamma (\epsilon )  \left(y_{IJk}\right)^{-\frac{\epsilon }{2}}
        \\=\,\,& \frac{\alpha_s}{4\pi}\left(\frac{\bar{m}^2}{\mu^2}\right)^{-\epsilon}\bigg\{\frac{2}{\epsilon^2}-\frac{\log (y_{IJk})}{\epsilon}+\frac{1}{4} \left(\log ^2(y_{IJk})+2 \pi ^2\right)
        \\&+\frac{1}{24} \epsilon \left(-\log ^3(y_{IJk})-6 \pi ^2 \log (y_{IJk})-16 \zeta (3)\right)\bigg\}+{\cal O}\left(\epsilon ^2\right)+{\cal O}(\lambda),
     \end{split}
     \\
     \begin{split}
          {\Y}_{(Ik)}^{\{C_{\IR}\}}=\,\,& \frac{\alpha_s}{4\pi}\left(\frac{m^2}{\mu^2\pi}\right)^{-\epsilon}2 \pi \csc(\pi  \epsilon )   \Gamma (\epsilon )  \left(y_{IJk}\right)^{\frac{\epsilon }{2}}
        \\=\,\,& \frac{\alpha_s}{4\pi}\left(\frac{\bar{m}^2}{\mu^2}\right)^{-\epsilon}\bigg\{\frac{2}{\epsilon^2}+\frac{\log (y_{IJk})}{\epsilon}+\frac{1}{4} \left(\log ^2(y_{IJk})+2 \pi ^2\right)
        \\&+\frac{1}{24} \epsilon \left(\log ^3(y_{IJk})+6 \pi ^2 \log (y_{IJk})-16 \zeta (3)\right)\bigg\}+{\cal O}\left(\epsilon ^2\right)+{\cal O}(\lambda).
     \end{split}
\end{align}
\end{subequations}

\section{Computation and result for IR regions at two loops}
\label{app:2loopIR}

In this appendix, we provide the computations and the results of the IR regions at two loops. The method we used at two loops is the same at one loop; it is described in the beginning of appendix~\ref{app:oneloop}. For each region $R$, we split the section into three subsections which contain the computations of the connected web, multi-gluon-exchanged webs, as well as a summary presenting the integration order of the non-trivial integrals and the result of the region function ${\cal R}^R$.  

For the most complicated IR region $\{C_{\IR},H\}$ in appendix~\ref{app:IRRegionHCIR}, we explain the steps in performing the integration. In the remaining sections, we will only provide the initial parametric representations and the order of integration. 

\subsection{\texorpdfstring{Region ${\cal R}^{C_{\mathrm{IR}},H}_{(IJk)}$}{R{CIR,H}}}
\label{app:IRRegionHCIR}

In this section, we will provide the main steps of the integration of the non-trivial integrals.

\subsubsection{Connected web}
\label{app:111webHCIR}

\begin{align}
\begin{split}
\label{111RCIRH}
{\Y}_{(IJk)}^{\left\{C_{\IR},H\right\}}=\,\,&\vcenter{\hbox{\includegraphics[width=1.5cm]{fig/111CIRH.pdf}}}
\\=\,\,&\left(\frac{\alpha_s}{4\pi}\right)^2{\cal N}^2\frac{\left(y_{IJk}-1\right) }{2 \sqrt{y_{IJk}}}I_{(IJk),1}^{\left\{C_{\IR},H\right\}}(\alpha_{IJ},y_{IJk})
\\&+\left(\frac{\alpha_s}{4\pi}\right)^2{\cal N}^2\frac{\left(\alpha _{IJ}^2+1\right) }{\epsilon  \alpha _{IJ}}\left[I_{(IJk),2}^{\left\{C_{\IR},H\right\}}(\alpha_{IJ},y_{IJk})-I_{(IJk),2}^{\left\{C_{\IR},H\right\}}(\alpha_{IJ},y_{IJk}^{-1})\right]
\\&+\left(\frac{\alpha_s}{4\pi}\right)^2{\cal N}^2\frac{9 (\epsilon -1) \epsilon +2 }{8 \epsilon ^2}\left[\sqrt{y_{IJk}}I_{(IJk),3}^{\left\{C_{\IR},H\right\}}(y_{IJk})-\frac{1}{\sqrt{y_{IJk}}}I_{(IJk),3}^{\left\{C_{\IR},H\right\}}(y_{IJk}^{-1})\right]
\\&+{\cal O}(\lambda),
\end{split}
 \end{align}
The master integrals in the Feynman parameter  representation are
\begin{subequations}
 \begin{align}
\begin{split}
I_{(IJk),1}^{\left\{C_{\IR},H\right\}}=\,\,&- 4^{2\epsilon+2}\Gamma(2+2\epsilon)\int_{a>0} [da] \left[\left(a_1+a_2\right) a_6\right]^{3 \epsilon }
\\&\bigg[4 a_1  a_6\left(a_3+a_4+a_5\right)+4 a_2  a_6\left(a_3+a_4+a_5\right)+\left(a_3^2+a_4^2\right) a_6
\\&\hspace{20mm}+a_3 a_4 a_6 \left(\alpha _{IJ}+\frac{1}{\alpha _{IJ}}\right)+a_2 a_3 a_5 \sqrt{y_{IJk}}+\frac{a_1 a_4 a_5}{\sqrt{y_{IJk}}}\bigg]^{-2 (\epsilon +1)},
\end{split}
 \\
\begin{split}
I_{(IJk),2}^{\left\{C_{\IR},H\right\}}= &- 4^{2\epsilon+2} \Gamma(2+2\epsilon)\int_{a>0} [da] a_5\left(a_1 a_5\right)^{3 \epsilon }
\\&\bigg[a_5 \left(a_2^2+a_3^2\right)+4 a_1a_5 \left(a_2+a_3+a_4\right)
\\&\hspace{20mm}+a_5a_3 a_2 \left(\alpha _{IJ}+\frac{1}{\alpha_{IJ}}\right)+\frac{a_1 a_3 a_4}{\sqrt{y_{IJk}}}\bigg]^{-2 (\epsilon +1)},
\end{split}
\\
     \begin{split}
     \label{commonInt1}
      I_{(IJk),3}^{\left\{C_{\IR},H\right\}}=\,\,&-4^{2\epsilon}  \Gamma (2 \epsilon )\int_{a>0}[da]\left(a_1 a_4\right)^{-2+3 \epsilon } \\&\left[a_2^2a_4+4 a_1 a_4\left(a_2+a_3\right)+a_1 a_2 a_3 \sqrt{y_{IJk}}\right]^{-2 \epsilon }.
     \end{split}
 \end{align}
 \end{subequations}

 For $I_{(IJk),1}^{\left\{C_{\IR},H\right\}}$, we set the parameter $a_4=1$, and by integrating $a_5$, $a_6$, $a_1$ and $a_2$ sequentially, we get the following integral with $a_3$ to be integrated,
\begin{align}
\label{regularInt1}
\begin{split}
I_{(IJk),1}^{\left\{C_{\IR},H\right\}}=\,\,&-  2^{1-4 \epsilon } 3\pi \csc (\pi  \epsilon )\sqrt{y_{IJk}} \Gamma (-\epsilon ) \Gamma (3 \epsilon )  \\&\int_0^{\infty}da_3\frac{ \left(a_3+1\right)^{-3 \epsilon -1} }{a_3 y_{IJk}-1}\left[a_3^{\epsilon } \left(y_{IJk}\right)^{\frac{\epsilon}{2}  }- \left(y_{IJk}\right)^{-\frac{\epsilon}{2} }\right]
 \left(a_3+\alpha _{IJ}\right)^{\epsilon } \left(a_3 +\frac{1}{\alpha _{IJ}}\right)^{\epsilon }.
\end{split}
 \end{align}
This integral is not easy to get a close form in $\epsilon$. Fortunately, the integrand has a regular behaviour for both large and small $a_3$. Therefore, it is safe to expand the integrand in $\epsilon$ before the integration. We then perform the expansion, together with the variable transformation $a_3\rightarrow \frac{x}{1-x}$, getting 
\begin{align}
\label{ExpIntegrand1}
\begin{split}
{\cal N}^2\frac{\left(y_{IJk}-1\right) }{2 \sqrt{y_{IJk}}}I_{(IJk),1}^{\left\{C_{\IR},H\right\}}=\,\,&\left(\frac{\bar{m}^2}{\mu^2}\right)^{-2\epsilon}\int_{0}^1dx\frac{ y_{IJk}-1}{x y_{IJk}+x-1} \log \left(\frac{x y_{IJk}}{1-x}\right)
\\&\bigg\{\frac{2}{\epsilon^2}+\frac{1}{\epsilon}\bigg[2 \log \left(x\alpha _{IJ}-x+1\right)+2 \log \left(1-x +\frac{x}{\alpha_{IJ}}\right)
\\&\hspace{40mm}+\log (1-x)+\log (x)\bigg]+{\cal O}(\epsilon^0)\bigg\}.
\end{split}
 \end{align}
Now the integral at each order of $\epsilon$ is evaluated to generalized polylogarithms. Notice that the corresponding prefactor in eq.~\eqref{111RCIRH} has been included in  eq.~\eqref{ExpIntegrand1}. Using the package {\tt PolyLogTools}~\cite{Duhr:2019tlz}, we obtained the result of the first two orders,
\begin{align}
\label{Result1}
\begin{split}
{\cal N}^2\frac{\left(y_{IJk}-1\right) }{2 \sqrt{y_{IJk}}}I_{(IJk),1}^{\left\{C_{\IR},H\right\}}=\,\,&\left(\frac{\bar{m}^2}{\mu^2}\right)^{-2\epsilon}\frac{ 1-y_{IJk} }{1+y_{IJk}}\bigg[\frac{2}{\epsilon ^2} U_1\left(y_{IJk}\right)+\frac{ 1}{2 \epsilon  }U_2\left(\alpha _{IJ},y_{IJk}\right)+O\left(\epsilon ^0\right)\bigg].
\end{split}
 \end{align}
The functions $U_1$ and $U_2$ can be found in appendix~\ref{app:convfunc}. 

For the integral $I_{(IJk),2}^{\left\{C_{\IR},H\right\}}$, we set $a_2=1$ and integrate $a_5$, $a_1$ and $a_4$  sequentially, leaving behind $a_3$ as the final variable to be integrated,
\begin{align}
    \begin{split}
       I_{(IJk),2}^{\left\{C_{\IR},H\right\}}=\,\,& 4^{1-2 \epsilon } \epsilon   \Gamma (-\epsilon )^2 \Gamma (3 \epsilon ) \Gamma (\epsilon +1)\left(y_{IJk}\right)^{\frac{\epsilon}{2}}
       \\&\int_0^{\infty}da_3 a_3^{\epsilon}\left(a_3+1\right)^{-3 \epsilon }\left(a_3+\alpha _{IJ}\right)^{\epsilon -1} \left(a_3 +\frac{1}{\alpha _{IJ}}\right)^{\epsilon -1}.
    \end{split}
\end{align}
 We then consider the $(I,J)$ antisymmetric combination, which is how 
 $I_{(IJk),2}^{\left\{C_{\IR},H\right\}}$ appears in eq.~\eqref{111RCIRH},
 \begin{align}
    \begin{split}
     {\cal N}^2 \frac{\left(\alpha _{IJ}^2+1\right) }{\epsilon  \alpha _{IJ}}\left[I_{(IJk),2}^{\left\{C_{\IR},H\right\}}(\alpha_{IJ},y_{IJk})-I_{(IJk),2}^{\left\{C_{\IR},H\right\}}(\alpha_{IJ},y_{IJk}^{-1})\right]=\,\,&
      \\&\hspace{-60mm}  {\cal N}^2 4^{1-2 \epsilon } \frac{\left(\alpha _{IJ}^2+1\right) }{ \alpha _{IJ}}  \Gamma (-\epsilon )^2 \Gamma (3 \epsilon ) \Gamma (\epsilon +1)\left(\left(y_{IJk}\right)^{\frac{\epsilon}{2}}-\left(y_{IJk}\right)^{-\frac{\epsilon}{2}}\right)
      \\&\hspace{-50mm}\int_0^{\infty}da_3 a_3^{\epsilon}\left(a_3+1\right)^{-3 \epsilon } \left(a_3+\alpha _{IJ}\right)^{\epsilon -1}\left(a_3 +\frac{1}{\alpha _{IJ}}\right)^{\epsilon -1}.
    \end{split}
\end{align}
Notice that, similarly to  eq.~\eqref{regularInt1}, the integrand is regular for both small and large values of~$a_3$. Next, we change the variables according to $a_3\rightarrow\frac{x}{1-x}$ and expand the integrand in~$\epsilon$. Finally, we get the result in terms of generalized polylogarithms,
\begin{align}
\label{Result2}
    \begin{split}
     {\cal N}^2 \frac{\left(\alpha _{IJ}^2+1\right) }{\epsilon  \alpha _{IJ}}\left[I_{(IJk),2}^{\left\{C_{\IR},H\right\}}(\alpha_{IJ},y_{IJk})-I_{(IJk),2}^{\left\{C_{\IR},H\right\}}(\alpha_{IJ},y_{IJk}^{-1})\right]=\,\,&
      \\&\hspace{-100mm}\left(\frac{\bar{m}^2}{\mu^2}\right)^{-2\epsilon}\frac{ 1+\alpha_{IJ}^2 }{1-\alpha_{IJ}^2}\log\left(y_{IJk}\right)
\bigg\{\frac{8 }{3 \epsilon ^2 }\log(\alpha_{IJ})-\frac{1}{\epsilon}\bigg[\frac{2  }{3 } M_{100}\left(\alpha _{IJ}\right)+2V_1\left(\alpha _{IJ}\right)\bigg]+{\cal O}(\epsilon^0)\bigg\}.
    \end{split}
\end{align}
The function $V_1$ can be found in appendix~\ref{app:convfunc}.

Compared to the two complicated integrals discussed above, the integral $I_{(IJk),3}^{\left\{C_{\IR},H\right\}}$ is straightforward to evaluate and we will present the result directly. Using the command {\tt AlphaIntEvaluate} provided in the package {\tt AmpRed}~\cite{Chen:2024xwt}, we get the following result containing simply ordinary logarithms, 
\begin{align}
\label{Result3}
    \begin{split}
       {\cal N}^2 \frac{9 (\epsilon -1) \epsilon +2 }{8 \epsilon ^2}\left[\sqrt{y_{IJk}}I_{(IJk),3}^{\left\{C_{\IR},H\right\}}(y_{IJk})-\frac{1}{\sqrt{y_{IJk}}}I_{(IJk),3}^{\left\{C_{\IR},H\right\}}(y_{IJk}^{-1})\right]=\,\,&
        \\&\hspace{-100mm}\left(\frac{\bar{m}^2}{\mu^2}\right)^{-2\epsilon}\bigg\{-\frac{2}{3\epsilon^3}\log(y_{IJk})-\frac{1}{36\epsilon}\log(y_{IJk})\left[\log^2(y_{IJk})+24\pi^2\right]+{\cal O}(\epsilon^0)\bigg\}
    \end{split}
\end{align}
Finally, by summing up the results in eqs.~\eqref{Result1},~\eqref{Result2} and~\eqref{Result3}, we get the kinematic function contributing to the region $\{H,C_{\IR}\}$ of the connected web,  
\begin{align}
\label{111HCIR}
\begin{split}
{\Y}_{(IJk)}^{\{H,C_{\IR}\}}=\,\,&\left(\frac{\alpha_s}{4\pi}\right)^2\left(\frac{\bar{m}^2}{\mu^2}\right)^{-2\epsilon}\bigg\{-\frac{2}{3\epsilon^3}\log(y_{IJk})
\\&+\frac{1}{\epsilon^2}\bigg[\frac{1+\alpha_{IJ}^2}{1-\alpha_{IJ}^2}\frac{8}{3}\log(y_{IJk})\log(\alpha_{IJ})+\frac{1-y_{IJk}}{1+y_{IJk}}2U_1(y_{IJk})\bigg]
\\&+\frac{1}{\epsilon}\bigg[-\frac{1+\alpha_{IJ}^2}{1-\alpha_{IJ}^2}\log\left(y_{IJk}\right)\bigg(\frac{4  }{3 } \log\left(\alpha _{IJ}\right)+2 V_1\left(\alpha _{IJ}\right)\bigg)
\\&-\log(y_{IJk})\bigg(\frac{1}{36}\log^2(y_{IJk})+\frac{2\pi^2}{3}\bigg)+\frac{1-y_{IJk}}{1+y_{IJk}}\frac{1}{2}U_{2}(\alpha_{IJ},y_{IJk})\bigg]\bigg\}+{\cal O}(\epsilon^0) \\
&+{\cal O}(\lambda).
\end{split}
 \end{align}

\subsubsection{Multiple-gluon-exchange webs}
\label{sec:TwoLoopMGEWs}

\begin{subequations}

\begin{align}
    \begin{split}
    \label{IJJKCIRH}
        {\Y}_{(Jk)(JI)}^{\left\{C_{\IR},H\right\}}=\,\,&\frac{1}{2}\left(\vcenter{\hbox{\includegraphics[width=1.5cm]{fig/121bCIRH.pdf}}} -\vcenter{\hbox{\includegraphics[width=1.5cm]{fig/121aCIRH.pdf}}}  \right)\\
=\,\,&\left(\frac{\alpha_s}{4\pi}\right)^2{\cal N}^2\frac{\left(\alpha _{IJ}^2+1\right) }{8 \alpha _{IJ}}\frac{1}{ \sqrt{y_{IJk}}}
\left[I_{(Jk)(JI),1}^{\left\{C_{\IR},H\right\}}(\alpha_{IJ},y_{IJk})-I_{(IJ) (Jk),2}^{\left\{C_{\IR},H\right\}}(\alpha_{IJ},y_{IJk})\right]
\\&+{\cal O}(\lambda),
    \end{split}
    \\
\begin{split}
    \label{IIJKCIRH}
      {\Y}_{(IJ)(Ik)}^{\left\{C_{\IR},H\right\}}=\,\,& \frac{1}{2}\left(\vcenter{\hbox{\includegraphics[width=1.5cm]{fig/211aCIRH.pdf}}} -\vcenter{\hbox{\includegraphics[width=1.5cm]{fig/211bCIRH.pdf}}}  \right)
      \\=\,\,&-\left(\frac{\alpha_s}{4\pi}\right)^2{\cal N}^2\frac{\left(\alpha _{IJ}^2+1\right) }{8 \alpha _{IJ}}\sqrt{y_{IJk}}\left[I_{(Jk)(JI),1}^{\left\{C_{\IR},H\right\}}(\alpha_{IJ},y_{IJk}^{-1})-I_{(IJ) (Jk),2}^{\left\{C_{\IR},H\right\}}(\alpha_{IJ},y_{IJk}^{-1})\right]
      \\&+{\cal O}(\lambda),
    \end{split}
\\
    \begin{split}
    \label{IJKKCIRH}
      {\Y}_{(kI)(kJ)}^{\left\{C_{\IR},H\right\}}= &\frac{1}{2}\left(\vcenter{\hbox{\includegraphics[width=1.5cm]{fig/112aCIRH.pdf}}} -\vcenter{\hbox{\includegraphics[width=1.5cm]{fig/112bHCIR.pdf}}}  \right)
      \\=\,\,&\left(\frac{\alpha_s}{4\pi}\right)^2{\cal N}^2\frac{2 \epsilon -1 }{8 \epsilon }\left[\sqrt{y_{IJk}}I_{(kI)(kJ),1}^{\left\{C_{\IR},H\right\}}(y_{IJk})-\frac{1}{\sqrt{y_{IJk}}}I_{(kI)(kJ),1}^{\left\{C_{\IR},H\right\}}(y_{IJk}^{-1})\right]
      \\&+{\cal O}(\lambda).
    \end{split}
\end{align}
\end{subequations}
We will only compute ${\Y}_{(Jk)(JI)}^{\left\{C_{\IR},H\right\}}$, since ${\Y}_{(IJ)(Ik)}^{\left\{C_{\IR},H\right\}}$, can be obtained from it by performing an $(I,J)$ interchange. The web in eq.~\eqref{IJKKCIRH}, where both gluons connect to the lightlike line, is of course different to the other two, and will be computed as well.

The Feynman parameter representation of the master integrals is given below.
\begin{subequations}
    \begin{align}
    \begin{split}
     I_{(Jk)(JI),1}^{\left\{C_{\IR},H\right\}}=\,\,&- 4^{2\epsilon+2} \Gamma (2 \epsilon +2) \int_{a>0}[da]\left(a_1 a_2\right)^{3 \epsilon } \\ &\bigg[ a_2a_3^2+a_2\left(a_4+a_6\right)^2+\left(a_4+a_6\right) a_2a_3 \left(\alpha _{IJ}+\frac{1}{\alpha _{IJ}}\right)
     \\&\hspace{20mm}+4a_1 a_2 \left(a_3+a_4+a_5+2 a_6\right)+\frac{1}{\sqrt{y_{IJk}}}a_1a_5 a_6\bigg]^{-2 (\epsilon +1)},
    \end{split}
\\
    \begin{split}
     I_{(Jk)(JI),2}^{\left\{C_{\IR},H\right\}}=\,\,&-4^{2\epsilon+2}\Gamma (2 \epsilon +2) \int_{a>0}[da]\left(a_1 a_2\right)^{3 \epsilon }  \\&\bigg[a_2 \left(a_3^2+a_6^2\right)+a_2a_6 a_3 \left(\alpha _{IJ}+\frac{1}{\alpha _{IJ}}\right)
     \\&\hspace{5mm}+ 4 a_1a_2 \left(a_3+a_4+a_5+2 a_6\right)+a_1a_5\left(a_4+a_6\right) \frac{1}{\sqrt{y_{IJk}}}\bigg]^{-2 (\epsilon +1)},
    \end{split}
\\
    \begin{split}
     I_{(kI)(kJ),1}^{\left\{C_{\IR},H\right\}}=\,\,&-4^{2 \epsilon +1} \Gamma (2 \epsilon +1) \int_{a>0}[da]\left(a_1 a_2\right)^{3 \epsilon -1} \\&\left[a_1 a_4^2+4 a_1a_2 \left(a_3+a_4+a_5\right)+a_2 a_3 a_5 \sqrt{y_{IJk}}\right]^{-2 \epsilon -1}.
    \end{split}
\end{align}
\end{subequations}

For $ I_{(Jk)(JI),1}^{\left\{C_{\IR},H\right\}}$, we set $a_4+a_6=1$ and integrate $a_5$, $a_1$, $a_2$ and $a_6$ sequentially. Note that the integral range over $a_6$ is from $0$ to $1$. The final integration over $a_3$, remains complicated due to the hypergeometric function appearing in the integrand, 
\begin{align}
\label{intgrandExp}
    \begin{split}
      I_{(Jk)(JI),1}^{\left\{C_{\IR},H\right\}}= & -4^{2-2 \epsilon }\frac{\pi ^2\csc ^2(\pi  \epsilon ) \Gamma (3 \epsilon )  }{\Gamma (\epsilon +1)} \left(y_{IJk}\right)^{\frac{1}{2}-\frac{\epsilon }{2}}
      \\&\int_{0}^{\infty}da_3\left(a_3+1\right)^{-3 \epsilon }  \left(a_3+\alpha _{IJ}\right)^{\epsilon -1} \left(a_3 +\frac{1}{\alpha_{IJ}}\right)^{\epsilon -1}
      \\&\hspace{20mm}\, _2F_1\left(\epsilon ,3 \epsilon ;\epsilon +1;-\frac{1}{a_3+1}\right).
    \end{split}
\end{align}
Let us examine the behaviour of the integrand near the integration boundary. We find that the hypergeometric function is finite at small and large values of $a_3$,
\begin{subequations}
\begin{align}
    \begin{split}
    \lim_{\alpha_{3}\rightarrow 0} \, _2F_1\left(\epsilon ,3 \epsilon ;\epsilon +1;-\frac{1}{a_3+1}\right)=\,\,&\, _2F_1\left(\epsilon ,3 \epsilon ;\epsilon +1;-1\right)+{\cal O}(a_3),
    \end{split}
\\
    \begin{split}
    \lim_{\alpha_{3}\rightarrow \infty} \, _2F_1\left(\epsilon ,3 \epsilon ;\epsilon +1;-\frac{1}{a_3+1}\right)=\,\,&1+O\left(\frac{1}{a_3}\right).
    \end{split}
\end{align}
\end{subequations}
Furthermore, the product of rational factors in the integrand of eq.~\eqref{intgrandExp} is also regular in these two limits of $a_3$. Thus, the entire integrand is finite in both limits, and therefore $\epsilon$ is not necessarily there as a regulator. We perform the variable transformation $a_3\rightarrow \frac{x}{1-x}$ and expand the integrand in $\epsilon$, similarly to how we proceeeded in appendix~\ref{app:111webHCIR}. The integral is then evaluated in terms of generalized polylogarithms. We will present the result after combining the two integrals $I_{(Jk)(JI),1}^{\left\{C_{\IR},H\right\}}$ and $I_{(Jk)(JI),2}^{\left\{C_{\IR},H\right\}}$, since it is simpler than each one of two individually. 

For the integral $I_{(Jk)(JI),2}^{\left\{C_{\IR},H\right\}}$, we set $a_6=1$. We integrate $a_5$, $a_1$, $a_2$ and $a_4$ sequentially, and the remaining integral of $a_3$ is 
\begin{align}
    \begin{split}
       I_{(Jk)(JI),2}^{\left\{C_{\IR},H\right\}}= &  -4^{2-2 \epsilon }\pi ^2\csc ^2(\pi  \epsilon ) \left(y_{IJk}\right)^{\frac{1}{2}-\frac{\epsilon }{2}}
       \\&\int_{0}^{\infty}da_3 \left(a_3+\alpha _{IJ}\right)^{\epsilon -1} \left(a_3 +\frac{1}{\alpha _{IJ}}\right)^{\epsilon -1}  \\&\left[\left(a_3+1\right)^{-2 \epsilon } \Gamma (2 \epsilon )-\frac{\left(a_3+2\right)^{-3 \epsilon } \Gamma (3 \epsilon ) }{\Gamma (\epsilon +1)}\, _2F_1\left(1,3 \epsilon ;\epsilon +1;\frac{1}{a_3+2}\right)\right].
    \end{split}
\end{align}
The asymptotic behaviour of the hypergeometric function is regular in both limits,
\begin{subequations}
\begin{align}
    \begin{split}
    \lim_{\alpha_{3}\rightarrow 0}\, _2F_1\left(1,3 \epsilon ;\epsilon +1;\frac{1}{a_3+2}\right)=\,\,&\, _2F_1\left(1,3 \epsilon ;\epsilon +1;\frac{1}{2}\right)+{\cal O}(a_3),
    \end{split}
\\
    \begin{split}
    \lim_{\alpha_{3}\rightarrow \infty}\, _2F_1\left(1,3 \epsilon ;\epsilon +1;\frac{1}{a_3+2}\right)=\,\,&1+O\left(\frac{1}{a_3}\right),
    \end{split}
\end{align}
\end{subequations}
so we can again expand the integrand in $\epsilon$ and perform the integration order by order. Finally, the result of the kinematic function contributing to the region $\{C_{\IR},H\}$ is given below,
\begin{align}
\label{HCIR211}
\begin{split}
   {\Y}_{(JI)(Jk)}^{\left\{C_{\IR},H\right\}} =\,\,&\left(\frac{\alpha_s}{4\pi}\right)^2\left(\frac{\bar{m}^2}{\mu^2}\right)^{-2\epsilon}\frac{1+\alpha_{IJ}^2}{1-\alpha_{IJ}^2}\bigg\{-\frac{4}{3\epsilon^3}\log\left(\alpha_{IJ}\right)
   \\&+\frac{1}{\epsilon^2}\left[\textcolor{red}{\frac{1}{3}\log \left(\alpha _{IJ}\right) \log \left(y_{IJk}\right)}+\frac{2}{3} M_{100}\left(\alpha _{IJ}\right)+ V_2\left(\alpha _{IJ}\right)\right]
   \\&+\frac{1}{\epsilon}\bigg[-\frac{8}{3} G\left(-1,-1,0,\alpha _{IJ}\right)-\frac{4}{3} G\left(-1,0,0,\alpha _{IJ}\right)+\frac{16}{3} G\left(-1,1,0,\alpha _{IJ}\right)
   \\&-\frac{4}{3} G\left(0,-1,0,\alpha _{IJ}\right)-\frac{16}{3} G\left(0,1,0,\alpha _{IJ}\right)+4 G\left(0,\frac{1}{2},0,\alpha _{IJ}\right)
   \\&+\frac{16}{3} G\left(1,-1,0,\alpha _{IJ}\right)-\frac{4}{3} G\left(1,0,0,\alpha _{IJ}\right)+\frac{16}{3} G\left(1,1,0,\alpha _{IJ}\right)
   \\&-4 G\left(1,2,0,\alpha _{IJ}\right)-4 G\left(1,\frac{1}{2},0,\alpha _{IJ}\right)+4 \log (2) G\left(0,\frac{1}{2},\alpha _{IJ}\right)
   \\&+4 \log (2) G\left(1,2,\alpha _{IJ}\right)-4 \log (2) G\left(1,\frac{1}{2},\alpha _{IJ}\right)-\frac{1}{9} \log ^3\left(\alpha _{IJ}\right)
   \\&-\log (2) \log ^2\left(\alpha _{IJ}\right)+2 \log ^2(2) \log \left(1-\alpha _{IJ}\right)-\frac{8}{9} \pi ^2 \log \left(\alpha _{IJ}\right)
   \\&+\frac{8}{9} \pi ^2 \log \left(1-\alpha _{IJ}\right)-\frac{10}{9} \pi ^2 \log \left(\alpha _{IJ}+1\right)
   \\&-\textcolor{red}{\frac{1}{6} \log \left(y_{IJk}\right) \left(2 M_{100}\left(\alpha _{IJ}\right)+3 V_1\left(\alpha _{IJ}\right)\right)}
   -\frac{1}{12} \log \left(\alpha _{IJ}\right) \log ^2\left(y_{IJk}\right)
   \\&+\log (2) \left(-M_{100}\left(\alpha _{IJ}\right)-V_1\left(\alpha _{IJ}\right)\right)+\frac{37 \zeta (3)}{6}-\frac{2}{3} \pi ^2 \log (2)\bigg]\bigg\}+{\cal O}(\epsilon^0)+{\cal O}(\lambda).
   \end{split}
\end{align}
The same region of the other web ${\Y}_{(IJ)(Ik)}^{\left\{C_{\IR},H\right\}} $ can be obtained by performing $(I,J)$ permutation,
\begin{align}
 {\Y}_{(IJ)(Ik)}^{\left\{C_{\IR},H\right\}}\left(\alpha_{IJ},y_{IJk}\right)=-{\Y}_{(JI)(Jk)}^{\left\{C_{\IR},H\right\}} \left(\alpha_{IJ},y_{IJk}^{-1}\right)\,.
\end{align}
Most of the terms appearing in eq.~\eqref{HCIR211} are $(I,J)$ symmetric, while only the two terms coloured in red are antisymmetric and will survive  in the sum of the two webs,
\begin{align}
    \begin{split}
       {\Y}_{(JK)(JI)}^{\left\{C_{\IR},H\right\}}+{\Y}_{(IJ)(Ik)}^{\left\{C_{\IR},H\right\}}=\,\,&\left(\frac{\alpha_s}{4\pi}\right)^2\left(\frac{\bar{m}^2}{\mu^2}\right)^{-2\epsilon}\frac{1+\alpha_{IJ}^2}{1-\alpha_{IJ}^2}\log(y_{IJk})\bigg\{-\frac{2}{3\epsilon^2}\log(\alpha_{IJ})
        \\&+\frac{1}{\epsilon}\bigg[\frac{2}{3}M_{100}(\alpha_{IJ})+V_1(\alpha_{IJ})\bigg]+{\cal O}(\epsilon^0)\bigg\}+{\cal O}(\lambda).
    \end{split}
\end{align}
 The integral $I_{(kI)(kJ),1}^{\left\{C_{\IR},H\right\}}$ is straightforward to evaluate, and the result of $ {\Y}_{(kI)(kJ)}^{\left\{C_{\IR},H\right\}}$ is given by
 \begin{align}
 \label{112CIRH}
    \begin{split}
   {\Y}_{(kI)(kJ)}^{\left\{C_{\IR},H\right\}}=\left(\frac{\alpha_s}{4\pi}\right)^2\left(\frac{\bar{m}^2}{\mu^2}\right)^{-2\epsilon}\log(y_{IJk})\bigg\{\!-\frac{1}{\epsilon^3}-\frac{1}{24\epsilon}\left[\log^2(y_{IJk})+16\pi^2\right]+{\cal O}(\epsilon^0)\bigg\}+{\cal O}(\lambda).
    \end{split}
\end{align}

\subsubsection{\texorpdfstring{Summary for ${\cal R}_{(IJk)}^{\left\{C_{\IR},H\right\}}$}{Summary {CIR,H}}}

The procedure we used to compute the non-trivial master integrals is summarized in table~\ref{IntCIRH}.
\renewcommand{\arraystretch}{1.5}
\begin{table}[htb]
\centering
\begin{tabular}{|c|c|c|}
\hline
  Integral & Cheng-Wu & Integration orders  \\
    \hhline{|===|}
  $I_{(IJk),1}^{\left\{C_{\IR},H\right\}}$ &$a_4=1$& $a_5\rightarrow a_6\rightarrow a_1\rightarrow a_2\rightarrow a_3$ \\
   \hline
 $I_{(IJk),2}^{\left\{C_{\IR},H\right\}}$ &$a_2=1$& $a_5\rightarrow a_1\rightarrow a_4\rightarrow a_3$  \\
  \hline
 $I_{(Jk)(JI),1}^{\left\{C_{\IR},H\right\}}$ &$a_4+a_6=1$& $a_5\rightarrow a_1\rightarrow a_2\rightarrow a_6\rightarrow a_3$  \\
 \hline
 $I_{(Jk)(JI),2}^{\left\{C_{\IR},H\right\}}$ &$a_6=1$& $a_5\rightarrow a_1\rightarrow a_2\rightarrow a_4\rightarrow a_3$  \\
   \hline
 \end{tabular}
  \caption{The set of master integrals of the $\{C_{\IR},H\}$ region associated with the tripole colour structure, specifying the integration measure in the middle column and the order of integration  in the rightmost column.
  The two remaining integrals $I_{(IJk),3}^{\left\{C_{\IR},H\right\}}$ and $I_{(kI)(kJ),1}^{\left\{C_{\IR},H\right\}}$ are simple, and have been evaluated directly in {\tt{AmpRed}}.}
    \label{IntCIRH}
 \end{table}
Upon summing the webs according to (\ref{eq:C_IRH}) the $\{C_{\IR},H\}$ region 
contribution to the correlators is given by
\begin{align}
\label{RegionCIRH}
\begin{split}
     {\cal R}_{(IJk)}^{\left\{C_{\IR},H\right\}}=\,\,&\left(\frac{\alpha_s}{4\pi}\right)^2\left(\frac{\bar{m}^2}{\mu^2}\right)^{-2\epsilon}\bigg\{-\frac{5}{3\epsilon^3}\log(y_{IJk})
     \\&+\frac{1}{\epsilon^2}\bigg[2\frac{1+\alpha_{IJ}^2}{1-\alpha_{IJ}^2}\log(y_{IJk})\log(\alpha_{IJ})+\frac{1-y_{IJk}}{1+y_{IJk}}2U_1(y_{IJk})\bigg]
     \\&+\frac{1}{\epsilon}\bigg[-\frac{1+\alpha_{IJ}^2}{1-\alpha_{IJ}^2}\log(y_{IJk})V_1(\alpha_{IJ})+\frac{1-y_{IJk}}{1+y_{IJk}}\frac{1}{2}U_{2}(\alpha_{IJ},y_{IJk})
     \\&\hspace{30mm}-\frac{5}{72}\log^3(y_{IJk})-\frac{4}{3}\pi^2\log(y_{IJk})\bigg]\bigg\}+{\cal O}(\epsilon^0)+{\cal O}(\lambda).
     \end{split}
\end{align}

\subsection{\texorpdfstring{Region ${\cal R}_{(IJk)}^{\left\{C_{\IR},C_{\IR}\right\}}$}{R{CIR,CIR}}}

 \subsubsection{Connected web}

 Decomposing the double collinear-IR region of the connected web into master integrals we get 
 \begin{align}
     \begin{split}
         {\Y}_{(IJk)}^{\left\{C_{\IR},C_{\IR}\right\}}=\,\,&\vcenter{\hbox{\includegraphics[width=1.5cm]{fig/111CIRCIR.pdf}}}
         \\=\,\,&\left(\frac{\alpha_s}{4\pi}\right)^2{\cal N}^2\frac{\sqrt{y_{IJk}}}{1+y_{IJk}}\left[I_{(IJk),1}^{\left\{C_{\IR},C_{\IR}\right\}}(y_{IJk})-I_{(IJk),1}^{\left\{C_{\IR},C_{\IR}\right\}}(y_{IJk}^{-1})\right]
         \\&+\left(\frac{\alpha_s}{4\pi}\right)^2{\cal N}^2\frac{2\epsilon-1}{8\epsilon}\frac{1-y_{IJk}}{1+y_{IJk}}I_{(IJk),2}^{\left\{C_{\IR},C_{\IR}\right\}}(y_{IJk})+{\cal O}(\lambda).
     \end{split}
 \end{align}
 The Feynman parameter representation of the master integrals is
 \begin{subequations}
 \begin{align}
     \begin{split}
       I_{(IJk),1}^{\left\{C_{\IR},C_{\IR}\right\}}(y_{IJk})= &-\Gamma (2 \epsilon +1)\int_{a>0}[da]\left(a_2 a_5+a_1a_2+a_1a_5\right)^{3 \epsilon -1} 
       \\&\bigg[ a_2 a_5\left(a_3+a_4\right)+a_1 \left(a_2+a_5\right)\left(a_3+a_4\right)
      +\frac{1}{4 \sqrt{y_{IJk}}}a_1 a_3 a_4\bigg]^{-2 \epsilon -1},
     \end{split}
 \\
     \begin{split}
       I_{(IJk),2}^{\left\{C_{\IR},C_{\IR}\right\}}(y_{IJk})= &-4^{2 \epsilon +1} \Gamma (2 \epsilon +1)\int_{a>0}[da]\left(a_1 a_2\right)^{3 \epsilon -1}
       \\&\left[a_2 a_3 a_5 \sqrt{y_{IJk}}+ 4 a_1a_2 \left(a_3+a_4+a_5\right)+\frac{1}{\sqrt{y_{IJk}}}a_1a_4 a_5\right]^{-2 \epsilon -1}.
     \end{split}
 \end{align}
 \end{subequations}
 The result of the function ${\Y}_{(IJk)}^{\{C_{\IR},C_{\IR}\}}$ is 
 \begin{align}
 \label{111CIRCIR}
     \begin{split}
        {\Y}_{(IJk)}^{\{C_{\IR},C_{\IR}\}} =\,\,&\left(\frac{\alpha_s}{4\pi}\right)^2\left(\frac{\bar{m}^2}{\mu^2}\right)^{-2\epsilon}\bigg\{\frac{1}{\epsilon^3}\log(y_{IJk})-\frac{1}{\epsilon^2}\frac{1-y_{IJk}}{1+y_{IJk}}U_1(y_{IJk})
         \\&+\frac{1}{\epsilon}\log(y_{IJk})\bigg[\frac{1}{6}\log^2(y_{IJk})+\frac{7}{6}\pi^2\bigg]\bigg\}+{\cal O}(\epsilon^0)+{\cal O}(\lambda)\,.
     \end{split}
 \end{align}
 
 \subsubsection{Multiple-gluon-exchange webs}
 
 Decomposing the double collinear-IR region of the multiple-gluon-exchange web $W_{(kI)(kJ)}$ into master integrals we get 
\begin{align}
\label{IRIR121int}
    \begin{split}
       {\Y}_{(kI)(kJ)}^{\left\{C_{\IR},C_{\IR}\right\}} =\,\,&\frac{1}{2}\left(\vcenter{\hbox{\includegraphics[width=1.5cm]{fig/112aCIRCIR.pdf}}} -\vcenter{\hbox{\includegraphics[width=1.5cm]{fig/112bCIRCIR.pdf}}}  \right)
       \\=\,\,&\left(\frac{\alpha_s}{4\pi}\right)^2{\cal N}^2\frac{1}{8}\left[I_{(kI)(kJ),1}^{\left\{C_{\IR},C_{\IR}\right\}}(y_{IJk})-I_{(kI)(kJ),1}^{\left\{C_{\IR},C_{\IR}\right\}}(y_{IJk}^{-1})\right]+{\cal O}(\lambda)\,.
    \end{split}
\end{align}
The Feynman parameter representation of the master integral is 
\begin{align}
    \begin{split}
        I_{(kI)(kJ),1}^{\left\{C_{\IR},C_{\IR}\right\}}(y_{IJk})=\,\,&-4^{2\epsilon+2}\Gamma (2 \epsilon +2)\int_{a>0}[da]\left(a_1 a_2\right)^{3 \epsilon }
        \\&
        \hspace{-20mm}\bigg[a_2 a_3 a_6 \sqrt{y_{IJk}}+4 a_1a_2 \left(a_3+a_4+a_5+2 a_6\right)
        +\frac{1}{\sqrt{y_{IJk}}}a_1a_4 \left(a_5+a_6\right)\bigg]^{-2 (\epsilon +1)}.
    \end{split}
\end{align}
By performing the variable transformation
\begin{align}
a_1\rightarrow b_3\equiv \sqrt{y_{IJk}}a_1\,,
\qquad
a_2\rightarrow b_2\equiv \frac{1}{\sqrt{y_{IJk}}}a_2\,,
\end{align}
one finds that the integrand is independent on $y_{IJk}$,
\begin{align}
    \begin{split}
        I_{(kI)(kJ),1}^{\left\{C_{\IR},C_{\IR}\right\}}(y_{IJk})=\,\,&-4^{2\epsilon+2}\Gamma (2 \epsilon +2)\int_0^{\infty}db_1\int_0^{\infty}db_2\int_0^{\infty}da_3\int_0^{\infty}da_4\int_0^{\infty}da_5\delta(1-a_5)
        \\&\left(b_1 b_2\right)^{3 \epsilon }\bigg[b_2 a_3 a_6 +4 b_1b_2 \left(a_3+a_4+a_5+2 a_6\right)+b_1a_4 \left(a_5+a_6\right)\bigg]^{-2 (\epsilon +1)}
        \\=\,\,&\text{constant}\,.
    \end{split}
\end{align}
Notice that we have used the Cheng-Wu theorem to replace the delta function $\delta(1-\sum_{i=1}^5a_i)$ by $\delta(1-a_5)$. Owing to the antisymmetry  in eq.~\eqref{IRIR121int}, we  conclude  that this web is trivial in this region,
\begin{align}
    \begin{split}
        {\Y}_{(kI)(kJ)}^{\left\{C_{\IR},C_{\IR}\right\}} =0.
    \end{split}
\end{align}

\subsubsection{\texorpdfstring{Summary of ${\cal R}_{(IJk)}^{\left\{C_{\IR},C_{\IR}\right\}}$}{Summary R{CIR,CIR}}}

The integration order for the master integrals appearing in this region is summarized in table~\ref{IntCIRCIR}.
\renewcommand{\arraystretch}{1.5}
\begin{table}[ht]
\centering
\begin{tabular}{|c|c|c|}
\hline
  Integral & Cheng-Wu & Integration orders  \\
    \hhline{|=|=|=|}
  $I_{(IJk),1}^{\left\{C_{\IR},C_{\IR}\right\}}$ &$a_5=1$&$a_3\rightarrow a_4\rightarrow a_1\rightarrow a_2$ \\
  \hline
  $I_{(kI)(kJ),1}^{\left\{C_{\IR},C_{\IR}\right\}}$&$a_5=1$&$a_3\rightarrow a_4\rightarrow a_1\rightarrow a_2$\\
   \hline
 \end{tabular}
  \caption{The set of master integrals of the $\{C_{\IR},C_{\IR}\}$ region associated with the tripole colour structure, specifying the integration measure in the middle column and the order of integration  in the rightmost column.
  Th remaining integral $I_{(IJk),2}^{\left\{C_{\IR},C_{\IR}\right\}}$ is simple, and have been evaluated directly in {\tt{AmpRed}}.}
    \label{IntCIRCIR}
 \end{table}
 
The region function ${\cal R}_{(IJk)}^{\left\{C_{\IR},C_{\IR}\right\}}$  equals the result of the connected web given in eq.~\eqref{111CIRCIR},
\begin{align}
\label{RegionCIRCIR}
     \begin{split}
        {\cal R}_{(IJk)}^{\left\{C_{\IR},C_{\IR}\right\}}=\,\,&\left(\frac{\alpha_s}{4\pi}\right)^2\left(\frac{\bar{m}^2}{\mu^2}\right)^{-2\epsilon}\bigg\{\frac{1}{\epsilon^3}\log(y_{IJk})-\frac{1}{\epsilon^2}\frac{1-y_{IJk}}{1+y_{IJk}}U_1(y_{IJk})
         \\&+\frac{1}{\epsilon}\log(y_{IJk})\bigg[\frac{1}{6}\log^2(y_{IJk})+\frac{7}{6}\pi^2\bigg]\bigg\}+{\cal O}(\epsilon^0)+{\cal O}(\lambda)\,.
     \end{split}
 \end{align}

\subsection{\texorpdfstring{Region ${\cal R}_{(IJk)}^{\left\{C_N,C_{\IR}\right\}}$}{R{CN,CIR}}}

 \subsubsection{Connected web}
 \begin{align}
     \begin{split}
         {\Y}_{(IJk)}^{\left\{C_N,C_{\IR}\right\}}=\,\,&\left(\vcenter{\hbox{\includegraphics[width=1.5cm]{fig/111CNCIRa.pdf}}}+\vcenter{\hbox{\includegraphics[width=1.5cm]{fig/111CNCIRb.pdf}}}\right)
         \\=\,\,&\left(\frac{\alpha_s}{4\pi}\right)^2{\cal N}^2\frac{9 (\epsilon -1) \epsilon +2 }{8 \epsilon ^2}\left[\sqrt{y_{IJk}}I_{(IJk),1}^{\left\{C_N,C_{\IR}\right\}}(y_{IJk})-\frac{1}{\sqrt{y_{IJk}}}I_{(IJk),1}^{\left\{C_N,C_{\IR}\right\}}(y_{IJk}^{-1})\right]
         \\&+{\cal O}(\lambda)
     \end{split}
 \end{align}
 The Feynman parameter representation of the master integral is 
  \begin{align}
     \begin{split}
       I_{(IJk),1}^{\left\{C_N,C_{\IR}\right\}}(y_{IJk})=\,\,&-4^{2\epsilon}  \Gamma (2 \epsilon )\int_{a>0}[da]\left(a_1 a_4\right)^{-2+3 \epsilon } \\&\left[a_2^2a_4+4 a_1a_4 \left(a_2+a_3\right)+a_1 a_2 a_3 \sqrt{y_{IJk}}\right]^{-2 \epsilon },
     \end{split}
 \end{align}
Note that this integral is the same as $I_{(IJk),3}^{\left\{C_{\IR},H\right\}}(y_{IJk})$; see eq.~\eqref{commonInt1}, so we directly write down the result:
 \begin{align}
 \label{111CNCIR}
     \begin{split}
          {\Y}_{(IJk)}^{\{C_N,C_{\IR}\}}=\,\,&\left(\frac{\alpha_s}{4\pi}\right)^2\left(\frac{\bar{m}^2}{\mu^2}\right)^{-2\epsilon}
          \log(y_{IJk})
          \bigg\{-\frac{2}{3\epsilon^3}-\frac{1}{36\epsilon}\left[\log^2(y_{IJk})+24\pi^2\right]\bigg\}
          \\
          &+{\cal O}(\epsilon^0)+{\cal O}(\lambda).
     \end{split}
 \end{align}
 
\subsubsection{Multiple-gluon-exchange webs}

\begin{align}
    \begin{split}
       {\Y}_{(kI)(kJ)}^{\left\{C_N,C_{\IR}\right\}}=\,\,& \frac{1}{2}\left(\vcenter{\hbox{\includegraphics[width=1.5cm]{fig/112aCNCIR.pdf}}} -\vcenter{\hbox{\includegraphics[width=1.5cm]{fig/112bCNCIR.pdf}}}  \right)+\frac{1}{2}\left(\vcenter{\hbox{\includegraphics[width=1.5cm]{fig/112aCIRCN.pdf}}} -\vcenter{\hbox{\includegraphics[width=1.5cm]{fig/112bCIRCN.pdf}}}  \right)
        \\=\,\,&\left(\frac{\alpha_s}{4\pi}\right)^2{\cal N}^2\frac{1}{8\epsilon}\bigg[\sqrt{y_{IJk}}I^{\left\{C_N,C_{\IR}\right\}}_{(kI)(kJ),1}(y_{IJk})-\frac{1}{\sqrt{y_{IJk}}}I^{\left\{C_N,C_{\IR}\right\}}_{(kI)(kJ),1}(y_{IJk}^{-1})\bigg]
        \\&+\left(\frac{\alpha_s}{4\pi}\right)^2{\cal N}^2\frac{2\epsilon-1}{8\epsilon^2}\bigg[\frac{1}{\sqrt{y_{IJk}}}I^{\left\{C_N,C_{\IR}\right\}}_{(kI)(kJ),2}(y_{IJk})-\sqrt{y_{IJk}}I^{\left\{C_N,C_{\IR}\right\}}_{(kI)(kJ),2}(y_{IJk}^{-1})\bigg]
        \\&+\left(\frac{\alpha_s}{4\pi}\right)^2{\cal N}^2\frac{9 (\epsilon -1) \epsilon +2}{16\epsilon^2}\bigg[\frac{1}{\sqrt{y_{IJk}}}I^{\left\{C_N,C_{\IR}\right\}}_{(kI)(kJ),3}(y_{IJk})-\sqrt{y_{IJk}}I^{\left\{C_N,C_{\IR}\right\}}_{(kI)(kJ),3}(y_{IJk}^{-1})\bigg]
        \\&+{\cal O}(\lambda).
    \end{split}
\end{align}
The Feynman parameter representations of these master integrals are  
\begin{subequations}
\begin{align}
    \begin{split}
        I_{(kI)(kJ),1}^{\left\{C_N,C_{\IR}\right\}}(y_{IJk})= &-4^{2\epsilon+2}\Gamma (2 \epsilon +2)\int_{a>0}[da] a_5\left(a_1 a_2\right)^{3 \epsilon } 
        \\&\hspace{-10mm}\left[ a_1a_5^2+4 a_1a_2 \left(a_3+a_4+2 a_5\right)+a_2 a_3 \left(a_4+a_5\right) \sqrt{y_{IJk}}\right]^{-2 (\epsilon +1)},
    \end{split}
\\
    \begin{split}
         I_{(kI)(kJ),2}^{\left\{C_N,C_{\IR}\right\}}(y_{IJk})= & -4^{2\epsilon+2} \Gamma (2 \epsilon +2)\int_{a>0}[da]a_5\left(a_1 a_2\right)^{3 \epsilon } 
         \\&\hspace{-10mm}\left[a_2 \left(a_4+a_5\right)^2+4 a_1 a_2 \left(a_3+a_4+2 a_5\right)+\frac{1 }{\sqrt{y_{IJk}}} a_1a_3 a_5\right]^{-2 (\epsilon +1)},
    \end{split}
\\
    \begin{split}
         I_{(kI)(kJ),3}^{\left\{C_N,C_{\IR}\right\}}(y_{IJk})= & -4^{2\epsilon}\Gamma (2 \epsilon ) \int_{a>0}[da]\left(a_1 a_2\right)^{-2+3 \epsilon }  \\&\left[a_2 a_4^2+4 a_1 a_2 \left(a_3+2 a_4\right)+\frac{1}{\sqrt{y_{IJk}}}a_1 a_3 a_4\right]^{-2 \epsilon }.
    \end{split}
\end{align}
\end{subequations}
The result of the function $ {\Y}_{(kI)(kJ)}^{\left\{C_N,C_{\IR}\right\}}$ is 
\begin{align}
\label{112CNCIR}
    \begin{split}
       {\Y}_{(kI)(kJ)}^{\left\{C_N,C_{\IR}\right\}}=\,\,&\left(\frac{\alpha_s}{4\pi}\right)^2\left(\frac{\bar{m}^2}{\mu^2}\right)^{-2\epsilon}\log(y_{IJk})\bigg\{\frac{1}{3\epsilon^3}+\frac{1}{\epsilon}\left[\frac{1}{72}\log^2(y_{IJk})+\frac{1}{3}\pi^2\right]\bigg\}
       \\&+{\cal O}(\epsilon^0)+{\cal O}(\lambda).
    \end{split}
\end{align}

\subsubsection{\texorpdfstring{Summary for ${\cal R}_{(IJk)}^{\left\{C_N,C_{\IR}\right\}}$}{Summary R{CN,CIR}}}

The order of integration used to compute the master integrals in this region is summarized in table~\ref{IntCNCIR}.
\renewcommand{\arraystretch}{1.5}
\begin{table}[ht]
\centering
\begin{tabular}{|c|c|c|}
\hline
  Integral & Cheng-Wu & Integration orders  \\
    \hhline{|===|}
  $I_{(kI)(kJ),1}^{\left\{C_N,C_{\IR}\right\}}$ &$a_5=1$& $a_1\rightarrow a_2\rightarrow a_3\rightarrow a_4$ \\
   \hline
  $I_{(kI)(kJ),2}^{\left\{C_N,C_{\IR}\right\}}$ &$a_5=1$& $a_1\rightarrow a_2\rightarrow a_3\rightarrow a_4$ \\
   \hline
 \end{tabular}
  \caption{The set of master integrals of the $\{C_N,C_{\IR}\}$ 
  region associated with the tripole colour structure, specifying the integration measure in the middle column and the order of integration  in the rightmost column.
  The two remaining integrals $I_{(IJk),1}^{\left\{C_N,C_{\IR}\right\}}$ and $I_{(kI)(kJ),3}^{\left\{C_N,C_{\IR}\right\}}$ are simple, and have been evaluated directly in {\tt{AmpRed}}.}
    \label{IntCNCIR}
 \end{table}

Finally, the result of the region function $ {\cal R}_{(IJk)}^{\left\{C_N,C_{\IR}\right\}}$ is  
\begin{align}
\label{RegionCNCIR}
    \begin{split}
   {\cal R}_{(IJk)}^{\left\{C_N,C_{\IR}\right\}}=\,\,&\left(\frac{\alpha_s}{4\pi}\right)^2\left(\frac{\bar{m}^2}{\mu^2}\right)^{-2\epsilon}\log(y_{IJk})\bigg\{-\frac{1}{3\epsilon^3}-\frac{1}{\epsilon}\left[\frac{1}{72}\log^2(y_{IJk})+\frac{1}{3}\pi^2\right]\bigg\}
   \\&+{\cal O}(\epsilon^0)+{\cal O}(\lambda).
    \end{split}
\end{align}

\section{Computation and result for neutral regions at two loops}
\label{app:2loopNe}

In this appendix, we present the computations and the results for neutral regions at two loops. We follow the method described in appendix~\ref{app:oneloop}, and organise the results similarly to appendix~\ref{app:2loopIR}. 

For the hard region of the connected webs summarized in appendix~\ref{app:comp111webHH}, we use the method of differential equations~\cite{Henn:2013pwa} to evaluate the master integrals. We use the packages {\tt Kira }~\cite{Maierhofer:2017gsa,Klappert:2020nbg,Lange:2025fba} and {\tt LiteRed}~\cite{Lee:2012cn,Lee:2013mka} to perform the IBP reduction and derive the differential equations respectively.

\subsection{\texorpdfstring{Region ${\cal R}_{(IJk)}^{\left\{C_N,C_N\right\}}$}{R{CN,CN}}}
 Each web in this region is trivial, because the neutral collinear modes will only extract the anti-lightcone direction of the external velocities; see section~\ref{sec:neutralregions}.
 \subsubsection{Connected web}
 \begin{subequations}
 \begin{align}
     \begin{split}
         {\Y}_{(IJk)}^{\left\{C_N,C_N\right\}}=  \vcenter{\hbox{\includegraphics[width=1.5cm]{fig/111CNCN.pdf}}}=0\,.
     \end{split}
      \end{align}
      
\subsubsection{Multiple-gluon-exchange webs}
 \begin{align}
 \label{IkJkCNCN}
 \begin{split}
        {\Y}_{(kI)(kJ)}^{\left\{C_N,C_N\right\}}= \frac{1}{2}\left(\vcenter{\hbox{\includegraphics[width=1.5cm]{fig/112aCNCN.pdf}}} -\vcenter{\hbox{\includegraphics[width=1.5cm]{fig/112bCNCN.pdf}}}  \right)=0\,.
     \end{split}
 \end{align}
 \end{subequations}
 
 \subsubsection{\texorpdfstring{Summary for ${\cal R}_{(IJk)}^{\left\{C_N,C_N\right\}}$}{Summary R{CN,CN}}}
The region function ${\cal R}_{(IJk)}^{\left\{C_N,C_N\right\}}$ is vanishing,
  \begin{align}
     \begin{split}
        {\cal R}_{(IJk)}^{\left\{C_N,C_N\right\}}=0\,.
     \end{split}
 \end{align}
 
 \subsection{\texorpdfstring{Region ${\cal R}_{(IJk)}^{\left\{C_N,H\right\}}$}{R{CN,H}}}
  \subsubsection{Connected web}
  The connected web if this region is trivial due to the $(I,J)$ antisymmetry,
 \begin{subequations}
 \begin{align}
     \begin{split}
         {\Y}_{(IJk)}^{\left\{C_N,H\right\}}  =\,\,&\left(\vcenter{\hbox{\includegraphics[width=1.5cm]{fig/111HCNap.pdf}}}+\vcenter{\hbox{\includegraphics[width=1.5cm]{fig/111HCNbp.pdf}}}\right)=0\,.
     \end{split}
     \end{align}
     
\subsubsection{Multiple-gluon-exchange webs}

 \begin{align}
     \begin{split}
         {\Y}_{(IJ)(Ik)}^{\left\{C_N,H\right\}}=\,\,&\frac{1}{2}\left(\vcenter{\hbox{\includegraphics[width=1.5cm]{fig/211aCNH.pdf}}} -\vcenter{\hbox{\includegraphics[width=1.5cm]{fig/211bCNH.pdf}}}  \right)=-\left(\frac{\alpha_s}{4\pi}\right)^2{\cal N}^2\frac{\alpha _{IJ}^2+1}{4 \epsilon  \alpha _{IJ}}I_{(IJ)(Ik),1}^{\{C_N,H\}}\left(\alpha_{IJ}\right)+{\cal O}(\lambda)\,,
     \end{split}
     \\
     \begin{split}
        {\Y}_{(Jk)(JI)}^{\left\{C_N,H\right\}}=\,\,&\frac{1}{2}\left(\vcenter{\hbox{\includegraphics[width=1.5cm]{fig/121bCNH.pdf}}} -\vcenter{\hbox{\includegraphics[width=1.5cm]{fig/121aCNH.pdf}}}  \right)=\left(\frac{\alpha_s}{4\pi}\right)^2{\cal N}^2\frac{\alpha _{IJ}^2+1}{4 \epsilon  \alpha _{IJ}}I_{(IJ)(Ik),1}^{\{C_N,H\}}\left(\alpha_{IJ}\right)+{\cal O}(\lambda)\,,
     \end{split}
 \\
     \begin{split}
          {\Y}_{(kI)(kJ)}^{\left\{C_N,H\right\}} = &\frac{1}{2}\vcenter{\hbox{\includegraphics[width=1.5cm]{fig/112aCNH.pdf}}} -\frac{1}{2}\vcenter{\hbox{\includegraphics[width=1.5cm]{fig/112bHCN.pdf}}} =0\,.
     \end{split}
 \end{align}
 \end{subequations}
 The Feynman parameter representation of the master integral is  
 \begin{align}
 \begin{split}
     I_{(IJ)(Ik),1}^{\{C_N,H\}}=\,\,&-4^{2\epsilon +2} \Gamma (2 \epsilon +2)\int_{a>0}\left[da\right]a_2 \left(a_1 a_2\right)^{3 \epsilon } 
     \\& \left[\frac{a_2 \left(a_3 \alpha _{IJ}+a_4\right) 
     \left(a_4 \alpha _{IJ}+a_3\right)}{\alpha _{IJ}}+a_1 a_5^2+4 a_1a_2 \left(a_3+a_4+a_5\right)\right]^{-2 (\epsilon +1)}\,.
     \end{split}
 \end{align}
  The result of the function ${\Y}_{(IJ)(Ik)}^{\left\{C_N,H\right\}}$ is 
 \begin{align}
 \label{211CNH}
 \begin{split}
      {\Y}_{(IJ)(Ik)}^{\left\{C_N,H\right\}}=\,\,&\left(\frac{\alpha_s}{4\pi}\right)^2\left(\frac{\bar{m}^2}{\mu^2}\right)^{-2\epsilon}\bigg\{\frac{1}{\epsilon^3}\log\left(\alpha_{IJ}\right)+\frac{1}{2\epsilon^2}V_1\left(\alpha_{IJ}\right)
      \\&+\frac{1}{\epsilon}\bigg[4 G\left(-1,-1,0,\alpha _{IJ}\right)-2 G\left(-1,0,0,\alpha _{IJ}\right)-2 G\left(0,-1,0,\alpha _{IJ}\right)
      \\&+\frac{1}{6} \log ^3\left(\alpha _{IJ}\right)+\frac{2}{3} \pi ^2 \log \left(\alpha _{IJ}\right)+\frac{1}{3} \pi ^2 \log \left(\alpha _{IJ}+1\right)-2 \zeta (3)\bigg]\bigg\}+{\cal O}(\epsilon^0)+{\cal O}(\lambda)\,.
      \end{split}
 \end{align}
 Notice that the result is symmetric under the $(I,J)$ interchange. Therefore, the sum of the two webs is trivial,
 \begin{align}
      {\Y}_{(IJ)(Ik)}^{\left\{C_N,H\right\}}+ {\Y}_{(Jk)(JI)}^{\left\{C_N,H\right\}}=0\,.
 \end{align}
 
 \subsubsection{\texorpdfstring{Summary for ${\cal R}_{(IJk)}^{\left\{C_N,H\right\}}$}{Summary R{CN,H}}}

The integration order of the master integral appearing in this region is summarized in table~\ref{IntCNH}.
 \begin{table}[ht]
\centering
\begin{tabular}{|c|c|c|}
\hline
  Integral & Cheng-Wu & Integration orders  \\
    \hhline{|===|}
  $ I_{(IJ)(Ik),1}^{\{C_N,H\}}$ &$a_5=1$& $a_1\rightarrow a_2\rightarrow a_4\rightarrow a_3$ \\
   \hline
 \end{tabular}
  \caption{The master integral of the $\{C_N,H\}$ region associated with the tripole colour structure, specifying the integration measure in the middle column and the order of integration  in the rightmost column.
  }
    \label{IntCNH}
 \end{table}
 
 The region function $ {\cal R}_{(IJk)}^{\left\{C_N,H\right\}}$ is also trivial due to the $(I,J)$ antisymmetry,
  \begin{align}
     \begin{split}
        {\cal R}_{(IJk)}^{\left\{C_N,H\right\}}=0.
     \end{split}
 \end{align}

 \subsection{\texorpdfstring{Region ${\cal R}_{(IJk)}^{\left\{H,H\right\}}$}{R{H,H}}}
 
 \subsubsection{Connected web}
 \label{app:comp111webHH}

 For the hard region of the connected web, we will use the method of differential equations~\cite{Henn:2013pwa} to compute the integrals.

 As the strict lightcone limit of the timelike web, we simply replace the time timelike velocity $\beta_K$ by the lightlike $\beta_k$ in the integrand in eq.~\eqref{Tripole_connected_web} at the leading order in $\lambda$,
 \begin{align}
 \label{111HHintegrand}
     \begin{split}
         {\Y}_{(IJk)}^{\left\{H,H\right\}}=\vcenter{\hbox{\includegraphics[width=1.5cm]{fig/111HH.pdf}}}=\,\,&\left(\frac{\alpha_s}{4\pi}\right)^2{\cal N}^2\int\frac{d^Dk_I}{i\pi^{\frac{D}{2}}}\int\frac{d^Dk_J}{i\pi^{\frac{D}{2}}}\int \frac{d^Dk_K}{i\pi^{\frac{D}{2}}} \frac{1}{k_I^2}\frac{1}{k_J^2}\frac{1}{(k_I+k_J)^2}
 \\&\frac{v_I^{\mu}}{v_I\cdot k_I-1}\frac{v_J^{\nu}}{v_J\cdot k_J-1}\frac{\tilde{\beta}_k^{\rho}}{-\tilde{\beta}_k\cdot (k_I+k_J)}
 \\&\left[g_{\mu\nu}\left(k_I-k_J\right)_{\rho}+g_{\nu\rho}\left(2k_J+k_I\right)_{\mu}-g_{\rho\mu}\left(2k_I+k_J\right)_{\nu}\right]+{\cal O}(\lambda).
         \end{split}
 \end{align}
For the lightlike velocity $\beta_k$, the normalized velocity is not well defined due to the vanishing virtuality. For convenience we then define $\tilde{\beta}_k$ as follows,
 \begin{align}
     \begin{split}
         \tilde{\beta}_k^\nu=\frac{\beta_k^\nu}{-\beta_k\cdot\beta_J}.
     \end{split}
 \end{align}
 With this definition, the scalar products are  
 
 \begin{align}
     \begin{split}
     v_I^2=v_J^2=1,
    \qquad
     v_I\cdot v_{J}=-\frac{1}{2}\left(\frac{1}{\alpha_{IJ}}+\alpha_{IJ}\right),
     \\
     v_I\cdot\tilde{\beta}_{k}=-y_{IJk},
    \qquad
     v_J\cdot\tilde{\beta}_{k}=-1,
     \qquad
     \tilde{\beta}_{k}^2=0.
     \end{split}
 \end{align}
Having defined the rescaled velocities as above, $\{\tilde{\beta}_k,v_I,v_J\}$,  rescaling invariance is manifest.
 
As the starting point for deriving the set of differential equations, we define the integral family which includes the propagators in eq.~\eqref{111HHintegrand} as well as three auxiliary scalar products,
 \begin{align}
     \begin{split}
         J_{\nu_1,\nu_{2},\nu_{3},\nu_4,\nu_{5},\nu_{6},\nu_7,\nu_{8},\nu_{9}}=\,\,&\left(\frac{\alpha_s}{4\pi}\right)^2{\cal N}^2\int\frac{d^Dk_I}{i\pi^{\frac{D}{2}}}\int\frac{d^Dk_J}{i\pi^{\frac{D}{2}}}
         \\&\hspace{-30mm}\frac{\left[v_J\cdot k_I\right]^{-\nu_7}\left[\tilde{\beta}_k\cdot k_J\right]^{-\nu_8}\left[v_I\cdot k_J\right]^{-\nu_9}}{ \left[k_I^2\right]^{\nu_1}\left[k_J^2\right]^{\nu_2}\left[(k_I+k_J)^2\right]^{\nu_3}\left[v_I\cdot k_I-1\right]^{\nu_4}\left[v_J\cdot k_J-1\right]^{\nu_5}\left[-\tilde{\beta}_k\cdot (k_I+k_J)\right]^{\nu_6}}.
     \end{split}
 \end{align}
 After the IBP reduction, the integrals in this family are expressed as linear combinations of eleven master integrals,
 \begin{align}
     \begin{split}
         \left( \begin{array}{c}
         J_1 \\
         J_2 \\
         J_3 \\
         J_4 \\
         J_5 \\
         J_6 \\
         J_7 \\
         J_8 \\
         J_9 \\
         J_{10} \\
         J_{11} 
 \end{array} \right)= \left( \begin{array}{c}
         J_{0, 1, 1, 1, 0, 0, 0, 0, 0} \\
         J_{1, 1, 0, 1, 1, 0, 0, 0, 0} \\
         J_{0, 1, 1, 1, 1, 0, 0, 0, 0} \\
         J_{0, 1, 1, 2, 1, 0, 0, 0, 0} \\
         J_{1, 1, 1, 1, 1, 0, 0, 0, 0} \\
         J_{1, 1, 0, 1, 1, 1, 0, 0, 0} \\
         J_{1, 1, 0, 1, 1, 2, 0, 0, 0} \\
         J_{1, 1, 1, 1, 1, 1, 0, 0, 0} \\
         J_{1, 1, 1, 1, 1, 1, -1, 0, 0} \\
         J_{1, 1, 1, 1, 1, 1, 0, -1, 0} \\
         J_{1, 1, 1, 1, 1, 1, 0, 0, -1} \,,
 \end{array} \right)
     \end{split}
 \end{align}
 with the top sector defined by the last four master integrals in this eleven dimensional basis.
 
 The subsectors have been transformed into canonical form for the fully timelike case in ref.~\cite{Milloy:2020hzi}. For the one-regulator case, the top sector can also be transformed into canonical form~\cite{Henn:2023pqn}. Making use of this previous work, along with the package {\tt CANONICA}~\cite{Meyer:2017joq}, we find the following Uniformly Transcendental (UT) basis for this system,
\begin{subequations}
     \begin{align}
     \begin{split}
         I_{(IJk),1}^{\left\{H,H\right\}}=\,\,&\frac{ (3-2 \epsilon  (8 \epsilon  (2 \epsilon -3)+11))}{32 \epsilon ^3}J_1,
          \end{split}
    \\
     \begin{split}
        \\I_{(IJk),2}^{\left\{H,H\right\}}=\,\,&\frac{ (3-2 \epsilon  (8 \epsilon  (2 \epsilon -3)+11))}{3 \epsilon ^3}J_1+\frac{(1-2 \epsilon )^2}{\epsilon ^2}J_2,
         \end{split}
    \\
     \begin{split}
       \\ I_{(IJk),3}^{\left\{H,H\right\}}=\,\,&-\frac{2 (2 \epsilon  (4 \epsilon -5)+3) \left(\alpha _{IJ}+1\right)}{3 \epsilon ^2 \left(\alpha _{IJ}-1\right)}J_1-\frac{2 (1-2 \epsilon )^2 \left(\alpha _{IJ}+1\right)}{3 \epsilon ^2 \left(\alpha _{IJ}-1\right)}J_3
       \\&+\frac{ (3 \epsilon -1) \left(\alpha _{IJ}^2-1\right)}{3 \epsilon ^2 \alpha _{IJ}}J_4,
        \end{split}
    \\
     \begin{split}
      \\ I_{(IJk),4}^{\left\{H,H\right\}}=\,\,&\frac{ (2 \epsilon  (4 \epsilon -5)+3) \left((4 \epsilon -1) \alpha _{IJ}-6 \epsilon +1\right)}{6 \epsilon ^3 \left(\alpha _{IJ}-1\right)}J_1
      \\&-\frac{(1-2 \epsilon )^2}{3 \epsilon ^2 \left(\alpha _{IJ}-1\right)}J_3-\frac{\left(\epsilon  \alpha _{IJ}+3 \epsilon -1\right)}{6 \epsilon ^2 \alpha _{IJ}}J_4,
       \end{split}
     \\
     \begin{split}
       \\  I_{(IJk),5}^{\left\{H,H\right\}}=\,\,&-\frac{ \left(1-\alpha _{IJ}^2\right)}{2 \alpha _{IJ}}J_5,
        \end{split}
   \\
     \begin{split}
       \\   I_{(IJk),6}^{\left\{H,H\right\}}=\,\,&\frac{ (1-2 \epsilon )^2 y_{IJk}}{8 \epsilon ^2 \left(y_{IJk}-1\right)}J_2+\frac{ (1-4 \epsilon ) y_{IJk}}{4 \epsilon  \left(y_{IJk}-1\right)}J_6+\frac{ y_{IJk}^2}{\epsilon  \left(y_{IJk}-1\right)}J_7,
 \end{split}
    \\
     \begin{split}
        \\ I_{(IJk),7}^{\left\{H,H\right\}}=\,\,&\frac{3  y_{IJk}}{16 \epsilon }J_7,
         \end{split}
    \\
     \begin{split}
        \\I_{(IJk),8}^{\left\{H,H\right\}}=\,\,&\left(y_{IJk}+1\right)J_8,
         \end{split}
     \\
     \begin{split}
        \\I_{(IJk),9}^{\left\{H,H\right\}}=\,\,& y_{IJk}J_8+ y_{IJk}J_9,
         \end{split}
    \\
     \begin{split}
        \\I_{(IJk),10}^{\left\{H,H\right\}}=\,\,&-\frac{ \left(1-\alpha _{IJ}^2\right)}{2 \alpha _{IJ}}J_{10},
         \end{split}
    \\
     \begin{split}
        \\I_{(IJk),11}^{\left\{H,H\right\}}=\,\,&J_{11}+J_8.
     \end{split}
 \end{align}
 \end{subequations}
With the above UT basis, the differential equation takes the canonical form,
\begin{align}
    \label{CForm}
d\overrightarrow{I}_{(IJk)}^{\left\{H,H\right\}}=\epsilon\sum_{i}d\log(\omega_i)\mathbf{C}_i\overrightarrow{I}_{(IJk)}^{\left\{H,H\right\}},
\end{align}
 where $\overrightarrow{I}_{(IJk)}^{\left\{H,H\right\}}$ is a vector whose components are the eleven UT basis elements above, and $\mathbf{C}_i$ is a constant matrix associated with the symbol letter $\omega_i$. The alphabet of this system is,
 \begin{align}
     \{\omega_i\}=\left\{\alpha _{IJ}+1,\alpha _{IJ},1-\alpha _{IJ},\alpha _{IJ}+y_{IJk},\alpha _{IJ} y_{IJk}+1,y_{IJk}+1,y_{IJk}\right\}.
 \end{align}
 The constant matrices $\mathbf{C}_i$ are stored in the {\tt Mathematica} notebook {\tt CanonicalForm.nb} given in ref.~\cite{MathematicaNoteBook}. In the notebook, the matrices are given by {\tt dlog2RegHH} following the order of the alphabet {\tt alpha2RegHH}.
 
The combination of integrals in which we are interested, namely the connected web of eq.~\eqref{111HHintegrand}, may be written in terms of UT basis as follows:
 \begin{align}
     \begin{split}
         {\Y}_{(IJk)}^{\left\{H,H\right\}}=\,\,&\frac{1-y_{IJk}}{1+y_{IJk}}I_{(IJk),8}^{\left\{H,H\right\}}(\alpha_{IJ},y_{IJk})
         \\&-\frac{1+\alpha_{IJ}^2}{1-\alpha_{IJ}^2}\left[I_{(IJk),5}^{\left\{H,H\right\}}(\alpha_{IJ})-2I_{(IJk),10}^{\left\{H,H\right\}}(\alpha_{IJ},y_{IJk})\right]
         \\&+2\left[I_{(IJk),9}^{\left\{H,H\right\}}(\alpha_{IJ},y_{IJk})-I_{(IJk),11}^{\left\{H,H\right\}}(\alpha_{IJ},y_{IJk})\right].
         \end{split}
 \end{align}
 
The solution of the differential equation presented in eq.~\eqref{CForm} can be formally written as the following path-ordered exponential,
\begin{align}
     \begin{split}
        \overrightarrow{I}_{(IJk)}^{\left\{H,H\right\}}(\alpha_{IJ},y_{IJk})=\text{P}\exp\left[\epsilon\sum_{i}\mathbf{C}_i\int_{\gamma_{\text{C}}}d\log(\omega_i)\right]\overrightarrow{I}_{(IJk)}^{\left\{H,H\right\}}(\bar{\alpha}_{IJ},\bar{y}_{IJk}),
         \end{split}
 \end{align}
where $\gamma_{\text{C}}$ is the integral contour defined in the two-dimensional space spanned by the two kinematic variables, starting from a certain point $(\bar{\alpha}_{IJ},\bar{y}_{IJk})$ where the boundary values are given, to the general point $(\alpha_{IJ},y_{IJk})$. In this calculation, the contour is chosen to be 
\begin{align}
     \begin{split}
        \gamma_{\text{C}}:(0,0)\rightarrow (0,y_{IJk})\rightarrow (\alpha_{IJ},y_{IJk}).
         \end{split}
 \end{align}
The next step is to determine the boundary values. We use the following three conditions to fix all the boundary values.

\subsubsection*{Single-pole condition}

The sum of all the regions must reproduce the lightcone expansion performed on the timelike result ${\cal T}_{\lambda}\left[{\cal Y}_{(IJK)}\right]$, which can only have a single pole in $\epsilon$, corresponding to the overall UV singularity generated by shrinking the entire web towards the interaction vertex of the Wilson lines. 
This requires that all the higher order poles appearing in the individual regions, cancel out upon summing up the region integrals,
\begin{align}
\label{PoleCancellation_boundary}
    \sum_{R}{\cal Y}_{(IJk)}^R={\cal O}(\epsilon^{-1})\,.
\end{align}
Defining the expansion coefficients of ${\cal Y}_{(IJk)}^{\{H,H\}}$ in $\epsilon$,
\begin{align}
   {\cal Y}_{(IJk)}^{\{H,H\}}=\sum_{l}{\cal Y}_{(IJk)}^{\{H,H\},(l)}\epsilon^l\,,
\end{align}
the above condition fixes the leading order of ${\cal Y}_{(IJk)}^{\{H,H\}}$, which is a double pole in $\epsilon$, as well as all full-depth terms in the single pole $ {\cal Y}_{(IJk)}^{\{H,H\},(-1)}$, leaving behind some undetermined coefficients of $\pi^2$ and $\zeta(3)$. 

\subsubsection*{Straight-line limit}

By taking the physical kinematic limit where the two timelike Wilson lines are along the same direction, but one is in the initial state and the other in the final state, 
$\beta_{I}\rightarrow-\beta_{J}$, the kinematic variables become
\begin{align}
\label{StraightLine_condition}
    \alpha_{IJ}\rightarrow 1\,,
    \qquad
    y_{IJk}\rightarrow -1\,.
\end{align}
We then perform the asymptotic expansion around this point by setting $\alpha_{IJ}\equiv  1+\alpha\delta$ and $y_{IJk}\equiv -1+y\delta$ with $\delta\to 0$. To this end we use the package {\tt AmpRed}, starting with the parametric representation of the integral~$ {\cal Y}_{(IJk)}^{\{H,H\}}$, finding that this  straight-line limit yields a pure power expansion in $\delta$, with no additional logarithms. We can therefore compute the integral given in eq.~\eqref{111HHintegrand}  in the strict straight-line limit, i.e., $\alpha_{IJ}=1$ and $y_{IJk}=-1$,
\begin{align}
     \begin{split}
         {\Y}_{(IJk)}^{\left\{H,H\right\}}\bigg|_{\alpha_{IJ}=1,\, y_{IJk}=-1}=\,\,&\left(\frac{\alpha_s}{4\pi}\right)^2{\cal N}^2\int\frac{d^Dk_I}{i\pi^{\frac{D}{2}}}\int\frac{d^Dk_J}{i\pi^{\frac{D}{2}}}\int \frac{d^Dk_K}{i\pi^{\frac{D}{2}}} \frac{1}{k_I^2}\frac{1}{k_J^2}\frac{1}{(k_I+k_J)^2}
 \\&\frac{-v_J^{\mu}}{-v_J\cdot k_I-1}\frac{v_J^{\nu}}{v_J\cdot k_J-1}\frac{\tilde{\beta}_k^{\rho}}{-\tilde{\beta}_k\cdot (k_I+k_J)}
 \\&\left[g_{\mu\nu}\left(k_I-k_J\right)_{\rho}+g_{\nu\rho}\left(2k_J+k_I\right)_{\mu}-g_{\rho\mu}\left(2k_I+k_J\right)_{\nu}\right]+{\cal O}(\lambda).
         \end{split}
 \end{align}
Moreover, we find that the integral vanishes in this limit after the IBP reduction is performed, i.e. we have
\begin{align}
    {\cal T}_{\delta}\left[{\cal Y}_{(IJk)}^{\{H,H\},(-1)}\right]=0\,.
\end{align}
This, together with eq.~(\ref{PoleCancellation_boundary}) allows us to fix all the boundary values, except for a single term proportional to $\pi^2$.

\subsubsection*{First-entry condition}

On the principal sheet, the unphysical symbol letters, $\alpha_{IJ}+y_{IJk}$ and $1+\alpha_{IJ}y_{IJk}$, should not appear as branch points. Given that the result is written in terms of generalized polylogarithms, this property can be checked by taking the discontinuity around the unphysical symbol letter $\omega_i$ using the coaction, see e.g.~\cite{Gaiotto:2011dt,Duhr:2012fh,Abreu:2014cla,Duhr:2019tlz}. We decompose the function $ {\cal Y}_{(IJk)}^{\{H,H\},(-1)}$ as follows,
\begin{align}
  {\cal Y}_{(IJk)}^{\{H,H\},(-1)}={\cal Y}_{(IJk),0}^{\{H,H\},(-1),}+\pi^2{\cal Y}_{(IJk),2}^{\{H,H\},(-1)}+\zeta(3){\cal Y}_{(IJk),3}^{\{H,H\},(-1)},
\end{align}
where ${\cal Y}_{(IJk),0}^{\{H,H\}}$ contains all the full-depth terms, while ${\cal Y}_{(IJk),2}^{\{H,H\}}$ and ${\cal Y}_{(IJk),3}^{\{H,H\}}$ are proportional to $\pi^2$ and $\zeta(3)$, respectively. Then, the discontinuity around the symbol letter $\omega_i$ is given by
\begin{align}
  \text{Disc}_{\omega_i}\left[{\cal Y}_{(IJk)}^{\{H,H\},(-1)}\right]=\mu\left\{\left(\text{Disc}_{\omega_i}\otimes \text{id}\right)\left[\Delta_{1,2}{\cal Y}_{(IJk),0}^{\{H,H\},(-1)}\right]\right\}+\pi^2\text{Disc}_{\omega_i}\left[{\cal Y}_{(IJk),2}^{\{H,H\},(-1)}\right],
\end{align}
where the operator $\mu$ maps the tensor product to the ordinary one. By requiring the vanishing of the discontinuities, 
\begin{align}
   \text{Disc}_{\alpha_{IJ}+y_{IJk}}\left[{\cal Y}_{(IJk)}^{\{H,H\},(-1)}\right]=\text{Disc}_{1+\alpha_{IJ}y_{IJk}}\left[{\cal Y}_{(IJk)}^{\{H,H\},(-1)}\right]=0\,,
\end{align}
we fix the remaining coefficient.

Finally, we obtain the result of the first two orders of  $ {\Y}_{(IJk)}^{\{H,H\}}$ in the $\epsilon$ expansion, 
 \begin{align}
     \begin{split}
       {\Y}_{(IJk)}^{\{H,H\}}=\,\,&\left(\frac{\alpha_s}{4\pi}\right)^2\left(\frac{\bar{m}^2}{\mu^2}\right)^{-2\epsilon}\bigg\{\frac{1}{\epsilon^2}\bigg[-\frac{1-y_{IJk}}{1+y_{IJk}}U_1(y_{IJk})-2\frac{1+\alpha_{IJ}^2}{1-\alpha_{IJ}^2}\log(y_{IJk})\log(\alpha_{IJ})\bigg]
       \\&+\frac{1}{\epsilon}\bigg[\frac{1+\alpha_{IJ}^2}{1-\alpha_{IJ}^2}\log(y_{IJk})\bigg(2V_1(\alpha_{IJ})+M_{100}(\alpha_{IJ})\bigg)-\frac{1}{2}\frac{1-y_{IJk}}{1+y_{IJk}}U_2(\alpha_{IJ},y_{IJk})
       \\&+2\log^2\left(\alpha _{IJ}\right) \log \left(y_{IJk}\right)
       -\frac{2}{3}  \log ^3\left(y_{IJk}\right)-\frac{2}{3} \pi ^2 \log \left(y_{IJk}\right)\bigg]\bigg\}+{\cal O}(\epsilon^0)+{\cal O}(\lambda)\,.
         \end{split}
 \end{align}
 
 \subsubsection{Multiple-gluon-exchange webs}
 
 \begin{subequations}
 \begin{align}
     \begin{split}
        {\Y}_{(IJ)(IK)}^{\left\{H,H\right\}}=\,\,&\frac{1}{2}\left(\vcenter{\hbox{\includegraphics[width=1.5cm]{fig/211aHH.pdf}}} -\vcenter{\hbox{\includegraphics[width=1.5cm]{fig/211bHH.pdf}}}  \right)
        \\=\,\,&\left(\frac{\alpha_s}{4\pi}\right)^2{\cal N}^2\frac{\alpha _{IJ}^2+1}{4 \epsilon  \alpha _{IJ}}\left[I_{(IJ)(Ik),1}^{\{H,H\}}\left(\alpha_{IJ}\right)-I_{(IJ)(Ik),2}^{\{H,H\}}\left(\alpha_{IJ}\right)+\frac{1}{\epsilon}I_{(IJ)(Ik),3}^{\{H,H\}}\left(\alpha_{IJ}\right)\right]
        \\&+{\cal O}(\lambda)\,,
     \end{split}
     \\
     \begin{split}
        {\Y}_{(Jk)(JI)}^{\left\{H,H\right\}}=\,\,&\frac{1}{2}\left(\vcenter{\hbox{\includegraphics[width=1.5cm]{fig/121bHH.pdf}}} -\vcenter{\hbox{\includegraphics[width=1.5cm]{fig/121aHH.pdf}}}  \right)
       \\ =\,\,&-\left(\frac{\alpha_s}{4\pi}\right)^2{\cal N}^2\frac{\alpha _{IJ}^2+1}{4 \epsilon  \alpha _{IJ}}\left[I_{(IJ)(Ik),1}^{\{H,H\}}\left(\alpha_{IJ}\right)-I_{(IJ)(Ik),2}^{\{H,H\}}\left(\alpha_{IJ}\right)+\frac{1}{\epsilon}I_{(IJ)(Ik),3}^{\{H,H\}}\left(\alpha_{IJ}\right)\right]
       \\&+{\cal O}(\lambda)\,,  
     \end{split}
 \\
     \begin{split}
       {\Y}_{(kI)(kJ)}^{\left\{H,H\right\}}=\,\,&\frac{1}{2}\left(\vcenter{\hbox{\includegraphics[width=1.5cm]{fig/112aHH.pdf}}} -\vcenter{\hbox{\includegraphics[width=1.5cm]{fig/112bHH.pdf}}}  \right) 
       \\=\,\,&-\left(\frac{\alpha_s}{4\pi}\right)^2{\cal N}^2\frac{y_{IJk}}{4\epsilon}I_{(kI)(kJ),1}^{\left\{H,H\right\}}(y_{IJk})+\left(\frac{\alpha_s}{4\pi}\right)^2{\cal N}^2\frac{1}{4\epsilon}I_{(kI)(kJ),2}^{\left\{H,H\right\}}(y_{IJk})+{\cal O}(\lambda)\,.
     \end{split}
 \end{align}
  \end{subequations}
 The Feynman parameter representation of the master integrals is  
\begin{subequations}
 \begin{align}
 \begin{split}
     I_{(IJ)(Ik),1}^{\{H,H\}}=\,\,&-4^{2\epsilon +2} \Gamma (2 \epsilon +2)\int_{a>0}\left[da\right]a_2  \left(a_1 a_2\right)^{3 \epsilon }
     \\&\hspace{-15mm}\left[\frac{a_2 \left(a_3 \alpha _{IJ}+a_5\right) \left(a_5 \alpha _{IJ}+a_3\right)}{\alpha _{IJ}}+a_1 \left(a_4+a_5\right)^2+4 a_1a_2 \left(a_3+a_4+2 a_5\right)\right]^{-2 (\epsilon +1)}\,,
 \end{split}
 \\
  \begin{split}
     I_{(IJ)(Ik),2}^{\{H,H\}}=\,\,&-4^{2\epsilon +2} \Gamma (2 \epsilon +2) \int_{a>0}\left[da\right]a_2\left(a_1 a_2\right)^{3 \epsilon }
     \\&\hspace{-15mm}\left[\frac{a_2 \left(a_3 \alpha _{IJ}+a_4+a_5\right) \left(a_4 \alpha _{IJ}+a_5 \alpha _{IJ}+a_3\right)}{\alpha _{IJ}}+ a_1a_5^2+4a_1 a_2 \left(a_3+a_4+2 a_5\right)\right]^{-2 (\epsilon +1)}\,,
 \end{split}
 \\
  \begin{split}
     I_{(IJ)(Ik),3}^{\{H,H\}}=\,\,&-2^{4 \epsilon +3}\Gamma (2 \epsilon +2)\int_{a>0}\left[da\right] a_4^2\left(a_1 a_2\right)^{3 \epsilon } 
     \\&\hspace{-15mm}\left[\frac{a_2 \left(a_3 \alpha _{IJ}+a_4\right) \left(a_4 \alpha _{IJ}+a_3\right)}{\alpha _{IJ}}+a_1 a_4^2+4a_1  a_2 \left(a_3+2 a_4\right)\right]^{-2 (\epsilon +1)}\,,
 \end{split}
 \\
     \begin{split}
       I_{(kI)(kJ),1}^{\left\{H,H\right\}}= &-4^{2\epsilon +2} \Gamma (2 \epsilon +2)  \int_{a>0}[da]a_4\left(a_1 a_2\right){}^{3 \epsilon }
       \\&\hspace{-15mm}[ a_1a_4 \left(2 a_5 +a_4\right)+4a_1 a_2 \left(a_3+a_4\right)
    +a_2 a_3 \left(2 a_5  y_{IJk}+a_3\right)]{}^{-2 (\epsilon +1)},  
     \end{split}
 \\
     \begin{split}
       I_{(kI)(kJ),2}^{\left\{H,H\right\}}= & -4^{2\epsilon +2}\Gamma (2 \epsilon +2)  \int_{a>0}[da]a_3\left(a_1 a_2\right){}^{3 \epsilon } 
       \\&\hspace{-15mm}[ a_1a_4 \left(2 a_5 +a_4\right)+4a_1 a_2 \left(a_3+a_4\right)
       +a_2 a_3 \left(2 a_5 y_{IJk}+a_3\right)]{}^{-2 (\epsilon +1)}\,.
     \end{split}
 \end{align}
 \end{subequations}
 The result of the function ${\Y}_{(IJ)(IK)}^{\left\{H,H\right\}}$ is
 \begin{align}
 \label{211Hard}
     \begin{split}
         {\Y}_{(IJ)(IK)}^{\left\{H,H\right\}}=\,\,&\left(\frac{\alpha_s}{4\pi}\right)^2\left(\frac{\bar{m}^2}{\mu^2}\right)^{-2\epsilon}\bigg\{-\frac{1}{2\epsilon^2}\left[M_{100}\left(\alpha _{IJ}\right)+V_1\left(\alpha _{IJ}\right)\right]
         \\&+\frac{1}{\epsilon}\bigg[2 G\left(-1,0,0,\alpha _{IJ}\right)-4 G\left(-1,1,0,\alpha _{IJ}\right)+2 G\left(0,-1,0,\alpha _{IJ}\right)
         \\&+4 G\left(0,1,0,\alpha _{IJ}\right)-4 G\left(0,\frac{1}{2},0,\alpha _{IJ}\right)-4 G\left(1,-1,0,\alpha _{IJ}\right)
         \\&-4 G\left(1,1,0,\alpha _{IJ}\right)+4 G\left(1,2,0,\alpha _{IJ}\right)+4 G\left(1,\frac{1}{2},0,\alpha _{IJ}\right)
         \\&-4 \log (2) G\left(0,\frac{1}{2},\alpha _{IJ}\right)-4 \log (2) G\left(1,2,\alpha _{IJ}\right)+4 \log (2) G\left(1,\frac{1}{2},\alpha _{IJ}\right)
         \\&+\frac{1}{6} \log ^3\left(\alpha _{IJ}\right)+\log (2) \log ^2\left(\alpha _{IJ}\right)-2 \log ^2(2) \log \left(1-\alpha _{IJ}\right)
         \\&+\frac{1}{2} \pi ^2 \log \left(\alpha _{IJ}\right)-\pi ^2 \log \left(1-\alpha _{IJ}\right)+\frac{2}{3} \pi ^2 \log \left(\alpha _{IJ}+1\right)
         \\&+\log (2) \left(M_{100}\left(\alpha _{IJ}\right)+V_1\left(\alpha _{IJ}\right)\right)-\frac{9 \zeta (3)}{2}+\frac{2}{3} \pi ^2 \log (2)\bigg]\bigg\}+{\cal O}(\epsilon^0)+{\cal O}(\lambda)\,.
     \end{split}
 \end{align}
 The function $ {\Y}_{(IJ)(IK)}^{\left\{H,H\right\}}$ is symmetric under the $(I,J)$ interchange, so the sum of the two webs is vanishing,
 \begin{align}
      {\Y}_{(IJ)(IK)}^{\left\{H,H\right\}}+ {\Y}_{(Jk)(JI)}^{\left\{H,H\right\}}=0\,.
 \end{align}
 The remaining function $ {\Y}_{(kI)(kJ)}^{\left\{H,H\right\}}$ is   
 \begin{align}
 \label{112Hard}
     \begin{split}
         {\Y}_{(kI)(kJ)}^{\left\{H,H\right\}}=\,\,&\left(\frac{\alpha_s}{4\pi}\right)^2\left(\frac{\bar{m}^2}{\mu^2}\right)^{-2\epsilon}\log(y_{IJk})\bigg\{\frac{1}{\epsilon^3}+\frac{1}{\epsilon}\left[\frac{1}{3}\log^2(y_{IJk})+\frac{11}{6}\pi^2\right]\bigg\}+{\cal O}(\epsilon^0)+{\cal O}(\lambda)\,.
     \end{split}
 \end{align}
 
 \subsubsection{\texorpdfstring{Summary for ${\cal R}_{(IJk)}^{\left\{H,H\right\}}$}{Summary R{H,H}}}

The integration order of the master integrals appearing in this region is summarized in table~\ref{IntHH}.
\renewcommand{\arraystretch}{1.5}
\begin{table}[ht]
\centering
\begin{tabular}{|c|c|c|}
\hline
  Integral & Cheng-Wu & Integration orders  \\
    \hhline{|===|}
  $I_{(IJ)(Ik),1}^{\{H,H\}}$ & $a_4=1-a_5$ & $a_1\rightarrow a_2\rightarrow a_5\rightarrow a_3$ \\
  \hline
  $I_{(IJ)(Ik),2}^{\{H,H\}}$ & $a_4=1-a_5$ & $a_1\rightarrow a_2\rightarrow a_5\rightarrow a_3$ \\
  \hline
  $I_{(IJ)(Ik),3}^{\{H,H\}}$ & $a_4=1$ & $a_2\rightarrow a_1\rightarrow  a_3$ \\
  \hline
    
  $I_{(kI)(kJ),1}^{\left\{H,H\right\}}$ &$a_4=1$&$a_5\rightarrow a_2\rightarrow a_3\rightarrow a_1$ \\
  \hline
  $I_{(kI)(kJ),2}^{\left\{H,H\right\}}$ &$a_3=1$&$a_5\rightarrow a_1\rightarrow a_4\rightarrow a_2$\\
   \hline
 \end{tabular}
  \caption{The set of master integrals of the $\{H,H\}$ region associated with the tripole colour structure, specifying the integration measure in the middle column and the order of integration  in the rightmost column.
  }
    \label{IntHH}
 \end{table}
 Note that the integrals of the connected web are computed using differential equations, and we do not collect them here.
 Finally, the result of the region function $ {\cal R}_{(IJk)}^{\left\{H,H\right\}}$ is   
 \begin{align}
 \label{RegionHH}
     \begin{split}
       {\cal R}_{(IJk)}^{\left\{H,H\right\}}=\,\,&\left(\frac{\alpha_s}{4\pi}\right)^2\left(\frac{\bar{m}^2}{\mu^2}\right)^{-2\epsilon}\bigg\{\frac{1}{\epsilon^3}\log(y_{IJk})
       \\&+\frac{1}{\epsilon^2}\bigg[-\frac{1-y_{IJk}}{1+y_{IJk}}U_1(y_{IJk})-2\frac{1+\alpha_{IJ}^2}{1-\alpha_{IJ}^2}\log(y_{IJk})\log(\alpha_{IJ})\bigg]
       \\&+\frac{1}{\epsilon}\bigg[\frac{1+\alpha_{IJ}^2}{1-\alpha_{IJ}^2}\log(y_{IJk})\bigg(2V_1(\alpha_{IJ})+M_{100}(\alpha_{IJ})\bigg)
       \\&\hspace{10mm}-\frac{1}{2}\frac{1-y_{IJk}}{1+y_{IJk}}U_2(\alpha_{IJ},y_{IJk})+2 \log^2\left(\alpha _{IJ}\right) \log \left(y_{IJk}\right)
       \\&\hspace{10mm}-\frac{1}{3}  \log ^3\left(y_{IJk}\right)+\frac{7}{6} \pi ^2 \log \left(y_{IJk}\right)\bigg]\bigg\}+{\cal O}(\epsilon^0)+{\cal O}(\lambda).
     \end{split}
 \end{align}
 
\section{Computation and result for UV regions at two loops}
\label{app:2loopUV}

In this section, we will summarize the results of the UV regions in the two-loop calculation. We will use the same method as we introduced in appendix~\ref{app:oneloop}, and follow the same structure as in appendix~\ref{app:2loopIR}. Regarding the calculation of the integrals, we will provide the Feynman parameter representation and the order of integration of the master integrals. We will also present the result of each web ${\cal Y}^R$ in the region $R$ as well as the region function ${\cal R}^R$, which is the sum of all the webs contributing to that region.
 
 \subsection{\texorpdfstring{Region ${\cal R}^{\left\{C_N,H_{\UV}\right\}}_{(IJk)}$}{R{CN,HUV}}}
 
 \subsubsection{Connected web}
 
 \begin{align}
     \begin{split}
        {\Y}_{(IJk)}^{\left\{C_N,H_{\UV}\right\}} =\,\,&\vcenter{\hbox{\includegraphics[width=1.5cm]{fig/111HUVCN.pdf}}}
        \\=\,\,&-\left(\frac{\alpha_s}{4\pi}\right)^2{\cal N}^2\frac{\alpha _{IJ}^2+1}{12 \epsilon  \alpha _{IJ}}\left[I^{\left\{C_N,H_{\UV}\right\}}_{(IJk),1}(\alpha_{IJ},y_{IJk})-I^{\left\{C_N,H_{\UV}\right\}}_{(IJk),1}(\alpha_{IJ},y_{IJk}^{-1})\right]
         \\&+\left(\frac{\alpha_s}{4\pi}\right)^2{\cal N}^2\frac{(1-2 \epsilon )^2 (4 \epsilon -1)}{12 \epsilon ^2 (3 \epsilon -1)}
         \\&\hspace{20mm}\left[ \sqrt{y_{IJk}}I^{\left\{C_N,H_{\UV}\right\}}_{(IJk),2}(y_{IJk})-\frac{1}{\sqrt{y_{IJk}}}I^{\left\{C_N,H_{\UV}\right\}}_{(IJk),2}(y_{IJk}^{-1})\right]+{\cal O}(\lambda)\,.
     \end{split}
 \end{align}
 The Feynman parameter representation of the master integrals is  
 \begin{subequations}
 \begin{align}
    \begin{split}
     \label{commonInt3}
        I^{\left\{C_N,H_{\UV}\right\}}_{(IJk),1}(\alpha_{IJ},y_{IJk})= & -4^{2\epsilon+2}\Gamma (2 \epsilon +2) \int_{a>0}[da]a_4\left(a_1 a_5\right)^{3 \epsilon } 
        \\&\bigg[a_1 \left(a_4+4 a_5\right) a_4+a_5 \left(a_2^2+a_3^2\right)
        \\&+ \left(\alpha _{IJ}+\frac{1}{\alpha_{IJ}}\right)a_5a_3 a_2+\frac{1}{\sqrt{y_{IJk}}}a_1 a_3 a_4\bigg]^{-2 (\epsilon +1)},
    \end{split}
\\
    \begin{split}
    \label{commonInt2}
        I^{\left\{C_N,H_{\UV}\right\}}_{(IJk),2}(y_{IJk})= & -4^{2\epsilon} \Gamma (2 \epsilon )\int_{a>0}[da]\left(a_1 a_4\right)^{-2+3 \epsilon } 
        \\&\left[a_4 a_2^2+a_1 a_3 \left(a_3+4 a_4\right)+\frac{1}{\sqrt{y_{IJk}}}a_1 a_3 a_2\right]^{-2 \epsilon }.
    \end{split}
\end{align}
\end{subequations}
The result of the function $ {\Y}_{(IJk)}^{\{C_{N},H_{\UV}\}}$ is 
\begin{align}
\label{111CNHUV}
     \begin{split}
          {\Y}_{(IJk)}^{\{C_{N},H_{\UV}\}}=\,\,&\left(\frac{\alpha_s}{4\pi}\right)^2\left(\frac{\bar{m}^2}{\mu^2}\right)^{-2\epsilon}
        \log(y_{IJk})
          \bigg\{\frac{1}{6\epsilon^3}-\frac{2}{3\epsilon^2}\frac{1+\alpha_{IJ}^2}{1-\alpha_{IJ}^2}\log(\alpha_{IJ})
         \\&+\frac{1}{12\epsilon}\bigg[-\frac{1+\alpha_{IJ}^2}{1-\alpha_{IJ}^2}4M_{100}(\alpha_{IJ})+\frac{1}{3}\log^2(y_{IJk})+5\pi^2\bigg]\bigg\}+{\cal O}(\epsilon^0)+{\cal O}(\lambda).
     \end{split}
 \end{align}
 
 \subsubsection{Multiple-gluon-exchange webs}
 
 \begin{subequations}
 \begin{align}
     \begin{split}
         {\Y}_{(Jk)(JI)}^{\left\{C_N,H_{\UV}\right\}}=\,\,&\frac{1} {2}\left(\vcenter{\hbox{\includegraphics[width=1.5cm]{fig/121aCNHUV.pdf}}} -\vcenter{\hbox{\includegraphics[width=1.5cm]{fig/121bCNHUV.pdf}}}\right)
         \\=\,\,&\left(\frac{\alpha_s}{4\pi}\right)^2{\cal N}^2\frac{\alpha _{IJ}^2+1}{12 \epsilon  \alpha _{IJ}}I^{\left\{C_N,H_{\UV}\right\}}_{(IJk),1}(\alpha_{IJ},y_{IJk})+{\cal O}(\lambda),
     \end{split}
 \\
     \begin{split}
           {\Y}_{(IJ)(Ik)}^{\left\{C_N,H_{\UV}\right\}}=\,\,& \frac{1} {2}\left(\vcenter{\hbox{\includegraphics[width=1.5cm]{fig/211bCNHUV.pdf}}} -\vcenter{\hbox{\includegraphics[width=1.5cm]{fig/211aCNHUV.pdf}}}\right)
           \\=\,\,&-\left(\frac{\alpha_s}{4\pi}\right)^2{\cal N}^2\frac{\alpha _{IJ}^2+1}{12 \epsilon  \alpha _{IJ}}I^{\left\{C_N,H_{\UV}\right\}}_{(IJk),1}(\alpha_{IJ},y_{IJk}^{-1})+{\cal O}(\lambda).
     \end{split}
 \end{align}
  \end{subequations}
 We expressed the result in terms of a master integral $I^{\left\{C_N,H_{\UV}\right\}}_{(IJk),1}$ that has been computed in the context of the connected web; see the parametric representation in eq.~\eqref{commonInt3}. We directly write down the result of the function ${\Y}_{(Jk)(JI)}^{\left\{C_N,H_{\UV}\right\}}$ here,
 \begin{align}
 \label{211CNHUV}
     \begin{split}
         {\Y}_{(Jk)(JI)}^{\left\{C_N,H_{\UV}\right\}}=\,\,&\left(\frac{\alpha_s}{4\pi}\right)^2\left(\frac{\bar{m}^2}{\mu^2}\right)^{-2\epsilon}\frac{1+\alpha_{IJ}^2}{1-\alpha_{IJ}^2}\log(y_{IJk})\bigg\{-\frac{1}{3\epsilon^3}\log\left(\alpha_{IJ}\right)
         \\&+\frac{1}{\epsilon^2}\left[\textcolor{red}{\frac{1}{3} \log \left(\alpha _{IJ}\right) \log \left(y_{IJk}\right)}-\frac{1}{6} M_{100}\left(\alpha _{IJ}\right)\right]
         \\&+\frac{1}{\epsilon}\bigg[-\frac{4}{3} G\left(-1,-1,0,\alpha _{IJ}\right)+\frac{4}{3} G\left(-1,0,0,\alpha _{IJ}\right)-\frac{4}{3} G\left(-1,1,0,\alpha _{IJ}\right)
         \\&+\frac{4}{3} G\left(0,-1,0,\alpha _{IJ}\right)+\frac{4}{3} G\left(0,1,0,\alpha _{IJ}\right)-\frac{4}{3} G\left(1,-1,0,\alpha _{IJ}\right)
         \\&+\frac{4}{3} G\left(1,0,0,\alpha _{IJ}\right)-\frac{4}{3} G\left(1,1,0,\alpha _{IJ}\right)-\frac{1}{9} 2 \log ^3\left(\alpha _{IJ}\right)
         \\&-\frac{17}{18} \pi ^2 \log \left(\alpha _{IJ}\right)+\frac{1}{9} \pi ^2 \log \left(1-\alpha _{IJ}\right)+\frac{1}{9} \pi ^2 \log \left(\alpha _{IJ}+1\right)
         \\&+\textcolor{red}{\frac{1}{6} M_{100}\left(\alpha _{IJ}\right) \log \left(y_{IJk}\right)}-\frac{1}{6} \log \left(\alpha _{IJ}\right) \log ^2\left(y_{IJk}\right)+\frac{\zeta (3)}{3}\bigg]\bigg\}+{\cal O}(\epsilon^0)+{\cal O}(\lambda)\,.
     \end{split}
 \end{align}
 Most of the terms in $ {\Y}_{(Jk)(JI)}^{\left\{C_N,H_{\UV}\right\}}$ are $(I,J)$ symmetric, while only the two terms in red are antisymmetric. Therefore, the sum of the two webs is  
 \begin{align}
     \begin{split}
       {\Y}_{(Jk)(JI)}^{\left\{C_N,H_{\UV}\right\}}+ {\Y}_{(IJ)(Ik)}^{\left\{C_N,H_{\UV}\right\}}=
       \\&\hspace{-40mm}\left(\frac{\alpha_s}{4\pi}\right)^2\left(\frac{\bar{m}^2}{\mu^2}\right)^{-2\epsilon}\frac{1+\alpha_{IJ}^2}{1-\alpha_{IJ}^2}\log(y_{IJk})\bigg\{\frac{2}{3\epsilon^2}\log(\alpha_{IJ})+\frac{1}{3\epsilon}M_{100}(\alpha_{IJ})\bigg\}+{\cal O}(\epsilon^0)+{\cal O}(\lambda)\,.
     \end{split}
 \end{align}
 
 \subsubsection{\texorpdfstring{Summary for ${\cal R}_{(IJk)}^{\left\{C_N,H_{\UV}\right\}}$}{Summary R{CN,HUV}}}

The integration order of the master integrals appearing in this region is summarized in table~\ref{IntCNHUV}.
\renewcommand{\arraystretch}{1.5}
\begin{table}[ht]
\centering
\begin{tabular}{|c|c|c|}
\hline
  Integral & Cheng-Wu & Integration orders  \\
    \hhline{|===|}
  $I_{(IJk),1}^{\left\{C_N,H_{\UV}\right\}}$ &$a_4=1$&$a_5\rightarrow a_1\rightarrow a_2\rightarrow a_3$ \\
   \hline
 \end{tabular}
  \caption{The set of master integrals of the $\{C_N,H_{\UV}\}$ region associated with the tripole colour structure, specifying the integration measure in the middle column and the order of integration  in the rightmost column.
  The remaining integral $I_{(IJk),2}^{\left\{C_N,H_{\UV}\right\}}$ is simple, and has been evaluated directly in {\tt{AmpRed}}.}
    \label{IntCNHUV}
 \end{table}
 
 Finally, the result of the region function $ {\cal R}_{(IJk)}^{\left\{C_N,H_{\UV}\right\}}$ is 
 \begin{align}
 \label{RegionCNHUV}
     \begin{split}
       {\cal R}_{(IJk)}^{\left\{C_N,H_{\UV}\right\}}=\,\,&\left(\frac{\alpha_s}{4\pi}\right)^2\left(\frac{\bar{m}^2}{\mu^2}\right)^{-2\epsilon}\log(y_{IJk})\bigg\{\frac{1}{6\epsilon^3}+\frac{1}{12\epsilon}\bigg[\frac{1}{3}\log^2(y_{IJk})+5\pi^2\bigg]\bigg\}
       \\&+{\cal O}(\epsilon^0)+{\cal O}(\lambda).
     \end{split}
 \end{align}
 Notice that the terms multiplied by $\frac{1+\alpha_{IJ}^2}{1-\alpha_{IJ}^2}$ are completely canceled between the two webs.

 \subsection{\texorpdfstring{Region ${\cal R}_{(IJk)}^{\left\{C_{\UV},H\right\}}$}{R{CUV,H}}}
 
 \subsubsection{Connected web}
 
 \begin{align}
     \begin{split}
     \label{UVconnected}
          {\Y}_{(IJk)}^{\left\{C_{\UV},H\right\}}=\,\,&\left(\vcenter{\hbox{\includegraphics[width=1.5cm]{fig/111CUVHa.pdf}}} + \vcenter{\hbox{\includegraphics[width=1.5cm]{fig/111CUVHb.pdf}}}\right)
         \\=\,\,&\left(\frac{\alpha_s}{4\pi}\right)^2{\cal N}^2\frac{(1-2 \epsilon )^2 (4 \epsilon -1)}{12 \epsilon ^2 (3 \epsilon -1)}\left[ \sqrt{y_{IJk}}I^{\left\{C_{\UV},H\right\}}_{(IJk),1}(y_{IJk})-\frac{1}{\sqrt{y_{IJk}}}I^{\left\{C_{\UV},H\right\}}_{(IJk),1}(y_{IJk}^{-1})\right]
         \\&+{\cal O}(\lambda)\,,
     \end{split}
 \end{align}
 The Feynman parameter representation of the master integral is  
 \begin{align}
 \label{commonInt4}
    \begin{split}
        I^{\left\{C_{\UV},H\right\}}_{(IJk),1}(y_{IJk})= & -4^{2\epsilon} \Gamma (2 \epsilon )\int_{a>0}[da]\left(a_1 a_4\right)^{-2+3 \epsilon } 
        \\&\left[a_4 a_2^2+a_1 a_3 \left(a_3+4 a_4\right)+\frac{1}{\sqrt{y_{IJk}}}a_1 a_3 a_2\right]^{-2 \epsilon }\,.
    \end{split}
\end{align}
Note that this integral is the same as $I^{\left\{C_N,H_{\UV}\right\}}_{(IJk),2}(y_{IJk})$; see eq.~\eqref{commonInt2}. The result of the function ${\Y}_{(IJk)}^{\{C_{\UV},H\}}$ is 
\begin{align}
\label{111CUVH}
     \begin{split}
        {\Y}_{(IJk)}^{\{C_{\UV},H\}}=\,\,&\left(\frac{\alpha_s}{4\pi}\right)^2\left(\frac{\bar{m}^2}{\mu^2}\right)^{-2\epsilon}\log(y_{IJk})\bigg\{\frac{1}{6\epsilon^3}+\frac{1}{12\epsilon}\bigg[\frac{1}{3}\log^2(y_{IJk})+5\pi^2\bigg]\bigg\}+{\cal O}(\epsilon^0)
        \\&+{\cal O}(\lambda).
     \end{split}
 \end{align}
 
\subsubsection{Multiple-gluon-exchange webs}

 \begin{align}
     \begin{split}
         {\Y}_{(kI)(kJ)}^{\left\{C_{\UV},H\right\}}=\,\,&\frac{1} {2}\left(\vcenter{\hbox{\includegraphics[width=1.5cm]{fig/112aCUVH.pdf}}} -\vcenter{\hbox{\includegraphics[width=1.5cm]{fig/112bCUVH.pdf}}}\right)+\frac{1} {2}\left(\vcenter{\hbox{\includegraphics[width=1.5cm]{fig/112aHCUV.pdf}}} -\vcenter{\hbox{\includegraphics[width=1.5cm]{fig/112bHCUV.pdf}}}\right)
      \\
      =\,\,&  \left(\frac{\alpha_s}{4\pi}\right)^2{\cal N}^2\frac{(1-2 \epsilon )^2 (4 \epsilon -1)}{6 \epsilon ^2 (3 \epsilon -1)}\left[
      \frac{1}{\sqrt{y_{IJk}}}I^{\left\{C_{\UV},H\right\}}_{(IJk),1}(y_{IJk}^{-1})
      -
      \sqrt{y_{IJk}}I^{\left\{C_{\UV},H\right\}}_{(IJk),1}(y_{IJk})
      \right]
      \\&+{\cal O}(\lambda).
     \end{split}
 \end{align}
The result is expressed in terms of the master integral, $I^{\left\{C_{\UV},H\right\}}_{(IJk),1}(y_{IJk})$, that has already appeared in the connected web in eq.~\eqref{UVconnected}; see the parametric representation in eq.~\eqref{commonInt4}. The result of the function ${\Y}_{(kI)(kJ)}^{\left\{C_{\UV},H\right\}}$ is
 \begin{align}
 \label{112CUVH}
    \begin{split}
     {\Y}_{(kI)(kJ)}^{\left\{C_{\UV},H\right\}}=\,\,&\left(\frac{\alpha_s}{4\pi}\right)^2\left(\frac{\bar{m}^2}{\mu^2}\right)^{-2\epsilon}\log(y_{IJk})\bigg\{-\frac{1}{3\epsilon^3}-\frac{1}{6\epsilon}\bigg[\frac{1}{3}\log^2(y_{IJk})+5\pi^2\bigg]\bigg\}
     \\&+{\cal O}(\epsilon^0)+{\cal O}(\lambda).
    \end{split}
\end{align}

\subsubsection{\texorpdfstring{Summary for ${\cal R}_{(IJk)}^{\left\{C_{\UV},H\right\}}$}{Summary R{CUV,H}}}

The master integral $I_{(IJk),1}^{\left\{C_{\UV},H\right\}}$ is simple, and has been evaluated directly in {\tt{AmpRed}}.
The result of the region function ${\cal R}_{(IJk)}^{\left\{C_{\UV},H\right\}}$ is   
\begin{align}
\label{RegionCUVH}
     \begin{split}
        {\cal R}_{(IJk)}^{\left\{C_{\UV},H\right\}}=\,\,&\left(\frac{\alpha_s}{4\pi}\right)^2\left(\frac{\bar{m}^2}{\mu^2}\right)^{-2\epsilon}\log(y_{IJk})\bigg\{-\frac{1}{6\epsilon^3}-\frac{1}{12\epsilon}\bigg[\frac{1}{3}\log^2(y_{IJk})+5\pi^2\bigg]\bigg\}
        \\&+{\cal O}(\epsilon^0)+{\cal O}(\lambda).
     \end{split}
 \end{align}
Notice that ${\cal R}_{(IJk)}^{\left\{C_{\UV},H\right\}}$ is exactly ${\cal R}_{(IJk)}^{\left\{C_{N},H_{\UV}\right\}}$ (eq.~\eqref{RegionCNHUV}) with an opposite sign. As a result, UV regions do not contribute at the correlator level.

\bibliographystyle{JHEP}
\bibliography{biblio.bib}

\end{document}